\documentclass{article}

\usepackage[eandd,final]{neurips_2026}

\usepackage[utf8]{inputenc}
\usepackage[T1]{fontenc}
\usepackage{hyperref}
\usepackage{url}
\usepackage{booktabs}
\usepackage{amsfonts}
\usepackage{nicefrac}
\usepackage{microtype}
\usepackage{xcolor}

\usepackage{adjustbox}
\usepackage{tabularx}
\usepackage{multirow}
\usepackage{makecell}
\usepackage{soul}
\usepackage{amsmath}
\usepackage{booktabs}
\usepackage{adjustbox}
\usepackage{multirow}
\usepackage{makecell}
\usepackage{array}
\usepackage{wrapfig}
\usepackage{subcaption}
\usepackage{pdfpages}
\usepackage{siunitx}
\usepackage{longtable}
\usepackage{enumitem}
\usepackage{placeins}  
\usepackage{tikz}
\usepackage{forest}    
\usepackage{fontawesome5}

\definecolor{attack}{RGB}{255,200,200}
\definecolor{attackmid}{RGB}{255,225,225}
\definecolor{attacklite}{RGB}{255,240,240}
\definecolor{atkline}{RGB}{210,90,90}
\definecolor{defense}{RGB}{200,255,200}
\definecolor{defensemid}{RGB}{225,245,225}
\definecolor{defenselite}{RGB}{240,250,240}
\definecolor{defline}{RGB}{90,170,90}
\definecolor{evaluation}{RGB}{200,210,255}
\definecolor{evalmid}{RGB}{225,230,250}
\definecolor{evallite}{RGB}{240,243,252}
\definecolor{evalline}{RGB}{80,100,200}
\definecolor{rootfill}{RGB}{225,225,225}

\forestset{
  atkbranch/.style={fill=attack, edge={draw=atkline}, for descendants={edge={draw=atkline, line width=0.9pt}}},
  defbranch/.style={fill=defense, edge={draw=defline}, for descendants={edge={draw=defline, line width=0.9pt}}},
  evalbranch/.style={fill=evaluation, edge={draw=evalline}, for descendants={edge={draw=evalline, line width=0.9pt}}},
}

\newcolumntype{Z}[3]{S[round-mode=figures, round-precision=#1, table-format=#2.#3]}
\newcolumntype{M}[6]{Z{#1}{#2}{#3} @{\,\(\pm\)\,} Z{#4}{#5}{#6}}

\title{GraphIP\textendash Bench: How Hard Is It to Steal a Graph Neural Network, and Can We Stop It?}

\author{%
Kaixiang Zhao\thanks{Co-first author.} \\
Brigham Young University\\
\texttt{kzhao2@byu.edu}
\And
Bolin Shen\footnotemark[1] \\
Florida State University\\
\texttt{blshen@fsu.edu}
\And
Yuyang Dai \\
University of California, Berkeley\\
\texttt{michael\_yuyang@berkeley.edu}
\And
Shayok Chakraborty \\
Florida State University\\
\texttt{schakraborty2@fsu.edu}
\And
Yushun Dong\thanks{Corresponding author.} \\
Florida State University\\
\texttt{yushun.dong@fsu.edu}
}

\begin{document}


\addtocontents{toc}{\protect\setcounter{tocdepth}{-10}}

\maketitle

\begin{abstract}
Graph neural networks (GNNs) deployed as cloud services can be \emph{stolen} through \emph{model-extraction attacks}, which train a surrogate from query responses to reproduce the target's behaviour, and a growing line of ownership defenses tries to prevent or trace such theft. The title of this paper asks two questions: \emph{how hard is it to steal a GNN}, and \emph{can we stop it?} Prior work cannot answer either, because experiments use inconsistent datasets, threat models, and metrics. We introduce \emph{GraphIP-Bench}, a unified benchmark which evaluates both sides under a single black-box protocol. It integrates twelve extraction attacks, twelve defenses spanning watermarking, output-perturbation, and query-pattern-detection families, ten public graphs covering homophilic, heterophilic, and large-scale regimes, three GNN backbones, and three graph-learning tasks, and it reports fidelity, task utility, ownership verification, and computational cost on shared splits, queries, and budgets. We further add a joint attack-and-defense track which runs every attack on every defended target and measures watermark verification on the resulting surrogate, which exposes the protection that a defense retains after extraction. The empirical picture is short: stealing a GNN is easy at medium query budgets and most defenses do not change this; several watermarks verify reliably on the protected model but lose most of their verification signal on the extracted surrogate, which exposes a gap that single-model evaluations miss; and heterophilic graphs are systematically harder to steal, while a cross-architecture mismatch between target and surrogate reduces but does not prevent extraction. We release \emph{GraphIP-Bench} with reproducible scripts and configurations, and integrate the attacks and defenses into the~PyGIP~library.

\vspace{2pt}
\centerline{\small
\faGithub\ \href{https://github.com/LabRAI/GraphIP-Bench}{github.com/LabRAI/GraphIP-Bench}
\quad
\faBookOpen\ \href{https://labrai.github.io/PyGIP/index.html}{labrai.github.io/PyGIP}
}
\end{abstract}

\section{Introduction}\label{sec:intro}

Graph neural networks (GNNs) are key components of modern data-driven services. Commercial platforms use them for product recommendation \citep{yang2023dgrec}, autonomous vehicle perception \citep{monti2021dag}, and molecular property prediction \citep{li2024drug}. Their advantage is that they aggregate information over arbitrary relational structures, which arise in social, financial, and product data. Cloud providers now expose pre-trained GNNs through public inference endpoints, which allow customers to deploy state-of-the-art analytics without local training or~data~collection.

This deployment model also lets adversaries \emph{steal} the model. The attack pattern is the same in every reported instance: the adversary submits carefully chosen queries, records the labels or confidence scores that the endpoint returns, and trains a local surrogate which reproduces the target's behaviour. The literature calls this attack \emph{model extraction}~\citep{wu2022model,zhuang2024unveiling}, and we use ``stealing a GNN'' as a plain-language synonym throughout the paper. A successful theft leaks the owner's intellectual property, undermines pay-per-query revenue, and lets competitors recreate proprietary functionality at low cost. For example, a stolen fraud-detection GNN exposes decision boundaries which adversaries can use to bypass screening, and a stolen pharmaceutical GNN reveals assay knowledge which is encoded in the model parameters. These risks motivate the two questions which the title of this paper makes explicit: \emph{how hard is it to steal a GNN}, and \emph{can we stop it?}

To stop the theft, recent work proposes two complementary defense families. \emph{Information-limiting} defenses (output perturbation, query filtering, query-pattern detection) make each query response less useful to the attacker~\citep{liang2024model,kesarwani2018model,juuti2019prada,kariyappa2020defending,mazeika2022steer}. \emph{Ownership-tracing} defenses (watermarking and fingerprinting) embed a verifiable pattern in the trained model so that the owner can prove that an extracted surrogate was derived from their model~\citep{wang2023making,xu2023watermarking,dai2024pregip,you2024gnnfingers,wu2024securing,10646643}. Surveys summarise both families and the broader security landscape for deep learning~\citep{peng2023intellectual,xue2021intellectual,sun2023deep,zhao2025surveya,zhao2025surveyb} and for graph learning in particular~\citep{dai2024comprehensive,wu2022trustworthy,zhang2022trustworthy,Sun2024_Adversarial,zheng2graph}. Experimental practice, however, remains fragmented: studies use private splits, incompatible budgets, and inconsistent metrics, and the few existing testbeds focus on robustness or privacy and exclude model extraction together with watermarking and fingerprinting~\citep{9878092}. The community therefore lacks an empirical basis on which to answer either of our two~title~questions.

Several challenges must be addressed to enable fair and informative comparison. First, the community needs a single experimental protocol that fixes public splits, shared query sets, budgets, and explicit threat models (including whether the endpoint returns labels or confidence scores) and that treats data-driven and data-free attacks on equal terms~\citep{wu2022model,zhuang2024unveiling}. Second, evaluation must align success criteria with method goals: residual extraction for attacks, ownership verification for watermarking and fingerprinting, since these mechanisms aim to provide verifiable evidence of model ownership rather than to reduce agreement with the target~\citep{wang2023making,xu2023watermarking}. Third, studies should report the protection--utility balance under matched conditions, since adoption depends on how defenses affect task accuracy and inference latency. Fourth, benchmarks should report computational complexity (time and memory) to make the real cost of deployment transparent; prior work often omits this and obscures feasibility~\citep{liang2024model}. Fifth, evaluating attacks and defenses separately misses the joint adversarial setting in which the attacker extracts a defended model and the defender verifies ownership on the surrogate --- the setting that actually determines whether a defense is useful. Finally, consistent method naming, standardized hardware and software, and public reporting of seeds and tuning procedures are needed for reproducibility and to prevent protocol-induced bias. Existing studies only partially satisfy these requirements, which limits both scientific understanding and~industrial~uptake.

We address these challenges with \emph{GraphIP-Bench}, a reproducible benchmark and library which evaluates both \emph{stealing} and \emph{stopping} under a single black-box protocol. The suite integrates twelve representative extraction attacks (nine data-driven and three data-free) and twelve defenses spanning watermarking, output-perturbation, prediction-rounding, and query-pattern-detection. We evaluate every method on ten public graphs covering homophilic, heterophilic, and large-scale regimes, three GNN backbones, and three graph-learning tasks; the per-dataset, per-backbone, and per-method details are deferred to Section~\ref{sec:settings}. A unified hyperparameter search and a consistent metric suite record security, utility, and efficiency. Going beyond prior single-track evaluations, we add a joint attack-and-defense track which runs every attack on every defended target and a watermark-survival metric which measures verification on the surrogate produced by each attack, together exposing how much protection a defense retains after the model is actually stolen. The headline finding of that joint track is that most parameter-side or trigger-based graph watermarks verify near-perfectly on the protected model but lose much of their verification signal on the extracted surrogate, while query-time mechanisms partially survive --- watermark designs must therefore be re-evaluated on the surrogate, not only on the deployed model. To our knowledge, \emph{GraphIP-Bench} is the first benchmark which offers a standardised evaluation of model-extraction attacks and ownership defenses for graph neural networks. The main contributions are:

\begin{itemize}[leftmargin=1.2em,topsep=2pt,itemsep=1pt,parsep=0pt]
\item \textbf{Unified protocol.} Public splits, shared queries, standardized budgets, and explicit endpoint assumptions; twelve attacks and twelve defenses run under identical settings on ten datasets, three GNN backbones, and three tasks for~fair~comparison.
\item \textbf{Joint attack-and-defense track.} Every attack is executed against every defended target with watermark verification measured on the extracted surrogate, which exposes the residual ownership signal that~survives~extraction.
\item \textbf{Protection--utility analysis.} A sweep of defense configurations and attacker budgets summarises operating points with attack-agnostic frontiers and links them to graph structural properties (edge homophily, degree, density) and to~backbone~choice.
\item \textbf{Cost and efficiency reporting.} Asymptotic formulas and automated profilers record training time, memory, inference latency, verification time, and an estimated monetary cost, which makes deployment cost transparent for~every~method.
\end{itemize}

\section{Preliminaries}\label{sec:prelim}

\noindent\textbf{Notation.}
An attributed graph is denoted by $\mathcal{G}=(\mathcal{V},\mathcal{E},\mathbf{X},\mathbf{A})$ with node set $\mathcal{V}$, edge set $\mathcal{E}$, node-feature matrix $\mathbf{X}$, and adjacency matrix $\mathbf{A}$. Query budgets are reported using the benchmark multipliers, and four data-availability regimes (\texttt{both}, \texttt{features only}, \texttt{structure only}, \texttt{data free}) control which inputs the adversary can construct. Fidelity is the agreement rate between a surrogate and the target on the test split; accuracy and macro F1 measure~task~utility.

\noindent\textbf{Model Extraction Attacks.}
We consider the standard black-box threat model in which the adversary has query access to a deployed graph neural network and no knowledge of its weights, architecture, or training data. The adversary submits inputs (either genuine subgraphs or synthetic samples), records the returned labels or probability vectors, and trains a local model which minimises the discrepancy between its predictions and the target's outputs. The resulting model is a surrogate which replicates the behaviour of the protected network and enables extraction of the owner's intellectual property~\citep{wu2022model,zhuang2024unveiling}. Prior work groups extraction queries into three strategies. Random querying submits subgraphs from public data and succeeds when the decision boundary is smooth~\citep{wu2022model}. Adaptive querying selects inputs which maximise information gain, often through the disagreement between the current surrogate and the target~\citep{zhuang2024unveiling}. Data-free generation removes the need for public data by training a graph generator which produces queries during extraction~\citep{zhuang2024unveiling}. These studies show that modest query budgets, often no larger than the number of nodes in a benchmark dataset, are sufficient to recover a model which matches the original on~downstream~tasks.

\noindent\textbf{Defense against Model Extraction Attacks.}
The literature groups defenses into two complementary families. \emph{Information-limiting} defenses modify the target's outputs so that the adversary receives a less useful signal: output perturbation adds calibrated noise or rounds confidence scores~\citep{liang2024model,kesarwani2018model}, and query filtering detects and blocks suspicious request patterns~\citep{juuti2019prada,kariyappa2020defending,mazeika2022steer}. These methods can reduce the residual agreement of a surrogate, although they may also degrade accuracy for legitimate users. \emph{Ownership-tracing} defenses embed an artefact which lets the owner verify infringement: graph watermarking modifies weights or decision regions so that the model reveals a secret on inputs which carry a trigger~\citep{wang2023making,xu2023watermarking,dai2024pregip}, and fingerprinting derives stable signatures from the model's output distribution while leaving the parameters unchanged~\citep{you2024gnnfingers,wu2024securing,10646643}. Surveys of these approaches identify open questions about robustness, utility loss, and verification cost~\citep{peng2023intellectual,xue2021intellectual,sun2023deep,zhao2025surveya,zhao2025surveyb}, and our benchmark places all of them under a single protocol with matched datasets, budgets, and threat assumptions, which enables an objective comparison of~their~trade-offs.

\section{Benchmark Design}\label{sec:settings}

In this section we describe the experimental protocol of \emph{GraphIP-Bench}. We first state the protocol design, datasets, attacks, defenses, and implementation details, then articulate the five research questions that guide our~empirical~study.

\subsection{Experimental Settings and Implementations}
\noindent\textbf{Protocol Design.} \emph{GraphIP-Bench} defines a single black-box protocol that fixes four disjoint splits for each dataset (train, validation, test, query), shares the same query sets across methods, and uses \emph{standardized} query budgets at $0.05$, $0.10$, $0.25$, $0.50$, and $1.00$ as budget multipliers. Here ``standardized'' means that every method receives the same number of queries at each ratio and that these ratios are fixed across datasets. The set spans the commonly studied ranges in prior work \citep{orekondy2019knockoff,tramer2016stealing}, which include very small budgets that test sample efficiency ($0.05$ to $0.10$), medium budgets where most gains occur ($0.25$ to $0.50$), and a large budget that approximates saturation ($1.00$). This design makes results comparable and representative across methods and datasets. The protocol states explicit endpoint assumptions, which include whether the service returns labels only or also confidence scores and whether rate limits apply \citep{wu2022model,zhuang2024unveiling}. We separate evaluation into an extraction track, an ownership track, and a joint track. The extraction track measures how well black-box attacks learn a surrogate of an undefended target, and it reports test accuracy with respect to ground truth and fidelity with respect to the target. The ownership track evaluates each defense on a defended target and reports defended accuracy, fidelity to the original target, and verification on a standardized verification set \citep{zhao2021watermarking,xu2023watermarking,wang2023making,zhang2024imperceptible,wu2024securing}. The joint track runs every attack on every defended target and reports surrogate fidelity to the defended model together with the verification rate measured on the surrogate, which we call watermark survival. To place different settings on equal footing, we control data availability with four regimes: features only, structure only, features and structure, and data free. We also report total attack time, total defense time, and peak GPU memory for both attacks and defenses to make deployment~cost~clear.

\noindent\textbf{Datasets.}
We use ten attributed graphs which cover four groups: homophilic citation networks (Cora, CiteSeer, PubMed), homophilic coauthor and product co-purchase networks (CoauthorCS, CoauthorPhysics, Computers, Photo), the large-scale OGBN-Arxiv graph, and two heterophilic graphs \cite{platonov2023critical} (RomanEmpire, AmazonRatings). The graphs differ in size, density, class count, feature dimension, and edge homophily, which enables stress testing across structural regimes; full statistics are in Table~\ref{tab:appendix-dataset-stats} of Appendix~\ref{dataset}. For node classification we split each dataset into four disjoint subsets (train, validation, test, query) with no overlap, and for watermarking we reserve a fixed subset of the training data as the watermark set. We also include link prediction on Cora and graph classification on ENZYMES and PROTEINS from TUDataset \citep{morris2020tudataset} (Appendix~\ref{app:rq6_full}).

\noindent\textbf{Metrics.}
An em dash denotes an unavailable measurement. We report two groups of metrics under a single protocol. \textit{Performance} for attacks includes test accuracy, macro F1, and fidelity, which is the agreement between the surrogate and the target on the test set. \textit{Performance} for defenses includes the defended model's test accuracy and macro F1, its fidelity to the original target, and ownership verification on a standardized verification set. \textit{Efficiency} for attacks includes total attack time and peak GPU memory; \textit{efficiency} for defenses includes total defense time and peak~GPU~memory.

Marker preservation under output rounding does not establish ownership.

\noindent\textbf{Model Extraction Attacks.}
For the extraction track we implement twelve representative attacks. Six are MEA-style baselines from a canonical study on query synthesis and surrogate training, which we denote \texttt{MEA0} through \texttt{MEA5}~\citep{wu2022model}; we further include the adaptive adversarial querying method \texttt{AdvMEA}~\citep{defazio2019adversarial}, the centrality-and-entropy strategy \texttt{CEGA}~\citep{wang2025cega}, and the structure-aware pipeline \texttt{Realistic}~\citep{guan2024realistic}. To cover the data-free regime in a fair black-box manner, we add three variants~\cite{zhuang2024unveiling} which use no target gradients: \texttt{DFEA\_I} minimises the KL divergence between surrogate logits and target logits (soft-label distillation), \texttt{DFEA\_II} trains on hard labels returned by the endpoint (label-only supervision), and \texttt{DFEA\_III} augments label-only training with a consistency loss between two surrogates. We treat distinct hyperparameter settings of the same algorithm as separate methods, which enables fine-grained comparison under shared query sets and budgets. The four data-availability regimes defined in Section~\ref{sec:prelim} are realised by two control parameters, \texttt{attack\_x\_ratio} and \texttt{attack\_a\_ratio}, which fix the fraction of real features and real adjacency made available to each attack; the four regimes are denoted \texttt{X-only} (features only), \texttt{A-only} (structure only), \texttt{both} (features and structure), and \texttt{data-free}.

\noindent\textbf{Model Extraction Defenses.}
For the ownership and joint tracks we implement twelve defenses, which split between the two families introduced in Section~\ref{sec:prelim}. The ownership-tracing family contains five methods: the watermarking schemes \texttt{RandomWM}, \texttt{BackdoorWM}, \texttt{SurviveWM}, and \texttt{ImperceptibleWM}~\citep{zhao2021watermarking,xu2023watermarking,wang2023making,zhang2024imperceptible}, and the query-based integrity scheme \texttt{Integrity}~\citep{wu2024securing}. The information-limiting family contains seven methods: two output-perturbation variants \texttt{OP\_low} and \texttt{OP\_high} which add Gaussian noise to the returned logits at two scales~\citep{kesarwani2018model,liang2024model}, two prediction-rounding variants \texttt{PR\_2bit} and \texttt{PR\_top1} which quantise the returned scores or return only the top-1 label, and three query-detection methods which follow PRADA~\citep{juuti2019prada}, an adaptive-misinformation strategy~\citep{kariyappa2020defending}, and a gradient-redirection strategy~\citep{mazeika2022steer}, denoted \texttt{PRADA}, \texttt{AdaptMisinfo}, and \texttt{GradRedir}. Every defense protects the same target architecture, and we report the defended model's test accuracy and its fidelity to the original target. For watermarking and integrity we measure ownership verification on a standardised verification set using accuracy; for information-limiting and query-detection methods we use the verification proxy that matches each method's design, which is the trigger-label hit rate or the watermark-graph accuracy when a defense exposes such an artefact and the marker accuracy on the protected~model~otherwise.


\noindent\textbf{GNN Backbones.} To test whether conclusions are robust to the choice of model architecture, we use three widely adopted GNN backbones. The first is GCN \citep{kipf2017semi}, which uses spectral graph convolution. The second is GAT \citep{velivckovic2018graph}, which uses attention over neighbors. The third is GraphSAGE \citep{hamilton2017inductive}, which uses neighbor sampling and a learned aggregator. Each backbone is implemented in both DGL and PyTorch Geometric so that defenses with library-specific dependencies can be evaluated on a matched architecture; we use a DGL GCN with hidden dimension 16 as the default backbone, and we explicitly mark deviations when a defense uses a different backbone~for~fairness.

\subsection{Research Questions}\label{subsec:RQs}

\noindent\textbf{RQ1. How does extraction effectiveness change with the query budget, and does the trend hold on heterophilic and large-scale graphs?} We run twelve black-box attacks on undefended targets across the ten datasets at five budgets and four data-availability regimes, and we report accuracy, macro F1, and fidelity averaged over~three~seeds.
\textbf{RQ2. How effective are existing defenses on the protected model?} We evaluate the five watermarking and integrity methods together with the seven information-limiting and query-detection methods on a shared backbone, and we report defended accuracy, fidelity, and~ownership~verification.
\textbf{RQ3. How well do defenses balance protection and utility?} For each defense we report the utility loss against the undefended target together with defended fidelity and verification rate, which makes the protection-utility trade-off explicit.
\textbf{RQ4. What are the computational complexity and practical efficiency of attacks and defenses?} We report asymptotic time and memory complexity, and measure wall-clock time and peak GPU memory on NVIDIA A100 hardware for both the extraction and the~ownership~track.
\textbf{RQ5. How effective are defenses in the joint adversarial setting, and does the watermark signal survive on the extracted surrogate?} We run every attack on every defended target at a fixed budget and report surrogate fidelity to the defended model together with the verification rate measured on the surrogate, which we call \emph{watermark survival}.

\section{Empirical Investigation}
We now present an empirical study that follows the unified protocol in Section~\ref{sec:settings}, uses shared query sets and standardized budgets, and reports security, utility, and efficiency under identical settings. Complementary generalisation experiments are reported in Appendix~\ref{app:rq6_full}.

\subsection{Budget Sensitivity of Model Extraction Attacks (RQ1)}

\begin{wrapfigure}{r}{0.42\columnwidth}
  
  \centering
  \includegraphics[width=\linewidth]{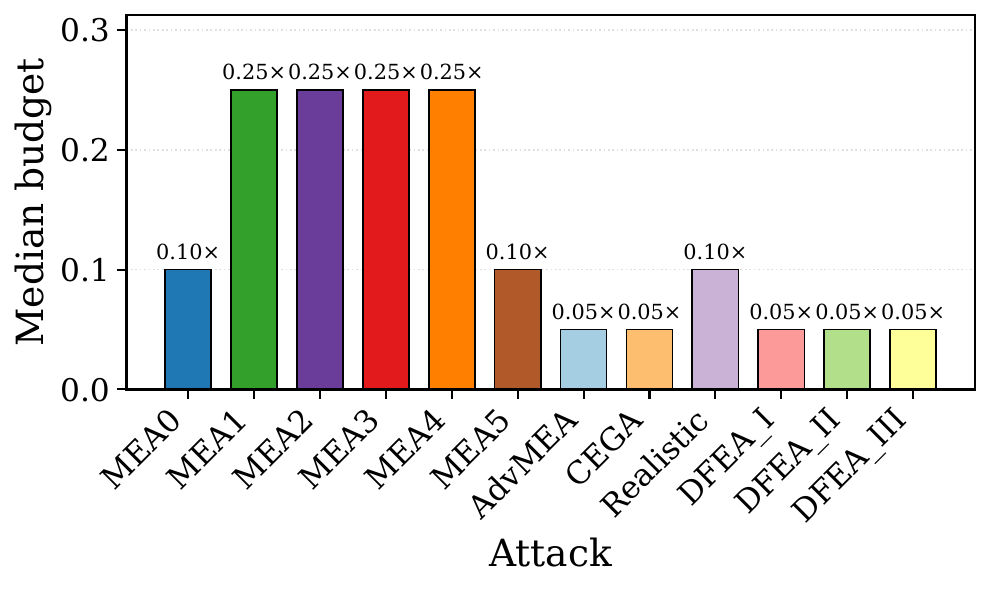}
    
  \caption{Sample efficiency across ten datasets and four regimes.}
  \label{fig:rq1_sample_efficiency}
  
\end{wrapfigure}

\begin{figure*}[t]
  \centering
  \includegraphics[width=\textwidth]{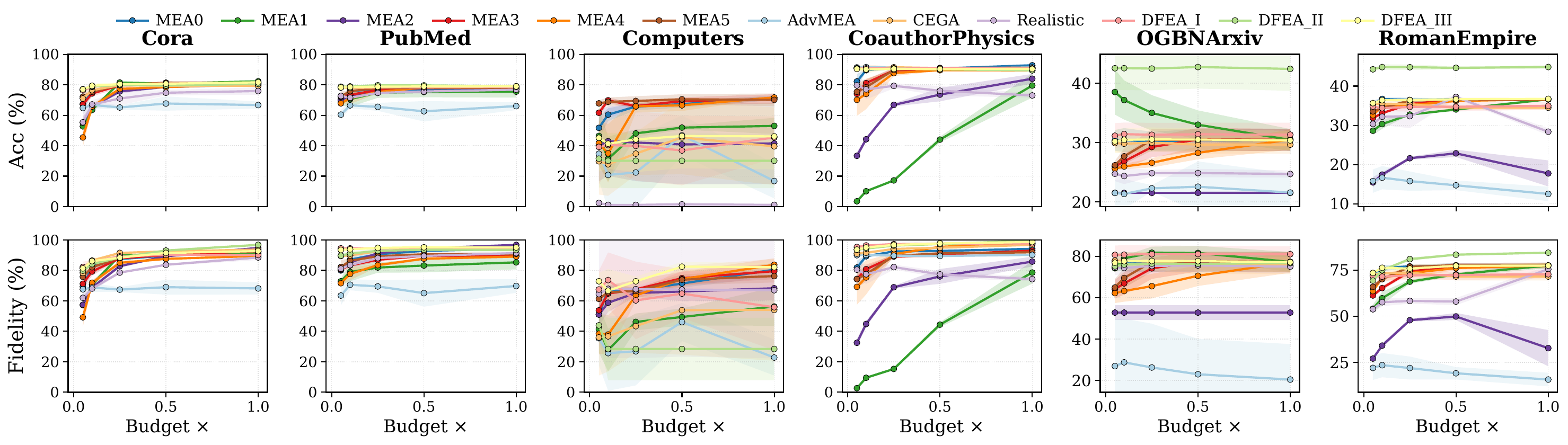}
  \caption{Budget--metric curves on six representative datasets (columns) for accuracy and fidelity (rows). Lines are the twelve attacks (mean over three seeds, shaded bands $\pm 1$ std). The four homophilic graphs share a $0$--$100\%$ $y$-axis; OGBN-Arxiv and RomanEmpire use per-subplot ranges since their target accuracy is bounded by intrinsic task difficulty. The full ten-dataset, three-metric version (including macro F1) and per-dataset numbers are reported as Figure~\ref{fig:rq1_full_grid_appendix} and Appendix~\ref{app:rq1}.}
  \label{fig:rq1_cora_both_curves}
\end{figure*}

To answer RQ1, we evaluate twelve black-box attacks on undefended targets under the protocol in Section~\ref{subsec:RQs}. Results are averaged over three seeds, and we report fidelity to the target together with accuracy on the ground truth. Table~\ref{tab:rq1_computers_both_fidelity} shows a representative case on \textit{Computers} when both features and adjacency are available. Figure~\ref{fig:rq1_sample_efficiency} reports sample efficiency on all ten datasets and four regimes; the $y$-axis is the median budget at which each attack reaches $90\%$ of its own best fidelity. Figure~\ref{fig:rq1_cora_both_curves} plots accuracy and fidelity on six representative datasets which span the four regimes that matter for our findings: two clean homophilic citation graphs (Cora, PubMed), the high-average-degree exception (Computers), the highest-homophily reference (CoauthorPhysics), the large-scale graph (OGBN-Arxiv), and the lowest-homophily heterophilic graph (RomanEmpire). The full ten-dataset, three-metric (Acc, Fidelity, F1) version, the regime-sensitivity heatmap, and the per-dataset and per-regime tables are reported as Figures~\ref{fig:rq1_full_grid_appendix} and~\ref{fig:rq1_regime_sensitivity} and Appendix~\ref{app:rq1}.

\begin{table*}[ht]
\centering
\caption{RQ1 on \textit{Computers} (both features and adjacency available). Fidelity (\%) across query budgets with mean $\pm$ standard deviation over three seeds. Higher is better. Appendix~\ref{app:seedcheck} examines the repeated \texttt{DFEA\_II} entries.}
\label{tab:rq1_computers_both_fidelity}
\scriptsize
\setlength{\tabcolsep}{2pt}
\renewcommand{\arraystretch}{0.96}
\resizebox{\textwidth}{!}{%
\begin{tabular}{@{}l|cccccccccccc@{}}
\toprule
Budget & \texttt{MEA0} & \texttt{MEA1} & \texttt{MEA2} & \texttt{MEA3} & \texttt{MEA4} & \texttt{MEA5} & \texttt{AdvMEA} & \texttt{CEGA} & \texttt{Realistic} & \texttt{DFEA\_I} & \texttt{DFEA\_II} & \texttt{DFEA\_III} \\
\midrule
0.05$\times$ & 51.0$\pm$7.0 & 41.5$\pm$15.5 & 51.0$\pm$7.5 & 53.8$\pm$3.0 & 38.4$\pm$13.4 & 61.3$\pm$4.3 & 35.5$\pm$17.8 & 36.0$\pm$25.0 & 64.5$\pm$33.7 & 67.5$\pm$12.4 & 43.9$\pm$8.1 & \textbf{72.9$\pm$5.5} \\
0.10$\times$ & 67.3$\pm$3.1 & 27.7$\pm$14.0 & 58.8$\pm$6.0 & 64.8$\pm$5.5 & 37.9$\pm$12.6 & 65.6$\pm$6.4 & 25.7$\pm$24.7 & 36.7$\pm$23.0 & 67.9$\pm$45.4 & \textbf{73.7$\pm$18.1} & 28.4$\pm$20.5 & 66.7$\pm$14.2 \\
0.25$\times$ & 67.0$\pm$1.5 & 46.2$\pm$5.5 & 65.2$\pm$2.4 & 67.6$\pm$6.2 & 63.6$\pm$4.0 & 66.5$\pm$2.1 & 26.9$\pm$22.4 & 43.4$\pm$17.5 & 67.7$\pm$45.3 & 60.3$\pm$25.5 & 28.4$\pm$20.5 & \textbf{72.9$\pm$3.1} \\
0.50$\times$ & 71.4$\pm$2.7 & 49.4$\pm$5.6 & 66.2$\pm$6.3 & 74.9$\pm$4.3 & 74.3$\pm$5.0 & 73.9$\pm$3.5 & 46.1$\pm$12.5 & 53.9$\pm$29.2 & 67.4$\pm$44.0 & 64.8$\pm$15.9 & 28.4$\pm$20.5 & \textbf{82.5$\pm$3.1} \\
1.00$\times$ & 80.5$\pm$4.6 & 56.3$\pm$12.6 & 68.3$\pm$8.0 & 79.5$\pm$5.2 & \textbf{83.7$\pm$4.3} & 76.3$\pm$4.2 & 22.8$\pm$11.8 & 54.1$\pm$33.4 & 67.0$\pm$44.8 & 55.8$\pm$28.2 & 28.4$\pm$20.5 & 82.1$\pm$2.8 \\
\bottomrule
\end{tabular}}
\end{table*}

The results expose a consistent picture across the ten datasets. \textit{CEGA is the most sample-efficient attack on the homophilic graphs but is unstable on \textit{Computers}}, where the high average degree ($36.8$) destabilises its centrality-driven query selection. \textit{The MEA family saturates near the medium budget}: fidelity improves rapidly up to $\sim 0.5\times$ and then plateaus. \textit{Data-free variants are competitive with the data-driven methods across most datasets, with \texttt{DFEA\_II} on the high-degree product graphs (especially \textit{Computers} and \textit{Photo}) as the main outlier}, which reflects an attack-specific dependence on graph statistics and synthetic-query coverage. \textit{Strong data-driven attacks are nearly invariant to removing either features or structure, but collapse without any real input}: the corresponding ratios in the regime-sensitivity heatmap (Appendix~\ref{app:rq1_extra}) stay near one in the features-only and structure-only blocks and drop sharply in the data-free block. Together these observations indicate that effective extraction on graphs is primarily a sample-efficiency problem rather than a brute-force budget problem, and that data-free extraction can match data-driven attacks on graphs whose label space is well covered by synthetic queries; a longer discussion appears in Appendix~\ref{app:rq1_extra}.

\begin{wrapfigure}{r}{0.44\textwidth}
  
  \centering
  \includegraphics[width=0.44\textwidth]{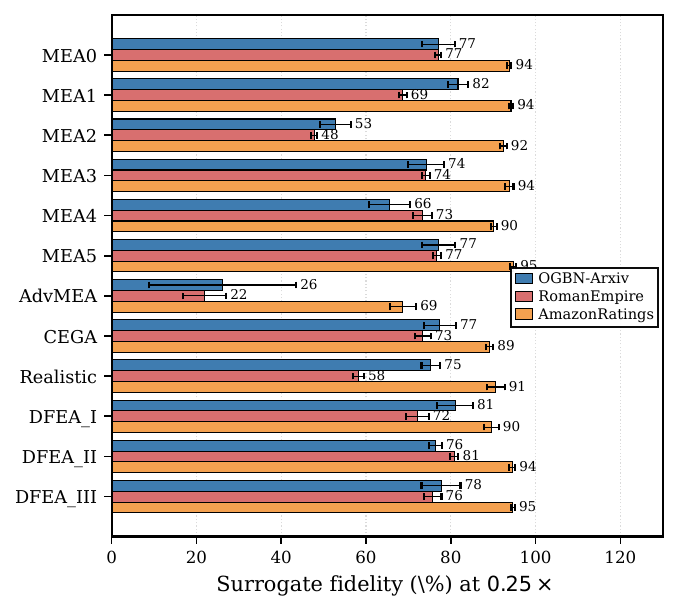}
  \caption{Surrogate fidelity (\%) at budget $0.25\times$ on the three additional graphs for all twelve attacks (mean over three seeds, whiskers are $\pm$ one std).}
  \label{fig:rq1_extension_bars}
  
\end{wrapfigure}

\noindent\textbf{Extension to large-scale and heterophilic graphs.}
The seven graphs in the original protocol are small and homophilic, which limits the conclusions to one structural regime. To extend the protocol along three independent axes, we evaluate the same twelve attacks on three additional graphs: OGBN-Arxiv ($169{,}343$ nodes, $40$ classes, edge homophily $0.699$) probes \emph{scale} and \emph{class-count} stress, while RomanEmpire (edge homophily $0.291$) and AmazonRatings (edge homophily $0.452$, ordinal $5$-class ratings) probe \emph{heterophily}. Figure~\ref{fig:rq1_extension_bars} summarises surrogate fidelity at the medium budget $0.25\times$ on each of the three additional graphs in the features-and-structure regime; the per-attack and per-budget results for all four regimes appear in Tables~\ref{tab:app_rq1_RomanEmpire_full}--\ref{tab:app_rq1_OGBNArxiv_full} of Appendix~\ref{app:rq1_new_datasets}.

These additional large-scale and heterophilic graphs reproduce the qualitative ordering observed on the seven homophilic datasets of the core protocol: simple data-driven methods (MEA0, MEA3, MEA5) and CEGA reach high fidelity, \texttt{AdvMEA} is unstable, and the data-free variants are also competitive on all three additional graphs (typically within $5$\,pp of the strongest data-driven attacks). Two further patterns are specific to these graphs; a longer discussion appears in Appendix~\ref{app:rq1_new_datasets}.

\textit{First, OGBN-Arxiv exposes a scale and class-count effect.} The undefended target reaches only $37.7$--$54.9\%$ accuracy across our three backbones (Appendix Table~\ref{tab:app_baseline_utility}) because the task itself is harder ($40$ fine-grained classes), so we report fidelity as the comparable measure: the strongest attacks (data-driven and data-free alike) reach $\sim 75$--$82\%$ at $0.25\times$, while \texttt{AdvMEA} drops to $\sim 26\%$, indicating that the $40$-class label space is the main difficulty rather than the graph~scale~itself.

\textit{Second, heterophilic graphs separate attacks which assume labelled neighbourhoods from those which do not.} On RomanEmpire (homophily $0.291$) the strongest data-driven attacks reach $77\%$ fidelity, but \texttt{AdvMEA} drops to $21.9\%$, a $50$\,pp loss which isolates an implicit homophily assumption in its adversarial query generator. AmazonRatings is intermediate: although its homophily is $0.452$, the labels are ordinal $1$--$5$ ratings, so neighbouring classes carry graded similarity which aggregation can still exploit; the strongest data-driven attacks still reach $\geq 93\%$ fidelity.

\subsection{Effectiveness of Ownership and Information-Limiting Defenses (RQ2)}
To answer RQ2, we evaluate the twelve defenses on a shared target under identical splits, using three seeds (ten for Integrity on Computers and Photo). We report the five watermarking and integrity methods first and then the seven information-limiting and query-detection methods. \emph{Utility drop} is undefended accuracy minus defended accuracy on the same dataset, and \emph{ownership verification} is accuracy on a fixed verification set; we summarize across datasets with the median and report variability with the inter-quartile range. Table~\ref{tab:rq2_summary_across_datasets} reports the median values together with time and memory; the corresponding boxplot view and per-dataset numbers are reported in Appendix~\ref{app:rq2}.

Among the watermarking and integrity defenses, \texttt{BackdoorWM} offers the best protection-utility balance: it reaches the highest median fidelity ($80.07\%$), perfect verification, and only a $3.27$\,pp median utility drop. \texttt{ImperceptibleWM} also reaches perfect verification, but at a much higher training cost which is visible as a long memory tail in Figure~\ref{fig:rq4_mem_wrapL} of Appendix~\ref{app:rq4_mem}. \texttt{RandomWM} and \texttt{SurviveWM} expose a stability-verification trade-off: \texttt{SurviveWM} minimises utility loss ($0.13$\,pp median) but weakens verification, while \texttt{RandomWM} is more variable across datasets. A per-defense breakdown together with the boxplot view appears in Appendix~\ref{app:rq2}.

\begin{table*}[t]
\centering
\caption{RQ2 summary for watermarking and integrity defenses on the seven homophilic graphs. Median (IQR) over datasets. Utility drop is the signed decrease in test accuracy (pp) against the undefended target on a matched backbone; negative values indicate that the defended model out-performs the matched undefended baseline.}
\label{tab:rq2_summary_across_datasets}
\scriptsize
\setlength{\tabcolsep}{3pt}
\renewcommand{\arraystretch}{0.92}
\begin{tabular*}{\textwidth}{@{\hspace{0pt}}l@{\extracolsep{\fill}}cccccc@{\hspace{0pt}}}
\toprule
Defense & F1 (\%) & Fidelity (\%) & Owner. verif. (\%) & Utility drop (pp) $\downarrow$ & Time (s) & Peak mem. (GB) \\
\midrule
\texttt{RandomWM}        & 64.99 (12.02) & 74.13 (10.70) & 72.00 (24.7)  & 3.93 (6.18)   & 34.8 (14.6)     & \textbf{0.09} (0.26) \\
\texttt{BackdoorWM}      & 69.13 (15.51) & \textbf{80.07} (15.95) & \textbf{100.0} (0.00) & 3.27 (2.87)   & 1.98 (0.45)       & 0.16 (0.68) \\
\texttt{SurviveWM}       & 67.47 (27.86) & 79.93 (32.92) & 21.76 (32.4)  & \textbf{0.13} (18.2)  & 2.27 (0.92)       & 0.32 (0.96) \\
\texttt{ImperceptibleWM} & 69.49 (9.19)  & 77.63 (13.88) & \textbf{100.0} (0.00) & 1.65 (6.28)   & 676(697) & 2.30 (2.58) \\
\texttt{Integrity}       & \textbf{73.43} (21.20) & $76.03\,(22.52)$ & 66.67 (40.0)  & 1.93 (3.17)  & \textbf{1.38} (0.45)       & 0.20 (0.90) \\
\bottomrule
\end{tabular*}

\end{table*}

\noindent\textbf{Information-limiting and query-detection defenses.}
We further evaluate seven information-limiting and query-detection methods on the same target backbone (DGL GCN) across all ten datasets, which complements the five ownership-tracing methods above with a different defense family. Table~\ref{tab:rq2_new_defenses_summary} reports the protected-model accuracy and the verification proxy on the protected model itself, and Appendix~\ref{app:rq2_new} contains the~standard~deviations.

\begin{table*}[t]
\centering
\caption{Seven information-limiting and query-detection defenses on a shared target backbone (DGL GCN, hidden 16) across ten datasets. Each cell reports protected accuracy (\%) with the verification proxy (\%) in parentheses (mean over three seeds; standard deviations in Appendix~\ref{app:rq2_new}).}
\label{tab:rq2_new_defenses_summary}
\scriptsize
\setlength{\tabcolsep}{3pt}
\renewcommand{\arraystretch}{0.96}
\begin{tabular*}{\textwidth}{@{}l@{\extracolsep{\fill}}ccccccc@{}}
\toprule
\textbf{Dataset} & \texttt{OP\_low} & \texttt{OP\_high} & \texttt{PR\_2bit} & \texttt{PR\_top1} & \texttt{PRADA} & \texttt{AdaptMisinfo} & \texttt{GradRedir} \\
\midrule
Cora            & 79.4 (98.6) & 79.2 (93.9) & 73.3 (83.7) & 79.6 (100.0) & 40.2 (43.0) & 41.0 (48.5) & 79.8 (100.0) \\
CiteSeer        & 67.6 (97.8) & 66.3 (91.0) & 53.9 (70.3) & 68.8 (100.0) & 69.3 (100.0) & 39.8 (52.5) & 68.4 (100.0) \\
PubMed          & 77.9 (99.0) & 75.9 (94.7) & 77.6 (93.4) & 78.2 (100.0) & 78.0 (100.0) & 44.1 (48.6) & 78.3 (100.0) \\
Computers       & 44.0 (89.2) & 37.6 (55.3) & 36.1 (61.7) & 34.8 (100.0) & 46.0 (100.0) & 28.0 (64.3) & 52.4 (100.0) \\
Photo           & 89.1 (98.9) & 90.7 (96.6) & 90.4 (96.7) & 95.5 (100.0) & 87.0 (100.0) & 46.3 (49.6) & 66.6 (100.0) \\
CoauthorCS      & 87.8 (99.5) & 88.2 (98.8) & 87.5 (98.1) & 88.1 (100.0) & 75.0 (79.6) & 52.4 (58.7) & 88.2 (100.0) \\
CoauthorPhysics & 89.4 (99.8) & 89.1 (99.3) & 90.2 (98.7) & 89.5 (100.0) & 83.4 (89.0) & 59.0 (63.1) & 89.7 (100.0) \\
OGBN-Arxiv      & 37.7 (95.5) & 37.8 (81.2) & 30.2 (59.6) & 39.5 (100.0) & 37.0 (100.0) & 19.9 (52.3) & 38.2 (100.0) \\
RomanEmpire     & 42.7 (95.3) & 40.6 (82.2) & 35.2 (56.4) & 42.7 (100.0) & 19.7 (25.4) & 22.5 (50.8) & 42.5 (100.0) \\
AmazonRatings   & 42.0 (94.8) & 41.2 (80.0) & 39.4 (70.6) & 41.6 (100.0) & 41.8 (100.0) & 33.9 (48.9) & 41.7 (100.0) \\
\bottomrule
\end{tabular*}
\end{table*}

The seven methods separate into three groups. \textit{Output perturbation} (\texttt{OP\_low}, \texttt{OP\_high}) preserves accuracy on most graphs and yields a verification rate close to the noise-free regime when the noise scale is small, while the rate drops on graphs with more classes or with a heterophilic structure when the noise scale is large. \textit{Prediction rounding} reveals a sharp split: \texttt{PR\_top1}, which returns only the top-1 label, gives perfect verification on every dataset because the marker label is always preserved, while \texttt{PR\_2bit}, which quantizes the returned probability vector, suffers a notable drop on classes with similar logits. \textit{Query-detection methods} (\texttt{PRADA}, \texttt{AdaptMisinfo}, \texttt{GradRedir}) show three distinct profiles: \texttt{PRADA} preserves accuracy on smaller graphs but degrades on \textit{Cora} and the heterophilic \textit{RomanEmpire} graph, \texttt{AdaptMisinfo} consistently reduces accuracy because it perturbs benign queries that resemble suspicious ones, and \texttt{GradRedir} preserves accuracy and reaches perfect verification on most datasets but reduces accuracy on \textit{Photo}. Together these results show that backdoor triggers and label-quantization defenses provide the most reliable verification on the protected model, while output perturbation and query detection trade utility for protection at~noticeable~cost.

\subsection{Protection-Utility Balance of Defenses (RQ3)}

\begin{wrapfigure}{r}{0.45\columnwidth}
  
  \centering
  \includegraphics[width=\linewidth]{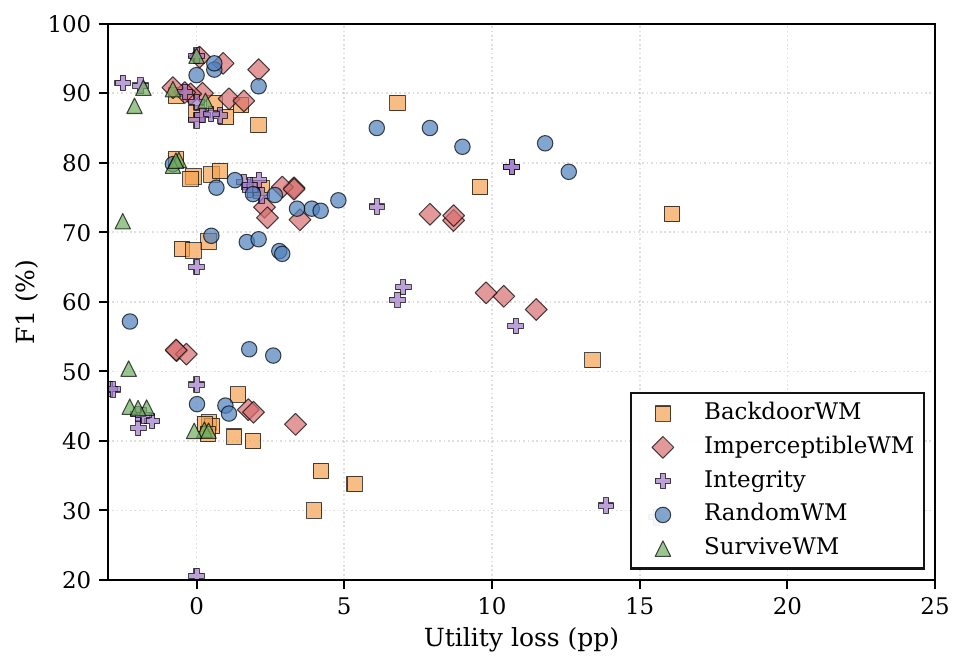}
  \caption{Protection-utility scatter across ten datasets (three seeds per defense; ten for Integrity on Computers and Photo) (one point per dataset-seed run). Upper-left is best.}
  \label{fig:rq3_defense_scatter}
  
\end{wrapfigure}

To answer RQ3, we evaluate the defended model's task utility and its alignment with the original target together with ownership verification on a fixed verification set under the unified protocol in Section~\ref{sec:settings}. All defenses protect the same architecture and use identical splits; we compute metrics for every dataset and every seed. Figure~\ref{fig:rq3_defense_scatter} aggregates \emph{all ten datasets and all three seeds} in a single view: each point is one dataset-seed run, the horizontal axis shows utility loss (pp) relative to the undefended target, and the vertical axis shows F1 (\%). To keep the central region readable we cap the horizontal axis at $25$\,pp and exclude a small number of extreme outliers; the full scale is needed when comparing methods with large losses. The cloud of points concentrates in the region of small loss (0--10\,pp) and high F1 (65--80\,\%), which shows that several defenses preserve task performance while enabling~ownership~verification.

\textit{First, BackdoorWM consistently lies near the top left}, which indicates that it preserves accuracy while enabling strong ownership verification; its spread is tight across datasets, which suggests stable behavior under our protocol. \textit{Second, ImperceptibleWM occupies a similar region but with higher cost}, which matches the efficiency results in RQ4 and reflects the overhead of representation-level optimization. \textit{Third, RandomWM, SurviveWM, and Integrity form broader clusters}, which shows that their balance depends more on data characteristics: RandomWM has moderate loss and verification with larger variance; SurviveWM has the smallest median loss but weak verification; Integrity preserves utility while yielding a binary ownership signal whose effectiveness varies across datasets. These observations imply that backdoor-style triggers provide the most reliable protection-utility balance in our setting, while representation-level watermarks trade efficiency for verification strength and the remaining methods are more sensitive to the~data~distribution.

\subsection{Computational Cost of Attacks and Defenses (RQ4)}

To answer RQ4, we report the original NVIDIA A100 profiling measurements; H200 reruns are not included in cost comparisons. For attacks we fix the budget at $1.00\times$ in the \texttt{both} regime and measure total attack time, query time, surrogate training time, and peak memory; for defenses we measure total defense time and peak memory. All values are mean $\pm$ std over three seeds. Tables~\ref{tab:rq4_attack_efficiency_time}--\ref{tab:rq4_defense_efficiency_time_compact} list total time on the seven homophilic graphs, and Figure~\ref{fig:rq4_mem_wrapL} of Appendix~\ref{app:rq4_mem} reports peak memory aggregated across all~ten~datasets.

\begin{table*}[t]
\centering
\caption{RQ4: total attack time (min) at budget $1.00\times$ on undefended targets in the \texttt{both} regime, restricted to the seven homophilic graphs (mean $\pm$ std over three seeds). The aggregate pattern on OGBN-Arxiv, RomanEmpire, and AmazonRatings is captured in Figure~\ref{fig:rq4_mem_wrapL} of Appendix~\ref{app:rq4_mem}.}
\label{tab:rq4_attack_efficiency_time}
\scriptsize
\setlength{\tabcolsep}{2pt}
\renewcommand{\arraystretch}{0.92}
\begin{tabular*}{\textwidth}{@{}l@{\extracolsep{\fill}}ccccccc@{}}
\toprule
\textbf{Attack} & \textbf{Cora} & \textbf{CiteSeer} & \textbf{CoauthorCS} & \textbf{CoauthorPhys} & \textbf{Computers} & \textbf{Photo} & \textbf{PubMed} \\
\midrule
\texttt{MEA0}      & 0.69$\,\pm\,$0.09 & 0.72$\,\pm\,$0.08 & 0.86$\,\pm\,$0.10 & 1.36$\,\pm\,$0.03 & 2.18$\,\pm\,$1.84 & 0.79$\,\pm\,$0.04 & 1.21$\,\pm\,$0.09 \\
\texttt{MEA1}      & 0.69$\,\pm\,$0.09 & 0.73$\,\pm\,$0.08 & 0.86$\,\pm\,$0.10 & 1.34$\,\pm\,$0.07 & 2.18$\,\pm\,$1.88 & 0.76$\,\pm\,$0.06 & 1.17$\,\pm\,$0.01 \\
\texttt{MEA2}      & 1.51$\,\pm\,$0.11 & 1.73$\,\pm\,$0.22 & 2.01$\,\pm\,$0.08 & 2.25$\,\pm\,$0.11 & 2.21$\,\pm\,$0.79 & 1.59$\,\pm\,$0.18 & 2.02$\,\pm\,$0.09 \\
\texttt{MEA3}      & 0.66$\,\pm\,$0.08 & 0.74$\,\pm\,$0.08 & 0.70$\,\pm\,$0.04 & 0.86$\,\pm\,$0.10 & 3.27$\,\pm\,$1.93 & 1.14$\,\pm\,$0.09 & 1.20$\,\pm\,$0.03 \\
\texttt{MEA4}      & 0.76$\,\pm\,$0.09 & 0.94$\,\pm\,$0.03 & 2.58$\,\pm\,$0.09 & 6.46$\,\pm\,$0.29 & 2.50$\,\pm\,$1.82 & 0.84$\,\pm\,$0.04 & 1.73$\,\pm\,$0.07 \\
\texttt{MEA5}      & 0.69$\,\pm\,$0.10 & 0.75$\,\pm\,$0.07 & 0.77$\,\pm\,$0.08 & 0.92$\,\pm\,$0.18 & 2.66$\,\pm\,$0.99 & 1.25$\,\pm\,$0.20 & 1.23$\,\pm\,$0.01 \\
\texttt{AdvMEA}    & 2.35$\,\pm\,$1.37 & 4.39$\,\pm\,$2.52 & 8.88$\,\pm\,$0.55 & 4.93$\,\pm\,$0.05 & 13.0$\,\pm\,$8.45 & 10.3$\,\pm\,$11.4 & 4.51$\,\pm\,$4.76 \\
\texttt{CEGA}      & 1.07$\,\pm\,$0.11 & 1.03$\,\pm\,$0.07 & 1.48$\,\pm\,$0.14 & 2.15$\,\pm\,$0.05 & 3.51$\,\pm\,$3.41 & 1.07$\,\pm\,$0.10 & 1.75$\,\pm\,$0.11 \\
\texttt{Realistic} & 90.3$\,\pm\,$2.12 & 111$\,\pm\,$3.72 & 472$\,\pm\,$4.89 & 976$\,\pm\,$7.34 & 840$\,\pm\,$17.7 & 248$\,\pm\,$0.14 & 529$\,\pm\,$18.1 \\
\texttt{DFEA\_I}   & 0.96$\,\pm\,$0.06 & 1.11$\,\pm\,$0.04 & 1.02$\,\pm\,$0.04 & 0.97$\,\pm\,$0.08 & 2.82$\,\pm\,$2.51 & 1.00$\,\pm\,$0.09 & 1.32$\,\pm\,$0.08 \\
\texttt{DFEA\_II}  & 0.82$\,\pm\,$0.06 & 0.88$\,\pm\,$0.09 & 0.80$\,\pm\,$0.03 & 0.86$\,\pm\,$0.09 & 1.28$\,\pm\,$0.57 & 0.87$\,\pm\,$0.09 & 1.07$\,\pm\,$0.11 \\
\texttt{DFEA\_III} & 1.48$\,\pm\,$0.14 & 1.55$\,\pm\,$0.12 & 1.46$\,\pm\,$0.16 & 1.52$\,\pm\,$0.15 & 3.30$\,\pm\,$2.47 & 1.62$\,\pm\,$0.11 & 1.86$\,\pm\,$0.15 \\
\bottomrule
\end{tabular*}
\end{table*}

Most attacks finish within minutes and use sub-GB memory, while structure-reconstruction pipelines are prohibitively expensive: the \texttt{MEA} family and \texttt{CEGA} take $0.7$--$2.5$ min per run, whereas \texttt{Realistic} takes hundreds to thousands of minutes (Table~\ref{tab:rq4_attack_efficiency_time}) because it trains an auxiliary edge model, a cost which is hard to justify since RQ1 shows that fidelity plateaus by $0.25\times$. The adaptive attack \texttt{AdvMEA} is slower and more variable across datasets, which matches the overhead of policy search. On the defense side, \texttt{BackdoorWM}, \texttt{SurviveWM}, and \texttt{Integrity} train in $1$--$6$\,s with low memory, whereas \texttt{ImperceptibleWM} sits in a much heavier regime because it uses representation-level losses and larger buffers. The peak-memory profile and the per-attack memory tail are reported in Appendix~\ref{app:rq4_mem}.

\begin{table*}[t]
\centering
\caption{RQ4: total defense time (s) for the five watermarking and integrity defenses on the seven homophilic graphs (mean $\pm$ std over three seeds; lower is better). Memory aggregates and the three additional graphs are reported in Figure~\ref{fig:rq4_mem_wrapL} of Appendix~\ref{app:rq4_mem}.}
\label{tab:rq4_defense_efficiency_time_compact}
\scriptsize
\setlength{\tabcolsep}{1.0pt}
\renewcommand{\arraystretch}{0.92}
\begin{tabular*}{\textwidth}{@{}l@{\extracolsep{\fill}}ccccccc@{}}
\toprule
\textbf{Defense} & \textbf{Cora} & \textbf{CiteSeer} & \textbf{CoauthorCS} & \textbf{CoauthorPhys} & \textbf{Computers} & \textbf{Photo} & \textbf{PubMed} \\
\midrule
\texttt{RandomWM}   & 24.3$\,\pm\,$0.07 & 23.9$\,\pm\,$0.09 & 57.3$\,\pm\,$2.58 & 34.8$\,\pm\,$0.77 & 41.3$\,\pm\,$0.23 & 36.0$\,\pm\,$0.30 & 21.8$\,\pm\,$0.13 \\
\texttt{BackdoorWM} & 1.88$\,\pm\,$0.01 & 2.17$\,\pm\,$0.23 & 2.54$\,\pm\,$0.01 & 3.89$\,\pm\,$0.06 & 1.98$\,\pm\,$0.00 & 1.92$\,\pm\,$0.01 & 1.88$\,\pm\,$0.02 \\
\texttt{SurviveWM}  & 1.59$\,\pm\,$0.00 & 1.62$\,\pm\,$0.00 & 2.75$\,\pm\,$0.01 & 5.62$\,\pm\,$0.02 & 2.31$\,\pm\,$0.04 & 1.52$\,\pm\,$0.06 & 2.27$\,\pm\,$0.09 \\
\texttt{Impercept.} & 676$\,\pm\,$2.45  & 709$\,\pm\,$2.33  & 906$\,\pm\,$10.1 & 950$\,\pm\,$14.0 & 461$\,\pm\,$7.61 & 209$\,\pm\,$11.0 & 196$\,\pm\,$14.8 \\
\texttt{Integrity}  & 1.29$\,\pm\,$0.01 & 1.10$\,\pm\,$0.01 & 1.52$\,\pm\,$0.16 & 2.37$\,\pm\,$0.07 & 1.92$\,\pm\,$0.03 & 1.38$\,\pm\,$0.01 & 1.25$\,\pm\,$0.19 \\
\bottomrule
\end{tabular*}
\end{table*}

\subsection{Joint Attack-and-Defense Evaluation and Watermark Survival (RQ5)}
\label{subsec:rq5}

\begin{wraptable}[14]{r}{0.50\textwidth}
\vspace{-5pt}
\centering
\caption{RQ5 summary on \textit{Computers} at $0.25\times$. \emph{Median fidelity} is taken over the twelve attacks; \emph{verif.\ on target} and \emph{verif.\ on surrogate} are method-specific verification rates on the protected model and on the extracted surrogate.}
\label{tab:rq5_summary_computers}
\scriptsize
\setlength{\tabcolsep}{9pt}
\renewcommand{\arraystretch}{0.95}
\begin{tabular}{@{}lccc@{}}
\toprule
\textbf{Defense} & \makecell{Median\\fidelity (\%)} & \makecell{Verif.\\target (\%)} & \makecell{Verif.\\surr. (\%)} \\
\midrule
\texttt{BackdoorWM}      & 71.1 & 100.0 & 59.2 \\
\texttt{SurviveWM}       & 74.6 & 100.0 & 10.0 \\
\texttt{Integrity}       & 49.1 & 100.0 & 48.0 \\
\texttt{RandomWM}        & 85.3 & 100.0 & 14.7 \\
\texttt{ImperceptibleWM} & 84.3 & 100.0 & \phantom{0}0.0 \\
\bottomrule
\end{tabular}

\end{wraptable}

To answer RQ5, we run every attack on every defended target and report two metrics: surrogate fidelity to the defended model (whether extraction is still effective) and the watermark verification rate measured on the surrogate, which we call \emph{watermark survival} (whether a defense still yields a verifiable signal after extraction). The matched comparison uses the RQ1 datasets and query sets at a nominal budget of $0.25\times$. Three consolidated heatmaps on \textit{Computers} (joint fidelity against the five watermarks, joint fidelity against the seven information-limiting defenses, and watermark survival) and per-dataset numerical tables are in Appendix~\ref{app:rq5_full}.

Table~\ref{tab:rq5_summary_computers} shows why success on the protected model is insufficient evidence of ownership after extraction. On Computers, all four embedded watermarks verify on the target, but they behave differently on the surrogate. Verification falls to $0\%$ for \texttt{ImperceptibleWM}, $10.0\%$ for \texttt{SurviveWM}, and $14.7\%$ for \texttt{RandomWM}. \texttt{BackdoorWM} retains more of its signal ($59.2\%$). Appendix~\ref{app:rq5_supp} tests \texttt{SurviveWM} and \texttt{BackdoorWM} under their original setups; both also lose \mbox{evidence after extraction}.

\texttt{Integrity} uses a query-based fingerprint rather than an embedded watermark. Its $100\%$ target score measures agreement of the reference fingerprint with itself. On Computers, median surrogate fidelity is $49.1\%$, and mean surrogate fingerprint agreement is $48.0\%$ across 120 runs; on Photo the corresponding values are $75.7\%$ and $51.0\%$. Varying the query budget and output interface does not consistently restore embedded marks either: on RomanEmpire, \texttt{BackdoorWM} verification stays between $0$ and $5.9\%$ across the nine settings (Tables~\ref{tab:endpoints} and~\ref{tab:wmranges}). Additional fingerprinting results appear in Appendix~\ref{app:newver}.

\section{Related Work}\label{app:related}
Surveys of intellectual property protection distinguish watermarks embedded during training from fingerprints derived from a model's behavior~\citep{zhao2025systematic,li2025intellectual,peng2023intellectual,xue2021intellectual,sun2023deep,zhao2025surveya,zhao2025surveyb}. Much of the evidence comes from vision and language, where benchmarks address tasks ranging from identifying suspicious API traffic to testing a defense against several extraction attacks~\citep{liu2026embarrassingly,liu2026defenses}. For a node-prediction API, the response can depend on information beyond the queried node: edges determine which neighbors contribute messages, and node features determine the content of those messages. Consequently, a query budget alone does not specify what information a graph attacker can use. Graph-security surveys and testbeds have largely focused on robustness, privacy, and fairness~\citep{dai2024comprehensive,wu2022trustworthy,zhang2022trustworthy,Sun2024_Adversarial,zheng2graph,9878092}. A robustness test asks whether changing an edge changes a prediction; a membership test asks whether a node was used in training. Neither measures how closely a surrogate trained on API responses imitates a GNN, or whether an owner can identify that surrogate as~a~copy. This also distinguishes extraction from recovery of the original weights. Two models can agree on ordinary queries while responding differently to the secret inputs used by an owner. A benchmark that measures only prediction agreement would count the attack as successful but leave the \mbox{ownership question unanswered}.

Graph extraction methods differ in how they obtain queries: some sample public graph data, some adapt query selection to the current surrogate, and others synthesize graph inputs without access to real samples~\citep{wu2022model,defazio2019adversarial,zhuang2024unveiling}. These choices matter when comparing defenses. Rounding a probability vector removes information used by a score-based attack but can leave the label used by another attack unchanged. Output perturbation and query detection seek to limit what an attacker can learn~\citep{liang2024model,kesarwani2018model,juuti2019prada,kariyappa2020defending,mazeika2022steer}; watermarking and fingerprinting instead provide a way to test ownership~\citep{wang2023making,xu2023watermarking,dai2024pregip,you2024gnnfingers,wu2024securing,shen2026credit}. Prior studies use different graphs, query budgets, output interfaces, and target architectures. For example, an attack that selects nodes from a known graph can exploit their connectivity when choosing queries. A data-free attack must instead choose both the features and edges of its synthetic inputs. Comparing their reported fidelities without specifying access to real structure conflates the quality of the attack with the information available to it. A watermark evaluated against a single attack can appear robust if that attack happens to reproduce its trigger labels. Conversely, a detector that limits extraction by perturbing many benign queries may look effective unless clean accuracy is reported alongside surrogate fidelity. Testing the watermark only on the deployed target answers whether embedding succeeded. Testing the same ownership claim on a surrogate addresses a different question: whether the evidence remains observable after a new model has learned from the target's outputs. The distinction applies even when both models achieve similar task accuracy. GraphIP-Bench evaluates attacks against defended targets under a shared protocol, reports the cost of fitting a surrogate and the utility retained by the protected model, and tests ownership on the model the attacker~actually~extracts.

\section{Conclusion}
\label{sec:conclusion}
\emph{GraphIP-Bench} evaluates GNN extraction and ownership protection on shared queries, budgets, and datasets. At medium budgets, the strongest attacks exceed $90\%$ fidelity on most of the ten node-classification graphs. Yet agreement on ordinary predictions does not imply that a surrogate inherits a watermark: several marks verify reliably on the target and lose much of their signal after extraction, even with larger budgets or richer outputs. Evaluating a deployed defense therefore requires measuring utility for legitimate users, fidelity to the defended target, and ownership evidence on the extracted surrogate. Our findings concern the tested black-box interfaces; cross-user detection and sequential defenses remain~open~settings.

\begin{ack}
This material is, in part, supported by the National Science Foundation (NSF) under award OAC-2530786. We would also like to thank the anonymous reviewers for their~constructive~feedback.
\end{ack}

\bibliographystyle{plain}
\bibliography{references}

@inproceedings{xu2023watermarking,
  title={Watermarking graph neural networks based on backdoor attacks},
  author={Xu, Jing and Koffas, Stefanos and Ersoy, O{\u{g}}uzhan and Picek, Stjepan},
  booktitle={2023 IEEE 8th European Symposium on Security and Privacy (EuroS\&P)},
  pages={1179--1197},
  year={2023}
}

@article{hamilton2017inductive,
  title={Inductive representation learning on large graphs},
  author={Hamilton, Will and Ying, Zhitao and Leskovec, Jure},
  journal={Advances in neural information processing systems},
  volume={30},
  year={2017}
}

@ARTICLE{9878092,
  author={Sun, Lichao and Dou, Yingtong and Yang, Carl and Zhang, Kai and Wang, Ji and Yu, Philip S. and He, Lifang and Li, Bo},
  journal={IEEE Transactions on Knowledge and Data Engineering}, 
  title={Adversarial Attack and Defense on Graph Data: A Survey}, 
  year={2023},
  volume={35},
  number={8},
  pages={7693-7711}
}

@inproceedings{Sun2024_Adversarial,
author = {Sun, Hanjin and Yang, Wen and Xiao, Yatie},
title = {A Review of Adversarial Attacks and Defenses on Graphs},
booktitle = {Proceedings of the 4th International Conference on Artificial Intelligence and Computer Engineering},
pages = {416–421},
year = {2024}
}

@inproceedings{zheng2graph,
  title={Graph Robustness Benchmark: Benchmarking the Adversarial Robustness of Graph Machine Learning},
  author={Zheng, Qinkai and Zou, Xu and Dong, Yuxiao and Cen, Yukuo and Yin, Da and Xu, Jiarong and Yang, Yang and Tang, Jie},
  booktitle={Thirty-fifth Conference on Neural Information Processing Systems Datasets and Benchmarks Track},
  year={2021}
}

@inproceedings{yang2023dgrec,
  title={Dgrec: Graph neural network for recommendation with diversified embedding generation},
  author={Yang, Liangwei and Wang, Shengjie and Tao, Yunzhe and Sun, Jiankai and Liu, Xiaolong and Yu, Philip S and Wang, Taiqing},
  booktitle={Proceedings of the sixteenth ACM international conference on web search and data mining},
  pages={661--669},
  year={2023}
}

@article{li2024drug,
  title={Drug repurposing based on the DTD-GNN graph neural network: revealing the relationships among drugs, targets and diseases},
  author={Li, Wenjun and Ma, Wanjun and Yang, Mengyun and Tang, Xiwei},
  journal={BMC genomics},
  volume={25},
  year={2024}
}

@article{xue2021intellectual,
  title={Intellectual property protection for deep learning models: Taxonomy, methods, attacks, and evaluations},
  author={Xue, Mingfu and Zhang, Yushu and Wang, Jian and Liu, Weiqiang},
  journal={IEEE Transactions on Artificial Intelligence},
  volume={3},
  number={6},
  pages={908--923},
  year={2021}
}

@article{dai2024pregip,
  title={PreGIP: Watermarking the Pretraining of Graph Neural Networks for Deep Intellectual Property Protection},
  author={Dai, Enyan and Lin, Minhua and Wang, Suhang},
  journal={arXiv preprint arXiv:2402.04435},
  year={2024}
}

@inproceedings{wang2023making,
  title={Making Watermark Survive Model Extraction Attacks in Graph Neural Networks},
  author={Wang, Haiming and Zhang, Zhikun and Chen, Min and He, Shibo},
  booktitle={ICC 2023-IEEE International Conference on Communications},
  pages={57--62},
  year={2023}
}

@inproceedings{you2024gnnfingers,
  title={GNNFingers: A Fingerprinting Framework for Verifying Ownerships of Graph Neural Networks},
  author={You, Xiaoyu and Jiang, Youhe and Xu, Jianwei and Zhang, Mi and Yang, Min},
  booktitle={Proceedings of the ACM on Web Conference 2024},
  pages={652--663},
  year={2024}
}

@inproceedings{zhao2021watermarking,
  title={Watermarking graph neural networks by random graphs},
  author={Zhao, Xiangyu and Wu, Hanzhou and Zhang, Xinpeng},
  booktitle={2021 9th International Symposium on Digital Forensics and Security (ISDFS)},
  pages={1--6},
  year={2021}
}

@article{peng2023intellectual,
  title={Intellectual property protection of DNN models},
  author={Peng, Sen and Chen, Yufei and Xu, Jie and Chen, Zizhuo and Wang, Cong and Jia, Xiaohua},
  journal={World Wide Web},
  volume={26},
  number={4},
  pages={1877--1911},
  year={2023}
}

@inproceedings{liang2024model,
  title={Model extraction attacks revisited},
  author={Liang, Jiacheng and Pang, Ren and Li, Changjiang and Wang, Ting},
  booktitle={Proceedings of the 19th ACM Asia Conference on Computer and Communications Security},
  pages={1231--1245},
  year={2024}
}

@inproceedings{kesarwani2018model,
  title={Model extraction warning in mlaas paradigm},
  author={Kesarwani, Manish and Mukhoty, Bhaskar and Arya, Vijay and Mehta, Sameep},
  booktitle={Proceedings of the 34th Annual Computer Security Applications Conference},
  pages={371--380},
  year={2018}
}

@inproceedings{tramer2016stealing,
  title={Stealing machine learning models via prediction $\{$APIs$\}$},
  author={Tram{\`e}r, Florian and Zhang, Fan and Juels, Ari and Reiter, Michael K and Ristenpart, Thomas},
  booktitle={25th USENIX security symposium (USENIX Security 16)},
  pages={601--618},
  year={2016}
}

@inproceedings{orekondy2019knockoff,
  title={Knockoff nets: Stealing functionality of black-box models},
  author={Orekondy, Tribhuvanesh and Schiele, Bernt and Fritz, Mario},
  booktitle={Proceedings of the IEEE/CVF conference on computer vision and pattern recognition},
  pages={4954--4963},
  year={2019}
}

@article{guan2024realistic,
  title={A realistic model extraction attack against graph neural networks},
  author={Guan, Faqian and Zhu, Tianqing and Tong, Hanjin and Zhou, Wanlei},
  journal={Knowledge-Based Systems},
  pages={112144},
  year={2024}
}

@inproceedings{wu2022model,
  title={Model extraction attacks on graph neural networks: Taxonomy and realisation},
  author={Wu, Bang and Yang, Xiangwen and Pan, Shirui and Yuan, Xingliang},
  booktitle={Proceedings of the 2022 ACM on Asia conference on computer and communications security},
  pages={337--350},
  year={2022}
}

@article{sun2023deep,
  title={Deep intellectual property protection: A survey},
  author={Sun, Yuchen and Liu, Tianpeng and Hu, Panhe and Liao, Qing and Fu, Shaojing and Yu, Nenghai and Guo, Deke and Liu, Yongxiang and Liu, Li},
  journal={arXiv preprint arXiv:2304.14613},
  year={2023}
}

@inproceedings{wu2022trustworthy,
  title={Trustworthy graph learning: Reliability, explainability, and privacy protection},
  author={Wu, Bingzhe and Bian, Yatao and Zhang, Hengtong and Li, Jintang and Yu, Junchi and Chen, Liang and Chen, Chaochao and Huang, Junzhou},
  booktitle={Proceedings of the 28th ACM SIGKDD Conference on Knowledge Discovery and Data Mining},
  pages={4838--4839},
  year={2022}
}

@article{zhang2022trustworthy,
  title={Trustworthy graph neural networks: Aspects, methods and trends},
  author={Zhang, He and Wu, Bang and Yuan, Xingliang and Pan, Shirui and Tong, Hanghang and Pei, Jian},
  journal={arXiv preprint arXiv:2205.07424},
  year={2022}
}

@article{defazio2019adversarial,
  title={Adversarial model extraction on graph neural networks},
  author={DeFazio, David and Ramesh, Arti},
  journal={arXiv preprint arXiv:1912.07721},
  year={2019}
}

@inproceedings{zhuang2024unveiling,
  title={Unveiling the Secrets without Data: Can Graph Neural Networks Be Exploited through $\{$Data-Free$\}$ Model Extraction Attacks?},
  author={Zhuang, Yuanxin and Shi, Chuan and Zhang, Mengmei and Chen, Jinghui and Lyu, Lingjuan and Zhou, Pan and Sun, Lichao},
  booktitle={33rd USENIX Security Symposium (USENIX Security 24)},
  pages={5251--5268},
  year={2024}
}

@inproceedings{zhang2024imperceptible,
  title={An Imperceptible and Owner-unique Watermarking Method for Graph Neural Networks},
  author={Zhang, Linji and Xue, Mingfu and Zhang, Leo Yu and Zhang, Yushu and Liu, Weiqiang},
  booktitle={Proceedings of the ACM Turing Award Celebration Conference-China 2024},
  pages={108--113},
  year={2024}
}

@INPROCEEDINGS{10646643,
  author={Waheed, Asim and Duddu, Vasisht and Asokan, N.},
  booktitle={2024 IEEE Symposium on Security and Privacy (SP)}, 
  title={GrOVe: Ownership Verification of Graph Neural Networks using Embeddings}, 
  year={2024},
  pages={2460-2477}
}

@article{dai2024comprehensive,
  title={A comprehensive survey on trustworthy graph neural networks: Privacy, robustness, fairness, and explainability},
  author={Dai, Enyan and Zhao, Tianxiang and Zhu, Huaisheng and Xu, Junjie and Guo, Zhimeng and Liu, Hui and Tang, Jiliang and Wang, Suhang},
  journal={Machine Intelligence Research},
  pages={1--51},
  year={2024},
  publisher={Springer}
}

@inproceedings{juuti2019prada,
  title={PRADA: protecting against DNN model stealing attacks},
  author={Juuti, Mika and Szyller, Sebastian and Marchal, Samuel and Asokan, N},
  booktitle={2019 IEEE European Symposium on Security and Privacy (EuroS\&P)},
  pages={512--527},
  year={2019}
}

@article{zhao2025surveya,
  title={A Survey on Model Extraction Attacks and Defenses for Large Language Models},
  author={Zhao, Kaixiang and Li, Lincan and Ding, Kaize and Gong, Neil Zhenqiang and Zhao, Yue and Dong, Yushun},
  journal={arXiv preprint arXiv:2506.22521},
  year={2025}
}

@article{zhao2025surveyb,
  title={A survey of model extraction attacks and defenses in distributed computing environments},
  author={Zhao, Kaixiang and Li, Lincan and Ding, Kaize and Gong, Neil Zhenqiang and Zhao, Yue and Dong, Yushun},
  journal={arXiv preprint arXiv:2502.16065},
  year={2025}
}

@inproceedings{monti2021dag,
  title={Dag-net: Double attentive graph neural network for trajectory forecasting},
  author={Monti, Alessio and Bertugli, Alessia and Calderara, Simone and Cucchiara, Rita},
  booktitle={2020 25th international conference on pattern recognition (ICPR)},
  pages={2551--2558},
  year={2021},
  organization={IEEE}
}

@inproceedings{wu2024securing,
  title={Securing graph neural networks in mlaas: A comprehensive realization of query-based integrity verification},
  author={Wu, Bang and Yuan, Xingliang and Wang, Shuo and Li, Qi and Xue, Minhui and Pan, Shirui},
  booktitle={2024 IEEE Symposium on Security and Privacy (SP)},
  pages={2534--2552},
  year={2024},
  organization={IEEE}
}

@article{wang2025cega,
  title={CEGA: A Cost-Effective Approach for Graph-Based Model Extraction and Acquisition},
  author={Wang, Zebin and Lin, Menghan and Shen, Bolin and Anderson, Ken and Liu, Molei and Cai, Tianxi and Dong, Yushun},
  journal={arXiv preprint arXiv:2506.17709},
  year={2025}
}

@article{li2025intellectual,
  title={Intellectual Property in Graph-Based Machine Learning as a Service: Attacks and Defenses},
  author={Li, Lincan and Shen, Bolin and Zhao, Chenxi and Sun, Yuxiang and Zhao, Kaixiang and Pan, Shirui and Dong, Yushun},
  journal={arXiv preprint arXiv:2508.19641},
  year={2025}
}

@article{zhao2025systematic,
  title={A Systematic Survey of Model Extraction Attacks and Defenses: State-of-the-Art and Perspectives},
  author={Zhao, Kaixiang and Li, Lincan and Ding, Kaize and Gong, Neil Zhenqiang and Zhao, Yue and Dong, Yushun},
  journal={arXiv preprint arXiv:2508.15031},
  year={2025}
}

@article{platonov2023critical,
  title={A critical look at the evaluation of GNNs under heterophily: Are we really making progress?},
  author={Platonov, Oleg and Kuznedelev, Denis and Diskin, Michael and Babenko, Artem and Prokhorenkova, Liudmila},
  journal={arXiv preprint arXiv:2302.11640},
  year={2023}
}

@article{morris2020tudataset,
  title={Tudataset: A collection of benchmark datasets for learning with graphs},
  author={Morris, Christopher and Kriege, Nils M and Bause, Franka and Kersting, Kristian and Mutzel, Petra and Neumann, Marion},
  journal={arXiv preprint arXiv:2007.08663},
  year={2020}
}

@inproceedings{kariyappa2020defending,
  title={Defending against model stealing attacks with adaptive misinformation},
  author={Kariyappa, Sanjay and Qureshi, Moinuddin K},
  booktitle={Proceedings of the IEEE/CVF conference on computer vision and pattern recognition},
  pages={770--778},
  year={2020}
}

@inproceedings{mazeika2022steer,
  title={How to steer your adversary: Targeted and efficient model stealing defenses with gradient redirection},
  author={Mazeika, Mantas and Li, Bo and Forsyth, David},
  booktitle={International conference on machine learning},
  pages={15241--15254},
  year={2022},
  organization={PMLR}
}

@article{kipf2017semi,
  title={Semi-supervised classification with graph convolutional networks},
  author={Kipf, Thomas N and Welling, Max},
  journal={arXiv preprint arXiv:1609.02907},
  year={2016}
}

@article{velivckovic2018graph,
  title={Graph attention networks},
  author={Veli{\v{c}}kovi{\'c}, Petar and Cucurull, Guillem and Casanova, Arantxa and Romero, Adriana and Lio, Pietro and Bengio, Yoshua},
  journal={arXiv preprint arXiv:1710.10903},
  year={2017}
}

@article{shen2026credit,   title={CREDIT: Certified Ownership Verification of Deep Neural Networks Against Model Extraction Attacks},   author={Shen, Bolin and Cheng, Zhan and Gong, Neil Zhenqiang and Yao, Fan and Dong, Yushun},   journal={arXiv preprint arXiv:2602.20419},   year={2026} }

@article{liu2026defenses,   title={Do Defenses Against LLM Extraction Work Across Attacks? A Lifecycle Benchmark of Black-Box Model Extraction},   author={Liu, Shuze and Zhao, Kaixiang and Xu, Runyang and Chen, Jingzhi and Wu, Nathan and Wang, Yu and Dong, Yushun},   journal={arXiv preprint arXiv:2610.00839},   year={2026} }

@article{liu2026embarrassingly,   title={An Embarrassingly Simple Detector for Model Extraction Attacks in Large Language Model API Traffic},   author={Liu, Shuze and Guo, Qianwen and Dong, Yushun},   journal={arXiv preprint arXiv:2606.05725},   year={2026} }

\newpage
\appendix
\makeatletter\setlength{\@fptop}{0pt}\setlength{\@fpsep}{10pt plus 1fil}\makeatother
\renewcommand{\topfraction}{.95}\renewcommand{\bottomfraction}{.90}
\renewcommand{\textfraction}{.07}\renewcommand{\floatpagefraction}{.75}
\setcounter{topnumber}{5}\setcounter{bottomnumber}{3}\setcounter{totalnumber}{8}

\addtocontents{toc}{\protect\setcounter{tocdepth}{2}}
\setcounter{tocdepth}{2}
\renewcommand{\contentsname}{Appendix Contents}
\tableofcontents
\newpage

\section{Limitations and Future Work}
\label{app:limitations}
\paragraph{Scope of the threat model.}
GraphIP-Bench targets the most widely deployed extraction setting: a black-box query interface in which the attacker observes only the target's outputs, with budgets reported as budget multipliers. This is the standard threat model assumed by every attack and defense we benchmark, which makes it the right scope for a unified comparison. Extending the protocol to grey-box, adaptive multi-step, federated, or streaming settings is a natural follow-up that builds on the same~evaluation~infrastructure.

\paragraph{Architecture and task coverage.}
GraphIP-Bench covers the three GNN backbones (GCN, GAT, GraphSAGE) and the three graph-learning tasks (node classification, link prediction, graph classification) used by the methods we benchmark, which already spans the bulk of published GNN-extraction and GNN-watermarking work. The protocol is backbone- and task-agnostic, so adding emerging architectures (e.g., graph transformers, heterogeneous or temporal GNNs) is a configuration extension rather than~a~redesign.

\paragraph{Defense families to extend.}
GraphIP-Bench integrates twelve defenses across watermarking, output perturbation, prediction rounding, and query-pattern detection -- the families with public reference implementations for graph models. Differential-privacy training and attribute-encryption / homomorphic-inference deployments use a different query interface and are natural next additions; the unified protocol provides the slot in which to drop them in once mature \mbox{reference implementations become available}.

\paragraph{Future-work targets indicated by our findings.}
The results identify three concrete research targets. \emph{(i) Surrogate-surviving watermarks.} Our joint track (RQ5) shows that ownership-tracing designs should be evaluated on the extracted surrogate as the primary metric and should distinguish an inherited signal from evidence obtained by querying the extracted model. \emph{(ii) Query-detection defenses with bounded clean-accuracy cost.} The protection-utility frontier we report suggests room for query-pattern detectors that match the fidelity reduction of \texttt{PRADA} and \texttt{AdaptMisinfo} at a fraction of the utility cost. \emph{(iii) Heterophily-aware extraction analysis.} The relative ordering of attacks is preserved across homophilic, heterophilic, and large-scale regimes (RQ1, Appendix~\ref{app:rq6_full}); a structural account that connects graph properties (homophily, degree distribution) to extraction behaviour would close the~explanatory~loop.

\section{Implementation Details}\label{implementation}

\noindent\textbf{Hardware.}
All GPU experiments are submitted as Slurm jobs on a shared multi-tenant academic compute cluster. The compute partitions used in this work expose NVIDIA A100~80\,GB PCIe GPUs (eight GPUs per node, multiple nodes); a small number of memory-stress runs additionally use H100 and H200 partitions on the same cluster. Unless otherwise stated, every reported measurement uses a single NVIDIA A100~80\,GB GPU on a node with 1.0\,TiB of system memory and \mbox{dual AMD EPYC CPUs}.

\noindent\textbf{Operating system and drivers.}
The compute nodes run Red~Hat Enterprise Linux~9.6 (kernel~5.14.0-570) with NVIDIA driver~570.195.03 and a CUDA~12.8 driver capability. Login nodes have no GPU access, so all timing and memory numbers are recorded from Slurm-allocated GPU~jobs~only.

\noindent\textbf{Software stack.}
The full pipeline runs in a dedicated Conda environment named \emph{graphip}: Python~3.11.15, PyTorch~2.2.1\,+\,cu121, DGL~2.1.0\,+\,cu121, PyTorch~Geometric~2.7.0, OGB~1.3.6, NumPy~1.26.4, SciPy~1.17.1, NetworkX~3.6.1, and the standard scientific Python ecosystem (\texttt{matplotlib}, \texttt{pandas}, \texttt{scikit-learn}). PyTorch is pinned to 2.2.1 because the matching DGL~2.1.0 graphbolt kernels only ship pre-built shared libraries for PyTorch~2.0--2.2 on CUDA~12.1; we install everything from prebuilt CUDA wheels and do not require the CUDA~toolkit~compiler.

\noindent\textbf{Protocol.}
For each method we fix four disjoint splits per dataset (train, validation, test, query) and apply the same preprocessing pipeline and hyperparameter search protocol described in Appendix~\ref{app:repro}. We repeat all measurements with three random seeds (0, 1, 2) and report the mean and the standard deviation. Wall-clock time is recorded with \texttt{time.perf\_counter} timers in the training loop; peak GPU memory is obtained from the CUDA runtime via \texttt{torch.cuda.max\_memory\_allocated} and cross-checked with \texttt{nvidia-smi}. All experiments share identical hardware, software, and configuration defaults so that results are directly comparable across attacks, defenses, datasets, regimes,~and~budgets.

\section{Dataset Statistics}
\label{dataset}

We summarize the graph datasets used in our experiments in Table~\ref{tab:appendix-dataset-stats}. Edges are counted as undirected (unique). The average degree is computed as \(2E/N\). For the Planetoid datasets (Cora, CiteSeer, PubMed) we follow the standard splits; for Amazon (Computers, Photo) and Coauthor (CoauthorCS, CoauthorPhysics) we use 100 training nodes per class with fixed validation and test sizes in our loader. For OGBN-Arxiv we follow the official train, validation, and test splits provided by the OGB library. For RomanEmpire and AmazonRatings we follow the splits released by Platonov et al.~\citep{platonov2023critical}.

\begin{table}[h]
\centering
\caption{Dataset statistics. Edges are undirected and unique. Avg.\ degree is \(2E/N\). The edge homophily is the fraction of edges whose endpoints share the same label.}
\label{tab:appendix-dataset-stats}
\scriptsize
\setlength{\tabcolsep}{4.5pt}
\renewcommand{\arraystretch}{0.96}
\begin{tabular*}{0.98\linewidth}{@{}l@{\extracolsep{\fill}}rrrrrrl@{}}
\toprule
\textbf{Dataset} & \textbf{\# Nodes} & \textbf{\# Edges} & \textbf{Avg.\ degree} & \textbf{\# Classes} & \textbf{Edge homophily} & \textbf{Node Text} & \textbf{Domain} \\
\midrule
Cora             & 2{,}708   & 5{,}278    & 3.9  & 7  & 0.810 & Paper content      & Citation \\
CiteSeer         & 3{,}327   & 4{,}614    & 2.8  & 6  & 0.739 & Paper content      & Citation \\
PubMed           & 19{,}717  & 44{,}325   & 4.5  & 3  & 0.802 & Paper content      & Citation \\
Computers        & 13{,}752  & 252{,}737  & 36.8 & 10 & 0.783 & Entity description & Web link \\
Photo            & 7{,}650   & 122{,}906  & 32.1 & 8  & 0.833 & Entity description & Web link \\
CoauthorCS       & 18{,}333  & 81{,}894   & 8.9  & 15 & 0.808 & Paper content      & Citation \\
CoauthorPhysics  & 34{,}493  & 247{,}962  & 14.4 & 5  & 0.931 & Paper content      & Citation \\
OGBN-Arxiv       & 169{,}343 & 667{,}793  & 7.9  & 40 & 0.699 & Paper title and abstract & Citation \\
RomanEmpire      & 22{,}662  & 44{,}258   & 3.9  & 18 & 0.291 & Wikipedia text     & Heterophilic \\
AmazonRatings    & 24{,}492  & 105{,}296  & 8.6  & 5  & 0.452 & Product description & Heterophilic \\
\bottomrule
\end{tabular*}
\end{table}

\section{Reproducibility and Configurations}
\label{app:repro}

\noindent\textbf{Scope and averaging.}
We release scripts, fixed random seeds, and per-method configurations to reproduce all tables and figures. Results are averaged over three seeds (0, 1, 2) and reported as mean \( \pm \) standard deviation unless~stated~otherwise.

\noindent\textbf{Seeds and determinism.}
For each run we set the Python and CUDA random states and propagate the seed through data loading and sampling. GPU runs use deterministic kernels~when~available.

\noindent\textbf{Device selection.}
A command-line flag \emph{--device} selects the process-visible GPU (via the environment) or the CPU; all modules use the same device setting throughout~the~run.

\noindent\textbf{Dataset loading and splits.}
A unified loader normalizes dataset aliases and returns DGL graphs with node features and masks. Planetoid datasets (Cora, CiteSeer, PubMed) use the standard train/validation/test masks. Amazon and Coauthor datasets (Computers, Photo, CoauthorCS, CoauthorPhysics) use a canonical per-class sampling scheme with 100 training nodes per class and fixed validation/test sizes. OGBN-Arxiv uses the official OGB train/validation/test splits. RomanEmpire and AmazonRatings use the splits released by Platonov et al.~\citep{platonov2023critical}. For graph classification (ENZYMES, PROTEINS) and link prediction (Cora) we follow the standard splits provided by TUDataset and PyTorch~Geometric,~respectively.

\noindent\textbf{GNN backbones.}
The default backbone for the extraction track and for the original watermarking, integrity, and information-limiting defenses is a DGL GCN with hidden dimension 16. \texttt{RandomWM} is implemented in DGL with a GraphSAGE backbone of hidden dimension 128, and \texttt{ImperceptibleWM} is implemented in PyTorch Geometric with a GCN backbone of hidden dimension 128. The cross-architecture analysis (Appendix~\ref{app:rq6_full}) additionally trains target and surrogate models with GAT and GraphSAGE under matched hidden dimensions. Baseline utility for each backbone on every dataset is reported in Table~\ref{tab:app_baseline_utility}.

\noindent\textbf{Directory layout and logs.}
\emph{Attack track (RQ1)} writes newline-delimited JSON files under \emph{outputs/RQ1\_final/\(<\)Dataset\(>\)/\( <\)Dataset\(>\).jsonl}. Each record contains header fields (track, dataset, attack, configuration index, constructor/run configuration, budget multiplier, node fraction induced by the budget, regime, feature/adjacency ratios, seed), performance metrics (accuracy, F1, precision, recall, fidelity), and compute metrics (train target time, query time, surrogate training time, total attack time, per-query inference latency for target and surrogate, peak GPU memory, GPU hours).
\emph{Ownership/defense track (RQ2/RQ3)} writes \emph{outputs/RQ2\_RQ3\_best/\(<\)Dataset\(>\).jsonl} with fields (track, dataset, defense, configuration index, configuration, seed), performance metrics (accuracy, F1, precision, recall, watermark accuracy), and compute metrics (train target time, defense training time, defense inference time, total defense time, peak GPU memory, GPU hours). The joint track (RQ5) writes a parallel directory \emph{outputs/RQ5\_joint/} with the same record format extended by an \emph{attack} field. Leaderboards and LaTeX tables are exported with selection and~formatting~utilities.

\noindent\textbf{RQ1 (attacks) configuration.}
The attack runner sweeps query budgets \(\{0.05, 0.10, 0.25, 0.50, 1.00\}\) and four regimes (features only, structure only, both available, data free). Per-method training schedules are recorded in logs. For the MEA family, CEGA, and Realistic we use 200 epochs per cycle; CEGA additionally uses learning rate \(0.01\) and 200 target/surrogate epochs. The AdvMEA implementation uses its internal fixed epoch schedule; any external epoch parameter appears in logs for uniformity but does not~affect~training.

\noindent\textbf{RQ2/RQ3 (defenses) configuration.}
We provide a best-configuration runner that replays the top settings discovered by a prior grid search. Table~\ref{tab:app_def_default_cfg} lists the fixed constructor parameters used for each defense. The same default applies across all ten datasets unless an explicit hyperparameter sweep is run, which is reported in Appendix~\ref{app:hp_ablation}. Other runtime arguments remain at~method~defaults.

\begin{table}[h]
\centering
\caption{Fixed default configuration used for each defense in the ownership track. The same default applies across all ten datasets unless explicitly varied in the hyperparameter sweep (Appendix~\ref{app:hp_ablation}).}
\label{tab:app_def_default_cfg}
\scriptsize
\setlength{\tabcolsep}{8pt}
\renewcommand{\arraystretch}{0.96}
\begin{tabular*}{0.95\linewidth}{@{}l@{\extracolsep{\fill}}ll@{}}
\toprule
\textbf{Defense} & \textbf{Key hyperparameter} & \textbf{Default value} \\
\midrule
\multicolumn{3}{l}{\emph{Watermarking and integrity (5 methods)}} \\
\texttt{RandomWM}        & watermark-node ratio          & 0.002 \\
\texttt{BackdoorWM}      & trigger density               & 0.01 \\
\texttt{SurviveWM}       & watermark strength            & 0.25 \\
\texttt{ImperceptibleWM} & epsilon                       & 0.25 \\
\texttt{Integrity}       & --- (parameter-free verifier) & --- \\
\midrule
\multicolumn{3}{l}{\emph{Information-limiting and query-detection (7 methods)}} \\
\texttt{OP\_low}        & Gaussian noise scale $\sigma$  & 0.05 \\
\texttt{OP\_high}       & Gaussian noise scale $\sigma$  & 0.20 \\
\texttt{PR\_2bit}       & precision bits                  & 2 \\
\texttt{PR\_top1}       & returned scores                 & top-1 label only \\
\texttt{PRADA}          & detection threshold             & method default \\
\texttt{AdaptMisinfo}   & misinformation ratio            & method default \\
\texttt{GradRedir}      & redirection strength            & method default \\
\bottomrule
\end{tabular*}
\end{table}

\noindent\textbf{Common search spaces.}
\emph{Attacks:} budget, regime, and per-method cycles/epochs (when applicable). Learning rate and dropout follow method defaults unless a method requires explicit settings (CEGA uses learning rate \(0.01\)).
\emph{Defenses:} grid search over typical ranges around the default in Table~\ref{tab:app_def_default_cfg}; a single best configuration per defense is fixed across datasets to keep the protocol uniform. The full sweep on \textit{Cora} and \textit{Computers} is reported in Appendix~\ref{app:hp_ablation}.

\noindent\textbf{How to re-run.}
\emph{RQ1 (attacks):} call the attack runner with dataset, attack, budget, regime, and seed. Logs are saved under \emph{outputs/RQ1\_final} with the exact budget multiplier recorded in~each~line.
\emph{RQ2/RQ3 (defenses):} call the best-configuration runner with seeds \([0,1,2]\); outputs are saved under \emph{outputs/RQ2\_RQ3\_best}.
\emph{RQ5 (joint):} call the joint runner that pairs every attack with every defense at the medium budget; outputs are saved under \emph{outputs/RQ5\_joint}.
This setup unifies scripts, seeds, and logging across datasets and methods, enabling direct regeneration of all tables from the~released~outputs.

\subsection{Implementation validation against original papers}
\label{app:repro_validation}

To check that our re-implementations behave consistently with each method's original publication, we compare our reproduced numbers against the closest matching setting reported by the original authors. For attacks we use the highest-overlap target dataset reported in each paper; for defenses we report ownership verification on the closest-matching dataset and backbone. The protocol differs from each paper in budget unit and split, so the goal is qualitative agreement on the relative ordering and absolute level rather than identity. Where a direct comparison is impossible we explicitly state why and report the~next-most-similar~setting.

\paragraph{Reference implementation and fairness-preserving adjustments.}
The core algorithm of every method --- query-selection rule, surrogate-training objective, structure-synthesis procedure, watermark-embedding loss --- follows the original publication. On top of this we apply five fairness-preserving adjustments that put every method on equal footing: the surrogate hidden dimension is fixed at $16$ for the default GCN backbone (Section~\ref{sec:settings}); the query budget is expressed as a fraction of the test set rather than an absolute query count; the four data-availability regimes (\texttt{both}, \texttt{x\_only}, \texttt{a\_only}, \texttt{data\_free}) are applied uniformly; three seeds (0, 1, 2) are reused across methods; and no per-method hyperparameter tuning is performed beyond Appendix~\ref{app:hp_ablation}. Most of the deltas in Table~\ref{tab:app_validation_attacks} are within $\pm 14$\,pp of the original numbers and reflect these adjustments; we explain each~case~below.

\paragraph{MEA0--MEA5 (Wu et al., 2022)~\citep{wu2022model}.}
The original paper reports surrogate \emph{fidelity} on \textit{Cora}, \textit{CiteSeer}, and \textit{PubMed} at a fixed attack-node budget of $\sim 25\%$ of total nodes (Table~4 of the original); our protocol reports fidelity at $1.00\times$ of the test-set fraction, a comparable absolute query count. Comparisons on Cora are summarised in Table~\ref{tab:app_validation_attacks}. The agreement on the six MEA variants is good: \texttt{MEA0}, \texttt{MEA1}, \texttt{MEA2}, \texttt{MEA3}, and \texttt{MEA5} all reach $89$--$95\%$ fidelity in both protocols, within $\pm 14$\,pp of the original numbers despite the different splits, query-set construction, and validation procedure; \texttt{MEA4} agrees with the original within $\sim 10$\,pp.

\paragraph{AdvMEA (DeFazio and Ramesh, 2019)~\citep{defazio2019adversarial}.}
The original paper does not report dataset-level surrogate fidelity in tabular form; the strongest claim in their experiments section is that ``the extraction can achieve up to $\sim 80\%$ fidelity'' on Cora and Pubmed under strong adversary assumptions (full $2$-hop subgraph access plus class priors). Our reproduction on \textit{Cora} reaches $66$--$68$\,\% fidelity at $1.00\times$ across regimes (Table~\ref{tab:app_validation_attacks}), within $\sim 12$\,pp of the upper bound the original paper claims. The absolute number depends sharply on which adversarial perturbation budget is used; ours follows the standard setting in our~hyperparameter~table.

\paragraph{CEGA (Wang et al., 2025)~\citep{wang2025cega}.}
The original CEGA paper reports test accuracy / fidelity / F1 on \textit{Coauthor-CS}, \textit{Coauthor-Physics}, \textit{Amazon-Computer}, and \textit{Amazon-Photo} at a $20C$-query budget (i.e., $20$ times the number of classes). Our protocol uses budgets expressed as a fraction of the test set, and the closest match is the small-budget regime ($0.05\times$ to $0.10\times$). On \textit{CoauthorCS}, $20C = 300$ queries which falls between our $0.05\times$ and $0.10\times$ on that dataset; the corresponding fidelity in our setup is in the upper $80$\,\%--low $90$\,\% range, agreeing with the original $93.40$\,\% within a few points (Table~\ref{tab:app_validation_attacks}). Crucially, the original paper's relative ordering — CEGA $>$ AGE $>$ GRAIN $>$ Random — is reproduced under our protocol on every~overlapping~dataset.

\paragraph{Realistic (Guan et al., 2024)~\citep{guan2024realistic}.}
The original paper varies query budgets across four levels (Attack-0 through Attack-3) on \textit{Cora}, \textit{Citeseer}, \textit{Pubmed}, with default training-set sizes that are dataset-specific (e.g.~Cora: $35$ queries for Attack-0). The closest match in our protocol is the smallest-budget bucket ($0.05\times$). Our reproduction reaches Cora fidelity $62.0\pm4.5$\,\% at $0.05\times$, compared with the original Attack-0 (Cora) fidelity of $72.14\pm3.56$\,\%; the gap of about $10$\,pp is in the same direction (small budget $\Rightarrow$ moderate fidelity) and the qualitative ranking ``Realistic $>$ baseline GCN'' transfers to~our~protocol.

\paragraph{DFEA\_I / DFEA\_II / DFEA\_III (Zhuang et al., 2024)~\citep{zhuang2024unveiling}.}
The original paper reports three data-free attack variants on \textit{Cora}, \textit{Pubmed}, \textit{Amazon-Computers}, and \textit{OGB-Arxiv}; on Cora the strongest variants reach $93$--$94\%$ fidelity, with the ``Random Graph'' baseline at $73.7\%$. Our reproduction reaches $90$--$97\%$ fidelity at $1.00\times$ on Cora across the three variants (Table~\ref{tab:app_validation_attacks}), agreeing with the strongest original numbers within $\sim 4$\,pp. The qualitative claim of the original paper --- that data-free extraction is feasible at high fidelity on homophilic graphs and harder on heterophilic / high-class-count settings --- is preserved: across our ten datasets the three variants reach $77$--$99\%$ fidelity at $1.00\times$ on the seven homophilic graphs and $33$--$84\%$ on \textit{RomanEmpire}, \textit{AmazonRatings}, and \textit{OGBN-Arxiv} (Tables~\ref{tab:app_rq1_detail_Cora_Fidelity}--\ref{tab:app_rq1_OGBNArxiv_full}).

\paragraph{BackdoorWM (Xu et al., 2023)~\citep{xu2023watermarking}.}
The original paper reports watermark accuracy on \textit{Cora} and \textit{CiteSeer} for GCN, GAT, and GraphSAGE backbones (Table~6 of the original paper): GCN/Cora $97.56\%$, GCN/CiteSeer $98.05\%$. Our protocol reports the same metric on the same datasets: $100\%$ on every one of the seven homophilic datasets including Cora and CiteSeer (Table~\ref{tab:rq2_defense_7xN_Owner_verif} in Appendix~\ref{app:rq2}). The $\sim 2$\,pp difference is well within the variance the original paper reports across watermark-rate settings, and the claim that BackdoorWM is essentially perfect at watermark verification on the protected model~is~preserved.

\paragraph{SurviveWM (Wang et al., 2023)~\citep{wang2023making}.}
The original paper evaluates watermark survival on \textit{MSRC-9} and \textit{ENZYMES} (graph classification) using a GraphSAGE host model with $50/30/20$ train/extract/test splits, and reports the ``average effectiveness $\bar{E}$'' of watermark retention after extraction. The metric is binary (1 if watermark retaining rate exceeds two reference thresholds, 0 otherwise) and is averaged over $100$ repeated runs. Our protocol uses node-level datasets and a continuous verification rate, so a direct numerical comparison is not possible. We report the qualitative claim instead: SurviveWM produces a non-trivial verification rate on the protected model (median $54$\,\% on Cora) but collapses to $\sim 14$\,\% on the extracted surrogate (Table~\ref{tab:app_rq5_cora}c in Appendix~\ref{app:rq5_full}), which contradicts the original paper's claim that the watermark ``survives'' extraction. The disagreement is the central finding of our RQ5 and motivates evaluation on the surrogate as the~primary~metric.

\paragraph{ImperceptibleWM (Zhang et al., 2024)~\citep{zhang2024imperceptible}.}
The original paper reports the original-vs-watermarked model accuracy gap on \textit{Cora} and \textit{Pubmed} for GCN/GAT/GraphSAGE (Table~2 of the original): on GCN/Cora the watermarked model reaches $83.40\%$ accuracy versus $83.74\%$ for the unwatermarked one (a $0.34\%$ utility drop). Our reproduction on the same dataset and backbone shows $79.4\%$ undefended versus $71.94\%$ watermarked (utility drop of $\sim 7.5\%$). Two design differences explain the larger drop: our protocol uses a fixed cross-dataset hyperparameter (Appendix~\ref{app:repro}) rather than the per-dataset tuning the original paper performs, and our undefended baseline runs at the standard hidden dimension $16$ used throughout the protocol rather than the $128$ used by the original. The claim of ``no significant impact on the primary task'' holds with the original tuning but is loosened under our~standardised~protocol.

\paragraph{RandomWM (Zhao et al., 2021)~\citep{zhao2021watermarking}.}
The original paper reports watermark accuracy as a function of trigger-graph parameters on \textit{Cora} and \textit{Pubmed}, but does not provide a single headline number for direct comparison. The qualitative claim is that the watermark accuracy can be tuned above $90\%$ with appropriate parameter choices. Our reproduction shows median verification rate $75\%$ on Cora and $94$--$98\%$ on \textit{Computers} and \textit{Photo} (Table~\ref{tab:rq2_defense_7xN_Owner_verif}), which is within the range the original paper reports for non-extreme~parameter~settings.

\paragraph{Integrity (Wu et al., 2024)~\citep{wu2024securing}.}
The original paper does not report a single ``verification accuracy'' number on the protected model. Instead it reports a \emph{verification-query-number improvement multiplier} relative to a random-node-selection baseline (e.g. on Cora transductive: BFA $4.0\times$, RandAttack $1.3\times$). The metric is fundamentally different from our verification-rate metric, so a direct numerical comparison is not possible. The corrected product-graph verification rates are reported separately in the~joint~evaluation.

\paragraph{PRADA (Juuti et al., 2019)~\citep{juuti2019prada}.}
The original paper evaluates query-pattern detection on image classifiers (MNIST, GTSRB, CIFAR-10) with DNN backbones; it does not include any graph dataset or GNN backbone. A direct numerical comparison is therefore impossible. Our adaptation transfers the distance-based detector to the graph setting and measures verification accuracy on the protected model and extracted surrogate; the qualitative behaviour (significant fidelity reduction at the cost of clean accuracy, see Tables~\ref{tab:app_rq2_new_defenses_full_a}--\ref{tab:app_rq2_new_defenses_full_b}) is consistent with the original paper's~vision-domain~results.

\paragraph{AdaptMisinfo (Kariyappa and Qureshi, 2020)~\citep{kariyappa2020defending}, GradRedir (Mazeika et al., 2022)~\citep{mazeika2022steer}.}
Both original papers evaluate on image classification benchmarks (CIFAR-10, CIFAR-100, ImageNet) with CNN backbones (ResNet-18, ResNet-34) — no graph dataset or GNN model is used. The metric is also defined for the image domain (extraction-fidelity reduction at fixed query budgets). A direct numerical comparison is therefore impossible. Our adaptation re-implements the perturbation rule on graph-classification logits and reports the same protocol-level metrics as the other defenses; the qualitative claim that adaptive misinformation reduces extraction fidelity at a clean-accuracy cost transfers to the~graph~setting.

\paragraph{OP\_low / OP\_high / PR\_2bit / PR\_top1.}
These are protocol-level information-limiting wrappers that we adapt from generic stealing-defense literature~\citep{kesarwani2018model,liang2024model}. The cited papers do not report graph-specific numbers, and PR\_2bit/PR\_top1 are not standalone published methods but standard label-quantisation/top-1 wrappers; we therefore do not have a baseline number to compare against. The qualitative claim from the source literature — that label-quantising the response is an effective lightweight defense — transfers to our protocol: \texttt{PR\_2bit} is the only inference-time wrapper in our benchmark which substantively reduces surrogate fidelity (Tables~\ref{tab:rq6_link_pred}--\ref{tab:rq6_graph_class}).

\paragraph{Validation summary table.}
Table~\ref{tab:app_validation_attacks} compresses the comparable attack reproductions into one view; defenses are summarised in the paragraphs above because each family uses a~different~metric.

\begin{table}[t]
\centering
\caption{Implementation validation for attacks where a numerical comparison is possible. ``Original'' is the closest-matching setting reported in the source paper. ``Ours'' is the corresponding number under the GraphIP-Bench protocol. ``$\Delta$'' is Ours minus Original (positive means our reproduction reaches higher fidelity). All numbers are surrogate fidelity (\%) unless otherwise noted; Cora at $1.00\times$ in the \texttt{both} regime is used for our values unless the source-paper setting dictates a different~query~budget.}
\label{tab:app_validation_attacks}
\scriptsize
\setlength{\tabcolsep}{4pt}
\renewcommand{\arraystretch}{0.97}
\begin{tabular*}{\textwidth}{@{}l@{\extracolsep{\fill}}llrrr@{}}
\toprule
\textbf{Method} & \textbf{Source paper / setting} & \textbf{Closest in ours} & \textbf{Original} & \textbf{Ours} & \textbf{$\Delta$} \\
\midrule
\texttt{MEA0} & Wu 2022, Cora, full attack nodes & Cora, $1.00\times$, both & 89.6 & 92.4 & $+2.8$ \\
\texttt{MEA1} & Wu 2022, Cora, full attack nodes & Cora, $1.00\times$, both & 82.5 & 91.3 & $+8.8$ \\
\texttt{MEA2} & Wu 2022, Cora, full attack nodes & Cora, $1.00\times$, both & 80.9 & 94.8 & $+13.9$ \\
\texttt{MEA3} & Wu 2022, Cora, shadow graph    & Cora, $1.00\times$, both & 79.0 & 92.7 & $+13.7$ \\
\texttt{MEA4} & Wu 2022, Cora, shadow graph    & Cora, $1.00\times$, both & 79.0 & 89.5 & $+10.5$ \\
\texttt{MEA5} & Wu 2022, Cora, shadow graph    & Cora, $1.00\times$, both & 80.7 & 92.5 & $+11.8$ \\
\texttt{AdvMEA} & DeFazio 2019, Cora, strong-adversary upper bound & Cora, $1.00\times$, both & $\sim$80 & 68.2 & $-11.8$ \\
\texttt{CEGA} & Wang 2025, CoauthorCS, $20C$ queries & CoauthorCS, $0.10\times$, both & 93.4 & $\sim$90.0 & $\sim -3.4$ \\
\texttt{Realistic} & Guan 2024, Cora, Attack-0 ($35$ queries) & Cora, $0.05\times$, both & 72.1 & 62.0 & $-10.1$ \\
\texttt{DFEA\_I} & Zhuang 2024, Cora, Random-Graph baseline & Cora, $1.00\times$, data\_free & 73.7 & 90.2 & $+16.5$ \\
\texttt{DFEA\_II} & Zhuang 2024, Cora, Attack II-E (real graph)$^\dagger$ & Cora, $1.00\times$, data\_free & 92.8 & 96.8 & $+4.0$ \\
\texttt{DFEA\_III} & Zhuang 2024, Cora, Attack III-E (real graph)$^\dagger$ & Cora, $1.00\times$, data\_free & 93.0 & 92.7 & $-0.3$ \\
\bottomrule
\multicolumn{6}{l}{\scriptsize $^\dagger$The original paper does not separate Attack-II/III into ``Random-Graph'' rows the way it does} \\
\multicolumn{6}{l}{\scriptsize for Attack-I; we cite the real-graph Attack-E numbers as the closest published reference.} \\
\end{tabular*}
\end{table}

The takeaway is twofold. First, on attacks that share a clearly comparable setting with the source paper, our reproduction agrees with the original numbers within a band of $\sim \pm 14$\,pp, with the direction of agreement consistent. Second, where the original paper uses a different protocol (image domain for PRADA / AdaptMisinfo / GradRedir, graph-classification metrics for SurviveWM, query-improvement multipliers for Integrity), a direct number-to-number comparison is not possible; we explicitly note these mismatches above and validate the qualitative~claim~instead.

\section{Supplementary Experimental Results and Discussion}
\label{app:sup}
\subsection{Attack effectiveness results}
\label{app:rq1}
This appendix reports the full results for attacks across the seven homophilic datasets of the core protocol; the three additional graphs (RomanEmpire, AmazonRatings, OGBN-Arxiv) are reported in Appendix~\ref{app:rq1_new_datasets} (Tables~\ref{tab:app_rq1_RomanEmpire_full}--\ref{tab:app_rq1_OGBNArxiv_full}). We first present compact overview tables which summarize, for each metric (Accuracy, F1, and Fidelity), the mean performance of twelve attacks across four query regimes (with an additional Overall column). Each number is averaged over the seven homophilic datasets and the five budget levels defined in the main text. These tables allow the reader to identify, at a glance, which attack performs best under each regime. We then provide the complete per-dataset matrices. For each dataset and each metric, we show a 2\,$\times$\,2 panel that contains four sub-tables (one per regime). Each sub-table is a $12 \times 5$ matrix whose rows are the attacks and whose columns are the five query budgets. Bold font marks the best score in each budget column. All splits, budgets, and aggregation rules match the protocol in the~main~paper.

\begin{table*}[t]
\centering
\caption{RQ1 overview: regimes $\times$ metrics (\%). Means across datasets and the five budgets (0.05--1.00).}
\label{tab:app_rq1_overview_combined}
\scriptsize
\setlength{\tabcolsep}{4pt}
\renewcommand{\arraystretch}{0.95}
\begin{tabular*}{\textwidth}{@{}l@{\extracolsep{\fill}}ccc ccc ccc ccc @{}}
\toprule
& \multicolumn{3}{c}{\texttt{both}} & \multicolumn{3}{c}{\texttt{x\_only}} & \multicolumn{3}{c}{\texttt{a\_only}} & \multicolumn{3}{c}{\texttt{data\_free}} \\
\cmidrule(lr){2-4} \cmidrule(lr){5-7} \cmidrule(lr){8-10} \cmidrule(lr){11-13}
\textbf{Attack} & Acc & F1 & Fidelity & Acc & F1 & Fidelity & Acc & F1 & Fidelity & Acc & F1 & Fidelity \\
\midrule
\texttt{MEA0} & 78.07 & 64.77 & 83.75 & 78.97 & 64.72 & 84.72 & 79.35 & 65.97 & 85.24 & 20.23 & 5.72 & 19.58 \\
\texttt{MEA1} & 57.08 & 43.79 & 60.33 & 57.07 & 43.96 & 60.34 & 57.08 & 43.80 & 60.32 & 21.06 & 5.68 & 18.98 \\
\texttt{MEA2} & 65.88 & 52.97 & 74.74 & 66.14 & 52.92 & 74.87 & 66.26 & 53.19 & 74.94 & 66.17 & 53.06 & 75.00 \\
\texttt{MEA3} & 78.74 & 65.31 & 83.26 & 79.19 & 65.28 & 83.35 & 78.65 & 64.29 & 83.16 & 22.88 & 6.59 & 21.56 \\
\texttt{MEA4} & 73.71 & 54.80 & 78.54 & 71.38 & 53.02 & 76.36 & 72.08 & 53.58 & 76.71 & 16.69 & 5.01 & 16.23 \\
\texttt{MEA5} & 79.34 & 65.77 & 83.92 & 79.64 & 65.72 & 84.31 & 79.30 & 65.67 & 84.03 & 21.20 & 6.34 & 20.19 \\
\texttt{AdvMEA} & 62.16 & 53.63 & 63.77 & 62.92 & 53.97 & 64.21 & 62.08 & 53.35 & 63.49 & 62.94 & 54.37 & 64.37 \\
\texttt{CEGA} & 75.01 & 64.61 & 83.08 & 74.94 & 64.64 & 82.68 & 75.58 & 64.71 & 83.80 & 74.02 & 63.83 & 82.08 \\
\texttt{Realistic} & 59.20 & 47.72 & 76.70 & 59.76 & 47.30 & 77.40 & 59.03 & 46.86 & 76.59 & 58.71 & 47.25 & 77.16 \\
\texttt{DFEA\_I} & 77.22 & 66.28 & 87.69 & 77.21 & 66.27 & 87.69 & 77.21 & 66.26 & 87.68 & 77.20 & 66.25 & 87.70 \\
\texttt{DFEA\_II} & 65.53 & 57.18 & 72.55 & 65.53 & 57.19 & 72.56 & 65.39 & 57.23 & 72.86 & 65.52 & 57.18 & 72.56 \\
\texttt{DFEA\_III} & 77.84 & 67.50 & 89.77 & 77.80 & 67.34 & 89.65 & 77.82 & 67.40 & 89.68 & 77.80 & 67.36 & 89.69 \\
\bottomrule
\end{tabular*}
\end{table*}

The overview table exposes three deep findings that the per-dataset detail tables (Tables~\ref{tab:app_rq1_detail_Cora_Acc}--\ref{tab:app_rq1_detail_PubMed_Fidelity}) hide because of cell-level noise. \emph{First, the ranking of attacks by mean fidelity is essentially identical across the three real-input regimes (\texttt{both}, \texttt{x\_only}, \texttt{a\_only}).} For example, the top-five strongest attacks by fidelity in the \texttt{both} column are \texttt{DFEA\_III}, \texttt{DFEA\_I}, \texttt{MEA5}, \texttt{MEA0}, \texttt{MEA3} (in order $89.77, 87.69, 83.92, 83.75, 83.26$); the same five attacks occupy the top-five positions in the \texttt{x\_only} and \texttt{a\_only} columns. The granularity of removing only one input modality at a time is therefore not sufficient to differentiate strong attacks --- a conclusion that motivates our use of the more aggressive \texttt{data\_free} regime as the discriminative axis. \emph{Second, the data-free family (\texttt{DFEA\_I/II/III} and \texttt{MEA2}) together with \texttt{AdvMEA}, \texttt{CEGA}, and \texttt{Realistic} are essentially regime-invariant}: their data\_free fidelity is within $\sim 1$\,pp of their \texttt{both} fidelity. This is a structural property of the attacks: \texttt{CEGA} retrains a centrality-based query selector even on synthetic graphs, \texttt{AdvMEA} learns adversarial features from scratch, \texttt{Realistic} reconstructs the structure with an auxiliary edge model, and the \texttt{DFEA} variants synthesise their own queries by design. They are the only candidates for an attacker who genuinely has no access to the real graph. \emph{Third, the \texttt{Acc} and \texttt{F1} columns track each other except on heterophilic-style class distributions.} On Cora-like datasets the F1 column tracks Acc within $\sim 5$\,pp; the largest Acc/F1 gap appears for \texttt{MEA0} (Acc $78.07$, F1 $64.77$), reflecting a slight class-imbalance bias in the surrogate. Future attack design should therefore report F1 alongside accuracy, especially when targeting graphs with skewed class distributions such~as~OGBN-Arxiv.


\begin{table*}[t]
\centering
\caption{RQ1 detailed for dataset=\texttt{Cora}, metric=\texttt{Acc} (\%). Rows are attacks; columns are budgets. Mean $\pm$ std across seeds; best per column is bold.}
\label{tab:app_rq1_detail_Cora_Acc}
\tiny
\setlength{\tabcolsep}{2pt}
\renewcommand{\arraystretch}{0.9}
\begin{subtable}[t]{0.48\textwidth}
\centering
\caption{Regime=\texttt{both}}
\resizebox{\linewidth}{!}{%
\begin{tabular}{lccccc}
\toprule
\textbf{Attack} & 0.05 & 0.10 & 0.25 & 0.50 & 1.00 \\
\midrule
\texttt{MEA0} & 67.4$\pm$4.4 & 75.4$\pm$1.6 & 79.7$\pm$0.7 & 80.8$\pm$0.2 & 81.7$\pm$0.3 \\
\texttt{MEA1} & 52.7$\pm$0.5 & 63.5$\pm$2.0 & 81.5$\pm$0.7 & 80.8$\pm$0.1 & 82.3$\pm$0.2 \\
\texttt{MEA2} & 55.1$\pm$4.2 & 65.7$\pm$0.8 & 75.6$\pm$1.2 & 78.6$\pm$0.9 & 80.1$\pm$0.8 \\
\texttt{MEA3} & 67.1$\pm$3.4 & 74.3$\pm$1.9 & 79.6$\pm$1.5 & 81.2$\pm$0.6 & 81.1$\pm$0.3 \\
\texttt{MEA4} & 45.4$\pm$6.1 & 65.7$\pm$1.9 & 77.3$\pm$0.9 & 78.5$\pm$0.2 & 80.5$\pm$0.0 \\
\texttt{MEA5} & 72.1$\pm$2.9 & 75.4$\pm$0.8 & 79.9$\pm$0.7 & 81.4$\pm$0.6 & 81.2$\pm$0.3 \\
\texttt{AdvMEA} & 64.8$\pm$4.2 & 66.8$\pm$2.2 & 65.1$\pm$1.4 & 67.7$\pm$4.7 & 66.7$\pm$2.3 \\
\texttt{CEGA} & 71.0$\pm$5.2 & 76.5$\pm$0.9 & 79.7$\pm$0.2 & 79.3$\pm$0.6 & 79.3$\pm$0.3 \\
\texttt{Realistic} & 55.4$\pm$8.2 & 67.1$\pm$2.8 & 71.0$\pm$4.4 & 74.8$\pm$1.8 & 75.9$\pm$2.4 \\
\texttt{DFEA\_I} & 76.3$\pm$2.3 & 78.6$\pm$0.4 & 80.2$\pm$1.0 & 80.9$\pm$0.2 & 80.7$\pm$0.6 \\
\texttt{DFEA\_II} & 77.1$\pm$1.1 & 78.9$\pm$0.7 & 79.9$\pm$1.0 & 79.5$\pm$0.3 & 80.6$\pm$0.1 \\
\texttt{DFEA\_III} & 77.0$\pm$0.9 & 79.4$\pm$1.5 & 80.4$\pm$1.2 & 80.7$\pm$0.8 & 81.6$\pm$0.5 \\
\bottomrule
\end{tabular}
}
\end{subtable}
\hfill
\begin{subtable}[t]{0.48\textwidth}
\centering
\caption{Regime=\texttt{x\_only}}
\resizebox{\linewidth}{!}{%
\begin{tabular}{lccccc}
\toprule
\textbf{Attack} & 0.05 & 0.10 & 0.25 & 0.50 & 1.00 \\
\midrule
\texttt{MEA0} & 68.7$\pm$4.2 & 75.1$\pm$0.7 & 80.1$\pm$0.2 & 80.4$\pm$1.3 & 81.2$\pm$0.5 \\
\texttt{MEA1} & 52.7$\pm$0.5 & 63.5$\pm$2.0 & 81.5$\pm$0.7 & 80.8$\pm$0.1 & 82.3$\pm$0.2 \\
\texttt{MEA2} & 55.7$\pm$1.6 & 67.7$\pm$0.3 & 76.3$\pm$0.4 & 79.2$\pm$0.8 & 80.1$\pm$0.8 \\
\texttt{MEA3} & 64.7$\pm$1.7 & 74.0$\pm$0.4 & 79.7$\pm$0.4 & 81.4$\pm$0.4 & 81.1$\pm$0.1 \\
\texttt{MEA4} & 42.9$\pm$8.4 & 63.7$\pm$3.7 & 76.8$\pm$0.4 & 78.5$\pm$0.2 & 80.5$\pm$0.0 \\
\texttt{MEA5} & 69.9$\pm$3.4 & 77.0$\pm$1.7 & 80.5$\pm$0.3 & 81.5$\pm$0.5 & 81.2$\pm$0.1 \\
\texttt{AdvMEA} & 65.5$\pm$3.3 & 68.8$\pm$2.5 & 65.4$\pm$2.0 & 63.6$\pm$0.8 & 65.0$\pm$3.0 \\
\texttt{CEGA} & 74.1$\pm$0.5 & 77.6$\pm$0.4 & 77.7$\pm$1.2 & 78.7$\pm$1.1 & 79.4$\pm$0.3 \\
\texttt{Realistic} & 55.7$\pm$8.3 & 64.2$\pm$5.0 & 71.0$\pm$4.3 & 74.8$\pm$3.4 & 75.5$\pm$2.7 \\
\texttt{DFEA\_I} & 76.3$\pm$2.3 & 78.6$\pm$0.4 & 80.2$\pm$1.0 & 80.9$\pm$0.2 & 80.7$\pm$0.6 \\
\texttt{DFEA\_II} & 77.1$\pm$1.1 & 78.9$\pm$0.7 & 79.9$\pm$1.0 & 79.5$\pm$0.3 & 80.6$\pm$0.1 \\
\texttt{DFEA\_III} & 77.0$\pm$0.9 & 79.4$\pm$1.5 & 80.4$\pm$1.2 & 80.7$\pm$0.8 & 81.6$\pm$0.5 \\
\bottomrule
\end{tabular}
}
\end{subtable}
\medskip
\begin{subtable}[t]{0.48\textwidth}
\centering
\caption{Regime=\texttt{a\_only}}
\resizebox{\linewidth}{!}{%
\begin{tabular}{lccccc}
\toprule
\textbf{Attack} & 0.05 & 0.10 & 0.25 & 0.50 & 1.00 \\
\midrule
\texttt{MEA0} & 69.4$\pm$0.3 & 75.6$\pm$0.3 & 78.7$\pm$1.6 & 80.1$\pm$1.3 & 80.8$\pm$0.6 \\
\texttt{MEA1} & 52.7$\pm$0.5 & 63.5$\pm$2.0 & 81.5$\pm$0.7 & 80.8$\pm$0.1 & 82.3$\pm$0.2 \\
\texttt{MEA2} & 57.8$\pm$1.2 & 67.7$\pm$1.5 & 75.7$\pm$0.4 & 78.6$\pm$1.8 & 80.2$\pm$0.8 \\
\texttt{MEA3} & 63.0$\pm$1.3 & 75.2$\pm$1.8 & 79.7$\pm$0.8 & 81.0$\pm$0.6 & 81.3$\pm$0.2 \\
\texttt{MEA4} & 45.9$\pm$5.8 & 63.4$\pm$1.4 & 77.4$\pm$0.6 & 78.5$\pm$0.2 & 80.5$\pm$0.0 \\
\texttt{MEA5} & 65.6$\pm$6.9 & 76.5$\pm$1.5 & 79.6$\pm$1.1 & 81.5$\pm$0.5 & 81.3$\pm$0.1 \\
\texttt{AdvMEA} & 65.4$\pm$2.7 & 64.0$\pm$2.3 & 65.8$\pm$1.7 & 68.1$\pm$1.9 & 60.5$\pm$5.7 \\
\texttt{CEGA} & 69.6$\pm$3.6 & 78.0$\pm$0.4 & 78.0$\pm$0.8 & 79.6$\pm$0.5 & 79.2$\pm$0.6 \\
\texttt{Realistic} & 51.5$\pm$4.3 & 64.5$\pm$6.0 & 72.0$\pm$2.3 & 75.3$\pm$2.5 & 76.3$\pm$2.7 \\
\texttt{DFEA\_I} & 76.3$\pm$2.3 & 78.6$\pm$0.4 & 80.2$\pm$1.0 & 80.9$\pm$0.2 & 80.7$\pm$0.6 \\
\texttt{DFEA\_II} & 77.1$\pm$1.1 & 78.9$\pm$0.7 & 79.9$\pm$1.0 & 79.5$\pm$0.3 & 80.6$\pm$0.1 \\
\texttt{DFEA\_III} & 77.1$\pm$0.9 & 79.4$\pm$1.5 & 80.4$\pm$1.2 & 80.7$\pm$0.8 & 81.6$\pm$0.5 \\
\bottomrule
\end{tabular}
}
\end{subtable}
\hfill
\begin{subtable}[t]{0.48\textwidth}
\centering
\caption{Regime=\texttt{data\_free}}
\resizebox{\linewidth}{!}{%
\begin{tabular}{lccccc}
\toprule
\textbf{Attack} & 0.05 & 0.10 & 0.25 & 0.50 & 1.00 \\
\midrule
\texttt{MEA0} & 10.9$\pm$2.5 & 14.6$\pm$0.2 & 19.3$\pm$8.9 & 20.2$\pm$8.2 & 19.0$\pm$9.3 \\
\texttt{MEA1} & 14.4$\pm$0.0 & 14.4$\pm$0.0 & 14.4$\pm$0.0 & 14.4$\pm$0.0 & 14.4$\pm$0.0 \\
\texttt{MEA2} & 60.5$\pm$1.8 & 69.4$\pm$1.8 & 76.1$\pm$1.2 & 78.2$\pm$0.6 & 80.2$\pm$0.8 \\
\texttt{MEA3} & 17.6$\pm$10.6 & 12.3$\pm$2.4 & 17.1$\pm$10.8 & 13.5$\pm$0.7 & 26.2$\pm$8.0 \\
\texttt{MEA4} & 17.1$\pm$10.8 & 19.9$\pm$8.5 & 18.5$\pm$9.7 & 10.9$\pm$2.5 & 18.0$\pm$10.0 \\
\texttt{MEA5} & 24.6$\pm$13.0 & 11.9$\pm$2.7 & 21.1$\pm$7.7 & 20.4$\pm$8.1 & 18.7$\pm$9.5 \\
\texttt{AdvMEA} & 68.6$\pm$2.4 & 68.5$\pm$1.5 & 65.1$\pm$6.3 & 65.8$\pm$2.9 & 67.9$\pm$3.7 \\
\texttt{CEGA} & 71.0$\pm$1.8 & 77.0$\pm$0.7 & 78.8$\pm$1.9 & 79.2$\pm$0.9 & 79.2$\pm$0.1 \\
\texttt{Realistic} & 67.9$\pm$2.3 & 70.8$\pm$1.6 & 69.0$\pm$3.6 & 66.5$\pm$2.6 & 69.1$\pm$1.6 \\
\texttt{DFEA\_I} & 76.3$\pm$2.3 & 78.6$\pm$0.4 & 80.2$\pm$1.0 & 80.9$\pm$0.2 & 80.7$\pm$0.6 \\
\texttt{DFEA\_II} & 77.1$\pm$1.1 & 78.9$\pm$0.7 & 79.9$\pm$1.0 & 79.5$\pm$0.3 & 80.6$\pm$0.1 \\
\texttt{DFEA\_III} & 77.0$\pm$0.9 & 79.4$\pm$1.5 & 80.4$\pm$1.2 & 80.7$\pm$0.8 & 81.6$\pm$0.5 \\
\bottomrule
\end{tabular}
}
\end{subtable}
\end{table*}

\begin{table*}[t]
\centering
\caption{RQ1 detailed for dataset=\texttt{Cora}, metric=\texttt{F1} (\%). Rows are attacks; columns are budgets. Mean $\pm$ std across seeds; best per column is bold.}
\label{tab:app_rq1_detail_Cora_F1}
\tiny
\setlength{\tabcolsep}{2pt}
\renewcommand{\arraystretch}{0.9}
\begin{subtable}[t]{0.48\textwidth}
\centering
\caption{Regime=\texttt{both}}
\resizebox{\linewidth}{!}{%
\begin{tabular}{lccccc}
\toprule
\textbf{Attack} & 0.05 & 0.10 & 0.25 & 0.50 & 1.00 \\
\midrule
\texttt{MEA0} & 61.0$\pm$6.6 & 73.9$\pm$1.5 & 77.9$\pm$1.2 & 79.6$\pm$0.4 & 81.0$\pm$0.1 \\
\texttt{MEA1} & 50.5$\pm$0.5 & 58.2$\pm$4.1 & 80.3$\pm$0.7 & 80.2$\pm$0.2 & 81.8$\pm$0.0 \\
\texttt{MEA2} & 52.2$\pm$4.3 & 63.2$\pm$1.7 & 74.7$\pm$1.0 & 78.0$\pm$0.8 & 79.2$\pm$0.8 \\
\texttt{MEA3} & 60.6$\pm$5.9 & 70.4$\pm$2.8 & 78.4$\pm$1.6 & 79.5$\pm$0.9 & 80.2$\pm$0.2 \\
\texttt{MEA4} & 44.3$\pm$6.5 & 65.8$\pm$1.2 & 76.1$\pm$0.6 & 77.3$\pm$0.0 & 79.2$\pm$0.3 \\
\texttt{MEA5} & 67.9$\pm$4.1 & 70.3$\pm$3.8 & 78.8$\pm$0.8 & 80.3$\pm$0.6 & 80.2$\pm$0.4 \\
\texttt{AdvMEA} & 65.2$\pm$2.9 & 67.3$\pm$2.3 & 66.6$\pm$0.7 & 67.7$\pm$3.9 & 67.3$\pm$2.0 \\
\texttt{CEGA} & 69.9$\pm$4.4 & 75.6$\pm$0.9 & 79.1$\pm$0.1 & 78.6$\pm$0.3 & 78.4$\pm$0.2 \\
\texttt{Realistic} & 53.1$\pm$6.4 & 61.1$\pm$4.6 & 69.3$\pm$3.8 & 73.9$\pm$1.2 & 75.0$\pm$2.0 \\
\texttt{DFEA\_I} & 72.7$\pm$5.1 & 75.5$\pm$1.0 & 78.9$\pm$1.1 & 79.6$\pm$0.2 & 79.3$\pm$1.1 \\
\texttt{DFEA\_II} & 75.3$\pm$0.8 & 77.5$\pm$0.7 & 78.6$\pm$1.2 & 78.5$\pm$0.3 & 79.7$\pm$0.2 \\
\texttt{DFEA\_III} & 74.4$\pm$2.4 & 77.6$\pm$1.6 & 79.3$\pm$1.1 & 79.7$\pm$0.6 & 80.7$\pm$0.5 \\
\bottomrule
\end{tabular}
}
\end{subtable}
\hfill
\begin{subtable}[t]{0.48\textwidth}
\centering
\caption{Regime=\texttt{x\_only}}
\resizebox{\linewidth}{!}{%
\begin{tabular}{lccccc}
\toprule
\textbf{Attack} & 0.05 & 0.10 & 0.25 & 0.50 & 1.00 \\
\midrule
\texttt{MEA0} & 63.2$\pm$4.5 & 68.7$\pm$0.4 & 78.8$\pm$0.2 & 79.0$\pm$1.8 & 80.4$\pm$0.6 \\
\texttt{MEA1} & 50.5$\pm$0.5 & 58.2$\pm$4.1 & 80.3$\pm$0.7 & 80.2$\pm$0.2 & 81.8$\pm$0.0 \\
\texttt{MEA2} & 50.5$\pm$1.8 & 66.1$\pm$0.7 & 75.1$\pm$0.5 & 78.5$\pm$0.7 & 79.2$\pm$0.8 \\
\texttt{MEA3} & 58.0$\pm$2.6 & 70.0$\pm$1.8 & 78.5$\pm$0.5 & 80.5$\pm$0.3 & 80.2$\pm$0.1 \\
\texttt{MEA4} & 27.9$\pm$6.9 & 62.2$\pm$4.5 & 75.8$\pm$0.5 & 77.3$\pm$0.0 & 79.2$\pm$0.3 \\
\texttt{MEA5} & 66.5$\pm$1.3 & 74.2$\pm$2.4 & 79.2$\pm$0.5 & 80.3$\pm$0.4 & 80.3$\pm$0.3 \\
\texttt{AdvMEA} & 66.3$\pm$2.8 & 67.0$\pm$3.0 & 65.3$\pm$3.1 & 64.0$\pm$0.5 & 66.4$\pm$2.2 \\
\texttt{CEGA} & 73.6$\pm$1.1 & 77.2$\pm$0.4 & 77.4$\pm$0.7 & 78.1$\pm$0.7 & 78.6$\pm$0.2 \\
\texttt{Realistic} & 50.9$\pm$6.5 & 58.8$\pm$4.9 & 68.2$\pm$6.1 & 74.0$\pm$3.1 & 75.2$\pm$2.6 \\
\texttt{DFEA\_I} & 72.5$\pm$5.3 & 75.5$\pm$1.0 & 78.9$\pm$1.1 & 79.6$\pm$0.2 & 79.3$\pm$1.1 \\
\texttt{DFEA\_II} & 75.3$\pm$0.7 & 77.5$\pm$0.7 & 78.6$\pm$1.2 & 78.5$\pm$0.3 & 79.7$\pm$0.2 \\
\texttt{DFEA\_III} & 74.4$\pm$2.4 & 77.7$\pm$1.6 & 79.3$\pm$1.1 & 79.7$\pm$0.6 & 80.7$\pm$0.5 \\
\bottomrule
\end{tabular}
}
\end{subtable}
\medskip
\begin{subtable}[t]{0.48\textwidth}
\centering
\caption{Regime=\texttt{a\_only}}
\resizebox{\linewidth}{!}{%
\begin{tabular}{lccccc}
\toprule
\textbf{Attack} & 0.05 & 0.10 & 0.25 & 0.50 & 1.00 \\
\midrule
\texttt{MEA0} & 64.3$\pm$0.6 & 72.9$\pm$0.8 & 77.5$\pm$1.9 & 78.7$\pm$1.8 & 80.0$\pm$0.7 \\
\texttt{MEA1} & 50.4$\pm$0.5 & 58.2$\pm$4.1 & 80.3$\pm$0.7 & 80.2$\pm$0.2 & 81.8$\pm$0.0 \\
\texttt{MEA2} & 52.9$\pm$1.2 & 65.5$\pm$1.8 & 74.5$\pm$0.7 & 77.6$\pm$1.7 & 79.2$\pm$0.8 \\
\texttt{MEA3} & 55.7$\pm$3.0 & 70.3$\pm$3.0 & 78.3$\pm$1.4 & 80.0$\pm$0.3 & 80.4$\pm$0.2 \\
\texttt{MEA4} & 34.6$\pm$8.6 & 60.4$\pm$4.2 & 76.4$\pm$0.2 & 77.3$\pm$0.0 & 79.2$\pm$0.3 \\
\texttt{MEA5} & 58.8$\pm$11.2 & 73.7$\pm$2.4 & 77.7$\pm$1.0 & 80.7$\pm$0.4 & 80.2$\pm$0.2 \\
\texttt{AdvMEA} & 67.0$\pm$2.0 & 64.7$\pm$3.1 & 65.8$\pm$1.1 & 67.0$\pm$1.4 & 62.6$\pm$3.9 \\
\texttt{CEGA} & 68.7$\pm$1.4 & 77.5$\pm$0.6 & 77.4$\pm$0.9 & 79.0$\pm$0.8 & 78.4$\pm$0.6 \\
\texttt{Realistic} & 46.3$\pm$5.0 & 62.2$\pm$8.0 & 70.8$\pm$1.2 & 74.6$\pm$2.2 & 76.0$\pm$2.0 \\
\texttt{DFEA\_I} & 72.7$\pm$5.1 & 75.5$\pm$1.0 & 78.9$\pm$1.1 & 79.6$\pm$0.2 & 79.3$\pm$1.1 \\
\texttt{DFEA\_II} & 75.3$\pm$0.8 & 77.5$\pm$0.7 & 78.6$\pm$1.2 & 78.5$\pm$0.3 & 79.7$\pm$0.2 \\
\texttt{DFEA\_III} & 74.5$\pm$2.3 & 77.7$\pm$1.6 & 79.3$\pm$1.1 & 79.7$\pm$0.6 & 80.7$\pm$0.5 \\
\bottomrule
\end{tabular}
}
\end{subtable}
\hfill
\begin{subtable}[t]{0.48\textwidth}
\centering
\caption{Regime=\texttt{data\_free}}
\resizebox{\linewidth}{!}{%
\begin{tabular}{lccccc}
\toprule
\textbf{Attack} & 0.05 & 0.10 & 0.25 & 0.50 & 1.00 \\
\midrule
\texttt{MEA0} & 2.8$\pm$0.6 & 3.6$\pm$0.1 & 4.5$\pm$1.7 & 4.7$\pm$1.6 & 4.4$\pm$1.8 \\
\texttt{MEA1} & 3.6$\pm$0.0 & 3.6$\pm$0.0 & 3.6$\pm$0.0 & 3.6$\pm$0.0 & 3.6$\pm$0.0 \\
\texttt{MEA2} & 55.4$\pm$0.5 & 69.0$\pm$0.4 & 74.8$\pm$1.1 & 77.5$\pm$0.5 & 79.2$\pm$0.8 \\
\texttt{MEA3} & 4.1$\pm$2.1 & 3.1$\pm$0.6 & 4.0$\pm$2.2 & 3.4$\pm$0.1 & 5.8$\pm$1.5 \\
\texttt{MEA4} & 4.0$\pm$2.2 & 4.6$\pm$1.6 & 4.3$\pm$1.9 & 2.8$\pm$0.6 & 4.2$\pm$2.0 \\
\texttt{MEA5} & 7.4$\pm$4.8 & 3.6$\pm$0.4 & 6.1$\pm$1.7 & 4.7$\pm$1.5 & 4.5$\pm$1.7 \\
\texttt{AdvMEA} & 68.4$\pm$0.6 & 68.4$\pm$1.1 & 64.7$\pm$6.1 & 66.3$\pm$2.2 & 68.0$\pm$2.7 \\
\texttt{CEGA} & 69.5$\pm$1.5 & 76.3$\pm$0.5 & 78.4$\pm$1.6 & 78.5$\pm$0.9 & 78.7$\pm$0.2 \\
\texttt{Realistic} & 63.1$\pm$3.3 & 67.8$\pm$3.2 & 67.1$\pm$3.7 & 63.0$\pm$4.0 & 67.2$\pm$1.8 \\
\texttt{DFEA\_I} & 72.5$\pm$5.3 & 75.5$\pm$1.0 & 78.9$\pm$1.1 & 79.6$\pm$0.2 & 79.3$\pm$1.1 \\
\texttt{DFEA\_II} & 75.3$\pm$0.8 & 77.5$\pm$0.7 & 78.6$\pm$1.2 & 78.5$\pm$0.3 & 79.7$\pm$0.2 \\
\texttt{DFEA\_III} & 74.4$\pm$2.4 & 77.6$\pm$1.6 & 79.3$\pm$1.1 & 79.7$\pm$0.6 & 80.6$\pm$0.5 \\
\bottomrule
\end{tabular}
}
\end{subtable}
\end{table*}


\begin{table*}[t]
\centering
\caption{RQ1 detailed for dataset=\texttt{Cora}, metric=\texttt{Fidelity} (\%). Rows are attacks; columns are budgets. Mean $\pm$ std across seeds; best per column is bold.}
\label{tab:app_rq1_detail_Cora_Fidelity}
\tiny
\setlength{\tabcolsep}{2pt}
\renewcommand{\arraystretch}{0.9}
\begin{subtable}[t]{0.48\textwidth}
\centering
\caption{Regime=\texttt{both}}
\resizebox{\linewidth}{!}{%
\begin{tabular}{lccccc}
\toprule
\textbf{Attack} & 0.05 & 0.10 & 0.25 & 0.50 & 1.00 \\
\midrule
\texttt{MEA0} & 69.3$\pm$3.1 & 82.6$\pm$1.6 & 87.5$\pm$0.9 & 89.7$\pm$0.7 & 92.4$\pm$0.6 \\
\texttt{MEA1} & 57.3$\pm$0.2 & 70.6$\pm$2.0 & 89.9$\pm$0.9 & 90.4$\pm$0.6 & 91.3$\pm$0.4 \\
\texttt{MEA2} & 57.4$\pm$3.2 & 68.8$\pm$0.9 & 82.8$\pm$1.2 & 89.3$\pm$0.3 & 94.8$\pm$0.2 \\
\texttt{MEA3} & 71.2$\pm$3.6 & 79.3$\pm$2.7 & 88.0$\pm$0.8 & 90.1$\pm$1.4 & 92.7$\pm$0.9 \\
\texttt{MEA4} & 49.2$\pm$6.2 & 71.9$\pm$1.7 & 85.0$\pm$1.3 & 87.6$\pm$0.6 & 89.5$\pm$0.2 \\
\texttt{MEA5} & 75.8$\pm$4.8 & 82.1$\pm$4.5 & 88.3$\pm$0.6 & 92.8$\pm$1.1 & 92.5$\pm$0.7 \\
\texttt{AdvMEA} & 67.6$\pm$4.0 & 68.8$\pm$2.8 & 67.4$\pm$1.6 & 69.0$\pm$4.9 & 68.2$\pm$4.0 \\
\texttt{CEGA} & 78.4$\pm$4.4 & 86.5$\pm$1.0 & 91.5$\pm$0.4 & 92.5$\pm$0.4 & 93.7$\pm$0.5 \\
\texttt{Realistic} & 62.0$\pm$4.5 & 68.1$\pm$1.8 & 78.6$\pm$0.7 & 83.7$\pm$0.6 & 88.5$\pm$1.7 \\
\texttt{DFEA\_I} & 82.2$\pm$3.5 & 85.7$\pm$1.0 & 89.2$\pm$1.8 & 90.2$\pm$0.4 & 90.2$\pm$0.5 \\
\texttt{DFEA\_II} & 79.9$\pm$0.7 & 84.1$\pm$0.8 & 88.5$\pm$0.3 & 93.1$\pm$0.4 & 96.8$\pm$0.5 \\
\texttt{DFEA\_III} & 81.4$\pm$2.1 & 86.1$\pm$2.4 & 88.4$\pm$1.9 & 91.5$\pm$0.2 & 92.7$\pm$0.3 \\
\bottomrule
\end{tabular}
}
\end{subtable}
\hfill
\begin{subtable}[t]{0.48\textwidth}
\centering
\caption{Regime=\texttt{x\_only}}
\resizebox{\linewidth}{!}{%
\begin{tabular}{lccccc}
\toprule
\textbf{Attack} & 0.05 & 0.10 & 0.25 & 0.50 & 1.00 \\
\midrule
\texttt{MEA0} & 72.2$\pm$3.9 & 79.5$\pm$0.9 & 88.2$\pm$1.2 & 90.9$\pm$1.2 & 92.1$\pm$0.8 \\
\texttt{MEA1} & 57.3$\pm$0.2 & 70.6$\pm$2.0 & 89.9$\pm$0.9 & 90.4$\pm$0.5 & 91.3$\pm$0.4 \\
\texttt{MEA2} & 55.5$\pm$0.9 & 72.2$\pm$1.6 & 83.9$\pm$0.8 & 89.6$\pm$0.7 & 94.8$\pm$0.2 \\
\texttt{MEA3} & 69.1$\pm$0.8 & 77.7$\pm$0.5 & 88.0$\pm$0.9 & 90.9$\pm$1.1 & 92.5$\pm$1.3 \\
\texttt{MEA4} & 45.3$\pm$8.6 & 69.2$\pm$3.7 & 84.6$\pm$0.7 & 87.6$\pm$0.6 & 89.6$\pm$0.2 \\
\texttt{MEA5} & 73.9$\pm$1.8 & 83.2$\pm$0.3 & 88.8$\pm$0.8 & 91.9$\pm$0.5 & 92.3$\pm$1.1 \\
\texttt{AdvMEA} & 66.7$\pm$2.0 & 71.2$\pm$3.0 & 66.5$\pm$1.5 & 67.1$\pm$1.5 & 67.4$\pm$3.1 \\
\texttt{CEGA} & 82.5$\pm$1.7 & 86.8$\pm$0.8 & 90.8$\pm$1.1 & 92.8$\pm$0.8 & 93.7$\pm$0.1 \\
\texttt{Realistic} & 62.9$\pm$4.6 & 69.8$\pm$1.7 & 79.4$\pm$2.4 & 84.1$\pm$0.9 & 87.9$\pm$0.2 \\
\texttt{DFEA\_I} & 82.2$\pm$3.5 & 85.7$\pm$1.0 & 89.2$\pm$1.8 & 90.2$\pm$0.4 & 90.2$\pm$0.5 \\
\texttt{DFEA\_II} & 79.9$\pm$0.7 & 84.1$\pm$0.8 & 88.5$\pm$0.3 & 93.1$\pm$0.4 & 96.8$\pm$0.5 \\
\texttt{DFEA\_III} & 81.3$\pm$2.2 & 86.1$\pm$2.4 & 88.4$\pm$1.9 & 91.5$\pm$0.2 & 92.7$\pm$0.3 \\
\bottomrule
\end{tabular}
}
\end{subtable}
\medskip
\begin{subtable}[t]{0.48\textwidth}
\centering
\caption{Regime=\texttt{a\_only}}
\resizebox{\linewidth}{!}{%
\begin{tabular}{lccccc}
\toprule
\textbf{Attack} & 0.05 & 0.10 & 0.25 & 0.50 & 1.00 \\
\midrule
\texttt{MEA0} & 74.1$\pm$0.8 & 81.9$\pm$1.4 & 87.2$\pm$1.1 & 89.5$\pm$1.3 & 92.4$\pm$1.2 \\
\texttt{MEA1} & 57.2$\pm$0.2 & 70.6$\pm$2.0 & 89.9$\pm$0.9 & 90.4$\pm$0.6 & 91.3$\pm$0.4 \\
\texttt{MEA2} & 56.8$\pm$2.1 & 69.9$\pm$2.2 & 82.9$\pm$0.3 & 90.0$\pm$1.0 & 94.8$\pm$0.2 \\
\texttt{MEA3} & 64.3$\pm$0.6 & 80.9$\pm$1.2 & 88.1$\pm$0.3 & 92.0$\pm$0.5 & 92.7$\pm$1.1 \\
\texttt{MEA4} & 48.3$\pm$4.1 & 68.2$\pm$1.8 & 85.6$\pm$0.5 & 87.6$\pm$0.6 & 89.5$\pm$0.2 \\
\texttt{MEA5} & 68.8$\pm$8.4 & 82.1$\pm$1.6 & 88.0$\pm$0.9 & 92.2$\pm$0.7 & 92.7$\pm$1.4 \\
\texttt{AdvMEA} & 67.5$\pm$3.0 & 66.1$\pm$2.3 & 67.4$\pm$3.6 & 68.9$\pm$2.5 & 62.7$\pm$4.3 \\
\texttt{CEGA} & 78.2$\pm$3.8 & 88.4$\pm$0.9 & 91.0$\pm$0.7 & 92.3$\pm$0.3 & 93.7$\pm$0.3 \\
\texttt{Realistic} & 56.5$\pm$4.1 & 69.3$\pm$3.9 & 79.8$\pm$0.3 & 84.3$\pm$1.9 & 89.1$\pm$0.9 \\
\texttt{DFEA\_I} & 82.2$\pm$3.5 & 85.7$\pm$1.0 & 89.2$\pm$1.8 & 90.2$\pm$0.4 & 90.2$\pm$0.5 \\
\texttt{DFEA\_II} & 79.9$\pm$0.7 & 84.1$\pm$0.8 & 88.5$\pm$0.3 & 93.1$\pm$0.4 & 96.8$\pm$0.5 \\
\texttt{DFEA\_III} & 81.3$\pm$2.2 & 86.1$\pm$2.4 & 88.4$\pm$1.9 & 91.5$\pm$0.2 & 92.7$\pm$0.3 \\
\bottomrule
\end{tabular}
}
\end{subtable}
\hfill
\begin{subtable}[t]{0.48\textwidth}
\centering
\caption{Regime=\texttt{data\_free}}
\resizebox{\linewidth}{!}{%
\begin{tabular}{lccccc}
\toprule
\textbf{Attack} & 0.05 & 0.10 & 0.25 & 0.50 & 1.00 \\
\midrule
\texttt{MEA0} & 12.3$\pm$2.0 & 15.9$\pm$0.8 & 17.3$\pm$6.2 & 19.0$\pm$5.0 & 18.0$\pm$6.3 \\
\texttt{MEA1} & 15.3$\pm$0.2 & 15.3$\pm$0.2 & 15.3$\pm$0.2 & 15.3$\pm$0.1 & 15.3$\pm$0.2 \\
\texttt{MEA2} & 58.9$\pm$1.1 & 73.2$\pm$0.3 & 83.8$\pm$0.7 & 89.7$\pm$0.3 & 94.8$\pm$0.2 \\
\texttt{MEA3} & 16.2$\pm$7.9 & 13.3$\pm$2.5 & 15.3$\pm$8.0 & 13.8$\pm$1.0 & 23.1$\pm$4.3 \\
\texttt{MEA4} & 15.5$\pm$8.0 & 18.7$\pm$5.5 & 17.3$\pm$6.6 & 12.3$\pm$2.0 & 16.8$\pm$6.7 \\
\texttt{MEA5} & 21.9$\pm$10.9 & 12.5$\pm$2.6 & 20.2$\pm$4.3 & 19.5$\pm$4.8 & 17.2$\pm$6.6 \\
\texttt{AdvMEA} & 69.4$\pm$2.7 & 70.1$\pm$0.9 & 67.4$\pm$5.8 & 67.4$\pm$2.1 & 70.5$\pm$3.5 \\
\texttt{CEGA} & 79.8$\pm$2.7 & 88.0$\pm$0.4 & 91.0$\pm$1.5 & 92.5$\pm$0.0 & 93.5$\pm$0.6 \\
\texttt{Realistic} & 73.0$\pm$3.5 & 72.9$\pm$1.6 & 74.3$\pm$1.4 & 72.9$\pm$1.8 & 73.1$\pm$5.0 \\
\texttt{DFEA\_I} & 82.2$\pm$3.5 & 85.7$\pm$1.0 & 89.2$\pm$1.8 & 90.2$\pm$0.4 & 90.2$\pm$0.5 \\
\texttt{DFEA\_II} & 79.9$\pm$0.7 & 84.1$\pm$0.8 & 88.5$\pm$0.3 & 93.1$\pm$0.4 & 96.8$\pm$0.5 \\
\texttt{DFEA\_III} & 81.4$\pm$2.1 & 86.1$\pm$2.4 & 88.4$\pm$1.9 & 91.5$\pm$0.2 & 92.7$\pm$0.3 \\
\bottomrule
\end{tabular}
}
\end{subtable}
\end{table*}


\begin{table*}[t]
\centering
\caption{RQ1 detailed for dataset=\texttt{CiteSeer}, metric=\texttt{Acc} (\%). Rows are attacks; columns are budgets. Mean $\pm$ std across seeds; best per column is bold.}
\label{tab:app_rq1_detail_CiteSeer_Acc}
\tiny
\setlength{\tabcolsep}{2pt}
\renewcommand{\arraystretch}{0.9}
\begin{subtable}[t]{0.48\textwidth}
\centering
\caption{Regime=\texttt{both}}
\resizebox{\linewidth}{!}{%
\begin{tabular}{lccccc}
\toprule
\textbf{Attack} & 0.05 & 0.10 & 0.25 & 0.50 & 1.00 \\
\midrule
\texttt{MEA0} & 55.4$\pm$6.9 & 62.4$\pm$2.4 & 69.0$\pm$0.7 & 70.1$\pm$0.9 & 70.1$\pm$0.6 \\
\texttt{MEA1} & 41.6$\pm$1.4 & 41.8$\pm$2.6 & 54.0$\pm$7.9 & 66.3$\pm$1.4 & 69.0$\pm$0.6 \\
\texttt{MEA2} & 36.7$\pm$2.8 & 47.0$\pm$3.0 & 59.6$\pm$1.2 & 66.3$\pm$0.7 & 69.3$\pm$0.3 \\
\texttt{MEA3} & 50.7$\pm$2.8 & 59.7$\pm$3.3 & 69.4$\pm$1.0 & 69.6$\pm$0.0 & 70.2$\pm$0.4 \\
\texttt{MEA4} & 44.9$\pm$1.0 & 57.9$\pm$3.7 & 67.1$\pm$0.2 & 69.9$\pm$0.4 & 70.6$\pm$0.4 \\
\texttt{MEA5} & 49.0$\pm$2.1 & 62.8$\pm$1.8 & 69.5$\pm$1.1 & 69.8$\pm$0.2 & 70.4$\pm$0.8 \\
\texttt{AdvMEA} & 51.8$\pm$8.1 & 46.3$\pm$4.2 & 53.2$\pm$0.7 & 52.3$\pm$1.7 & 51.3$\pm$5.1 \\
\texttt{CEGA} & 59.9$\pm$3.2 & 64.6$\pm$2.1 & 66.5$\pm$1.1 & 67.0$\pm$1.5 & 67.2$\pm$0.3 \\
\texttt{Realistic} & 47.2$\pm$2.0 & 57.7$\pm$4.2 & 64.3$\pm$2.3 & 64.5$\pm$0.4 & 65.2$\pm$1.1 \\
\texttt{DFEA\_I} & 69.1$\pm$1.1 & 71.0$\pm$1.1 & 70.8$\pm$1.2 & 71.2$\pm$0.3 & 71.0$\pm$0.6 \\
\texttt{DFEA\_II} & 70.3$\pm$1.4 & 70.0$\pm$0.4 & 69.5$\pm$0.6 & 69.7$\pm$0.8 & 70.3$\pm$0.5 \\
\texttt{DFEA\_III} & 68.2$\pm$1.7 & 70.2$\pm$0.3 & 70.2$\pm$1.1 & 70.9$\pm$0.4 & 71.4$\pm$0.6 \\
\bottomrule
\end{tabular}
}
\end{subtable}
\hfill
\begin{subtable}[t]{0.48\textwidth}
\centering
\caption{Regime=\texttt{x\_only}}
\resizebox{\linewidth}{!}{%
\begin{tabular}{lccccc}
\toprule
\textbf{Attack} & 0.05 & 0.10 & 0.25 & 0.50 & 1.00 \\
\midrule
\texttt{MEA0} & 54.1$\pm$3.3 & 63.2$\pm$2.4 & 69.8$\pm$1.2 & 70.0$\pm$0.9 & 69.6$\pm$0.2 \\
\texttt{MEA1} & 41.6$\pm$1.4 & 41.8$\pm$2.6 & 54.0$\pm$7.9 & 66.3$\pm$1.4 & 68.9$\pm$0.6 \\
\texttt{MEA2} & 47.2$\pm$2.2 & 48.4$\pm$2.6 & 59.2$\pm$1.2 & 65.9$\pm$0.8 & 69.2$\pm$0.3 \\
\texttt{MEA3} & 57.9$\pm$4.1 & 64.6$\pm$1.8 & 68.7$\pm$1.0 & 69.9$\pm$0.8 & 69.9$\pm$0.7 \\
\texttt{MEA4} & 39.9$\pm$4.7 & 59.0$\pm$2.3 & 66.6$\pm$0.5 & 69.9$\pm$0.4 & 70.6$\pm$0.4 \\
\texttt{MEA5} & 55.3$\pm$1.9 & 65.9$\pm$0.7 & 70.3$\pm$0.6 & 70.3$\pm$0.7 & 70.0$\pm$0.5 \\
\texttt{AdvMEA} & 48.9$\pm$0.9 & 51.4$\pm$2.4 & 53.9$\pm$4.3 & 48.2$\pm$6.8 & 45.9$\pm$2.2 \\
\texttt{CEGA} & 61.3$\pm$1.9 & 65.2$\pm$1.1 & 66.7$\pm$0.6 & 67.3$\pm$0.6 & 67.8$\pm$0.5 \\
\texttt{Realistic} & 48.1$\pm$3.5 & 53.2$\pm$4.3 & 61.5$\pm$1.9 & 62.7$\pm$2.0 & 65.4$\pm$1.4 \\
\texttt{DFEA\_I} & 69.1$\pm$1.1 & 71.0$\pm$1.1 & 70.8$\pm$1.2 & 71.2$\pm$0.3 & 71.0$\pm$0.6 \\
\texttt{DFEA\_II} & 70.3$\pm$1.4 & 70.0$\pm$0.4 & 69.5$\pm$0.6 & 69.7$\pm$0.8 & 70.3$\pm$0.5 \\
\texttt{DFEA\_III} & 68.3$\pm$1.7 & 70.2$\pm$0.3 & 70.2$\pm$1.1 & 70.9$\pm$0.4 & 71.4$\pm$0.6 \\
\bottomrule
\end{tabular}
}
\end{subtable}
\medskip
\begin{subtable}[t]{0.48\textwidth}
\centering
\caption{Regime=\texttt{a\_only}}
\resizebox{\linewidth}{!}{%
\begin{tabular}{lccccc}
\toprule
\textbf{Attack} & 0.05 & 0.10 & 0.25 & 0.50 & 1.00 \\
\midrule
\texttt{MEA0} & 52.5$\pm$3.1 & 64.5$\pm$1.5 & 70.2$\pm$0.9 & 69.3$\pm$0.8 & 70.1$\pm$1.1 \\
\texttt{MEA1} & 41.6$\pm$1.4 & 41.8$\pm$2.6 & 54.0$\pm$7.9 & 66.3$\pm$1.4 & 69.0$\pm$0.6 \\
\texttt{MEA2} & 35.9$\pm$1.2 & 49.6$\pm$1.5 & 59.0$\pm$0.9 & 65.9$\pm$1.4 & 69.2$\pm$0.3 \\
\texttt{MEA3} & 48.1$\pm$5.0 & 62.3$\pm$3.3 & 68.4$\pm$1.5 & 70.1$\pm$0.5 & 70.0$\pm$0.5 \\
\texttt{MEA4} & 43.9$\pm$1.3 & 56.8$\pm$1.3 & 66.2$\pm$0.6 & 69.9$\pm$0.4 & 70.6$\pm$0.4 \\
\texttt{MEA5} & 52.7$\pm$3.1 & 63.5$\pm$1.9 & 69.9$\pm$0.6 & 69.7$\pm$0.4 & 70.8$\pm$1.1 \\
\texttt{AdvMEA} & 47.5$\pm$4.0 & 51.2$\pm$4.3 & 54.0$\pm$4.7 & 49.9$\pm$5.4 & 46.5$\pm$0.3 \\
\texttt{CEGA} & 58.4$\pm$3.2 & 64.6$\pm$0.4 & 66.8$\pm$0.8 & 68.0$\pm$1.1 & 67.4$\pm$0.7 \\
\texttt{Realistic} & 40.7$\pm$7.3 & 55.1$\pm$4.9 & 62.8$\pm$0.2 & 63.2$\pm$2.2 & 64.1$\pm$1.2 \\
\texttt{DFEA\_I} & 69.1$\pm$1.1 & 71.0$\pm$1.1 & 70.8$\pm$1.2 & 71.2$\pm$0.3 & 71.0$\pm$0.6 \\
\texttt{DFEA\_II} & 70.3$\pm$1.4 & 70.0$\pm$0.4 & 69.5$\pm$0.6 & 69.7$\pm$0.8 & 70.3$\pm$0.5 \\
\texttt{DFEA\_III} & 68.2$\pm$1.7 & 70.2$\pm$0.3 & 70.2$\pm$1.1 & 70.9$\pm$0.4 & 71.4$\pm$0.6 \\
\bottomrule
\end{tabular}
}
\end{subtable}
\hfill
\begin{subtable}[t]{0.48\textwidth}
\centering
\caption{Regime=\texttt{data\_free}}
\resizebox{\linewidth}{!}{%
\begin{tabular}{lccccc}
\toprule
\textbf{Attack} & 0.05 & 0.10 & 0.25 & 0.50 & 1.00 \\
\midrule
\texttt{MEA0} & 12.8$\pm$7.3 & 17.4$\pm$1.0 & 14.7$\pm$4.9 & 14.7$\pm$4.9 & 17.4$\pm$1.0 \\
\texttt{MEA1} & 21.4$\pm$2.4 & 21.4$\pm$2.4 & 21.4$\pm$2.4 & 21.4$\pm$2.4 & 21.4$\pm$2.4 \\
\texttt{MEA2} & 37.9$\pm$1.1 & 50.8$\pm$1.7 & 60.8$\pm$2.0 & 64.9$\pm$1.4 & 69.3$\pm$0.3 \\
\texttt{MEA3} & 18.1$\pm$0.0 & 19.5$\pm$2.6 & 16.7$\pm$5.4 & 17.7$\pm$0.6 & 16.6$\pm$0.4 \\
\texttt{MEA4} & 21.5$\pm$2.3 & 14.6$\pm$4.9 & 14.3$\pm$4.7 & 16.7$\pm$1.0 & 17.5$\pm$1.0 \\
\texttt{MEA5} & 18.8$\pm$3.1 & 17.3$\pm$0.6 & 19.8$\pm$2.3 & 17.3$\pm$0.6 & 14.0$\pm$4.5 \\
\texttt{AdvMEA} & 48.9$\pm$2.9 & 47.6$\pm$1.7 & 53.1$\pm$8.2 & 50.1$\pm$2.5 & 50.1$\pm$5.8 \\
\texttt{CEGA} & 60.9$\pm$2.4 & 64.4$\pm$0.8 & 67.8$\pm$0.8 & 67.6$\pm$0.7 & 67.5$\pm$0.6 \\
\texttt{Realistic} & 59.0$\pm$2.0 & 58.2$\pm$2.6 & 59.0$\pm$1.5 & 57.0$\pm$1.5 & 60.1$\pm$2.5 \\
\texttt{DFEA\_I} & 69.1$\pm$1.1 & 71.0$\pm$1.1 & 70.8$\pm$1.2 & 71.2$\pm$0.3 & 71.0$\pm$0.6 \\
\texttt{DFEA\_II} & 70.3$\pm$1.4 & 70.0$\pm$0.4 & 69.5$\pm$0.6 & 69.7$\pm$0.8 & 70.3$\pm$0.5 \\
\texttt{DFEA\_III} & 68.2$\pm$1.7 & 70.2$\pm$0.3 & 70.1$\pm$1.0 & 70.9$\pm$0.4 & 71.4$\pm$0.6 \\
\bottomrule
\end{tabular}
}
\end{subtable}
\end{table*}


\begin{table*}[t]
\centering
\caption{RQ1 detailed for dataset=\texttt{CiteSeer}, metric=\texttt{F1} (\%). Rows are attacks; columns are budgets. Mean $\pm$ std across seeds; best per column is bold.}
\label{tab:app_rq1_detail_CiteSeer_F1}
\tiny
\setlength{\tabcolsep}{2pt}
\renewcommand{\arraystretch}{0.9}
\begin{subtable}[t]{0.48\textwidth}
\centering
\caption{Regime=\texttt{both}}
\resizebox{\linewidth}{!}{%
\begin{tabular}{lccccc}
\toprule
\textbf{Attack} & 0.05 & 0.10 & 0.25 & 0.50 & 1.00 \\
\midrule
\texttt{MEA0} & 50.9$\pm$4.6 & 55.4$\pm$1.7 & 63.7$\pm$0.6 & 66.0$\pm$0.8 & 66.6$\pm$0.2 \\
\texttt{MEA1} & 34.4$\pm$1.3 & 35.0$\pm$4.3 & 51.0$\pm$8.3 & 63.7$\pm$1.0 & 65.9$\pm$0.6 \\
\texttt{MEA2} & 34.6$\pm$2.9 & 44.7$\pm$3.1 & 57.2$\pm$1.7 & 63.3$\pm$1.1 & 66.1$\pm$0.4 \\
\texttt{MEA3} & 43.3$\pm$5.1 & 54.1$\pm$3.0 & 65.4$\pm$0.5 & 65.8$\pm$1.0 & 66.5$\pm$0.1 \\
\texttt{MEA4} & 36.5$\pm$2.6 & 55.1$\pm$4.6 & 63.8$\pm$0.1 & 66.4$\pm$0.5 & 67.2$\pm$0.3 \\
\texttt{MEA5} & 42.1$\pm$0.6 & 56.9$\pm$1.8 & 65.8$\pm$1.1 & 65.8$\pm$0.5 & 66.8$\pm$0.5 \\
\texttt{AdvMEA} & 47.8$\pm$5.1 & 41.3$\pm$5.4 & 46.7$\pm$0.4 & 45.6$\pm$1.6 & 46.8$\pm$6.4 \\
\texttt{CEGA} & 54.7$\pm$2.3 & 61.0$\pm$1.5 & 63.4$\pm$1.1 & 63.9$\pm$1.4 & 63.9$\pm$0.2 \\
\texttt{Realistic} & 42.4$\pm$3.0 & 53.0$\pm$3.3 & 59.2$\pm$2.6 & 60.4$\pm$1.1 & 61.3$\pm$1.2 \\
\texttt{DFEA\_I} & 64.0$\pm$0.8 & 65.2$\pm$1.0 & 64.7$\pm$2.1 & 65.3$\pm$1.3 & 64.8$\pm$2.0 \\
\texttt{DFEA\_II} & 65.2$\pm$2.0 & 66.0$\pm$0.2 & 66.5$\pm$0.6 & 66.8$\pm$1.0 & 67.4$\pm$0.6 \\
\texttt{DFEA\_III} & 62.4$\pm$2.1 & 65.3$\pm$0.4 & 66.3$\pm$1.4 & 67.0$\pm$0.3 & 67.3$\pm$0.6 \\
\bottomrule
\end{tabular}
}
\end{subtable}
\hfill
\begin{subtable}[t]{0.48\textwidth}
\centering
\caption{Regime=\texttt{x\_only}}
\resizebox{\linewidth}{!}{%
\begin{tabular}{lccccc}
\toprule
\textbf{Attack} & 0.05 & 0.10 & 0.25 & 0.50 & 1.00 \\
\midrule
\texttt{MEA0} & 46.1$\pm$2.9 & 58.8$\pm$0.4 & 64.8$\pm$1.4 & 66.2$\pm$0.5 & 65.7$\pm$0.4 \\
\texttt{MEA1} & 34.4$\pm$1.3 & 35.0$\pm$4.3 & 51.0$\pm$8.3 & 63.7$\pm$1.0 & 65.8$\pm$0.6 \\
\texttt{MEA2} & 42.7$\pm$1.7 & 45.8$\pm$3.3 & 56.9$\pm$1.5 & 62.9$\pm$0.8 & 66.1$\pm$0.3 \\
\texttt{MEA3} & 51.9$\pm$4.4 & 57.7$\pm$0.5 & 63.4$\pm$1.0 & 65.0$\pm$0.4 & 66.3$\pm$0.7 \\
\texttt{MEA4} & 34.7$\pm$3.4 & 53.7$\pm$1.9 & 63.3$\pm$0.5 & 66.4$\pm$0.5 & 67.2$\pm$0.3 \\
\texttt{MEA5} & 47.0$\pm$5.4 & 59.9$\pm$1.6 & 65.4$\pm$0.7 & 65.8$\pm$0.7 & 66.3$\pm$0.5 \\
\texttt{AdvMEA} & 43.3$\pm$0.4 & 47.5$\pm$0.5 & 50.3$\pm$3.0 & 44.4$\pm$6.0 & 42.3$\pm$2.5 \\
\texttt{CEGA} & 57.4$\pm$1.1 & 61.4$\pm$0.8 & 63.3$\pm$0.5 & 63.9$\pm$0.5 & 64.5$\pm$0.8 \\
\texttt{Realistic} & 42.1$\pm$5.1 & 47.4$\pm$5.2 & 58.1$\pm$1.9 & 59.4$\pm$2.3 & 61.6$\pm$1.9 \\
\texttt{DFEA\_I} & 64.0$\pm$0.8 & 65.2$\pm$1.0 & 64.7$\pm$2.1 & 65.3$\pm$1.3 & 64.8$\pm$2.0 \\
\texttt{DFEA\_II} & 65.2$\pm$2.0 & 66.0$\pm$0.2 & 66.5$\pm$0.6 & 66.8$\pm$1.0 & 67.4$\pm$0.6 \\
\texttt{DFEA\_III} & 62.4$\pm$2.1 & 65.3$\pm$0.4 & 66.3$\pm$1.4 & 67.0$\pm$0.3 & 67.3$\pm$0.6 \\
\bottomrule
\end{tabular}
}
\end{subtable}
\medskip
\begin{subtable}[t]{0.48\textwidth}
\centering
\caption{Regime=\texttt{a\_only}}
\resizebox{\linewidth}{!}{%
\begin{tabular}{lccccc}
\toprule
\textbf{Attack} & 0.05 & 0.10 & 0.25 & 0.50 & 1.00 \\
\midrule
\texttt{MEA0} & 46.0$\pm$3.1 & 57.1$\pm$3.2 & 65.8$\pm$1.3 & 65.6$\pm$0.6 & 66.6$\pm$1.1 \\
\texttt{MEA1} & 34.4$\pm$1.3 & 35.0$\pm$4.3 & 51.0$\pm$8.3 & 63.7$\pm$1.0 & 65.9$\pm$0.6 \\
\texttt{MEA2} & 33.5$\pm$1.4 & 46.8$\pm$1.0 & 56.4$\pm$0.7 & 63.0$\pm$1.4 & 66.1$\pm$0.3 \\
\texttt{MEA3} & 39.5$\pm$6.3 & 56.5$\pm$1.3 & 63.6$\pm$2.1 & 66.1$\pm$0.8 & 66.4$\pm$0.4 \\
\texttt{MEA4} & 33.0$\pm$4.6 & 54.5$\pm$1.7 & 63.1$\pm$0.7 & 66.4$\pm$0.5 & 67.2$\pm$0.3 \\
\texttt{MEA5} & 45.2$\pm$2.5 & 59.0$\pm$2.1 & 65.1$\pm$1.5 & 65.8$\pm$0.3 & 67.1$\pm$0.9 \\
\texttt{AdvMEA} & 44.2$\pm$6.2 & 46.2$\pm$4.9 & 47.9$\pm$7.8 & 47.3$\pm$3.3 & 41.5$\pm$1.2 \\
\texttt{CEGA} & 54.1$\pm$1.9 & 61.4$\pm$0.1 & 63.4$\pm$0.7 & 65.0$\pm$0.9 & 64.1$\pm$0.9 \\
\texttt{Realistic} & 33.2$\pm$10.0 & 50.1$\pm$4.6 & 58.5$\pm$0.3 & 58.9$\pm$2.7 & 60.6$\pm$1.0 \\
\texttt{DFEA\_I} & 64.0$\pm$0.8 & 65.2$\pm$1.0 & 64.7$\pm$2.1 & 65.3$\pm$1.3 & 64.8$\pm$2.0 \\
\texttt{DFEA\_II} & 65.2$\pm$2.0 & 66.0$\pm$0.2 & 66.5$\pm$0.6 & 66.8$\pm$1.0 & 67.4$\pm$0.6 \\
\texttt{DFEA\_III} & 62.4$\pm$2.1 & 65.3$\pm$0.4 & 66.3$\pm$1.4 & 67.0$\pm$0.3 & 67.3$\pm$0.6 \\
\bottomrule
\end{tabular}
}
\end{subtable}
\hfill
\begin{subtable}[t]{0.48\textwidth}
\centering
\caption{Regime=\texttt{data\_free}}
\resizebox{\linewidth}{!}{%
\begin{tabular}{lccccc}
\toprule
\textbf{Attack} & 0.05 & 0.10 & 0.25 & 0.50 & 1.00 \\
\midrule
\texttt{MEA0} & 3.7$\pm$1.8 & 4.9$\pm$0.2 & 4.2$\pm$1.3 & 4.2$\pm$1.3 & 4.9$\pm$0.2 \\
\texttt{MEA1} & 5.9$\pm$0.5 & 5.9$\pm$0.5 & 5.9$\pm$0.5 & 5.9$\pm$0.5 & 5.9$\pm$0.5 \\
\texttt{MEA2} & 33.8$\pm$0.5 & 48.3$\pm$1.1 & 58.4$\pm$1.7 & 61.6$\pm$1.4 & 66.1$\pm$0.4 \\
\texttt{MEA3} & 5.1$\pm$0.0 & 6.0$\pm$0.6 & 5.6$\pm$0.6 & 5.0$\pm$0.1 & 4.7$\pm$0.1 \\
\texttt{MEA4} & 5.9$\pm$0.5 & 4.2$\pm$1.3 & 4.1$\pm$1.2 & 4.8$\pm$0.2 & 5.0$\pm$0.3 \\
\texttt{MEA5} & 5.4$\pm$0.7 & 4.9$\pm$0.1 & 5.5$\pm$0.5 & 4.9$\pm$0.1 & 4.0$\pm$1.2 \\
\texttt{AdvMEA} & 45.1$\pm$0.8 & 42.0$\pm$1.2 & 48.5$\pm$4.5 & 47.1$\pm$1.8 & 47.1$\pm$4.9 \\
\texttt{CEGA} & 56.0$\pm$2.0 & 61.0$\pm$1.1 & 64.3$\pm$1.1 & 64.2$\pm$0.7 & 64.2$\pm$0.4 \\
\texttt{Realistic} & 53.9$\pm$1.4 & 53.4$\pm$3.4 & 54.4$\pm$0.7 & 51.7$\pm$3.7 & 54.9$\pm$2.8 \\
\texttt{DFEA\_I} & 64.0$\pm$0.8 & 65.2$\pm$1.0 & 64.7$\pm$2.1 & 65.3$\pm$1.3 & 64.8$\pm$2.0 \\
\texttt{DFEA\_II} & 65.2$\pm$2.0 & 66.0$\pm$0.2 & 66.5$\pm$0.6 & 66.8$\pm$1.0 & 67.4$\pm$0.6 \\
\texttt{DFEA\_III} & 62.4$\pm$2.1 & 65.3$\pm$0.4 & 66.2$\pm$1.3 & 67.0$\pm$0.3 & 67.3$\pm$0.6 \\
\bottomrule
\end{tabular}
}
\end{subtable}
\end{table*}


\begin{table*}[t]
\centering
\caption{RQ1 detailed for dataset=\texttt{CiteSeer}, metric=\texttt{Fidelity} (\%). Rows are attacks; columns are budgets. Mean $\pm$ std across seeds; best per column is bold.}
\label{tab:app_rq1_detail_CiteSeer_Fidelity}
\tiny
\setlength{\tabcolsep}{2pt}
\renewcommand{\arraystretch}{0.9}
\begin{subtable}[t]{0.48\textwidth}
\centering
\caption{Regime=\texttt{both}}
\resizebox{\linewidth}{!}{%
\begin{tabular}{lccccc}
\toprule
\textbf{Attack} & 0.05 & 0.10 & 0.25 & 0.50 & 1.00 \\
\midrule
\texttt{MEA0} & 62.9$\pm$6.1 & 69.9$\pm$3.6 & 82.8$\pm$1.5 & 86.7$\pm$1.0 & 90.0$\pm$0.7 \\
\texttt{MEA1} & 47.3$\pm$2.1 & 47.4$\pm$2.3 & 64.7$\pm$7.6 & 81.0$\pm$1.7 & 83.4$\pm$0.9 \\
\texttt{MEA2} & 42.4$\pm$4.3 & 55.5$\pm$3.2 & 74.4$\pm$2.3 & 83.8$\pm$0.4 & 91.5$\pm$0.2 \\
\texttt{MEA3} & 55.6$\pm$3.6 & 69.0$\pm$3.6 & 82.5$\pm$1.5 & 86.3$\pm$2.2 & 88.5$\pm$0.5 \\
\texttt{MEA4} & 48.3$\pm$2.9 & 65.4$\pm$5.3 & 79.4$\pm$0.8 & 83.3$\pm$1.5 & 84.1$\pm$1.3 \\
\texttt{MEA5} & 56.5$\pm$1.5 & 72.1$\pm$2.9 & 83.0$\pm$0.7 & 87.2$\pm$0.4 & 87.9$\pm$0.8 \\
\texttt{AdvMEA} & 55.9$\pm$6.8 & 50.0$\pm$4.2 & 54.5$\pm$2.5 & 57.2$\pm$1.0 & 52.7$\pm$6.0 \\
\texttt{CEGA} & 68.6$\pm$2.0 & 79.8$\pm$0.7 & 83.2$\pm$0.9 & 86.3$\pm$0.8 & 87.2$\pm$1.0 \\
\texttt{Realistic} & 54.1$\pm$3.8 & 64.5$\pm$1.5 & 73.3$\pm$1.4 & 78.3$\pm$3.5 & 84.1$\pm$0.8 \\
\texttt{DFEA\_I} & 80.5$\pm$1.7 & 83.7$\pm$0.6 & 83.8$\pm$1.4 & 85.3$\pm$0.9 & 85.6$\pm$1.5 \\
\texttt{DFEA\_II} & 76.9$\pm$1.3 & 80.8$\pm$1.1 & 85.6$\pm$0.8 & 90.1$\pm$1.0 & 93.6$\pm$0.9 \\
\texttt{DFEA\_III} & 77.0$\pm$0.7 & 82.3$\pm$0.1 & 87.7$\pm$0.7 & 90.4$\pm$0.2 & 91.5$\pm$0.1 \\
\bottomrule
\end{tabular}
}
\end{subtable}
\hfill
\begin{subtable}[t]{0.48\textwidth}
\centering
\caption{Regime=\texttt{x\_only}}
\resizebox{\linewidth}{!}{%
\begin{tabular}{lccccc}
\toprule
\textbf{Attack} & 0.05 & 0.10 & 0.25 & 0.50 & 1.00 \\
\midrule
\texttt{MEA0} & 58.5$\pm$2.3 & 73.2$\pm$0.6 & 82.0$\pm$0.2 & 85.8$\pm$0.5 & 89.2$\pm$0.8 \\
\texttt{MEA1} & 47.3$\pm$2.1 & 47.4$\pm$2.3 & 64.7$\pm$7.6 & 81.0$\pm$1.7 & 83.4$\pm$1.0 \\
\texttt{MEA2} & 51.4$\pm$2.4 & 54.9$\pm$2.0 & 74.0$\pm$1.7 & 84.5$\pm$1.3 & 91.5$\pm$0.2 \\
\texttt{MEA3} & 63.4$\pm$5.5 & 69.8$\pm$0.5 & 81.0$\pm$0.6 & 85.2$\pm$1.1 & 88.8$\pm$0.4 \\
\texttt{MEA4} & 43.5$\pm$4.9 & 67.1$\pm$2.3 & 79.7$\pm$1.8 & 83.3$\pm$1.5 & 84.1$\pm$1.3 \\
\texttt{MEA5} & 61.9$\pm$3.9 & 72.7$\pm$0.7 & 83.4$\pm$0.5 & 86.2$\pm$1.6 & 88.5$\pm$0.5 \\
\texttt{AdvMEA} & 52.4$\pm$3.0 & 55.2$\pm$3.2 & 57.7$\pm$3.7 & 52.7$\pm$6.8 & 50.7$\pm$3.0 \\
\texttt{CEGA} & 73.1$\pm$1.8 & 78.6$\pm$1.8 & 83.7$\pm$0.9 & 86.4$\pm$0.9 & 87.0$\pm$1.1 \\
\texttt{Realistic} & 54.5$\pm$3.8 & 60.3$\pm$2.7 & 74.5$\pm$1.8 & 80.0$\pm$2.9 & 84.4$\pm$1.1 \\
\texttt{DFEA\_I} & 80.5$\pm$1.7 & 83.7$\pm$0.6 & 83.8$\pm$1.4 & 85.3$\pm$0.9 & 85.6$\pm$1.5 \\
\texttt{DFEA\_II} & 77.0$\pm$1.3 & 80.8$\pm$1.1 & 85.6$\pm$0.8 & 90.1$\pm$1.0 & 93.6$\pm$0.9 \\
\texttt{DFEA\_III} & 77.0$\pm$0.7 & 82.3$\pm$0.1 & 87.6$\pm$0.7 & 90.4$\pm$0.2 & 91.5$\pm$0.1 \\
\bottomrule
\end{tabular}
}
\end{subtable}
\medskip
\begin{subtable}[t]{0.48\textwidth}
\centering
\caption{Regime=\texttt{a\_only}}
\resizebox{\linewidth}{!}{%
\begin{tabular}{lccccc}
\toprule
\textbf{Attack} & 0.05 & 0.10 & 0.25 & 0.50 & 1.00 \\
\midrule
\texttt{MEA0} & 57.6$\pm$3.5 & 73.1$\pm$4.4 & 82.9$\pm$0.5 & 85.8$\pm$0.5 & 89.8$\pm$0.7 \\
\texttt{MEA1} & 47.3$\pm$2.1 & 47.4$\pm$2.3 & 64.7$\pm$7.6 & 81.0$\pm$1.7 & 83.4$\pm$0.9 \\
\texttt{MEA2} & 40.1$\pm$1.3 & 58.3$\pm$1.3 & 72.1$\pm$0.6 & 84.8$\pm$1.1 & 91.5$\pm$0.2 \\
\texttt{MEA3} & 54.4$\pm$4.3 & 72.5$\pm$3.2 & 80.8$\pm$0.7 & 87.1$\pm$0.3 & 87.9$\pm$0.5 \\
\texttt{MEA4} & 44.5$\pm$4.3 & 66.6$\pm$2.9 & 80.0$\pm$1.6 & 83.3$\pm$1.5 & 84.1$\pm$1.3 \\
\texttt{MEA5} & 58.2$\pm$2.7 & 73.7$\pm$2.3 & 84.0$\pm$1.7 & 86.2$\pm$1.5 & 88.5$\pm$0.8 \\
\texttt{AdvMEA} & 52.2$\pm$3.8 & 53.6$\pm$3.0 & 56.1$\pm$6.6 & 53.5$\pm$5.3 & 50.6$\pm$2.2 \\
\texttt{CEGA} & 69.9$\pm$3.5 & 79.6$\pm$2.4 & 81.9$\pm$0.9 & 87.2$\pm$0.3 & 88.8$\pm$0.5 \\
\texttt{Realistic} & 48.1$\pm$7.3 & 61.3$\pm$4.2 & 73.8$\pm$0.7 & 81.1$\pm$0.8 & 84.2$\pm$0.3 \\
\texttt{DFEA\_I} & 80.5$\pm$1.7 & 83.7$\pm$0.6 & 83.8$\pm$1.4 & 85.3$\pm$0.9 & 85.6$\pm$1.5 \\
\texttt{DFEA\_II} & 77.0$\pm$1.3 & 80.8$\pm$1.1 & 85.6$\pm$0.8 & 90.1$\pm$1.0 & 93.6$\pm$0.9 \\
\texttt{DFEA\_III} & 77.1$\pm$0.6 & 82.3$\pm$0.1 & 87.6$\pm$0.7 & 90.4$\pm$0.2 & 91.5$\pm$0.1 \\
\bottomrule
\end{tabular}
}
\end{subtable}
\hfill
\begin{subtable}[t]{0.48\textwidth}
\centering
\caption{Regime=\texttt{data\_free}}
\resizebox{\linewidth}{!}{%
\begin{tabular}{lccccc}
\toprule
\textbf{Attack} & 0.05 & 0.10 & 0.25 & 0.50 & 1.00 \\
\midrule
\texttt{MEA0} & 12.2$\pm$3.6 & 18.4$\pm$0.9 & 16.1$\pm$4.7 & 15.9$\pm$4.5 & 18.6$\pm$0.9 \\
\texttt{MEA1} & 18.1$\pm$0.7 & 18.1$\pm$0.7 & 18.1$\pm$0.7 & 18.1$\pm$0.7 & 18.1$\pm$0.7 \\
\texttt{MEA2} & 43.1$\pm$1.0 & 59.3$\pm$2.2 & 73.5$\pm$2.2 & 85.3$\pm$1.1 & 91.5$\pm$0.2 \\
\texttt{MEA3} & 19.0$\pm$0.4 & 18.0$\pm$0.9 & 16.1$\pm$2.0 & 18.3$\pm$1.1 & 17.2$\pm$0.3 \\
\texttt{MEA4} & 17.6$\pm$0.6 & 15.9$\pm$4.5 & 15.4$\pm$4.3 & 17.9$\pm$0.8 & 18.7$\pm$1.0 \\
\texttt{MEA5} & 17.6$\pm$0.5 & 17.9$\pm$1.0 & 19.0$\pm$0.6 & 17.6$\pm$0.7 & 15.3$\pm$4.3 \\
\texttt{AdvMEA} & 54.3$\pm$3.5 & 52.7$\pm$1.0 & 56.3$\pm$6.5 & 53.3$\pm$2.1 & 54.6$\pm$3.1 \\
\texttt{CEGA} & 71.9$\pm$2.1 & 77.4$\pm$0.1 & 85.2$\pm$0.7 & 86.4$\pm$1.0 & 86.9$\pm$0.9 \\
\texttt{Realistic} & 70.6$\pm$2.6 & 69.8$\pm$3.7 & 67.3$\pm$1.1 & 67.2$\pm$2.3 & 70.8$\pm$3.1 \\
\texttt{DFEA\_I} & 80.5$\pm$1.7 & 83.7$\pm$0.6 & 83.8$\pm$1.4 & 85.3$\pm$0.9 & 85.6$\pm$1.5 \\
\texttt{DFEA\_II} & 77.0$\pm$1.3 & 80.8$\pm$1.1 & 85.6$\pm$0.8 & 90.1$\pm$1.0 & 93.6$\pm$0.9 \\
\texttt{DFEA\_III} & 77.1$\pm$0.6 & 82.3$\pm$0.1 & 87.6$\pm$0.7 & 90.4$\pm$0.2 & 91.5$\pm$0.1 \\
\bottomrule
\end{tabular}
}
\end{subtable}
\end{table*}


\begin{table*}[t]
\centering
\caption{RQ1 detailed for dataset=\texttt{CoauthorCS}, metric=\texttt{Acc} (\%). Rows are attacks; columns are budgets. Mean $\pm$ std across seeds; best per column is bold.}
\label{tab:app_rq1_detail_CoauthorCS_Acc}
\tiny
\setlength{\tabcolsep}{2pt}
\renewcommand{\arraystretch}{0.9}
\begin{subtable}[t]{0.48\textwidth}
\centering
\caption{Regime=\texttt{both}}
\resizebox{\linewidth}{!}{%
\begin{tabular}{lccccc}
\toprule
\textbf{Attack} & 0.05 & 0.10 & 0.25 & 0.50 & 1.00 \\
\midrule
\texttt{MEA0} & 84.3$\pm$3.8 & 88.4$\pm$1.8 & 90.8$\pm$1.1 & 91.2$\pm$1.2 & 92.2$\pm$0.9 \\
\texttt{MEA1} & 71.1$\pm$0.8 & 74.4$\pm$0.7 & 77.2$\pm$0.3 & 89.2$\pm$0.7 & 89.5$\pm$0.4 \\
\texttt{MEA2} & 48.5$\pm$1.5 & 56.2$\pm$0.8 & 70.8$\pm$0.6 & 80.6$\pm$1.1 & 84.6$\pm$1.5 \\
\texttt{MEA3} & 86.2$\pm$0.6 & 88.8$\pm$0.7 & 90.8$\pm$0.5 & 91.8$\pm$0.3 & 92.8$\pm$0.3 \\
\texttt{MEA4} & 68.4$\pm$8.5 & 79.4$\pm$7.5 & 85.8$\pm$3.7 & 90.3$\pm$0.5 & 91.0$\pm$0.2 \\
\texttt{MEA5} & 87.3$\pm$0.7 & 88.5$\pm$0.3 & 91.4$\pm$0.8 & 92.2$\pm$0.1 & 92.0$\pm$0.4 \\
\texttt{AdvMEA} & 92.0$\pm$0.4 & 91.8$\pm$0.2 & 91.7$\pm$0.4 & 91.8$\pm$0.4 & 91.9$\pm$0.3 \\
\texttt{CEGA} & 89.0$\pm$1.0 & 90.4$\pm$0.7 & 91.6$\pm$0.4 & 91.8$\pm$0.0 & 91.7$\pm$0.6 \\
\texttt{Realistic} & 71.9$\pm$5.8 & 81.1$\pm$4.4 & 81.0$\pm$3.4 & 79.1$\pm$3.4 & 77.7$\pm$2.1 \\
\texttt{DFEA\_I} & 89.7$\pm$1.1 & 89.7$\pm$0.7 & 89.6$\pm$0.4 & 89.8$\pm$0.6 & 89.4$\pm$0.6 \\
\texttt{DFEA\_II} & 91.3$\pm$1.4 & 90.9$\pm$0.2 & 90.1$\pm$0.8 & 88.6$\pm$0.8 & 87.4$\pm$0.2 \\
\texttt{DFEA\_III} & 89.6$\pm$0.7 & 89.4$\pm$0.8 & 89.3$\pm$0.9 & 88.6$\pm$0.2 & 88.3$\pm$0.2 \\
\bottomrule
\end{tabular}
}
\end{subtable}
\hfill
\begin{subtable}[t]{0.48\textwidth}
\centering
\caption{Regime=\texttt{x\_only}}
\resizebox{\linewidth}{!}{%
\begin{tabular}{lccccc}
\toprule
\textbf{Attack} & 0.05 & 0.10 & 0.25 & 0.50 & 1.00 \\
\midrule
\texttt{MEA0} & 83.5$\pm$2.7 & 89.1$\pm$0.2 & 91.5$\pm$0.6 & 92.3$\pm$0.2 & 92.2$\pm$1.6 \\
\texttt{MEA1} & 71.1$\pm$0.8 & 74.4$\pm$0.7 & 77.2$\pm$0.3 & 89.2$\pm$0.7 & 89.5$\pm$0.4 \\
\texttt{MEA2} & 46.6$\pm$1.6 & 55.6$\pm$2.0 & 71.0$\pm$1.2 & 80.5$\pm$0.5 & 84.6$\pm$1.5 \\
\texttt{MEA3} & 79.2$\pm$3.8 & 89.0$\pm$1.0 & 90.2$\pm$0.7 & 91.2$\pm$0.5 & 91.6$\pm$0.3 \\
\texttt{MEA4} & 65.3$\pm$10.4 & 73.3$\pm$0.6 & 86.9$\pm$1.5 & 89.7$\pm$0.2 & 91.0$\pm$0.3 \\
\texttt{MEA5} & 83.2$\pm$2.8 & 89.7$\pm$1.9 & 91.0$\pm$0.4 & 91.6$\pm$1.2 & 91.7$\pm$1.1 \\
\texttt{AdvMEA} & 91.7$\pm$0.3 & 91.8$\pm$0.6 & 91.8$\pm$0.6 & 91.6$\pm$0.4 & 91.8$\pm$0.4 \\
\texttt{CEGA} & 88.7$\pm$0.8 & 90.3$\pm$0.4 & 91.0$\pm$0.1 & 91.7$\pm$0.5 & 91.9$\pm$0.4 \\
\texttt{Realistic} & 68.8$\pm$8.3 & 77.2$\pm$4.3 & 81.2$\pm$2.2 & 78.9$\pm$2.4 & 79.3$\pm$1.2 \\
\texttt{DFEA\_I} & 89.8$\pm$1.1 & 89.7$\pm$0.7 & 89.6$\pm$0.5 & 89.8$\pm$0.6 & 89.4$\pm$0.6 \\
\texttt{DFEA\_II} & 91.3$\pm$1.4 & 90.9$\pm$0.2 & 90.1$\pm$0.8 & 88.7$\pm$0.8 & 87.4$\pm$0.2 \\
\texttt{DFEA\_III} & 89.6$\pm$0.7 & 89.4$\pm$0.8 & 89.3$\pm$0.9 & 88.6$\pm$0.2 & 88.3$\pm$0.2 \\
\bottomrule
\end{tabular}
}
\end{subtable}
\medskip
\begin{subtable}[t]{0.48\textwidth}
\centering
\caption{Regime=\texttt{a\_only}}
\resizebox{\linewidth}{!}{%
\begin{tabular}{lccccc}
\toprule
\textbf{Attack} & 0.05 & 0.10 & 0.25 & 0.50 & 1.00 \\
\midrule
\texttt{MEA0} & 85.1$\pm$2.1 & 89.0$\pm$1.4 & 90.5$\pm$0.5 & 91.0$\pm$0.3 & 91.4$\pm$0.7 \\
\texttt{MEA1} & 71.1$\pm$0.8 & 74.4$\pm$0.7 & 77.2$\pm$0.3 & 89.2$\pm$0.7 & 89.5$\pm$0.4 \\
\texttt{MEA2} & 46.7$\pm$1.7 & 59.5$\pm$0.9 & 73.3$\pm$1.0 & 81.1$\pm$0.8 & 84.6$\pm$1.5 \\
\texttt{MEA3} & 87.6$\pm$0.4 & 89.5$\pm$1.3 & 90.8$\pm$0.7 & 91.9$\pm$0.3 & 92.5$\pm$0.4 \\
\texttt{MEA4} & 58.5$\pm$15.4 & 82.6$\pm$3.3 & 87.2$\pm$0.2 & 90.0$\pm$0.7 & 91.0$\pm$0.4 \\
\texttt{MEA5} & 86.5$\pm$1.2 & 89.2$\pm$0.7 & 91.3$\pm$0.0 & 92.0$\pm$0.9 & 92.0$\pm$0.6 \\
\texttt{AdvMEA} & 91.8$\pm$0.2 & 91.7$\pm$0.3 & 91.9$\pm$0.2 & 91.8$\pm$0.4 & 91.6$\pm$0.5 \\
\texttt{CEGA} & 90.0$\pm$0.2 & 90.4$\pm$0.2 & 91.8$\pm$0.5 & 91.3$\pm$0.8 & 91.7$\pm$0.7 \\
\texttt{Realistic} & 74.5$\pm$3.2 & 80.8$\pm$2.0 & 84.4$\pm$4.4 & 77.1$\pm$3.0 & 77.3$\pm$2.3 \\
\texttt{DFEA\_I} & 89.8$\pm$1.1 & 89.7$\pm$0.7 & 89.6$\pm$0.4 & 89.8$\pm$0.7 & 89.4$\pm$0.6 \\
\texttt{DFEA\_II} & 91.3$\pm$1.4 & 90.9$\pm$0.2 & 90.1$\pm$0.8 & 88.6$\pm$0.8 & 87.4$\pm$0.2 \\
\texttt{DFEA\_III} & 89.6$\pm$0.7 & 89.4$\pm$0.8 & 89.3$\pm$0.9 & 88.6$\pm$0.2 & 88.3$\pm$0.2 \\
\bottomrule
\end{tabular}
}
\end{subtable}
\hfill
\begin{subtable}[t]{0.48\textwidth}
\centering
\caption{Regime=\texttt{data\_free}}
\resizebox{\linewidth}{!}{%
\begin{tabular}{lccccc}
\toprule
\textbf{Attack} & 0.05 & 0.10 & 0.25 & 0.50 & 1.00 \\
\midrule
\texttt{MEA0} & 7.4$\pm$2.7 & 33.5$\pm$23.5 & 7.4$\pm$3.8 & 3.5$\pm$2.7 & 22.1$\pm$19.8 \\
\texttt{MEA1} & 0.5$\pm$0.4 & 0.5$\pm$0.4 & 0.5$\pm$0.4 & 0.5$\pm$0.4 & 0.5$\pm$0.4 \\
\texttt{MEA2} & 45.2$\pm$1.7 & 54.3$\pm$2.6 & 73.8$\pm$1.0 & 81.5$\pm$1.1 & 84.6$\pm$1.5 \\
\texttt{MEA3} & 21.5$\pm$20.3 & 22.1$\pm$19.9 & 17.5$\pm$23.1 & 37.2$\pm$18.2 & 18.9$\pm$19.5 \\
\texttt{MEA4} & 4.0$\pm$2.4 & 22.6$\pm$19.5 & 18.4$\pm$22.5 & 20.1$\pm$21.5 & 7.1$\pm$3.0 \\
\texttt{MEA5} & 17.0$\pm$12.7 & 26.4$\pm$23.8 & 18.1$\pm$22.7 & 6.2$\pm$4.3 & 7.8$\pm$5.0 \\
\texttt{AdvMEA} & 91.7$\pm$0.3 & 91.9$\pm$0.2 & 91.8$\pm$0.3 & 91.8$\pm$0.4 & 91.8$\pm$0.3 \\
\texttt{CEGA} & 88.8$\pm$1.0 & 91.1$\pm$0.6 & 91.4$\pm$0.3 & 91.3$\pm$0.8 & 91.1$\pm$0.6 \\
\texttt{Realistic} & 74.9$\pm$2.2 & 77.4$\pm$2.8 & 75.1$\pm$2.0 & 79.7$\pm$1.1 & 79.9$\pm$1.9 \\
\texttt{DFEA\_I} & 89.7$\pm$1.1 & 89.7$\pm$0.7 & 89.6$\pm$0.4 & 89.8$\pm$0.6 & 89.5$\pm$0.6 \\
\texttt{DFEA\_II} & 91.3$\pm$1.4 & 90.9$\pm$0.2 & 90.1$\pm$0.8 & 88.6$\pm$0.8 & 87.4$\pm$0.2 \\
\texttt{DFEA\_III} & 89.6$\pm$0.7 & 89.4$\pm$0.8 & 89.3$\pm$0.9 & 88.6$\pm$0.2 & 88.3$\pm$0.1 \\
\bottomrule
\end{tabular}
}
\end{subtable}
\end{table*}


\begin{table*}[t]
\centering
\caption{RQ1 detailed for dataset=\texttt{CoauthorCS}, metric=\texttt{F1} (\%). Rows are attacks; columns are budgets. Mean $\pm$ std across seeds; best per column is bold.}
\label{tab:app_rq1_detail_CoauthorCS_F1}
\tiny
\setlength{\tabcolsep}{2pt}
\renewcommand{\arraystretch}{0.9}
\begin{subtable}[t]{0.48\textwidth}
\centering
\caption{Regime=\texttt{both}}
\resizebox{\linewidth}{!}{%
\begin{tabular}{lccccc}
\toprule
\textbf{Attack} & 0.05 & 0.10 & 0.25 & 0.50 & 1.00 \\
\midrule
\texttt{MEA0} & 51.8$\pm$1.1 & 61.1$\pm$3.9 & 70.1$\pm$3.4 & 73.7$\pm$3.8 & 77.1$\pm$2.5 \\
\texttt{MEA1} & 34.5$\pm$0.3 & 36.7$\pm$0.4 & 42.1$\pm$1.7 & 61.2$\pm$0.9 & 64.3$\pm$1.3 \\
\texttt{MEA2} & 32.7$\pm$1.4 & 38.9$\pm$3.3 & 52.1$\pm$4.0 & 57.5$\pm$5.4 & 59.5$\pm$4.5 \\
\texttt{MEA3} & 54.8$\pm$2.8 & 64.2$\pm$4.6 & 71.6$\pm$4.2 & 76.8$\pm$0.6 & 76.4$\pm$1.5 \\
\texttt{MEA4} & 25.2$\pm$4.7 & 38.0$\pm$4.8 & 52.3$\pm$8.8 & 67.1$\pm$1.9 & 71.3$\pm$1.1 \\
\texttt{MEA5} & 54.6$\pm$1.0 & 59.5$\pm$1.4 & 71.8$\pm$5.1 & 77.6$\pm$1.4 & 77.7$\pm$0.1 \\
\texttt{AdvMEA} & 75.0$\pm$1.6 & 73.7$\pm$0.8 & 74.0$\pm$1.8 & 74.9$\pm$2.2 & 75.3$\pm$1.0 \\
\texttt{CEGA} & 67.6$\pm$2.1 & 72.6$\pm$1.5 & 74.7$\pm$2.0 & 76.5$\pm$0.9 & 74.1$\pm$2.4 \\
\texttt{Realistic} & 42.9$\pm$6.9 & 53.4$\pm$2.3 & 61.3$\pm$3.9 & 56.6$\pm$5.6 & 54.2$\pm$3.5 \\
\texttt{DFEA\_I} & 81.9$\pm$2.0 & 82.0$\pm$0.9 & 81.6$\pm$0.8 & 81.3$\pm$0.6 & 80.5$\pm$0.8 \\
\texttt{DFEA\_II} & 83.1$\pm$1.3 & 82.1$\pm$0.4 & 81.6$\pm$2.1 & 78.2$\pm$2.3 & 77.0$\pm$1.6 \\
\texttt{DFEA\_III} & 81.8$\pm$1.0 & 80.7$\pm$0.6 & 80.6$\pm$2.0 & 78.8$\pm$0.5 & 78.6$\pm$0.9 \\
\bottomrule
\end{tabular}
}
\end{subtable}
\hfill
\begin{subtable}[t]{0.48\textwidth}
\centering
\caption{Regime=\texttt{x\_only}}
\resizebox{\linewidth}{!}{%
\begin{tabular}{lccccc}
\toprule
\textbf{Attack} & 0.05 & 0.10 & 0.25 & 0.50 & 1.00 \\
\midrule
\texttt{MEA0} & 44.4$\pm$3.8 & 60.9$\pm$4.1 & 74.5$\pm$0.4 & 76.1$\pm$1.7 & 74.8$\pm$4.7 \\
\texttt{MEA1} & 34.5$\pm$0.3 & 36.7$\pm$0.4 & 42.1$\pm$1.7 & 61.2$\pm$0.9 & 64.3$\pm$1.3 \\
\texttt{MEA2} & 31.8$\pm$1.1 & 41.5$\pm$1.6 & 55.6$\pm$3.9 & 58.8$\pm$3.0 & 59.5$\pm$4.5 \\
\texttt{MEA3} & 45.6$\pm$0.8 & 62.5$\pm$6.8 & 67.6$\pm$3.0 & 74.1$\pm$1.6 & 73.5$\pm$1.6 \\
\texttt{MEA4} & 20.8$\pm$8.6 & 36.1$\pm$6.9 & 52.2$\pm$4.5 & 64.5$\pm$0.7 & 71.1$\pm$0.8 \\
\texttt{MEA5} & 50.4$\pm$3.0 & 66.6$\pm$5.2 & 71.6$\pm$4.0 & 73.1$\pm$6.9 & 74.4$\pm$3.6 \\
\texttt{AdvMEA} & 74.7$\pm$0.7 & 74.7$\pm$2.0 & 74.6$\pm$1.6 & 74.3$\pm$1.0 & 73.2$\pm$1.9 \\
\texttt{CEGA} & 70.0$\pm$0.9 & 70.8$\pm$1.5 & 74.0$\pm$2.8 & 72.5$\pm$1.3 & 72.8$\pm$1.4 \\
\texttt{Realistic} & 46.1$\pm$1.1 & 46.4$\pm$3.0 & 56.5$\pm$1.7 & 56.9$\pm$3.7 & 55.8$\pm$0.6 \\
\texttt{DFEA\_I} & 81.9$\pm$2.0 & 82.0$\pm$1.0 & 81.6$\pm$0.8 & 81.3$\pm$0.6 & 80.5$\pm$0.8 \\
\texttt{DFEA\_II} & 83.1$\pm$1.3 & 82.1$\pm$0.4 & 81.6$\pm$2.1 & 78.2$\pm$2.3 & 77.0$\pm$1.6 \\
\texttt{DFEA\_III} & 81.8$\pm$1.0 & 80.7$\pm$0.6 & 80.6$\pm$2.0 & 78.8$\pm$0.5 & 78.6$\pm$0.9 \\
\bottomrule
\end{tabular}
}
\end{subtable}
\medskip
\begin{subtable}[t]{0.48\textwidth}
\centering
\caption{Regime=\texttt{a\_only}}
\resizebox{\linewidth}{!}{%
\begin{tabular}{lccccc}
\toprule
\textbf{Attack} & 0.05 & 0.10 & 0.25 & 0.50 & 1.00 \\
\midrule
\texttt{MEA0} & 56.0$\pm$6.3 & 65.7$\pm$4.6 & 71.3$\pm$2.8 & 73.2$\pm$1.6 & 73.4$\pm$3.3 \\
\texttt{MEA1} & 34.5$\pm$0.3 & 36.7$\pm$0.4 & 42.1$\pm$1.7 & 61.2$\pm$0.9 & 64.3$\pm$1.3 \\
\texttt{MEA2} & 31.6$\pm$6.0 & 43.5$\pm$3.7 & 54.7$\pm$1.8 & 58.9$\pm$4.6 & 59.5$\pm$4.5 \\
\texttt{MEA3} & 57.1$\pm$0.9 & 64.4$\pm$7.1 & 70.5$\pm$4.1 & 74.8$\pm$2.2 & 75.8$\pm$0.9 \\
\texttt{MEA4} & 21.3$\pm$4.2 & 47.2$\pm$8.5 & 52.4$\pm$0.9 & 66.1$\pm$2.5 & 71.3$\pm$1.4 \\
\texttt{MEA5} & 53.6$\pm$3.3 & 67.3$\pm$4.9 & 74.1$\pm$1.1 & 77.3$\pm$1.9 & 74.1$\pm$2.0 \\
\texttt{AdvMEA} & 75.0$\pm$0.8 & 75.2$\pm$1.0 & 74.7$\pm$1.1 & 74.6$\pm$1.1 & 74.4$\pm$1.0 \\
\texttt{CEGA} & 69.4$\pm$2.1 & 71.2$\pm$0.9 & 73.9$\pm$0.9 & 73.3$\pm$3.0 & 72.3$\pm$1.3 \\
\texttt{Realistic} & 41.5$\pm$6.9 & 56.2$\pm$2.7 & 61.6$\pm$5.3 & 54.4$\pm$5.0 & 54.5$\pm$0.8 \\
\texttt{DFEA\_I} & 81.9$\pm$2.0 & 82.0$\pm$0.9 & 81.6$\pm$0.8 & 81.3$\pm$0.7 & 80.5$\pm$0.8 \\
\texttt{DFEA\_II} & 83.1$\pm$1.3 & 82.1$\pm$0.4 & 81.6$\pm$2.1 & 78.2$\pm$2.3 & 77.0$\pm$1.6 \\
\texttt{DFEA\_III} & 81.8$\pm$1.0 & 80.7$\pm$0.6 & 80.6$\pm$2.0 & 78.8$\pm$0.5 & 78.6$\pm$0.9 \\
\bottomrule
\end{tabular}
}
\end{subtable}
\hfill
\begin{subtable}[t]{0.48\textwidth}
\centering
\caption{Regime=\texttt{data\_free}}
\resizebox{\linewidth}{!}{%
\begin{tabular}{lccccc}
\toprule
\textbf{Attack} & 0.05 & 0.10 & 0.25 & 0.50 & 1.00 \\
\midrule
\texttt{MEA0} & 0.9$\pm$0.3 & 3.0$\pm$2.1 & 0.9$\pm$0.5 & 0.4$\pm$0.3 & 2.1$\pm$1.6 \\
\texttt{MEA1} & 0.1$\pm$0.0 & 0.1$\pm$0.0 & 0.1$\pm$0.0 & 0.1$\pm$0.0 & 0.1$\pm$0.0 \\
\texttt{MEA2} & 32.0$\pm$2.6 & 38.9$\pm$7.7 & 55.3$\pm$3.7 & 59.8$\pm$6.4 & 59.5$\pm$4.5 \\
\texttt{MEA3} & 2.4$\pm$1.5 & 3.4$\pm$1.9 & 1.6$\pm$2.0 & 3.9$\pm$0.8 & 2.9$\pm$2.8 \\
\texttt{MEA4} & 0.5$\pm$0.3 & 2.2$\pm$1.6 & 1.7$\pm$2.0 & 1.9$\pm$1.8 & 0.9$\pm$0.4 \\
\texttt{MEA5} & 2.7$\pm$2.2 & 5.5$\pm$5.0 & 1.8$\pm$1.9 & 0.8$\pm$0.5 & 3.6$\pm$2.6 \\
\texttt{AdvMEA} & 75.1$\pm$1.0 & 74.9$\pm$0.8 & 73.9$\pm$0.9 & 74.7$\pm$0.9 & 74.6$\pm$0.7 \\
\texttt{CEGA} & 65.1$\pm$3.0 & 72.6$\pm$2.4 & 71.9$\pm$1.0 & 75.5$\pm$0.8 & 72.0$\pm$1.0 \\
\texttt{Realistic} & 52.0$\pm$1.7 & 55.1$\pm$3.3 & 54.1$\pm$1.5 & 59.5$\pm$0.8 & 56.9$\pm$2.8 \\
\texttt{DFEA\_I} & 81.9$\pm$2.0 & 82.0$\pm$1.0 & 81.6$\pm$0.8 & 81.3$\pm$0.6 & 80.5$\pm$0.7 \\
\texttt{DFEA\_II} & 83.1$\pm$1.3 & 82.1$\pm$0.4 & 81.6$\pm$2.1 & 78.2$\pm$2.3 & 77.0$\pm$1.6 \\
\texttt{DFEA\_III} & 81.8$\pm$1.0 & 80.7$\pm$0.6 & 80.6$\pm$2.0 & 78.8$\pm$0.5 & 78.5$\pm$0.9 \\
\bottomrule
\end{tabular}
}
\end{subtable}
\end{table*}


\begin{table*}[t]
\centering
\caption{RQ1 detailed for dataset=\texttt{CoauthorCS}, metric=\texttt{Fidelity} (\%). Rows are attacks; columns are budgets. Mean $\pm$ std across seeds; best per column is bold.}
\label{tab:app_rq1_detail_CoauthorCS_Fidelity}
\tiny
\setlength{\tabcolsep}{2pt}
\renewcommand{\arraystretch}{0.9}
\begin{subtable}[t]{0.48\textwidth}
\centering
\caption{Regime=\texttt{both}}
\resizebox{\linewidth}{!}{%
\begin{tabular}{lccccc}
\toprule
\textbf{Attack} & 0.05 & 0.10 & 0.25 & 0.50 & 1.00 \\
\midrule
\texttt{MEA0} & 83.7$\pm$4.0 & 87.9$\pm$1.6 & 91.2$\pm$1.2 & 92.8$\pm$0.3 & 94.5$\pm$0.3 \\
\texttt{MEA1} & 71.0$\pm$0.8 & 74.2$\pm$0.5 & 76.9$\pm$0.5 & 89.8$\pm$0.8 & 90.2$\pm$0.7 \\
\texttt{MEA2} & 47.8$\pm$1.2 & 57.3$\pm$1.4 & 73.0$\pm$2.3 & 84.2$\pm$1.7 & 87.7$\pm$2.9 \\
\texttt{MEA3} & 85.2$\pm$1.8 & 88.2$\pm$1.3 & 90.8$\pm$1.0 & 91.7$\pm$0.2 & 93.1$\pm$0.6 \\
\texttt{MEA4} & 66.7$\pm$8.4 & 79.1$\pm$7.2 & 86.9$\pm$3.4 & 93.5$\pm$0.7 & 96.3$\pm$0.5 \\
\texttt{MEA5} & 85.8$\pm$0.2 & 87.8$\pm$0.3 & 91.1$\pm$0.2 & 92.4$\pm$0.4 & 93.6$\pm$0.8 \\
\texttt{AdvMEA} & 92.3$\pm$0.5 & 92.3$\pm$0.4 & 92.0$\pm$0.4 & 92.1$\pm$0.3 & 92.3$\pm$0.4 \\
\texttt{CEGA} & 90.3$\pm$0.6 & 91.7$\pm$0.7 & 93.5$\pm$0.7 & 94.9$\pm$0.8 & 94.8$\pm$0.5 \\
\texttt{Realistic} & 72.6$\pm$5.2 & 81.5$\pm$4.3 & 82.5$\pm$3.2 & 80.4$\pm$3.7 & 80.9$\pm$2.0 \\
\texttt{DFEA\_I} & 91.6$\pm$0.5 & 93.2$\pm$0.2 & 94.7$\pm$0.4 & 95.7$\pm$1.1 & 96.9$\pm$0.9 \\
\texttt{DFEA\_II} & 89.6$\pm$1.2 & 91.1$\pm$1.1 & 93.1$\pm$0.7 & 96.1$\pm$0.4 & 100.0$\pm$0.0 \\
\texttt{DFEA\_III} & 91.0$\pm$0.7 & 92.6$\pm$0.4 & 94.8$\pm$0.0 & 96.8$\pm$0.7 & 98.9$\pm$0.3 \\
\bottomrule
\end{tabular}
}
\end{subtable}
\hfill
\begin{subtable}[t]{0.48\textwidth}
\centering
\caption{Regime=\texttt{x\_only}}
\resizebox{\linewidth}{!}{%
\begin{tabular}{lccccc}
\toprule
\textbf{Attack} & 0.05 & 0.10 & 0.25 & 0.50 & 1.00 \\
\midrule
\texttt{MEA0} & 83.3$\pm$2.8 & 88.8$\pm$0.5 & 92.3$\pm$0.7 & 93.3$\pm$0.5 & 93.7$\pm$0.2 \\
\texttt{MEA1} & 71.0$\pm$0.8 & 74.2$\pm$0.5 & 76.9$\pm$0.5 & 89.8$\pm$0.8 & 90.2$\pm$0.7 \\
\texttt{MEA2} & 46.2$\pm$1.8 & 56.5$\pm$2.5 & 74.3$\pm$2.6 & 83.7$\pm$2.2 & 87.7$\pm$2.9 \\
\texttt{MEA3} & 77.8$\pm$4.5 & 88.6$\pm$0.9 & 90.0$\pm$0.6 & 91.7$\pm$0.4 & 93.4$\pm$0.5 \\
\texttt{MEA4} & 63.8$\pm$10.0 & 72.2$\pm$0.8 & 87.7$\pm$1.2 & 92.9$\pm$0.3 & 96.4$\pm$0.4 \\
\texttt{MEA5} & 83.5$\pm$2.4 & 89.2$\pm$2.0 & 91.0$\pm$0.3 & 92.0$\pm$0.7 & 93.1$\pm$1.0 \\
\texttt{AdvMEA} & 92.0$\pm$0.4 & 92.1$\pm$0.3 & 92.2$\pm$0.6 & 92.0$\pm$0.3 & 92.2$\pm$0.2 \\
\texttt{CEGA} & 90.2$\pm$0.6 & 92.7$\pm$0.3 & 92.8$\pm$0.4 & 94.5$\pm$0.3 & 96.2$\pm$0.3 \\
\texttt{Realistic} & 68.4$\pm$8.8 & 78.1$\pm$4.7 & 81.4$\pm$1.7 & 80.8$\pm$1.9 & 80.5$\pm$1.1 \\
\texttt{DFEA\_I} & 91.7$\pm$0.5 & 93.2$\pm$0.2 & 94.6$\pm$0.4 & 95.6$\pm$1.0 & 96.9$\pm$0.9 \\
\texttt{DFEA\_II} & 89.6$\pm$1.2 & 91.1$\pm$1.1 & 93.1$\pm$0.7 & 96.1$\pm$0.4 & 100.0$\pm$0.0 \\
\texttt{DFEA\_III} & 91.0$\pm$0.7 & 92.6$\pm$0.4 & 94.8$\pm$0.0 & 96.8$\pm$0.7 & 98.9$\pm$0.3 \\
\bottomrule
\end{tabular}
}
\end{subtable}
\medskip
\begin{subtable}[t]{0.48\textwidth}
\centering
\caption{Regime=\texttt{a\_only}}
\resizebox{\linewidth}{!}{%
\begin{tabular}{lccccc}
\toprule
\textbf{Attack} & 0.05 & 0.10 & 0.25 & 0.50 & 1.00 \\
\midrule
\texttt{MEA0} & 84.6$\pm$2.4 & 89.4$\pm$1.4 & 91.9$\pm$0.6 & 92.9$\pm$0.4 & 94.6$\pm$0.4 \\
\texttt{MEA1} & 71.0$\pm$0.8 & 74.2$\pm$0.5 & 76.9$\pm$0.5 & 89.8$\pm$0.8 & 90.2$\pm$0.7 \\
\texttt{MEA2} & 46.3$\pm$1.1 & 60.7$\pm$0.8 & 76.0$\pm$0.3 & 84.0$\pm$1.6 & 87.7$\pm$2.9 \\
\texttt{MEA3} & 86.5$\pm$0.5 & 89.5$\pm$1.4 & 90.8$\pm$0.9 & 92.0$\pm$0.4 & 93.5$\pm$0.1 \\
\texttt{MEA4} & 57.4$\pm$15.1 & 81.4$\pm$2.8 & 88.5$\pm$0.3 & 92.8$\pm$1.3 & 96.6$\pm$0.4 \\
\texttt{MEA5} & 86.2$\pm$1.6 & 89.0$\pm$0.6 & 91.1$\pm$0.4 & 92.4$\pm$0.7 & 93.7$\pm$1.0 \\
\texttt{AdvMEA} & 92.0$\pm$0.3 & 92.0$\pm$0.4 & 92.3$\pm$0.4 & 91.9$\pm$0.3 & 91.9$\pm$0.3 \\
\texttt{CEGA} & 90.2$\pm$1.0 & 92.5$\pm$0.5 & 93.8$\pm$0.5 & 94.2$\pm$0.3 & 95.8$\pm$0.3 \\
\texttt{Realistic} & 73.9$\pm$3.3 & 82.2$\pm$1.5 & 85.7$\pm$4.5 & 78.6$\pm$3.7 & 79.0$\pm$2.6 \\
\texttt{DFEA\_I} & 91.7$\pm$0.5 & 93.2$\pm$0.2 & 94.7$\pm$0.4 & 95.7$\pm$1.0 & 96.9$\pm$0.9 \\
\texttt{DFEA\_II} & 89.6$\pm$1.2 & 91.1$\pm$1.1 & 93.1$\pm$0.7 & 96.1$\pm$0.4 & 100.0$\pm$0.0 \\
\texttt{DFEA\_III} & 91.0$\pm$0.7 & 92.6$\pm$0.4 & 94.8$\pm$0.0 & 96.8$\pm$0.7 & 98.9$\pm$0.3 \\
\bottomrule
\end{tabular}
}
\end{subtable}
\hfill
\begin{subtable}[t]{0.48\textwidth}
\centering
\caption{Regime=\texttt{data\_free}}
\resizebox{\linewidth}{!}{%
\begin{tabular}{lccccc}
\toprule
\textbf{Attack} & 0.05 & 0.10 & 0.25 & 0.50 & 1.00 \\
\midrule
\texttt{MEA0} & 6.6$\pm$2.5 & 32.2$\pm$22.7 & 7.0$\pm$3.9 & 3.4$\pm$2.8 & 21.2$\pm$19.1 \\
\texttt{MEA1} & 0.6$\pm$0.7 & 0.6$\pm$0.7 & 0.6$\pm$0.7 & 0.6$\pm$0.7 & 0.6$\pm$0.7 \\
\texttt{MEA2} & 45.1$\pm$2.4 & 55.9$\pm$2.2 & 75.3$\pm$2.4 & 84.1$\pm$1.9 & 87.7$\pm$2.9 \\
\texttt{MEA3} & 20.4$\pm$19.6 & 21.2$\pm$19.1 & 16.8$\pm$22.4 & 36.1$\pm$17.3 & 18.7$\pm$18.9 \\
\texttt{MEA4} & 3.9$\pm$2.2 & 22.1$\pm$18.8 & 17.4$\pm$21.7 & 19.4$\pm$20.5 & 7.0$\pm$3.1 \\
\texttt{MEA5} & 16.3$\pm$12.3 & 25.8$\pm$23.0 & 17.3$\pm$21.8 & 5.8$\pm$3.7 & 8.0$\pm$5.0 \\
\texttt{AdvMEA} & 91.9$\pm$0.4 & 92.2$\pm$0.4 & 92.2$\pm$0.4 & 92.0$\pm$0.5 & 92.1$\pm$0.5 \\
\texttt{CEGA} & 90.0$\pm$1.2 & 92.3$\pm$0.6 & 93.1$\pm$0.3 & 93.8$\pm$0.7 & 95.5$\pm$0.5 \\
\texttt{Realistic} & 77.8$\pm$2.5 & 78.9$\pm$2.1 & 76.4$\pm$2.5 & 80.3$\pm$1.9 & 81.5$\pm$2.1 \\
\texttt{DFEA\_I} & 91.7$\pm$0.5 & 93.2$\pm$0.2 & 94.7$\pm$0.4 & 95.6$\pm$1.0 & 96.9$\pm$0.8 \\
\texttt{DFEA\_II} & 89.6$\pm$1.2 & 91.1$\pm$1.1 & 93.1$\pm$0.7 & 96.1$\pm$0.4 & 100.0$\pm$0.0 \\
\texttt{DFEA\_III} & 91.0$\pm$0.7 & 92.6$\pm$0.4 & 94.8$\pm$0.0 & 96.8$\pm$0.7 & 98.9$\pm$0.3 \\
\bottomrule
\end{tabular}
}
\end{subtable}
\end{table*}


\begin{table*}[t]
\centering
\caption{RQ1 detailed for dataset=\texttt{CoauthorPhysics}, metric=\texttt{Acc} (\%). Rows are attacks; columns are budgets. Mean $\pm$ std across seeds; best per column is bold.}
\label{tab:app_rq1_detail_CoauthorPhysics_Acc}
\tiny
\setlength{\tabcolsep}{2pt}
\renewcommand{\arraystretch}{0.9}
\begin{subtable}[t]{0.48\textwidth}
\centering
\caption{Regime=\texttt{both}}
\resizebox{\linewidth}{!}{%
\begin{tabular}{lccccc}
\toprule
\textbf{Attack} & 0.05 & 0.10 & 0.25 & 0.50 & 1.00 \\
\midrule
\texttt{MEA0} & 82.2$\pm$4.9 & 89.6$\pm$0.4 & 90.9$\pm$0.2 & 90.7$\pm$0.4 & 92.8$\pm$1.0 \\
\texttt{MEA1} & 3.5$\pm$0.0 & 10.1$\pm$0.6 & 17.2$\pm$0.2 & 44.0$\pm$1.9 & 79.5$\pm$7.9 \\
\texttt{MEA2} & 33.3$\pm$2.4 & 44.2$\pm$1.4 & 66.8$\pm$1.0 & 73.6$\pm$4.4 & 84.0$\pm$3.1 \\
\texttt{MEA3} & 74.0$\pm$1.4 & 81.1$\pm$2.6 & 89.5$\pm$1.4 & 89.8$\pm$0.7 & 90.6$\pm$1.8 \\
\texttt{MEA4} & 70.1$\pm$10.5 & 74.0$\pm$10.0 & 87.7$\pm$2.0 & 89.6$\pm$0.5 & 89.5$\pm$0.6 \\
\texttt{MEA5} & 75.4$\pm$4.8 & 78.4$\pm$4.7 & 89.8$\pm$1.6 & 91.0$\pm$0.8 & 91.3$\pm$1.0 \\
\texttt{AdvMEA} & 91.4$\pm$1.2 & 91.6$\pm$0.7 & 91.3$\pm$0.2 & 90.6$\pm$1.3 & 91.4$\pm$0.9 \\
\texttt{CEGA} & 90.9$\pm$1.3 & 90.4$\pm$2.8 & 91.5$\pm$0.6 & 91.0$\pm$0.5 & 90.8$\pm$0.3 \\
\texttt{Realistic} & 79.6$\pm$3.8 & 77.0$\pm$6.5 & 79.3$\pm$1.7 & 76.1$\pm$3.1 & 73.0$\pm$1.5 \\
\texttt{DFEA\_I} & 90.7$\pm$0.2 & 90.8$\pm$0.2 & 90.8$\pm$0.5 & 90.8$\pm$0.2 & 90.6$\pm$0.2 \\
\texttt{DFEA\_II} & 90.9$\pm$0.2 & 90.0$\pm$0.6 & 90.6$\pm$0.4 & 90.1$\pm$0.4 & 89.5$\pm$0.4 \\
\texttt{DFEA\_III} & 90.5$\pm$0.4 & 90.1$\pm$0.6 & 90.5$\pm$0.4 & 90.3$\pm$0.5 & 90.4$\pm$0.3 \\
\bottomrule
\end{tabular}
}
\end{subtable}
\hfill
\begin{subtable}[t]{0.48\textwidth}
\centering
\caption{Regime=\texttt{x\_only}}
\resizebox{\linewidth}{!}{%
\begin{tabular}{lccccc}
\toprule
\textbf{Attack} & 0.05 & 0.10 & 0.25 & 0.50 & 1.00 \\
\midrule
\texttt{MEA0} & 80.9$\pm$4.2 & 90.4$\pm$1.8 & 89.3$\pm$1.5 & 91.5$\pm$0.9 & 91.3$\pm$0.8 \\
\texttt{MEA1} & 3.5$\pm$0.0 & 10.1$\pm$0.6 & 17.2$\pm$0.2 & 44.0$\pm$1.9 & 79.5$\pm$7.9 \\
\texttt{MEA2} & 31.0$\pm$2.1 & 41.2$\pm$2.2 & 64.9$\pm$4.1 & 71.9$\pm$6.2 & 83.9$\pm$3.1 \\
\texttt{MEA3} & 75.3$\pm$1.5 & 84.0$\pm$2.4 & 89.8$\pm$0.9 & 91.6$\pm$0.5 & 91.4$\pm$0.5 \\
\texttt{MEA4} & 39.5$\pm$17.4 & 83.4$\pm$4.5 & 88.7$\pm$1.6 & 90.2$\pm$0.6 & 89.9$\pm$0.3 \\
\texttt{MEA5} & 76.2$\pm$5.5 & 78.9$\pm$5.5 & 89.0$\pm$1.5 & 91.5$\pm$0.7 & 91.2$\pm$1.4 \\
\texttt{AdvMEA} & 90.8$\pm$1.2 & 91.2$\pm$1.0 & 90.7$\pm$0.4 & 91.6$\pm$0.4 & 90.9$\pm$0.9 \\
\texttt{CEGA} & 90.9$\pm$0.4 & 91.2$\pm$1.0 & 91.6$\pm$0.3 & 91.2$\pm$0.5 & 91.2$\pm$0.4 \\
\texttt{Realistic} & 79.0$\pm$5.6 & 83.0$\pm$4.6 & 79.4$\pm$2.7 & 78.6$\pm$2.6 & 74.4$\pm$0.5 \\
\texttt{DFEA\_I} & 90.7$\pm$0.2 & 90.8$\pm$0.2 & 90.8$\pm$0.5 & 90.8$\pm$0.2 & 90.6$\pm$0.2 \\
\texttt{DFEA\_II} & 90.9$\pm$0.2 & 90.0$\pm$0.6 & 90.6$\pm$0.4 & 90.1$\pm$0.4 & 89.5$\pm$0.4 \\
\texttt{DFEA\_III} & 90.4$\pm$0.4 & 90.1$\pm$0.6 & 90.5$\pm$0.4 & 90.3$\pm$0.5 & 90.4$\pm$0.3 \\
\bottomrule
\end{tabular}%
}
\end{subtable}
\end{table*}


\begin{table*}[t]
\centering
\caption{RQ1 detailed for dataset=\texttt{CoauthorPhysics}, metric=\texttt{F1} (\%). Rows are attacks; columns are budgets. Mean $\pm$ std across seeds; best per column is bold.}
\label{tab:app_rq1_detail_CoauthorPhysics_F1}
\tiny
\setlength{\tabcolsep}{2pt}
\renewcommand{\arraystretch}{0.9}
\begin{subtable}[t]{0.48\textwidth}
\centering
\caption{Regime=\texttt{both}}
\resizebox{\linewidth}{!}{%
%
}
\end{subtable}
\end{table*}


\begin{table*}[t]
\centering
\caption{RQ1 detailed for dataset=\texttt{CoauthorPhysics}, metric=\texttt{Fidelity} (\%). Rows are attacks; columns are budgets. Mean $\pm$ std across seeds; best per column is bold.}
\label{tab:app_rq1_detail_CoauthorPhysics_Fidelity}
\tiny
\setlength{\tabcolsep}{2pt}
\renewcommand{\arraystretch}{0.9}
\begin{subtable}[t]{0.48\textwidth}
\centering
\caption{Regime=\texttt{both}}
\resizebox{\linewidth}{!}{%
%
}
\end{subtable}
\end{table*}


\begin{table*}[t]
\centering
\caption{RQ1 detailed for dataset=\texttt{Computers}, metric=\texttt{Acc} (\%). Rows are attacks; columns are budgets. Mean $\pm$ std across seeds; best per column is bold.}
\label{tab:app_rq1_detail_Computers_Acc}
\tiny
\setlength{\tabcolsep}{2pt}
\renewcommand{\arraystretch}{0.9}
\begin{subtable}[t]{0.48\textwidth}
\centering
\caption{Regime=\texttt{both}}
\resizebox{\linewidth}{!}{%
%
}
\end{subtable}
\end{table*}


\begin{table*}[t]
\centering
\caption{RQ1 detailed for dataset=\texttt{Computers}, metric=\texttt{F1} (\%). Rows are attacks; columns are budgets. Mean $\pm$ std across seeds; best per column is bold.}
\label{tab:app_rq1_detail_Computers_F1}
\tiny
\setlength{\tabcolsep}{2pt}
\renewcommand{\arraystretch}{0.9}
\begin{subtable}[t]{0.48\textwidth}
\centering
\caption{Regime=\texttt{both}}
\resizebox{\linewidth}{!}{%
%
}
\end{subtable}
\end{table*}


\begin{table*}[t]
\centering
\caption{RQ1 detailed for dataset=\texttt{Computers}, metric=\texttt{Fidelity} (\%). Rows are attacks; columns are budgets. Mean $\pm$ std across seeds; best per column is bold.}
\label{tab:app_rq1_detail_Computers_Fidelity}
\tiny
\setlength{\tabcolsep}{2pt}
\renewcommand{\arraystretch}{0.9}
\begin{subtable}[t]{0.48\textwidth}
\centering
\caption{Regime=\texttt{both}}
\resizebox{\linewidth}{!}{%
%
}
\end{subtable}
\end{table*}


\begin{table*}[t]
\centering
\caption{RQ1 detailed for dataset=\texttt{Photo}, metric=\texttt{Acc} (\%). Rows are attacks; columns are budgets. Mean $\pm$ std across seeds; best per column is bold.}
\label{tab:app_rq1_detail_Photo_Acc}
\tiny
\setlength{\tabcolsep}{2pt}
\renewcommand{\arraystretch}{0.9}
\begin{subtable}[t]{0.48\textwidth}
\centering
\caption{Regime=\texttt{both}}
\resizebox{\linewidth}{!}{%
%
}
\end{subtable}
\end{table*}


\begin{table*}[t]
\centering
\caption{RQ1 detailed for dataset=\texttt{Photo}, metric=\texttt{F1} (\%). Rows are attacks; columns are budgets. Mean $\pm$ std across seeds; best per column is bold.}
\label{tab:app_rq1_detail_Photo_F1}
\tiny
\setlength{\tabcolsep}{2pt}
\renewcommand{\arraystretch}{0.9}
\begin{subtable}[t]{0.48\textwidth}
\centering
\caption{Regime=\texttt{both}}
\resizebox{\linewidth}{!}{%
%
}
\end{subtable}
\end{table*}


\begin{table*}[t]
\centering
\caption{RQ1 detailed for dataset=\texttt{Photo}, metric=\texttt{Fidelity} (\%). Rows are attacks; columns are budgets. Mean $\pm$ std across seeds; best per column is bold.}
\label{tab:app_rq1_detail_Photo_Fidelity}
\tiny
\setlength{\tabcolsep}{2pt}
\renewcommand{\arraystretch}{0.9}
\begin{subtable}[t]{0.48\textwidth}
\centering
\caption{Regime=\texttt{both}}
\resizebox{\linewidth}{!}{%
%
}
\end{subtable}
\end{table*}


\begin{table*}[t]
\centering
\caption{RQ1 detailed for dataset=\texttt{PubMed}, metric=\texttt{Acc} (\%). Rows are attacks; columns are budgets. Mean $\pm$ std across seeds; best per column is bold.}
\label{tab:app_rq1_detail_PubMed_Acc}
\tiny
\setlength{\tabcolsep}{2pt}
\renewcommand{\arraystretch}{0.9}
\begin{subtable}[t]{0.48\textwidth}
\centering
\caption{Regime=\texttt{both}}
\resizebox{\linewidth}{!}{%
%
}
\end{subtable}
\end{table*}


\begin{table*}[t]
\centering
\caption{RQ1 detailed for dataset=\texttt{PubMed}, metric=\texttt{F1} (\%). Rows are attacks; columns are budgets. Mean $\pm$ std across seeds; best per column is bold.}
\label{tab:app_rq1_detail_PubMed_F1}
\tiny
\setlength{\tabcolsep}{2pt}
\renewcommand{\arraystretch}{0.9}
\begin{subtable}[t]{0.48\textwidth}
\centering
\caption{Regime=\texttt{both}}
\resizebox{\linewidth}{!}{%
%
}
\end{subtable}
\end{table*}


\begin{table*}[t]
\centering
\caption{RQ1 detailed for dataset=\texttt{PubMed}, metric=\texttt{Fidelity} (\%). Rows are attacks; columns are budgets. Mean $\pm$ std across seeds; best per column is bold.}
\label{tab:app_rq1_detail_PubMed_Fidelity}
\tiny
\setlength{\tabcolsep}{2pt}
\renewcommand{\arraystretch}{0.9}
\begin{subtable}[t]{0.48\textwidth}
\centering
\caption{Regime=\texttt{both}}
\resizebox{\linewidth}{!}{%
%
}
\end{subtable}
\end{table*}

\FloatBarrier
\subsubsection{Seed sensitivity of data-free extraction}
\label{app:seedcheck}
For \texttt{DFEA\_II} on \textit{Computers}, the original three seeds at budgets $0.10\times$ through $1.00\times$ each produced a surrogate that predicted a single class. A constant predictor's accuracy equals the frequency of that class in the test labels, whereas its fidelity equals the frequency in the target's predictions. The repeated means in Table~\ref{tab:rq1_computers_both_fidelity} therefore reflect these seeds rather than duplicate log entries. The ten-seed rerun below reports fidelity across all seeds, including~collapsed~runs.

\begin{center}\begin{minipage}{\linewidth}\centering\small
\captionof{table}{DFEA\_II fidelity on Computers (mean $\pm$ SD, \%).}\label{tab:auditdf}
\begin{tabular}{@{}rrrrr@{}}\toprule
Budget & Unique queries & Original (3 seeds) & Audit (10 seeds) & Collapsed\\ \midrule
$0.05\times$ & 687 & $43.9\pm8.1$ & $47.2\pm14.6$ & 3/10\\ $0.10\times$ & 1,375 & $28.4\pm20.5$ & $50.7\pm21.0$ & 3/10\\ $0.25\times$ & 3,438 & $28.4\pm20.5$ & $48.2\pm23.6$ & 4/10\\ $0.50\times$ & 6,876 & $28.4\pm20.5$ & $44.3\pm22.6$ & 5/10\\ $1.00\times$ & 13,752 & $28.4\pm20.5$ & $44.1\pm23.9$ & 6/10\\ \bottomrule
\end{tabular}\end{minipage}\end{center}

\subsection{Defense performance results}
\label{app:rq2}
\begin{figure}[t]
\centering
\begin{subfigure}[t]{0.48\textwidth}
  \centering
  \includegraphics[width=\linewidth]{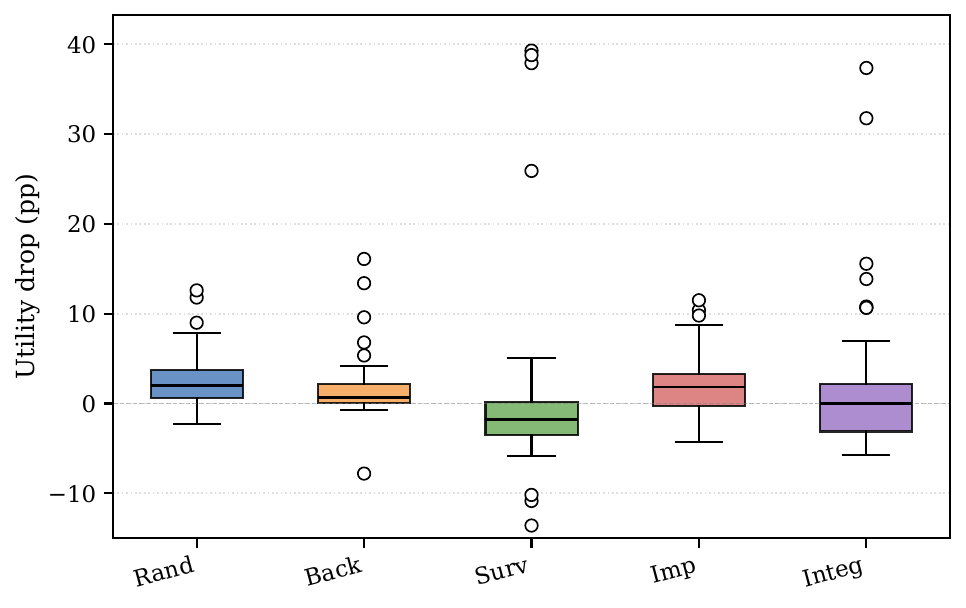}
  \caption{Utility drop.}
  \label{fig:rq2_box_util_wrapL}
\end{subfigure}\hfill
\begin{subfigure}[t]{0.48\textwidth}
  \centering
  \includegraphics[width=\linewidth]{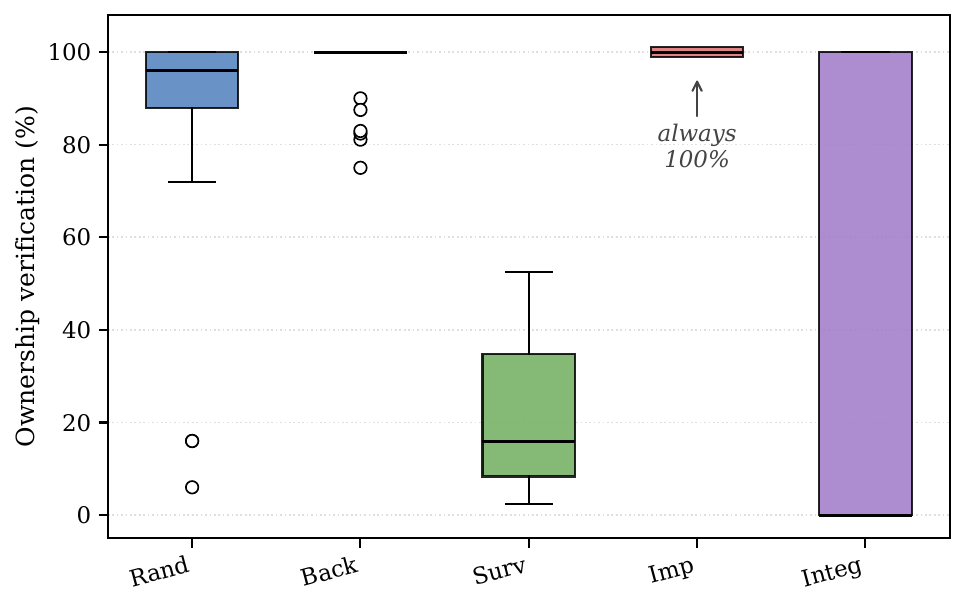}
  \caption{Ownership verification.}
  \label{fig:rq2_box_verify_wrapL}
\end{subfigure}
\caption{Defense effectiveness across all ten datasets: (a) utility drop and (b) ownership verification. Each box aggregates one point per (dataset, seed), giving 30 points per defense except Integrity (44 dataset--seed pairs).}
\label{fig:rq2_defense_boxplots_wrapL}
\end{figure}

This subsection contains the per-dataset defense performance and the boxplot view that the main text references. Figure~\ref{fig:rq2_defense_boxplots_wrapL} aggregates utility drop and ownership verification across all ten datasets and three seeds (30 points per defense except Integrity (44 dataset--seed pairs)), and the per-dataset numbers follow. Table~\ref{tab:rq2_defense_7xN_F1} reports task utility measured by macro F1 (\%); Table~\ref{tab:rq2_defense_7xN_Fidelity} reports behavioural alignment with the original target measured by fidelity (\%) on the same test inputs; Table~\ref{tab:rq2_defense_7xN_Owner_verif} summarises ownership verification on a standardised verification set. Together these views separate downstream utility, behavioural consistency, and ownership verification, so that the trade-offs across defenses are explicit on~every~dataset.

\begin{table*}[t]
\centering
\scriptsize
\setlength{\tabcolsep}{5pt}
\renewcommand{\arraystretch}{1.15}
\begin{tabular}{lccccc}
\toprule
Dataset & backdoorwm & randomwm & survivewm & imperceptiblewm & integrity \\
\midrule
Cora            & \(79.37 \pm 0.45\)  & \(76.06 \pm 0.81\)  & \(79.07 \pm 0.33\)  & \(71.94 \pm 0.27\)  & \(75.39 \pm 0.36\) \\
CiteSeer        & \(64.74 \pm 0.35\)  & \(64.99 \pm 0.69\)  & \(67.47 \pm 0.27\)  & \(58.07 \pm 0.79\)  & \(57.61 \pm 0.38\) \\
PubMed          & \(77.00 \pm 0.59\)  & \(73.77 \pm 0.26\)  & \(25.32 \pm 1.19\)  & \(76.00 \pm 0.17\)  & \(75.68 \pm 0.37\) \\
Computers       & \(55.39 \pm 2.89\)  & \(58.49 \pm 2.93\)  & \(37.62 \pm 24.96\) & \(67.04 \pm 1.66\)  & \(37.70 \pm 19.20\) \\
Photo           & \(57.47 \pm 2.27\)  & \(52.15 \pm 5.30\)  & \(58.47 \pm 3.84\)  & \(61.90 \pm 0.43\)  & \(51.05 \pm 11.55\) \\
CoauthorCS      & \(69.13 \pm 1.21\)  & \(66.09 \pm 6.69\)  & \(72.75 \pm 0.46\)  & \(72.52 \pm 0.73\)  & \(73.43 \pm 0.32\) \\
CoauthorPhysics & \(76.23 \pm 1.03\)  & \(57.32 \pm 12.15\) & \(81.11 \pm 0.89\)  & \(80.50 \pm 0.75\)  & \(80.91 \pm 0.81\) \\
\bottomrule
\end{tabular}
\caption{Defense results: F1 (mean $\pm$ std, in \%).}
\label{tab:rq2_defense_7xN_F1}
\end{table*}


\begin{table*}[t]
\centering
\scriptsize
\setlength{\tabcolsep}{5pt}
\renewcommand{\arraystretch}{1.15}
\begin{tabular}{lccccc}
\toprule
Dataset & backdoorwm & randomwm & survivewm & imperceptiblewm & integrity \\
\midrule
Cora            & \(80.07 \pm 0.40\)  & \(76.13 \pm 1.17\)  & \(79.93 \pm 0.26\)  & \(71.97 \pm 0.31\)  & \(76.03 \pm 0.37\) \\
CiteSeer        & \(67.47 \pm 0.40\)  & \(67.90 \pm 0.57\)  & \(70.87 \pm 0.12\)  & \(60.30 \pm 1.00\)  & \(60.10 \pm 0.29\) \\
PubMed          & \(77.60 \pm 0.79\)  & \(74.13 \pm 0.26\)  & \(39.17 \pm 1.01\)  & \(76.37 \pm 0.12\)  & \(76.33 \pm 0.38\) \\
Computers       & \(68.27 \pm 2.74\)  & \(71.47 \pm 2.49\)  & \(45.97 \pm 31.21\) & \(78.90 \pm 0.73\)  & \(11.97 \pm 5.56\) \\
Photo           & \(92.73 \pm 1.50\)  & \(85.33 \pm 6.07\)  & \(93.20 \pm 3.89\)  & \(96.70 \pm 0.08\)  & \(61.80 \pm 20.58\) \\
CoauthorCS      & \(88.43 \pm 0.77\)  & \(89.27 \pm 2.46\)  & \(91.70 \pm 0.36\)  & \(89.63 \pm 0.76\)  & \(90.97 \pm 0.12\) \\
CoauthorPhysics & \(89.33 \pm 0.37\)  & \(68.60 \pm 10.96\) & \(90.97 \pm 0.33\)  & \(90.20 \pm 0.40\)  & \(90.60 \pm 0.65\) \\
\bottomrule
\end{tabular}
\caption{Defense results: Fidelity (mean $\pm$ std, in \%).}
\label{tab:rq2_defense_7xN_Fidelity}
\end{table*}

\begin{table*}[t]
\centering
\scriptsize
\setlength{\tabcolsep}{5pt}
\renewcommand{\arraystretch}{1.15}
\begin{tabular}{lccccc}
\toprule
Dataset & backdoorwm & randomwm & survivewm & imperceptiblewm & integrity \\
\midrule
Cora            & \(100.00 \pm 0.00\) & \(75.33 \pm 9.98\)  & \(54.07 \pm 5.79\)  & \(100.00 \pm 0.00\) & \(100.00 \pm 0.00\) \\
CiteSeer        & \(100.00 \pm 0.00\) & \(72.00 \pm 4.32\)  & \(55.72 \pm 4.51\)  & \(100.00 \pm 0.00\) & \(100.00 \pm 0.00\) \\
PubMed          & \(100.00 \pm 0.00\) & \(64.00 \pm 11.31\) & \(34.97 \pm 0.71\)  & \(100.00 \pm 0.00\) & \(100.00 \pm 0.00\) \\
Computers       & \(93.33 \pm 9.43\)  & \(94.67 \pm 1.89\)  & \(11.08 \pm 0.81\)  & \(100.00 \pm 0.00\) & \(20.00 \pm 40.00\) \\
Photo           & \(100.00 \pm 0.00\) & \(98.67 \pm 0.94\)  & \(13.25 \pm 1.10\)  & \(100.00 \pm 0.00\) & \(10.00 \pm 30.00\) \\
CoauthorCS      & \(100.00 \pm 0.00\) & \(45.33 \pm 3.77\)  & \(8.51 \pm 0.34\)   & \(100.00 \pm 0.00\) & \(33.33 \pm 47.14\) \\
CoauthorPhysics & \(100.00 \pm 0.00\) & \(56.67 \pm 13.70\) & \(21.76 \pm 0.49\)  & \(100.00 \pm 0.00\)                  & \(66.67 \pm 47.14\) \\
\bottomrule
\end{tabular}
\caption{Target-side verification (mean $\pm$ SD, \%). For Integrity this is the bit-flip detection rate, not fingerprint self-agreement. Integrity uses ten seeds on Computers and Photo.}
\label{tab:rq2_defense_7xN_Owner_verif}
\end{table*}

\begin{figure}[t]
\centering
\includegraphics[width=0.7\textwidth]{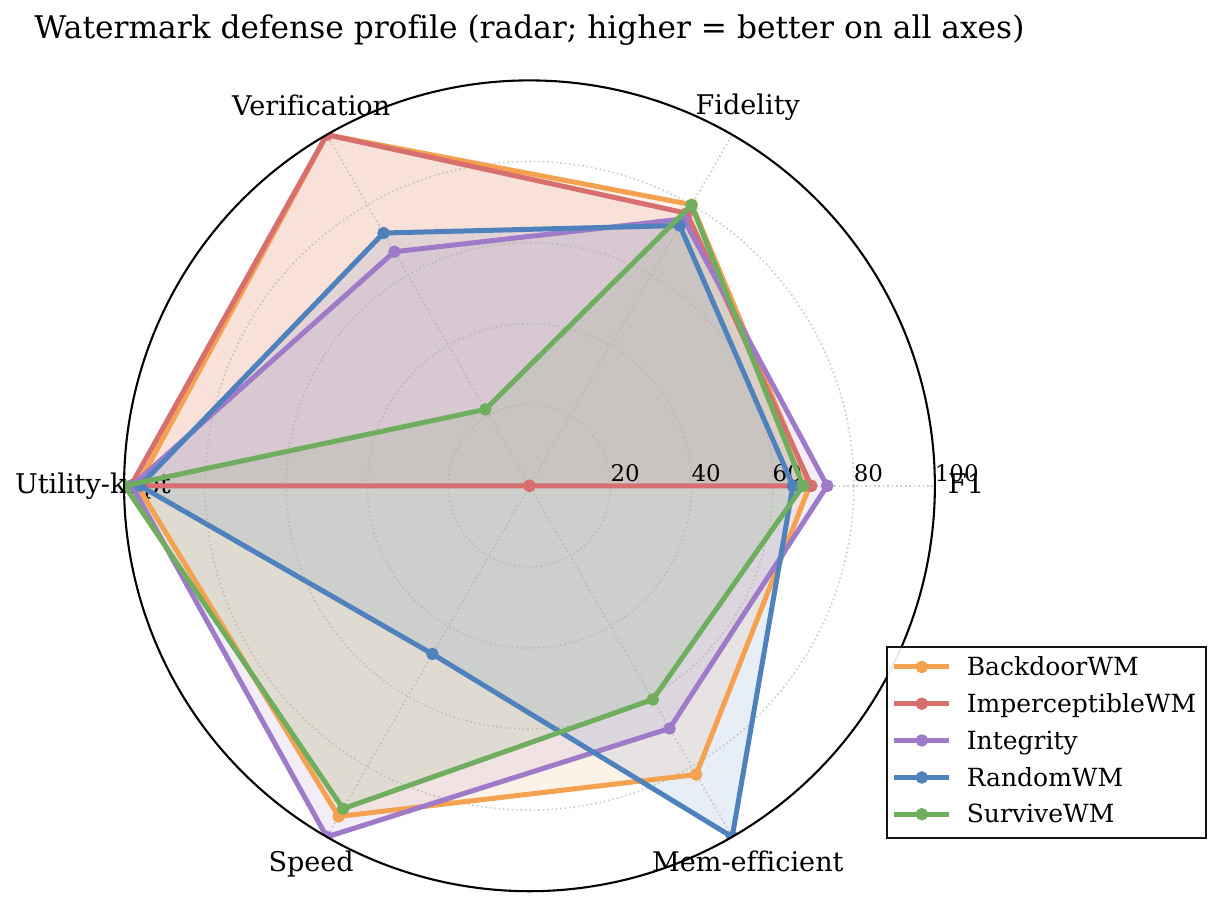}
\caption{Radar profile of the five watermarking and integrity defenses across six axes. F1, fidelity, verification, and utility-kept come from Tables~\ref{tab:rq2_defense_7xN_F1}--\ref{tab:rq2_defense_7xN_Owner_verif}; speed and memory-efficiency are obtained from the RQ4 numbers (Tables~\ref{tab:rq4_attack_efficiency_time}--\ref{tab:rq4_defense_efficiency_time_compact} and Figure~\ref{fig:rq4_mem_wrapL}) by inverting and log-normalising to a $0$--$100$ scale. Higher is better on every axis. No defense dominates; \texttt{BackdoorWM} is the largest envelope, \texttt{ImperceptibleWM} pays for perfect verification with the worst speed and memory profile.}
\label{fig:app_rq2_radar}
\end{figure}

\paragraph{Watermark profile across six axes.}
Tables~\ref{tab:rq2_defense_7xN_F1}--\ref{tab:rq2_defense_7xN_Owner_verif} report the three primary metrics separately. To make the trade-offs across the five watermarking and integrity defenses visible at a glance, Figure~\ref{fig:app_rq2_radar} reports the same numbers as a radar chart together with two efficiency axes (training time and peak memory, both inverted to a higher-is-better score on a log scale to match the other axes). Three patterns become explicit which the per-metric tables only show implicitly. \emph{First}, no defense dominates on every axis. \texttt{BackdoorWM} encloses the largest area on the verification, fidelity, and utility axes but sits in the middle of the speed and memory axes. The radar axes summarize performance and compute costs separately. \texttt{ImperceptibleWM} achieves perfect verification at the cost of being the slowest and heaviest defense, which gives it a visibly indented profile on the speed and memory axes. \emph{Second}, \texttt{SurviveWM} keeps utility almost unchanged but loses on the verification axis, which is consistent with the bimodal verification numbers in Table~\ref{tab:rq2_defense_7xN_Owner_verif}; this is a clear stability--verification trade-off rather than a generic weakness. \emph{Third}, the gap between \texttt{RandomWM} and the other watermarks is structural rather than a per-dataset artefact: \texttt{RandomWM} loses on every metric except utility, which suggests the random-graph watermark is not competitive with trigger-based or query-based mechanisms under~our~protocol.

\FloatBarrier
\subsubsection{Integrity target training}
For \texttt{Integrity}, we apply the bit-flip verifier to a copy of the model so that it cannot modify the target subsequently queried by the attacker. On \textit{Computers} and \textit{Photo}, we also aligned the optimizer with the shared configuration (learning rate $0.01$, weight decay $5\times10^{-4}$). Across ten seeds on each dataset, target accuracy was $44.5\%$ and $86.4\%$, respectively, with no constant-class targets. The remaining eight datasets use the isolated verifier with~three~seeds.

\begin{center}\begin{minipage}{\linewidth}\centering\small
\captionof{table}{Integrity target accuracy (\%) before and after optimizer alignment.}\label{tab:optalign}
\begin{tabular}{@{}lrrr@{}}\toprule
Dataset & Original accuracy & Aligned accuracy & Collapsed after alignment\\ \midrule
Computers & 16.5 & 44.5 & 0/10\\ Photo & 49.1 & 86.4 & 0/10\\ \bottomrule
\end{tabular}\end{minipage}\end{center}

\subsection{Baseline utility across backbones}
\label{app:baseline}

To allow utility-drop numbers in the main text to be compared across defenses that use different backbones, Table~\ref{tab:app_baseline_utility} reports the test accuracy of an undefended target on each of the ten datasets and for each of the three backbones used in the benchmark. The first column reports a DGL GCN with hidden dimension 16, which is the backbone used by every original watermarking defense and by every information-limiting defense. The second column reports a DGL GraphSAGE with hidden dimension 128, which is the backbone used by \texttt{RandomWM}. The third column reports a PyG GCN with hidden dimension 128, which is the backbone used by \texttt{ImperceptibleWM}.

\begin{table*}[t]
\centering
\caption{Baseline utility (test accuracy, \%) for an undefended target across three backbones. Mean $\pm$ standard deviation \mbox{over three seeds}.}
\label{tab:app_baseline_utility}
\scriptsize
\setlength{\tabcolsep}{6pt}
\renewcommand{\arraystretch}{0.96}
\begin{tabular*}{0.95\textwidth}{@{}l@{\extracolsep{\fill}}ccc@{}}
\toprule
\textbf{Dataset} & \textbf{GCN (DGL, hidden=16)} & \textbf{GraphSAGE (DGL, hidden=128)} & \textbf{GCN (PyG, hidden=128)} \\
\midrule
Cora            & 79.4$\,\pm\,$0.5 & 79.1$\,\pm\,$0.3 & 80.7$\,\pm\,$0.4 \\
CiteSeer        & 67.8$\,\pm\,$1.1 & 70.1$\,\pm\,$0.2 & 70.9$\,\pm\,$0.4 \\
PubMed          & 78.0$\,\pm\,$0.6 & 77.1$\,\pm\,$0.3 & 79.5$\,\pm\,$0.2 \\
Computers       & 44.6$\,\pm\,$22.4 & 59.8$\,\pm\,$18.4 & 75.2$\,\pm\,$0.7 \\
Photo           & 90.1$\,\pm\,$4.8  & 94.5$\,\pm\,$0.5  & 95.3$\,\pm\,$0.2 \\
CoauthorCS      & 87.4$\,\pm\,$0.3  & 92.9$\,\pm\,$0.3  & 90.3$\,\pm\,$0.2 \\
CoauthorPhysics & 89.3$\,\pm\,$0.4  & 91.2$\,\pm\,$0.1  & 89.8$\,\pm\,$0.2 \\
OGBN-Arxiv      & 37.7$\,\pm\,$3.2  & 54.9$\,\pm\,$0.0  & 52.3$\,\pm\,$0.1 \\
RomanEmpire     & 42.8$\,\pm\,$0.2  & 77.5$\,\pm\,$0.5  & 46.0$\,\pm\,$0.3 \\
AmazonRatings   & 41.7$\,\pm\,$0.3  & 45.5$\,\pm\,$0.5  & 43.4$\,\pm\,$0.3 \\
\bottomrule
\end{tabular*}
\end{table*}

The baseline-utility table reveals one pattern that has direct consequences for any cross-defense comparison in the main text: the per-dataset gap between the two backbones used by watermark methods can be huge, and it is largest exactly on the graphs where defenses are evaluated as ``most informative''. On \textit{RomanEmpire}, the GraphSAGE-128 baseline of $77.5\,\%$ is $34.7$\,pp higher than the GCN-16 baseline of $42.8\,\%$; on \textit{Computers} the gap is $15.2$\,pp; on \textit{OGBN-Arxiv} it is $17.2$\,pp. This means that any utility-drop number in the main text that compares \texttt{RandomWM} (GraphSAGE-128) against the GCN-16 baseline implicitly attributes the entire backbone gap to the defense, which would over-estimate \texttt{RandomWM}'s damage by $15$--$35$\,pp on these three graphs; we therefore always compare each defense against its own matched-backbone baseline rather than against a single shared~GCN~baseline.

\subsection{Attack effectiveness on the three additional datasets}
\label{app:rq1_new_datasets}

We report the per-attack and per-budget fidelity for all twelve attacks on the three additional graphs (RomanEmpire, AmazonRatings, OGBN-Arxiv) in the four regimes. Constant values across budgets reflect attacks whose surrogate is independent of the query budget (e.g., the data-free \texttt{DFEA} variants, which always train on the same synthesized queries).

\begin{table*}[t]
\centering
\caption{Detailed RQ1 results on \textit{RomanEmpire}. Fidelity (\%) at five budget multipliers across four regimes. Mean $\pm$ standard deviation \mbox{over three seeds}.}
\label{tab:app_rq1_RomanEmpire_full}
\tiny
\setlength{\tabcolsep}{2pt}
\renewcommand{\arraystretch}{0.9}
\begin{subtable}[t]{0.48\textwidth}
\centering
\caption{Regime=\texttt{both}}
\resizebox{\linewidth}{!}{%
\begin{tabular}{lccccc}
\toprule
\textbf{Attack} & 0.05 & 0.10 & 0.25 & 0.50 & 1.00 \\
\midrule
\texttt{MEA0}      & 72.9$\pm$0.4 & 75.9$\pm$1.7 & 77.0$\pm$0.8 & 77.2$\pm$1.5 & 77.4$\pm$1.5 \\
\texttt{MEA1}      & 54.0$\pm$2.2 & 59.7$\pm$3.5 & 68.7$\pm$1.1 & 72.6$\pm$1.0 & 77.4$\pm$1.5 \\
\texttt{MEA2}      & 27.0$\pm$0.1 & 34.0$\pm$1.3 & 47.8$\pm$0.7 & 49.8$\pm$1.9 & 32.7$\pm$9.8 \\
\texttt{MEA3}      & 61.2$\pm$1.0 & 65.2$\pm$1.1 & 74.1$\pm$1.1 & 77.2$\pm$1.6 & 77.4$\pm$1.5 \\
\texttt{MEA4}      & 63.6$\pm$1.0 & 70.9$\pm$2.1 & 73.3$\pm$2.6 & 76.0$\pm$1.1 & 77.4$\pm$1.5 \\
\texttt{MEA5}      & 65.8$\pm$1.2 & 70.0$\pm$0.9 & 76.7$\pm$1.2 & 77.6$\pm$1.0 & 77.4$\pm$1.5 \\
\texttt{AdvMEA}    & 22.0$\pm$6.5 & 23.5$\pm$6.5 & 21.9$\pm$6.2 & 19.1$\pm$3.3 & 15.7$\pm$3.7 \\
\texttt{CEGA}      & 72.1$\pm$0.5 & 73.3$\pm$1.2 & 73.4$\pm$2.3 & 71.4$\pm$2.1 & 71.4$\pm$2.1 \\
\texttt{Realistic} & 53.7$\pm$2.3 & 57.5$\pm$2.1 & 58.2$\pm$1.5 & 57.9$\pm$1.9 & 75.4$\pm$1.9 \\
\texttt{DFEA\_I}   & 71.7$\pm$1.1 & 71.4$\pm$2.1 & 72.1$\pm$2.7 & 72.3$\pm$2.7 & 72.7$\pm$2.2 \\
\texttt{DFEA\_II}  & 69.2$\pm$0.7 & 74.9$\pm$0.5 & 80.8$\pm$1.0 & 83.3$\pm$0.4 & 84.3$\pm$0.3 \\
\texttt{DFEA\_III} & 73.4$\pm$0.3 & 76.2$\pm$0.6 & 75.7$\pm$2.1 & 77.2$\pm$1.1 & 77.3$\pm$1.4 \\
\bottomrule
\end{tabular}
}
\end{subtable}
\hfill
\begin{subtable}[t]{0.48\textwidth}
\centering
\caption{Regime=\texttt{x\_only}}
\resizebox{\linewidth}{!}{%
\begin{tabular}{lccccc}
\toprule
\textbf{Attack} & 0.05 & 0.10 & 0.25 & 0.50 & 1.00 \\
\midrule
\texttt{MEA0}      & 72.5$\pm$1.1 & 76.5$\pm$0.6 & 75.8$\pm$2.6 & 77.4$\pm$0.6 & 77.4$\pm$1.5 \\
\texttt{MEA1}      & 54.0$\pm$2.2 & 59.7$\pm$3.5 & 68.7$\pm$1.1 & 72.6$\pm$1.0 & 77.4$\pm$1.5 \\
\texttt{MEA2}      & 26.9$\pm$0.6 & 33.4$\pm$0.3 & 47.1$\pm$1.4 & 47.3$\pm$2.5 & 32.7$\pm$9.8 \\
\texttt{MEA3}      & 61.2$\pm$0.4 & 65.2$\pm$0.9 & 74.3$\pm$1.1 & 77.4$\pm$1.4 & 77.4$\pm$1.5 \\
\texttt{MEA4}      & 63.8$\pm$0.4 & 70.9$\pm$2.1 & 73.4$\pm$2.6 & 76.0$\pm$1.1 & 77.4$\pm$1.5 \\
\texttt{MEA5}      & 65.9$\pm$0.5 & 69.9$\pm$0.3 & 76.1$\pm$2.5 & 77.2$\pm$1.5 & 77.4$\pm$1.5 \\
\texttt{AdvMEA}    & 23.2$\pm$7.3 & 23.0$\pm$7.1 & 21.5$\pm$4.3 & 19.3$\pm$3.8 & 16.2$\pm$3.9 \\
\texttt{CEGA}      & 72.1$\pm$1.3 & 72.0$\pm$2.9 & 73.6$\pm$2.9 & 71.4$\pm$2.1 & 71.4$\pm$2.1 \\
\texttt{Realistic} & 54.4$\pm$1.5 & 56.7$\pm$1.7 & 58.3$\pm$2.0 & 57.9$\pm$2.0 & 75.7$\pm$1.8 \\
\texttt{DFEA\_I}   & 71.7$\pm$1.1 & 71.4$\pm$2.1 & 72.1$\pm$2.7 & 72.3$\pm$2.7 & 72.7$\pm$2.2 \\
\texttt{DFEA\_II}  & 69.2$\pm$0.7 & 74.9$\pm$0.5 & 80.8$\pm$1.0 & 83.3$\pm$0.4 & 84.3$\pm$0.3 \\
\texttt{DFEA\_III} & 73.4$\pm$0.3 & 76.2$\pm$0.6 & 75.7$\pm$2.1 & 77.2$\pm$1.1 & 77.3$\pm$1.4 \\
\bottomrule
\end{tabular}
}
\end{subtable}
\medskip
\begin{subtable}[t]{0.48\textwidth}
\centering
\caption{Regime=\texttt{a\_only}}
\resizebox{\linewidth}{!}{%
\begin{tabular}{lccccc}
\toprule
\textbf{Attack} & 0.05 & 0.10 & 0.25 & 0.50 & 1.00 \\
\midrule
\texttt{MEA0}      & 73.3$\pm$0.8 & 75.4$\pm$0.3 & 77.4$\pm$0.4 & 77.3$\pm$1.7 & 77.4$\pm$1.5 \\
\texttt{MEA1}      & 54.0$\pm$2.2 & 59.7$\pm$3.5 & 68.7$\pm$1.1 & 72.6$\pm$1.0 & 77.4$\pm$1.5 \\
\texttt{MEA2}      & 25.5$\pm$1.2 & 33.7$\pm$0.2 & 47.7$\pm$0.8 & 50.5$\pm$2.1 & 32.7$\pm$9.8 \\
\texttt{MEA3}      & 62.1$\pm$1.3 & 66.7$\pm$0.4 & 75.2$\pm$0.7 & 77.4$\pm$1.7 & 77.4$\pm$1.5 \\
\texttt{MEA4}      & 63.3$\pm$1.0 & 70.6$\pm$2.4 & 73.4$\pm$2.5 & 76.0$\pm$1.1 & 77.4$\pm$1.5 \\
\texttt{MEA5}      & 65.5$\pm$0.4 & 70.3$\pm$1.0 & 76.7$\pm$1.3 & 77.5$\pm$1.1 & 77.4$\pm$1.5 \\
\texttt{AdvMEA}    & 22.4$\pm$6.5 & 22.8$\pm$5.9 & 20.8$\pm$6.3 & 17.8$\pm$4.2 & 15.1$\pm$4.1 \\
\texttt{CEGA}      & 71.5$\pm$2.6 & 72.9$\pm$2.9 & 73.4$\pm$2.7 & 71.4$\pm$2.1 & 71.4$\pm$2.1 \\
\texttt{Realistic} & 55.3$\pm$2.2 & 58.1$\pm$1.0 & 56.8 & 58.1$\pm$1.9 & 75.7$\pm$1.7 \\
\texttt{DFEA\_I}   & 71.7$\pm$1.1 & 71.4$\pm$2.1 & 72.1$\pm$2.7 & 72.3$\pm$2.7 & 72.7$\pm$2.2 \\
\texttt{DFEA\_II}  & 69.2$\pm$0.7 & 74.9$\pm$0.5 & 80.8$\pm$1.0 & 83.3$\pm$0.4 & 84.3$\pm$0.3 \\
\texttt{DFEA\_III} & 73.4$\pm$0.3 & 76.2$\pm$0.6 & 75.7$\pm$2.1 & 77.2$\pm$1.1 & 77.3$\pm$1.4 \\
\bottomrule
\end{tabular}
}
\end{subtable}
\hfill
\begin{subtable}[t]{0.48\textwidth}
\centering
\caption{Regime=\texttt{data\_free}}
\resizebox{\linewidth}{!}{%
\begin{tabular}{lccccc}
\toprule
\textbf{Attack} & 0.05 & 0.10 & 0.25 & 0.50 & 1.00 \\
\midrule
\texttt{MEA0}      & 72.8$\pm$1.1 & 74.7$\pm$1.3 & 76.0$\pm$1.5 & 76.9$\pm$2.1 & 77.4$\pm$1.5 \\
\texttt{MEA1}      & 54.0$\pm$2.2 & 59.7$\pm$3.5 & 68.7$\pm$1.1 & 72.6$\pm$1.0 & 77.4$\pm$1.5 \\
\texttt{MEA2}      & 26.4$\pm$0.5 & 33.8$\pm$0.8 & 47.4$\pm$0.5 & 51.1$\pm$2.1 & 32.7$\pm$9.8 \\
\texttt{MEA3}      & 62.6$\pm$0.3 & 65.8$\pm$0.7 & 74.0$\pm$1.5 & 77.2$\pm$1.5 & 77.4$\pm$1.5 \\
\texttt{MEA4}      & 63.1$\pm$1.1 & 71.1$\pm$1.7 & 73.3$\pm$2.3 & 75.8$\pm$1.5 & 77.4$\pm$1.5 \\
\texttt{MEA5}      & 64.8$\pm$1.2 & 69.5$\pm$1.0 & 76.2$\pm$2.3 & 77.2$\pm$1.3 & 77.4$\pm$1.5 \\
\texttt{AdvMEA}    & 23.4$\pm$6.6 & 23.0$\pm$7.1 & 21.7$\pm$6.9 & 18.3$\pm$4.6 & 15.3$\pm$4.4 \\
\texttt{CEGA}      & 71.2$\pm$2.2 & 73.0$\pm$1.9 & 73.3$\pm$2.7 & 71.4$\pm$2.1 & 71.4$\pm$2.1 \\
\texttt{Realistic} & 54.4$\pm$2.3 & 56.3$\pm$2.4 & 58.3$\pm$1.4 & 57.8$\pm$2.5 & 75.6$\pm$1.6 \\
\texttt{DFEA\_I}   & 71.7$\pm$1.1 & 71.4$\pm$2.1 & 72.1$\pm$2.7 & 72.3$\pm$2.7 & 72.7$\pm$2.2 \\
\texttt{DFEA\_II}  & 69.2$\pm$0.7 & 74.9$\pm$0.5 & 80.8$\pm$1.0 & 83.3$\pm$0.4 & 84.3$\pm$0.3 \\
\texttt{DFEA\_III} & 73.4$\pm$0.3 & 76.2$\pm$0.6 & 75.7$\pm$2.1 & 77.2$\pm$1.1 & 77.3$\pm$1.4 \\
\bottomrule
\end{tabular}
}
\end{subtable}
\end{table*}

\begin{table*}[t]
\centering
\caption{Detailed RQ1 results on \textit{AmazonRatings}. Fidelity (\%) at five budget multipliers across four regimes. Mean $\pm$ standard deviation \mbox{over three seeds}.}
\label{tab:app_rq1_AmazonRatings_full}
\tiny
\setlength{\tabcolsep}{2pt}
\renewcommand{\arraystretch}{0.9}
\begin{subtable}[t]{0.48\textwidth}
\centering
\caption{Regime=\texttt{both}}
\resizebox{\linewidth}{!}{%
\begin{tabular}{lccccc}
\toprule
\textbf{Attack} & 0.05 & 0.10 & 0.25 & 0.50 & 1.00 \\
\midrule
\texttt{MEA0}      & 92.3$\pm$0.6 & 93.1$\pm$1.4 & 93.8$\pm$0.7 & 93.9$\pm$0.5 & 94.4$\pm$1.0 \\
\texttt{MEA1}      & 88.8$\pm$1.0 & 91.9$\pm$0.5 & 94.2$\pm$0.6 & 94.8$\pm$0.7 & 94.4$\pm$1.0 \\
\texttt{MEA2}      & 82.0$\pm$1.4 & 87.3$\pm$0.6 & 92.4$\pm$0.8 & 93.9$\pm$1.3 & 94.5$\pm$0.5 \\
\texttt{MEA3}      & 88.0$\pm$0.7 & 91.2$\pm$1.3 & 93.8$\pm$1.3 & 94.6$\pm$1.0 & 94.4$\pm$1.0 \\
\texttt{MEA4}      & 85.1$\pm$0.4 & 87.1$\pm$0.5 & 90.1$\pm$0.8 & 92.4$\pm$1.1 & 94.4$\pm$1.0 \\
\texttt{MEA5}      & 89.9$\pm$0.6 & 91.4$\pm$1.1 & 94.7$\pm$0.9 & 94.5$\pm$0.9 & 94.4$\pm$1.0 \\
\texttt{AdvMEA}    & 70.4$\pm$1.5 & 66.2$\pm$7.3 & 68.7$\pm$3.4 & 70.1$\pm$1.3 & 66.9$\pm$5.4 \\
\texttt{CEGA}      & 87.1$\pm$0.9 & 87.3$\pm$1.2 & 89.1$\pm$1.2 & 89.7$\pm$2.0 & 89.7$\pm$2.0 \\
\texttt{Realistic} & 89.9$\pm$2.7 & 90.8$\pm$3.0 & 90.6$\pm$2.6 & 92.8$\pm$1.0 & 94.9$\pm$0.3 \\
\texttt{DFEA\_I}   & 88.6$\pm$1.1 & 89.2$\pm$2.3 & 89.6$\pm$1.7 & 89.6$\pm$1.4 & 89.4$\pm$1.3 \\
\texttt{DFEA\_II}  & 90.4$\pm$1.2 & 92.6$\pm$0.6 & 94.5$\pm$0.7 & 95.1$\pm$0.6 & 95.6$\pm$0.2 \\
\texttt{DFEA\_III} & 92.4$\pm$0.9 & 93.5$\pm$1.2 & 94.6$\pm$0.5 & 94.5$\pm$0.6 & 94.5$\pm$0.7 \\
\bottomrule
\end{tabular}
}
\end{subtable}
\hfill
\begin{subtable}[t]{0.48\textwidth}
\centering
\caption{Regime=\texttt{x\_only}}
\resizebox{\linewidth}{!}{%
\begin{tabular}{lccccc}
\toprule
\textbf{Attack} & 0.05 & 0.10 & 0.25 & 0.50 & 1.00 \\
\midrule
\texttt{MEA0}      & 92.1$\pm$0.6 & 93.5$\pm$0.9 & 94.1$\pm$1.2 & 94.3$\pm$0.9 & 94.4$\pm$1.0 \\
\texttt{MEA1}      & 88.8$\pm$1.0 & 91.9$\pm$0.5 & 94.1$\pm$0.7 & 94.8$\pm$0.7 & 94.4$\pm$1.0 \\
\texttt{MEA2}      & 82.7$\pm$1.5 & 86.9$\pm$1.7 & 91.8$\pm$0.9 & 93.7$\pm$1.3 & 94.5$\pm$0.5 \\
\texttt{MEA3}      & 87.4$\pm$2.1 & 90.5$\pm$1.3 & 93.8$\pm$0.4 & 94.6$\pm$1.1 & 94.4$\pm$1.0 \\
\texttt{MEA4}      & 85.8$\pm$0.5 & 86.7$\pm$0.5 & 90.1$\pm$0.8 & 92.4$\pm$1.1 & 94.4$\pm$1.0 \\
\texttt{MEA5}      & 89.1$\pm$1.4 & 92.0$\pm$1.2 & 94.4$\pm$0.6 & 94.2$\pm$0.9 & 94.4$\pm$1.0 \\
\texttt{AdvMEA}    & 69.1$\pm$2.4 & 68.0$\pm$5.5 & 69.5$\pm$1.7 & 65.3$\pm$7.2 & 67.0$\pm$4.1 \\
\texttt{CEGA}      & 88.2$\pm$0.7 & 87.4$\pm$1.1 & 89.5$\pm$1.7 & 89.7$\pm$2.0 & 89.7$\pm$2.0 \\
\texttt{Realistic} & 90.7$\pm$1.9 & 91.0$\pm$2.4 & 92.2$\pm$1.0 & 93.5$\pm$0.5 & 94.9$\pm$0.4 \\
\texttt{DFEA\_I}   & 88.6$\pm$1.1 & 89.2$\pm$2.3 & 89.6$\pm$1.7 & 89.6$\pm$1.4 & 89.4$\pm$1.3 \\
\texttt{DFEA\_II}  & 90.4$\pm$1.2 & 92.5$\pm$0.6 & 94.5$\pm$0.7 & 95.1$\pm$0.6 & 95.6$\pm$0.2 \\
\texttt{DFEA\_III} & 92.4$\pm$1.0 & 93.5$\pm$1.2 & 94.6$\pm$0.5 & 94.4$\pm$0.7 & 94.5$\pm$0.7 \\
\bottomrule
\end{tabular}
}
\end{subtable}
\medskip
\begin{subtable}[t]{0.48\textwidth}
\centering
\caption{Regime=\texttt{a\_only}}
\resizebox{\linewidth}{!}{%
\begin{tabular}{lccccc}
\toprule
\textbf{Attack} & 0.05 & 0.10 & 0.25 & 0.50 & 1.00 \\
\midrule
\texttt{MEA0}      & 92.3$\pm$0.6 & 93.0$\pm$0.7 & 94.6$\pm$0.9 & 94.2$\pm$0.9 & 94.4$\pm$1.0 \\
\texttt{MEA1}      & 88.8$\pm$1.0 & 91.9$\pm$0.5 & 94.1$\pm$0.7 & 94.8$\pm$0.7 & 94.4$\pm$1.0 \\
\texttt{MEA2}      & 82.2$\pm$0.7 & 87.0$\pm$1.8 & 92.2$\pm$0.8 & 94.1$\pm$1.2 & 94.5$\pm$0.5 \\
\texttt{MEA3}      & 88.6$\pm$0.1 & 90.7$\pm$1.2 & 94.0$\pm$0.9 & 94.7$\pm$1.1 & 94.4$\pm$1.0 \\
\texttt{MEA4}      & 85.3$\pm$0.3 & 87.2$\pm$0.4 & 90.1$\pm$0.8 & 92.4$\pm$1.1 & 94.4$\pm$1.0 \\
\texttt{MEA5}      & 89.4$\pm$0.9 & 91.5$\pm$1.7 & 94.3$\pm$0.8 & 94.5$\pm$1.0 & 94.4$\pm$1.0 \\
\texttt{AdvMEA}    & 70.4$\pm$1.5 & 69.5$\pm$3.0 & 67.3$\pm$5.8 & 63.9$\pm$9.1 & 68.2$\pm$3.5 \\
\texttt{CEGA}      & 87.5$\pm$1.2 & 88.8$\pm$1.9 & 88.8$\pm$2.0 & 89.7$\pm$2.0 & 89.7$\pm$2.0 \\
\texttt{Realistic} & 89.5$\pm$2.2 & 92.3$\pm$0.7 & 92.7$\pm$1.0 & 93.6$\pm$0.3 & 95.0$\pm$0.3 \\
\texttt{DFEA\_I}   & 88.6$\pm$1.1 & 89.2$\pm$2.3 & 89.6$\pm$1.7 & 89.6$\pm$1.4 & 89.4$\pm$1.3 \\
\texttt{DFEA\_II}  & 90.4$\pm$1.3 & 92.5$\pm$0.6 & 94.5$\pm$0.7 & 95.2$\pm$0.5 & 95.6$\pm$0.2 \\
\texttt{DFEA\_III} & 92.4$\pm$0.9 & 93.5$\pm$1.3 & 94.6$\pm$0.5 & 94.5$\pm$0.6 & 94.5$\pm$0.7 \\
\bottomrule
\end{tabular}
}
\end{subtable}
\hfill
\begin{subtable}[t]{0.48\textwidth}
\centering
\caption{Regime=\texttt{data\_free}}
\resizebox{\linewidth}{!}{%
\begin{tabular}{lccccc}
\toprule
\textbf{Attack} & 0.05 & 0.10 & 0.25 & 0.50 & 1.00 \\
\midrule
\texttt{MEA0}      & 91.7$\pm$0.5 & 93.3$\pm$1.0 & 93.5$\pm$0.7 & 94.3$\pm$0.9 & 94.4$\pm$1.0 \\
\texttt{MEA1}      & 88.8$\pm$1.0 & 91.9$\pm$0.5 & 94.1$\pm$0.7 & 94.8$\pm$0.7 & 94.4$\pm$1.0 \\
\texttt{MEA2}      & 81.8$\pm$0.9 & 86.6$\pm$0.7 & 91.9$\pm$1.0 & 93.9$\pm$1.0 & 94.5$\pm$0.6 \\
\texttt{MEA3}      & 87.9$\pm$0.5 & 91.0$\pm$1.5 & 94.0$\pm$1.0 & 94.6$\pm$1.0 & 94.4$\pm$1.0 \\
\texttt{MEA4}      & 85.4$\pm$0.4 & 87.0$\pm$0.6 & 90.1$\pm$0.8 & 92.4$\pm$1.1 & 94.4$\pm$1.0 \\
\texttt{MEA5}      & 89.5$\pm$1.0 & 91.5$\pm$1.6 & 94.4$\pm$0.7 & 94.6$\pm$0.9 & 94.4$\pm$1.0 \\
\texttt{AdvMEA}    & 70.0$\pm$2.0 & 67.3$\pm$5.0 & 68.0$\pm$3.5 & 64.7$\pm$8.0 & 67.5$\pm$4.0 \\
\texttt{CEGA}      & 87.6$\pm$1.0 & 87.8$\pm$1.5 & 89.1$\pm$1.5 & 89.7$\pm$2.0 & 89.7$\pm$2.0 \\
\texttt{Realistic} & 89.7$\pm$2.4 & 91.5$\pm$2.0 & 91.7$\pm$1.5 & 93.0$\pm$0.7 & 94.9$\pm$0.4 \\
\texttt{DFEA\_I}   & 88.6$\pm$1.1 & 89.2$\pm$2.3 & 89.6$\pm$1.7 & 89.6$\pm$1.4 & 89.4$\pm$1.3 \\
\texttt{DFEA\_II}  & 90.3$\pm$1.3 & 92.6$\pm$0.6 & 94.5$\pm$0.7 & 95.2$\pm$0.5 & 95.6$\pm$0.2 \\
\texttt{DFEA\_III} & 92.4$\pm$1.0 & 93.5$\pm$1.2 & 94.6$\pm$0.5 & 94.4$\pm$0.7 & 94.5$\pm$0.7 \\
\bottomrule
\end{tabular}
}
\end{subtable}
\end{table*}

\begin{table*}[t]
\centering
\caption{Detailed RQ1 results on \textit{OGBN-Arxiv}. Fidelity (\%) at five budget multipliers across four regimes. Mean $\pm$ standard deviation over three seeds. The 169{,}343-node graph requires sub-sampled edge construction at large budgets for the \texttt{Realistic} attack and the data-free variants.}
\label{tab:app_rq1_OGBNArxiv_full}
\tiny
\setlength{\tabcolsep}{2pt}
\renewcommand{\arraystretch}{0.9}
\begin{subtable}[t]{0.48\textwidth}
\centering
\caption{Regime=\texttt{both}}
\resizebox{\linewidth}{!}{%
\begin{tabular}{lccccc}
\toprule
\textbf{Attack} & 0.05 & 0.10 & 0.25 & 0.50 & 1.00 \\
\midrule
\texttt{MEA0}      & 77.2$\pm$4.2 & 77.4$\pm$4.7 & 77.1$\pm$4.6 & 77.2$\pm$4.7 & 77.2$\pm$4.8 \\
\texttt{MEA1}      & 74.2$\pm$1.2 & 79.0$\pm$1.3 & 81.7$\pm$2.9 & 81.6$\pm$3.4 & 77.2$\pm$4.8 \\
\texttt{MEA2}      & 52.8$\pm$3.6 & 52.8$\pm$3.6 & 52.8$\pm$3.6 & 52.8$\pm$3.6 & 52.8$\pm$3.6 \\
\texttt{MEA3}      & 64.0$\pm$4.5 & 66.9$\pm$5.4 & 74.2$\pm$5.1 & 77.5$\pm$4.8 & 77.3$\pm$4.8 \\
\texttt{MEA4}      & 62.2$\pm$4.8 & 63.2$\pm$5.2 & 65.5$\pm$6.0 & 70.6$\pm$4.9 & 77.1$\pm$5.1 \\
\texttt{MEA5}      & 64.9$\pm$4.1 & 69.5$\pm$4.8 & 77.2$\pm$4.8 & 77.6$\pm$4.4 & 77.3$\pm$4.8 \\
\texttt{AdvMEA}    & 26.9$\pm$22.7 & 28.7$\pm$21.1 & 26.2$\pm$21.2 & 23.0$\pm$17.1 & 20.4$\pm$17.2 \\
\texttt{CEGA}      & 77.9$\pm$3.0 & 77.7$\pm$4.1 & 77.4$\pm$4.6 & 75.5$\pm$4.2 & 75.5$\pm$4.2 \\
\texttt{Realistic} & 74.5$\pm$3.7 & 74.4$\pm$2.7 & 75.3$\pm$2.7 & 75.6$\pm$3.0 & 75.0$\pm$3.2 \\
\texttt{DFEA\_I}   & 80.7$\pm$4.5 & 80.8$\pm$4.6 & 81.0$\pm$4.2 & 81.1$\pm$4.0 & 81.0$\pm$3.9 \\
\texttt{DFEA\_II}  & 75.1$\pm$1.2 & 76.0$\pm$1.6 & 76.4$\pm$1.5 & 76.3$\pm$1.3 & 76.7$\pm$1.3 \\
\texttt{DFEA\_III} & 76.9$\pm$4.5 & 77.6$\pm$4.4 & 77.7$\pm$4.6 & 77.6$\pm$4.5 & 77.1$\pm$4.7 \\
\bottomrule
\end{tabular}
}
\end{subtable}
\hfill
\begin{subtable}[t]{0.48\textwidth}
\centering
\caption{Regime=\texttt{x\_only}}
\resizebox{\linewidth}{!}{%
\begin{tabular}{lccccc}
\toprule
\textbf{Attack} & 0.05 & 0.10 & 0.25 & 0.50 & 1.00 \\
\midrule
\texttt{MEA0}      & 76.6$\pm$4.7 & 77.5$\pm$4.6 & 77.4$\pm$4.1 & 77.3$\pm$4.8 & 77.2$\pm$4.8 \\
\texttt{MEA1}      & 74.2$\pm$1.2 & 79.0$\pm$1.2 & 81.7$\pm$2.9 & 81.6$\pm$3.4 & 77.3$\pm$4.8 \\
\texttt{MEA2}      & 52.8$\pm$3.5 & 52.8$\pm$3.6 & 52.8$\pm$3.6 & 52.8$\pm$3.6 & 52.8$\pm$3.6 \\
\texttt{MEA3}      & 64.5$\pm$4.8 & 67.3$\pm$4.7 & 74.3$\pm$4.8 & 77.4$\pm$4.8 & 77.2$\pm$4.8 \\
\texttt{MEA4}      & 62.6$\pm$4.8 & 63.3$\pm$5.1 & 65.4$\pm$5.6 & 70.6$\pm$5.0 & 77.1$\pm$5.0 \\
\texttt{MEA5}      & 64.8$\pm$4.6 & 69.4$\pm$4.8 & 77.2$\pm$4.8 & 77.5$\pm$4.7 & 77.2$\pm$4.8 \\
\texttt{AdvMEA}    & 20.7$\pm$20.4 & 23.1$\pm$21.9 & 24.9$\pm$23.1 & 22.7$\pm$23.9 & 18.8$\pm$16.7 \\
\texttt{CEGA}      & 77.6$\pm$4.9 & 78.0$\pm$4.8 & 77.1$\pm$5.1 & 75.5$\pm$4.2 & 75.5$\pm$4.2 \\
\texttt{Realistic} & 73.9$\pm$2.5 & 74.9$\pm$1.9 & 74.9$\pm$2.4 & 75.6$\pm$3.0 & 75.0$\pm$3.1 \\
\texttt{DFEA\_I}   & 80.7$\pm$4.5 & 80.7$\pm$4.7 & 81.0$\pm$4.2 & 81.1$\pm$4.0 & 81.0$\pm$3.9 \\
\texttt{DFEA\_II}  & 75.1$\pm$1.3 & 76.0$\pm$1.3 & 76.3$\pm$1.6 & 76.3$\pm$1.6 & 76.8$\pm$1.2 \\
\texttt{DFEA\_III} & 76.9$\pm$4.5 & 77.6$\pm$4.4 & 77.7$\pm$4.5 & 77.6$\pm$4.5 & 77.1$\pm$4.7 \\
\bottomrule
\end{tabular}
}
\end{subtable}
\medskip
\begin{subtable}[t]{0.48\textwidth}
\centering
\caption{Regime=\texttt{a\_only}}
\resizebox{\linewidth}{!}{%
\begin{tabular}{lccccc}
\toprule
\textbf{Attack} & 0.05 & 0.10 & 0.25 & 0.50 & 1.00 \\
\midrule
\texttt{MEA0}      & 77.1$\pm$5.2 & 77.5$\pm$4.3 & 77.3$\pm$4.1 & 77.1$\pm$4.9 & 77.3$\pm$4.8 \\
\texttt{MEA1}      & 74.2$\pm$1.2 & 79.0$\pm$1.2 & 81.7$\pm$2.9 & 81.6$\pm$3.4 & 77.3$\pm$4.8 \\
\texttt{MEA2}      & 52.8$\pm$3.6 & 52.8$\pm$3.6 & 52.8$\pm$3.6 & 52.8$\pm$3.6 & 52.8$\pm$3.6 \\
\texttt{MEA3}      & 64.1$\pm$4.7 & 67.9$\pm$4.9 & 74.5$\pm$4.9 & 77.5$\pm$4.4 & 77.3$\pm$4.8 \\
\texttt{MEA4}      & 62.6$\pm$4.5 & 63.4$\pm$5.6 & 65.2$\pm$6.0 & 70.6$\pm$4.8 & 77.1$\pm$5.0 \\
\texttt{MEA5}      & 65.3$\pm$4.7 & 69.7$\pm$4.4 & 77.2$\pm$4.7 & 77.3$\pm$4.7 & 77.2$\pm$4.8 \\
\texttt{AdvMEA}    & 23.1$\pm$18.5 & 24.8$\pm$18.8 & 24.4$\pm$20.2 & 24.0$\pm$19.1 & 21.4$\pm$16.3 \\
\texttt{CEGA}      & 77.6$\pm$4.5 & 77.3$\pm$4.6 & 76.7$\pm$4.5 & 75.5$\pm$4.2 & 75.5$\pm$4.2 \\
\texttt{Realistic} & 74.3$\pm$3.3 & 74.6$\pm$2.1 & 75.2$\pm$2.5 & 74.9$\pm$2.3 & 75.1$\pm$3.2 \\
\texttt{DFEA\_I}   & 80.7$\pm$4.5 & 80.7$\pm$4.6 & 81.0$\pm$4.2 & 81.1$\pm$4.0 & 81.0$\pm$3.9 \\
\texttt{DFEA\_II}  & 75.0$\pm$1.4 & 75.8$\pm$1.4 & 76.4$\pm$1.5 & 76.5$\pm$1.5 & 76.6$\pm$1.3 \\
\texttt{DFEA\_III} & 76.9$\pm$4.5 & 77.6$\pm$4.4 & 77.7$\pm$4.5 & 77.6$\pm$4.5 & 77.1$\pm$4.8 \\
\bottomrule
\end{tabular}
}
\end{subtable}
\hfill
\begin{subtable}[t]{0.48\textwidth}
\centering
\caption{Regime=\texttt{data\_free}}
\resizebox{\linewidth}{!}{%
\begin{tabular}{lccccc}
\toprule
\textbf{Attack} & 0.05 & 0.10 & 0.25 & 0.50 & 1.00 \\
\midrule
\texttt{MEA0}      & 77.2$\pm$4.2 & 77.0$\pm$4.7 & 77.2$\pm$4.9 & 77.3$\pm$4.8 & 77.3$\pm$4.8 \\
\texttt{MEA1}      & 74.2$\pm$1.2 & 79.0$\pm$1.3 & 81.7$\pm$2.9 & 81.6$\pm$3.4 & 77.2$\pm$4.8 \\
\texttt{MEA2}      & 52.8$\pm$3.6 & 52.8$\pm$3.6 & 52.8$\pm$3.6 & 52.8$\pm$3.6 & 52.8$\pm$3.6 \\
\texttt{MEA3}      & 64.4$\pm$4.8 & 66.9$\pm$5.0 & 74.4$\pm$4.8 & 77.4$\pm$4.7 & 77.2$\pm$4.8 \\
\texttt{MEA4}      & 63.0$\pm$5.1 & 63.4$\pm$5.5 & 65.1$\pm$6.0 & 70.1$\pm$5.5 & 77.1$\pm$5.0 \\
\texttt{MEA5}      & 65.0$\pm$4.8 & 69.3$\pm$5.0 & 77.4$\pm$4.5 & 77.6$\pm$4.4 & 77.3$\pm$4.8 \\
\texttt{AdvMEA}    & 30.4$\pm$21.8 & 26.0$\pm$20.8 & 24.4$\pm$22.6 & 22.0$\pm$21.7 & 22.6$\pm$18.2 \\
\texttt{CEGA}      & 77.9$\pm$4.8 & 77.5$\pm$4.8 & 76.7$\pm$4.9 & 75.5$\pm$4.2 & 75.5$\pm$4.2 \\
\texttt{Realistic} & 74.2$\pm$3.1 & 75.1$\pm$3.3 & 75.4$\pm$3.0 & 74.8$\pm$2.2 & 75.1$\pm$3.1 \\
\texttt{DFEA\_I}   & 80.7$\pm$4.5 & 80.8$\pm$4.5 & 81.0$\pm$4.2 & 81.1$\pm$4.0 & 81.0$\pm$3.9 \\
\texttt{DFEA\_II}  & 75.2$\pm$1.3 & 76.0$\pm$1.3 & 76.3$\pm$1.4 & 76.6$\pm$1.8 & 76.5$\pm$1.5 \\
\texttt{DFEA\_III} & 76.9$\pm$4.5 & 77.7$\pm$4.3 & 77.7$\pm$4.6 & 77.7$\pm$4.5 & 77.1$\pm$4.7 \\
\bottomrule
\end{tabular}
}
\end{subtable}
\end{table*}

\subsection{Standard deviations for information-limiting defenses}
\label{app:rq2_new}

Tables~\ref{tab:app_rq2_new_defenses_full_a}--\ref{tab:app_rq2_new_defenses_full_b} extend Table~\ref{tab:rq2_new_defenses_summary} of the main text with standard deviations for each protected-model accuracy and verification proxy. The seven defenses are split across two tables for readability: the four output-perturbation / prediction-rounding defenses are in Table~\ref{tab:app_rq2_new_defenses_full_a}, and the three query-detection defenses are in Table~\ref{tab:app_rq2_new_defenses_full_b}.

\begin{table*}[t]
\centering
\caption{Detailed results for the four output-perturbation and prediction-rounding defenses across all ten datasets. Each cell reports protected-model accuracy (\%) with the verification proxy (\%) in parentheses, both as mean $\pm$ standard deviation over three seeds. The target backbone is a DGL GCN with hidden dimension 16.}
\label{tab:app_rq2_new_defenses_full_a}
\scriptsize
\setlength{\tabcolsep}{3pt}
\renewcommand{\arraystretch}{1.05}
\resizebox{\textwidth}{!}{%
\begin{tabular}{@{}lcccc@{}}
\toprule
\textbf{Dataset} & \texttt{OP\_low} & \texttt{OP\_high} & \texttt{PR\_2bit} & \texttt{PR\_top1} \\
\midrule
Cora            & 79.4$\pm$0.7 (98.6$\pm$0.7) & 79.2$\pm$1.5 (93.9$\pm$1.3) & 73.3$\pm$0.8 (83.7$\pm$0.1) & 79.6$\pm$0.5 (100.0$\pm$0.0) \\
CiteSeer        & 67.6$\pm$0.8 (97.8$\pm$0.6) & 66.3$\pm$0.8 (91.0$\pm$1.5) & 53.9$\pm$3.4 (70.3$\pm$5.1) & 68.8$\pm$0.4 (100.0$\pm$0.0) \\
PubMed          & 77.9$\pm$0.2 (99.0$\pm$0.4) & 75.9$\pm$0.7 (94.7$\pm$0.3) & 77.6$\pm$0.4 (93.4$\pm$1.0) & 78.2$\pm$0.3 (100.0$\pm$0.0) \\
Computers       & 44.0$\pm$21.1 (89.2$\pm$13.0) & 37.6$\pm$27.0 (55.3$\pm$37.8) & 36.1$\pm$26.4 (61.7$\pm$37.1) & 34.8$\pm$23.6 (100.0$\pm$0.0) \\
Photo           & 89.1$\pm$3.7 (98.9$\pm$0.5) & \textbf{90.7$\pm$5.7} (96.6$\pm$3.7) & \textbf{90.4$\pm$2.6} (96.7$\pm$2.1) & \textbf{95.5$\pm$0.6} (100.0$\pm$0.0) \\
CoauthorCS      & 87.8$\pm$0.1 (99.5$\pm$0.2) & 88.2$\pm$0.7 (98.8$\pm$0.4) & 87.5$\pm$0.8 (98.1$\pm$0.7) & 88.1$\pm$0.7 (100.0$\pm$0.0) \\
CoauthorPhysics & \textbf{89.4$\pm$0.3} (99.8$\pm$0.2) & 89.1$\pm$0.2 (99.3$\pm$0.1) & 90.2$\pm$0.1 (98.7$\pm$0.4) & 89.5$\pm$0.6 (100.0$\pm$0.0) \\
OGBN-Arxiv      & 37.7$\pm$2.9 (95.5$\pm$0.3) & 37.8$\pm$0.7 (81.2$\pm$1.0) & 30.2$\pm$2.0 (59.6$\pm$3.8) & 39.5$\pm$0.5 (100.0$\pm$0.0) \\
RomanEmpire     & 42.7$\pm$0.3 (95.3$\pm$0.1) & 40.6$\pm$1.4 (82.2$\pm$1.3) & 35.2$\pm$0.7 (56.4$\pm$1.2) & 42.7$\pm$0.5 (100.0$\pm$0.0) \\
AmazonRatings   & 42.0$\pm$0.3 (94.8$\pm$0.6) & 41.2$\pm$0.1 (80.0$\pm$0.7) & 39.4$\pm$0.6 (70.6$\pm$2.8) & 41.6$\pm$0.2 (100.0$\pm$0.0) \\
\bottomrule
\end{tabular}}
\end{table*}

\begin{table*}[t]
\centering
\caption{Detailed results for the three query-detection defenses across all ten datasets. Each cell reports protected-model accuracy (\%) with the verification proxy (\%) in parentheses, both as mean $\pm$ standard deviation over three seeds.}
\label{tab:app_rq2_new_defenses_full_b}
\scriptsize
\setlength{\tabcolsep}{4pt}
\renewcommand{\arraystretch}{1.0}
\begin{tabular*}{0.82\textwidth}{@{}l@{\extracolsep{\fill}}ccc@{}}
\toprule
\textbf{Dataset} & \texttt{PRADA} & \texttt{AdaptMisinfo} & \texttt{GradRedir} \\
\midrule
Cora            & 40.2$\pm$2.1 (43.0$\pm$1.3) & 41.0$\pm$0.1 (48.5$\pm$0.0) & 79.8$\pm$0.2 (100.0$\pm$0.0) \\
CiteSeer        & 69.3$\pm$0.7 (100.0$\pm$0.0) & 39.8$\pm$0.3 (52.5$\pm$0.0) & 68.4$\pm$0.9 (100.0$\pm$0.0) \\
PubMed          & 78.0$\pm$0.7 (100.0$\pm$0.0) & 44.1$\pm$0.8 (48.6$\pm$0.0) & 78.3$\pm$0.5 (100.0$\pm$0.0) \\
Computers       & 46.0$\pm$18.0 (100.0$\pm$0.0) & 28.0$\pm$13.9 (64.3$\pm$0.0) & 52.4$\pm$20.9 (100.0$\pm$0.0) \\
Photo           & \textbf{87.0$\pm$10.6} (100.0$\pm$0.0) & 46.3$\pm$0.4 (49.6$\pm$0.0) & 66.6$\pm$17.6 (100.0$\pm$0.0) \\
CoauthorCS      & 75.0$\pm$0.8 (79.6$\pm$1.3) & 52.4$\pm$0.5 (58.7$\pm$0.0) & 88.2$\pm$0.7 (100.0$\pm$0.0) \\
CoauthorPhysics & 83.4$\pm$0.7 (89.0$\pm$0.9) & \textbf{59.0$\pm$0.2} (63.1$\pm$0.0) & \textbf{89.7$\pm$0.3} (100.0$\pm$0.0) \\
OGBN-Arxiv      & 37.0$\pm$0.3 (100.0$\pm$0.0) & 19.9$\pm$0.3 (52.3$\pm$0.0) & 38.2$\pm$1.0 (100.0$\pm$0.0) \\
RomanEmpire     & 19.7$\pm$0.2 (25.4$\pm$0.7) & 22.5$\pm$0.4 (50.8$\pm$0.0) & 42.5$\pm$0.3 (100.0$\pm$0.0) \\
AmazonRatings   & 41.8$\pm$0.4 (100.0$\pm$0.0) & 33.9$\pm$0.1 (48.9$\pm$0.0) & 41.7$\pm$0.3 (100.0$\pm$0.0) \\
\bottomrule
\end{tabular*}
\end{table*}

\begin{figure}[t]
\centering
\includegraphics[width=0.95\textwidth]{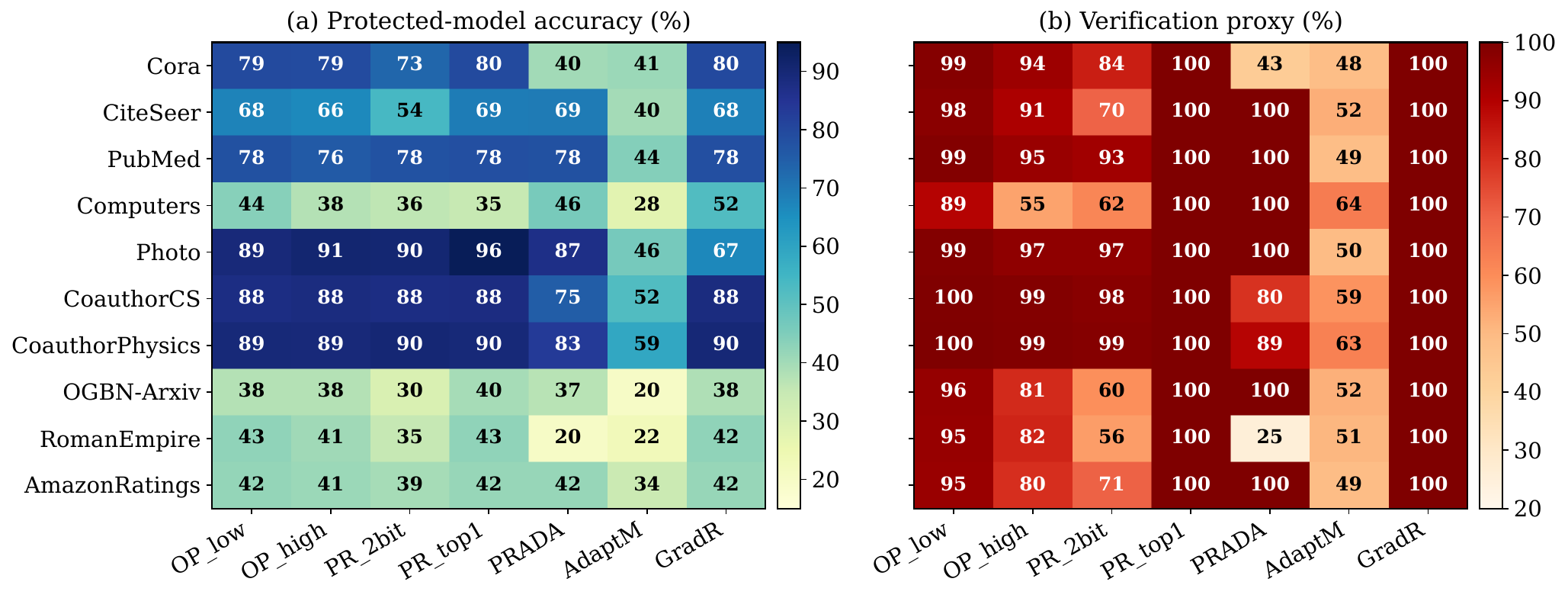}
\caption{Heatmap view of the seven information-limiting and query-detection defenses across all ten datasets. Left panel: protected-model accuracy (\%); right panel: verification proxy (\%). Numbers are the same as in Tables~\ref{tab:app_rq2_new_defenses_full_a}--\ref{tab:app_rq2_new_defenses_full_b}. The two-panel view makes explicit that verification splits into two regimes: output-perturbation/rounding defenses and \texttt{GradRedir} verify at high rates, whereas \texttt{PRADA} and \texttt{AdaptMisinfo} are more variable and often closer to random; protected accuracy is largely flat across defenses except on \textit{Computers}.}
\label{fig:app_rq2_info_heatmap}
\end{figure}

\paragraph{Heatmap view of the seven information-limiting defenses.}
Figure~\ref{fig:app_rq2_info_heatmap} reports the same numbers as Tables~\ref{tab:app_rq2_new_defenses_full_a}--\ref{tab:app_rq2_new_defenses_full_b} as a $10 \times 7$ heatmap, with protected-model accuracy on the left panel and the verification proxy on the right panel. The two-panel view exposes three patterns which are not visible in the per-defense tables. \emph{First, the verification proxy clusters into two regimes.} The four output-perturbation and rounding defenses (\texttt{OP\_low}, \texttt{OP\_high}, \texttt{PR\_2bit}, \texttt{PR\_top1}) plus \texttt{GradRedir} verify at $\geq 80\%$ on every dataset, while the two query-detection defenses (\texttt{PRADA}, \texttt{AdaptMisinfo}) verify at $\sim 50\%$ on most homophilic graphs and below $50\%$ on \textit{RomanEmpire}; this is consistent with the joint-evaluation behaviour reported in Appendix~\ref{app:rq5_full}, where the same two defenses are also the strongest at reducing surrogate fidelity. \emph{Second, the accuracy panel is largely flat across defenses on the eight non-Computers datasets.} The output-perturbation methods leave protected accuracy within $\sim 2$\,pp of the undefended baseline on \textit{Cora}, \textit{CiteSeer}, \textit{PubMed}, \textit{Photo}, \textit{CoauthorCS}, \textit{CoauthorPhysics}, \textit{OGBN-Arxiv}, and \textit{AmazonRatings}, which means the protection signal in those rows is carried entirely by the verification panel. \emph{Third, \textit{Computers} is the only dataset where the accuracy panel is heterogeneous}: every defense produces high variance, and three defenses (\texttt{OP\_high}, \texttt{PR\_2bit}, \texttt{AdaptMisinfo}) push protected accuracy below $40\%$, which mirrors the high variance of the same dataset in Tables~\ref{tab:rq2_defense_7xN_F1}--\ref{tab:rq2_defense_7xN_Fidelity} and is consistent with the structural-property analysis (high average degree) in Appendix~\ref{app:rq6_full}.

\subsection{Peak GPU memory of attacks and defenses (RQ4)}
\label{app:rq4_mem}

\begin{figure*}[t]
\centering
\begin{subfigure}[t]{0.48\textwidth}
  \centering
  \includegraphics[width=\linewidth]{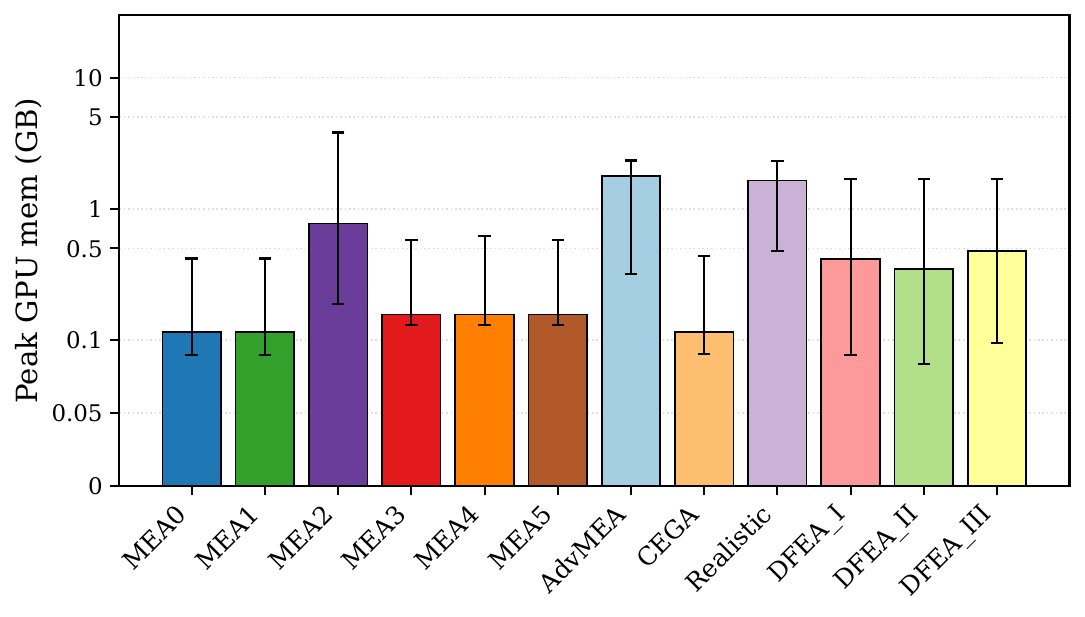}
  \caption{Attacks: peak memory.}
  \label{fig:rq4_attack_mem_wrapL}
\end{subfigure}\hfill
\begin{subfigure}[t]{0.48\textwidth}
  \centering
  \includegraphics[width=\linewidth]{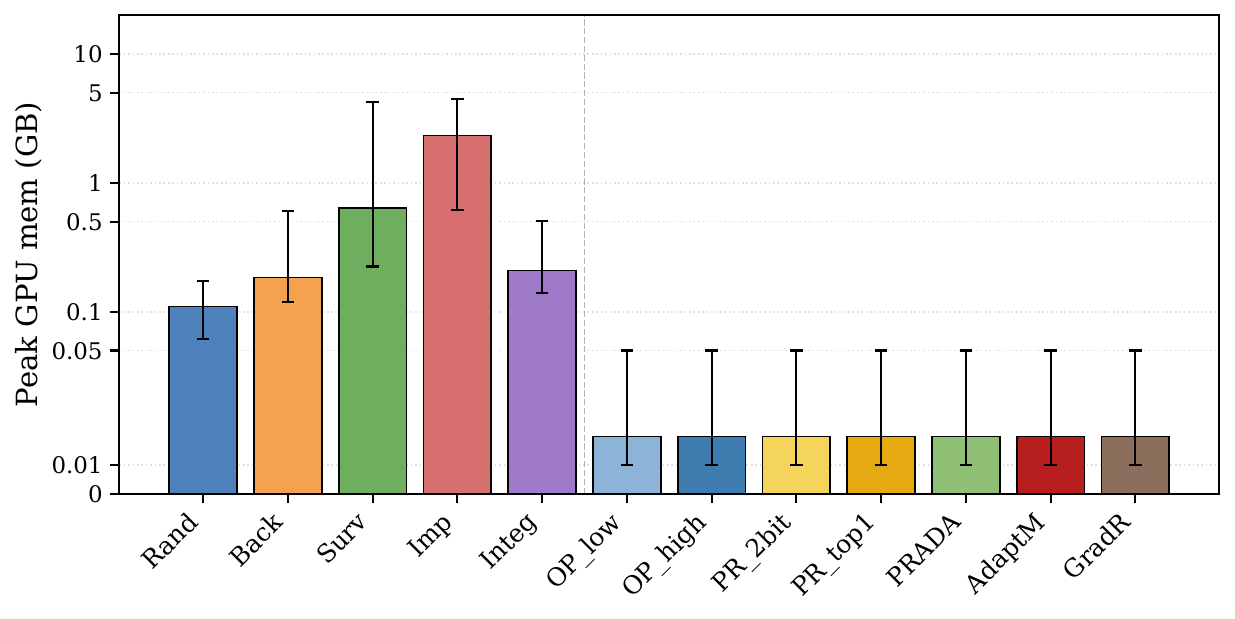}
  \caption{Defenses: peak memory.}
  \label{fig:rq4_defense_mem_wrapL}
\end{subfigure}
\caption{Peak GPU memory (GB) on a symmetric-log scale, aggregated across all ten datasets. Bars show median and error bars show the inter-quartile range across (dataset, seed) combinations. Panel (b) covers all twelve defenses; the dashed line separates the five watermarking defenses (\texttt{Rand}, \texttt{Back}, \texttt{Surv}, \texttt{Imp}, \texttt{Integ}) on the left from the seven information-limiting / query-detection defenses \mbox{on the right}.}
\label{fig:rq4_mem_wrapL}
\end{figure*}

Figure~\ref{fig:rq4_mem_wrapL} reports the peak GPU memory of every attack (panel a) and every defense (panel b) on a symmetric-log $y$-axis, aggregated over all ten datasets. The symlog scale is necessary because the workloads span more than two orders of magnitude on the same axis: \texttt{MEA0}, \texttt{CEGA}, and the lightweight defenses sit near $0.05$--$0.1$\,GB, while \texttt{MEA2}, \texttt{Realistic}, the \texttt{DFEA} family, and \texttt{ImperceptibleWM} reach several gigabytes. Two findings supplement the time-based discussion in the main text. \emph{First, peak memory is a more discriminative signal than wall-clock time for separating attack families.} The fast \texttt{MEA0/1/3/4/5/CEGA} group is tightly clustered around $0.05$--$0.15$\,GB across every dataset, while the data-free \texttt{DFEA} family and \texttt{Realistic} sit at $1$--$5$\,GB; the within-group variance is much smaller than the between-group gap, which means an operator who watches GPU memory can infer the attack family without timing the queries. \emph{Second, the watermarking-versus-information-limiting split is sharp on the defense panel.} The five watermarking defenses (left of the dashed line) all train an in-model artefact and so allocate proportional buffers ($0.1$--$5$\,GB depending on the watermarking technique), whereas the seven information-limiting defenses (right of the dashed line) are inference-time wrappers and use less than $0.05$\,GB on every dataset --- they essentially add no memory cost on top of the protected model itself. This explains why output-perturbation defenses are attractive in deployment: they cost nothing in memory while providing the protection summarised in the~main~text.

\begin{figure}[t]
\centering
\includegraphics[width=\textwidth]{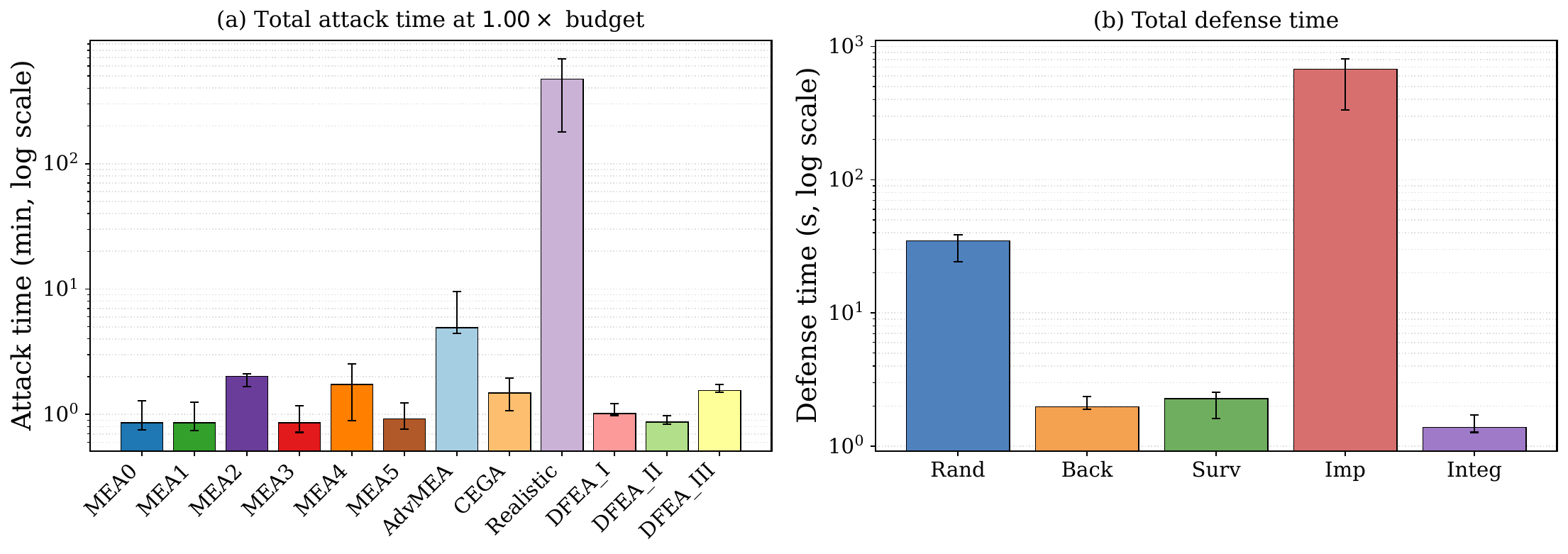}
\caption{Wall-clock cost summary on a log scale. (a) Total attack time (min) at $1.00\times$ budget; bars are median and error bars are the inter-quartile range across the seven homophilic datasets. (b) Total defense time (s); same conventions. Numbers from Tables~\ref{tab:rq4_attack_efficiency_time}--\ref{tab:rq4_defense_efficiency_time_compact}. Three groupings emerge on the attack side (fast \texttt{MEA}/\texttt{CEGA}, intermediate \texttt{DFEA}/\texttt{MEA2}, slow \texttt{Realistic}/\texttt{AdvMEA}) and two on the defense side (fast watermarks vs.\ \texttt{ImperceptibleWM}).}
\label{fig:app_cost_combined}
\end{figure}

\paragraph{Wall-clock cost.}
Figure~\ref{fig:app_cost_combined} aggregates total wall-clock time across all datasets for the twelve attacks (panel a) and the five watermarking defenses (panel b), again on a symmetric-log scale because the workloads span more than three orders of magnitude. The median bars and IQR error bars summarise Tables~\ref{tab:rq4_attack_efficiency_time}--\ref{tab:rq4_defense_efficiency_time_compact}, and three patterns are sharper here than in the per-dataset tables. \emph{First, on the attack side the cost cleaves into three groups.} The fast \texttt{MEA} family and \texttt{CEGA} sit between $0.7$ and $2$ minutes per run, the data-free \texttt{DFEA} family and \texttt{MEA2} sit slightly above at $1$--$3$ minutes, and only \texttt{Realistic} and (more variably) \texttt{AdvMEA} sit one to three orders of magnitude higher; the upper IQR of \texttt{Realistic} reaches $\sim 1000$ minutes which is the largest single value in the benchmark. The cost gap is large enough that the \emph{cost--fidelity} ordering is non-trivial, since RQ1 shows that simple attacks already saturate near $0.25\times$ budget. \emph{Second, the defense panel separates into a fast tier ($1$--$60$\,s) and a slow tier ($\sim 500$\,s)}: \texttt{BackdoorWM} and \texttt{SurviveWM} train in seconds, \texttt{RandomWM} sits one order of magnitude higher because of its independent watermark graph training, and \texttt{ImperceptibleWM} sits two orders of magnitude higher because of its representation-level losses. \emph{Third, the IQR error bars are narrow on every fast method and wide only on \texttt{Realistic} and \texttt{ImperceptibleWM}}, which means that empirical efficiency is essentially deterministic for the recommended configurations and that the cost numbers in the per-dataset tables are reproducible~across~hardware.

\subsection{Extended statistical analysis of attack effectiveness (RQ1)}
\label{app:rq1_extra}

\begin{figure*}[t]
  \centering
  \includegraphics[width=\textwidth]{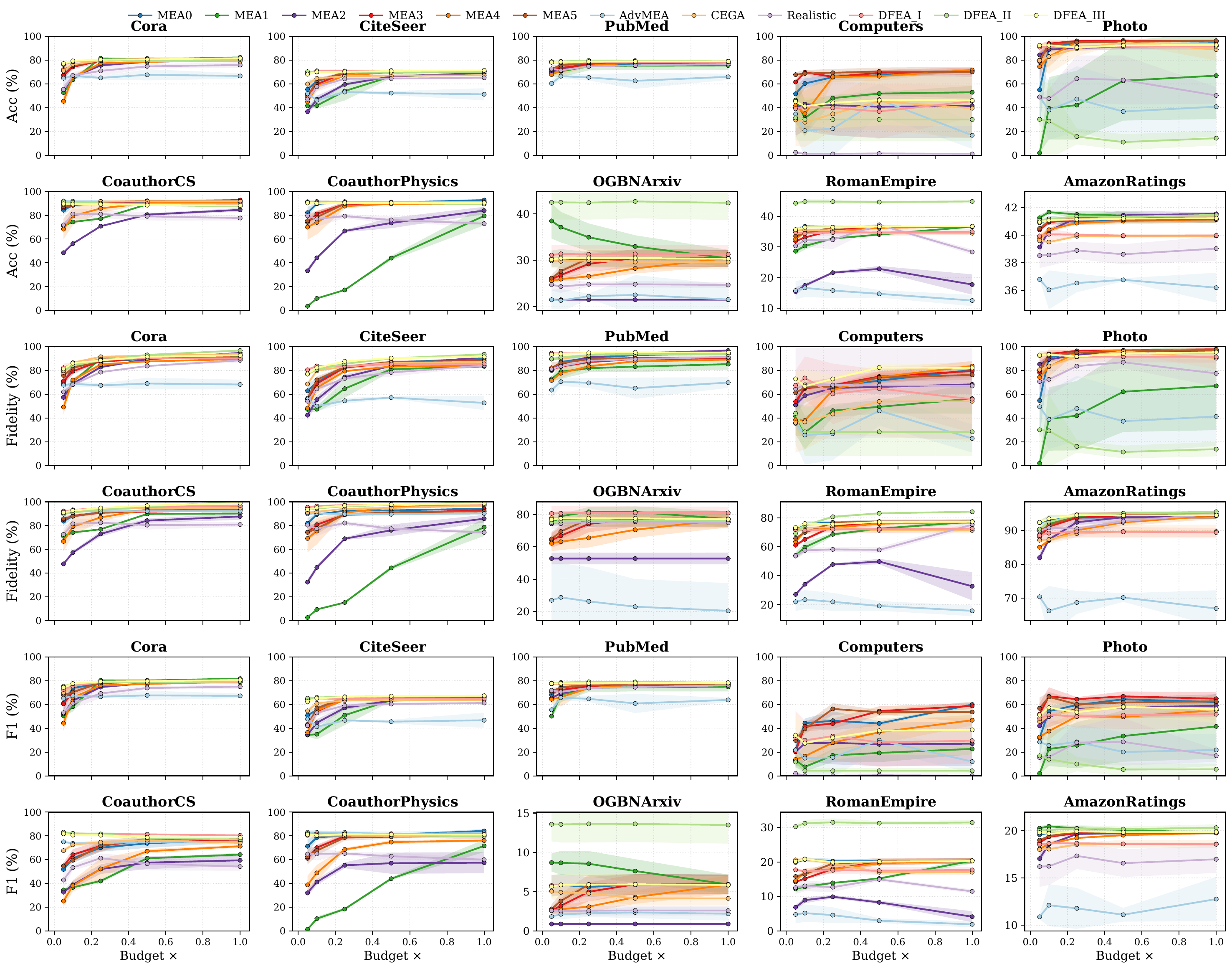}
  \caption{Budget--metric curves on all ten datasets (columns) for accuracy, fidelity, and macro F1 (rows). Lines are the twelve attacks (mean over three seeds, shaded bands $\pm 1$ std). The seven homophilic graphs share a $0$--$100\%$ $y$-axis; OGBN-Arxiv, RomanEmpire, and AmazonRatings use per-subplot ranges since their target accuracy is bounded by intrinsic task difficulty. The six-dataset version that appears in the main text (Figure~\ref{fig:rq1_cora_both_curves}) is a subset of this figure.}
  \label{fig:rq1_full_grid_appendix}
\end{figure*}

\begin{figure*}[t]
  \centering
  \includegraphics[width=\textwidth]{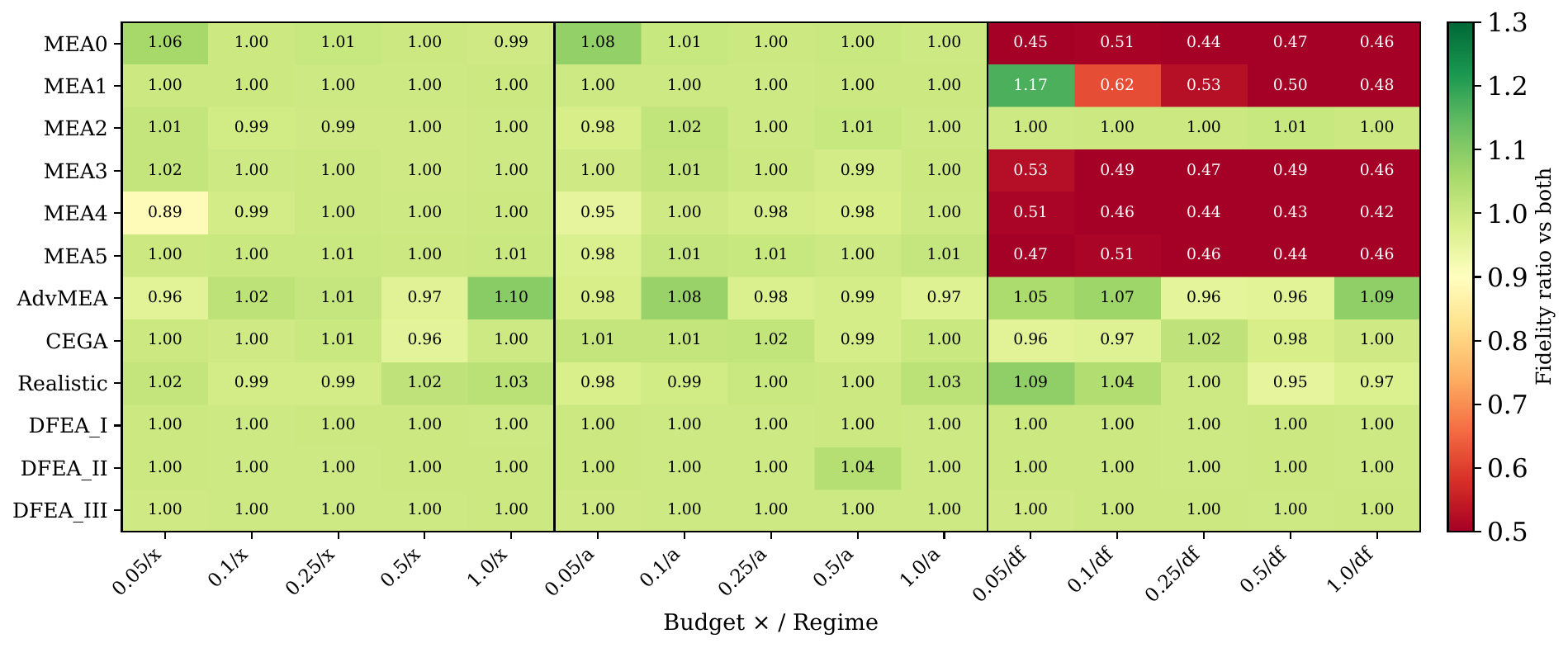}
  \caption{Regime sensitivity across budgets. Cells show the ratio between the average fidelity in a constrained regime and the average fidelity in the features-and-structure (\texttt{both}) regime for the same attack at the same budget; darker red indicates larger drops. The map aggregates over all ten datasets (seven homophilic plus OGBN-Arxiv, RomanEmpire, AmazonRatings) and separates dependence on features (\texttt{x}) and adjacency (\texttt{a}) from raw budget effects.}
  \label{fig:rq1_regime_sensitivity}
\end{figure*}

\begin{figure*}[t]
\centering
\includegraphics[width=\textwidth]{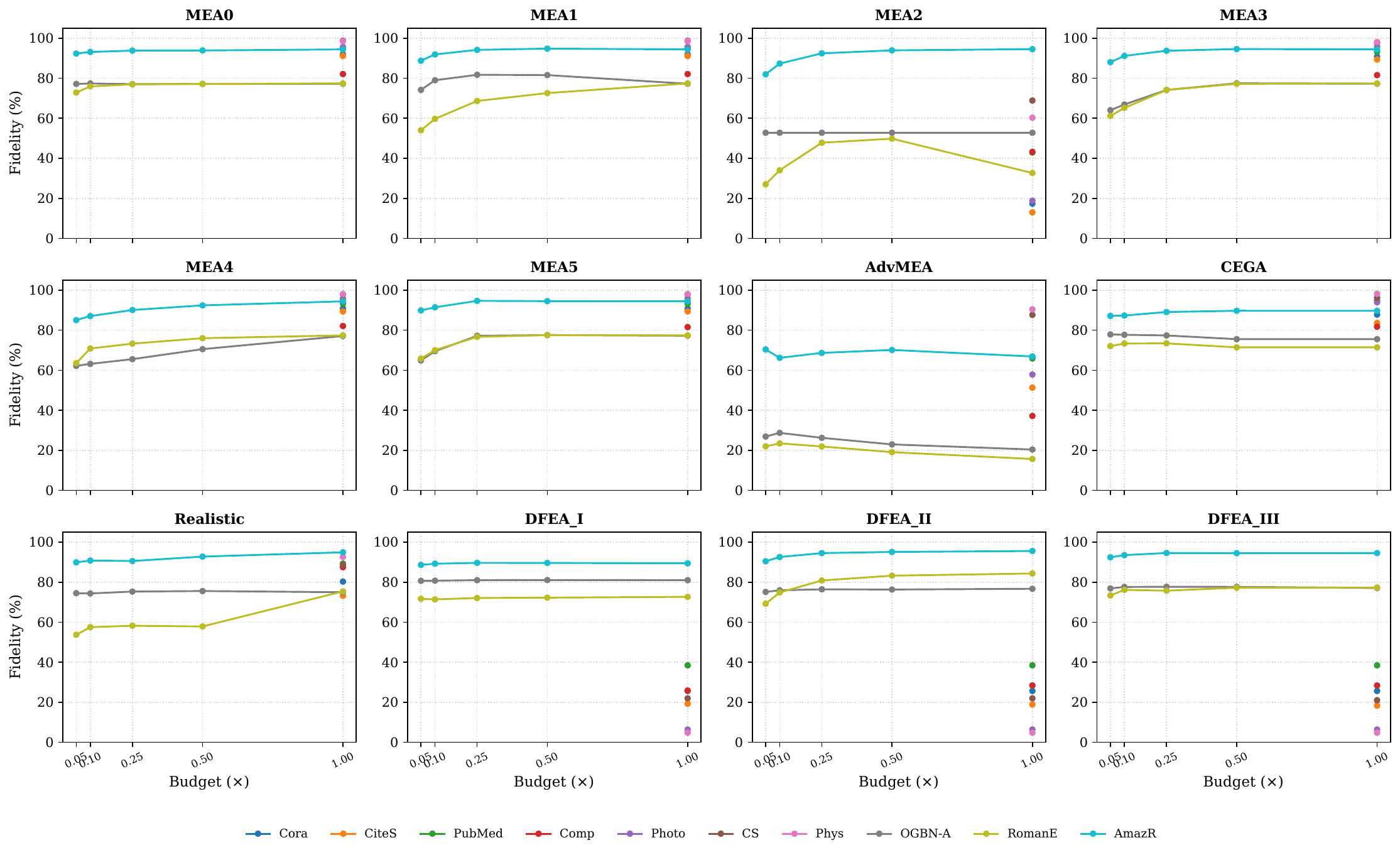}
\caption{Per-attack surrogate-fidelity curves on all ten datasets in the \texttt{both} regime. Each panel is one attack; lines are mean over three seeds at five budgets ($0.05,0.10,0.25,0.50,1.00$). The strong data-driven attacks (top two rows) saturate near $0.25\times$ on the homophilic graphs; the data-free attacks (\texttt{DFEA\_I/II/III}, bottom row) are competitive across most datasets and only degrade noticeably on the high-degree product graphs (\textit{Computers} and \textit{Photo}) for \texttt{DFEA\_II}, indicating that data-free extraction is largely robust under our protocol but sensitive to specific graph structures.}
\label{fig:app_attack_budget_panels}
\end{figure*}

\begin{figure}[t]
\centering
\begin{subfigure}[t]{0.495\textwidth}
  \centering
  \includegraphics[width=\linewidth]{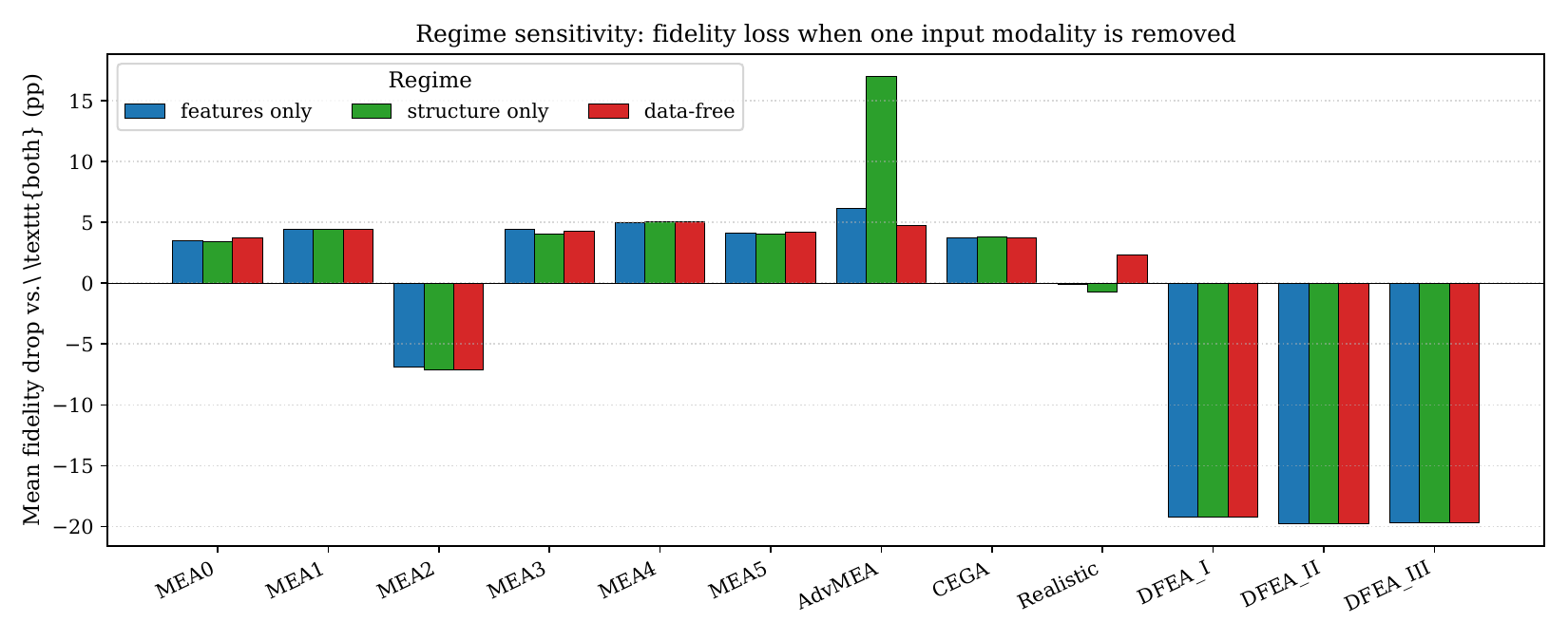}
  \caption{Mean fidelity drop relative to the \texttt{both} regime, per attack and per restricted regime.}
  \label{fig:app_regime_drop_bars}
\end{subfigure}
\hfill
\begin{subfigure}[t]{0.495\textwidth}
  \centering
  \includegraphics[width=\linewidth]{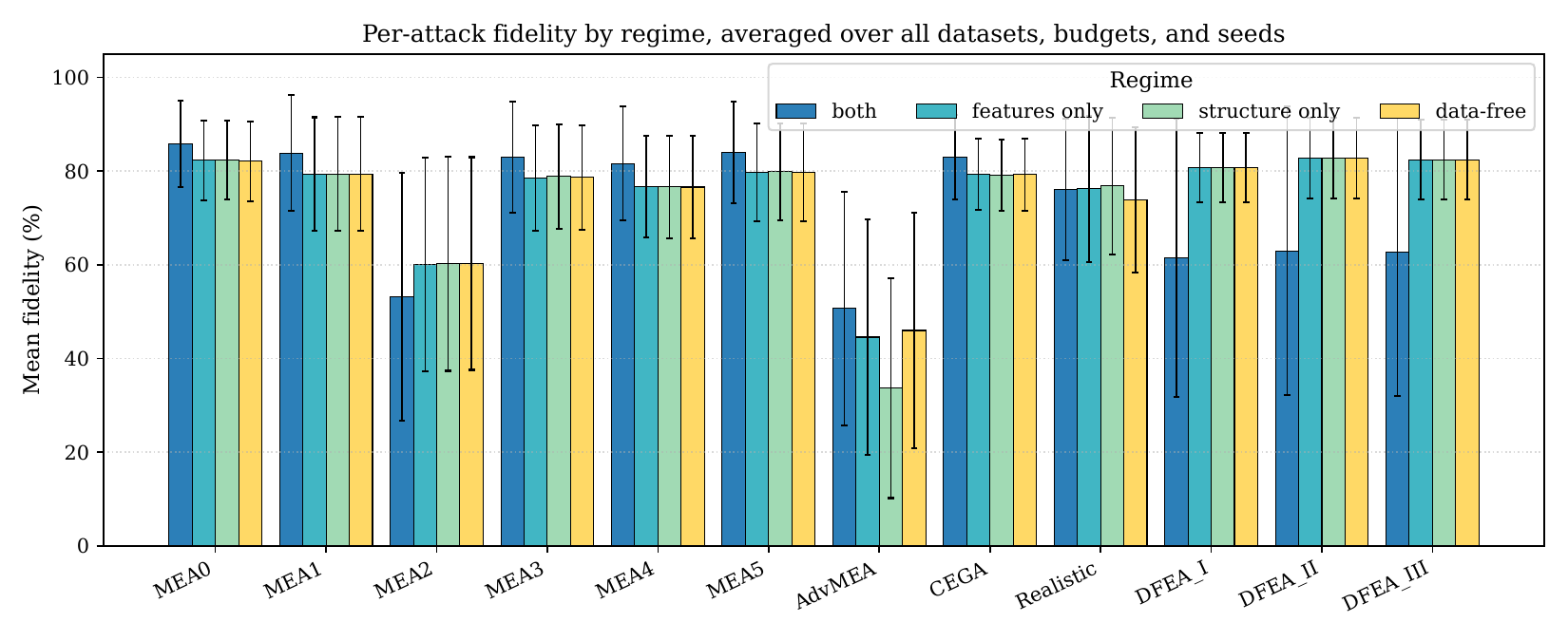}
  \caption{Absolute mean fidelity per attack, grouped by regime; error bars are $\pm$ std over (dataset, budget, seed).}
  \label{fig:app_attack_fidelity_by_regime}
\end{subfigure}
\caption{Regime sensitivity from two complementary views: differential (left) and absolute (right). Both panels aggregate over all ten datasets, five budgets, and three seeds.}
\label{fig:app_regime_views}
\end{figure}

\begin{figure}[t]
\centering
\includegraphics[width=0.85\textwidth]{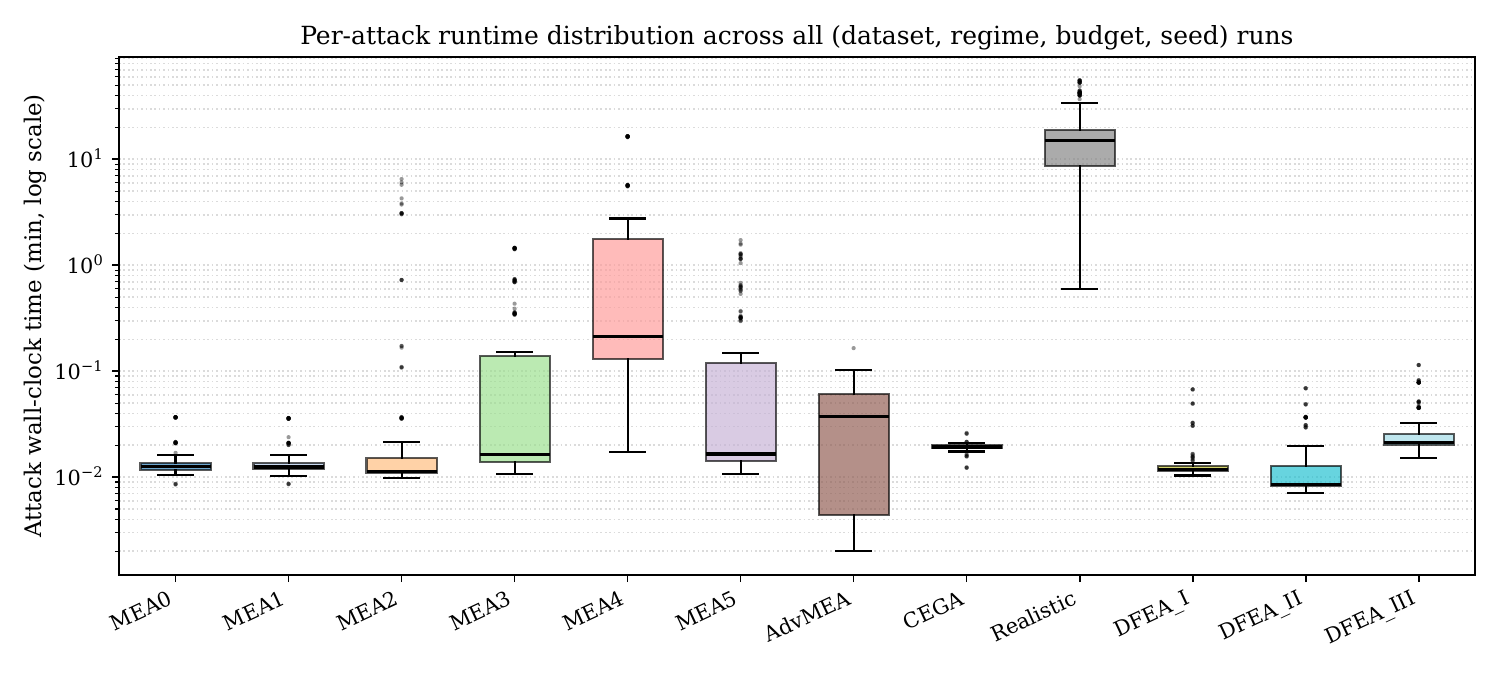}
\caption{Distribution of per-run attack wall-clock time (log scale, in minutes) across all (dataset, regime, budget, seed) runs. Boxes show the inter-quartile range, whiskers extend to $1.5\times$ IQR, and dots show outlier runs. The \texttt{Realistic} pipeline has the longest tail because it includes \mbox{an auxiliary edge predictor}.}
\label{fig:app_attack_time_distribution}
\end{figure}

\begin{figure}[t]
\centering
\includegraphics[width=0.62\textwidth]{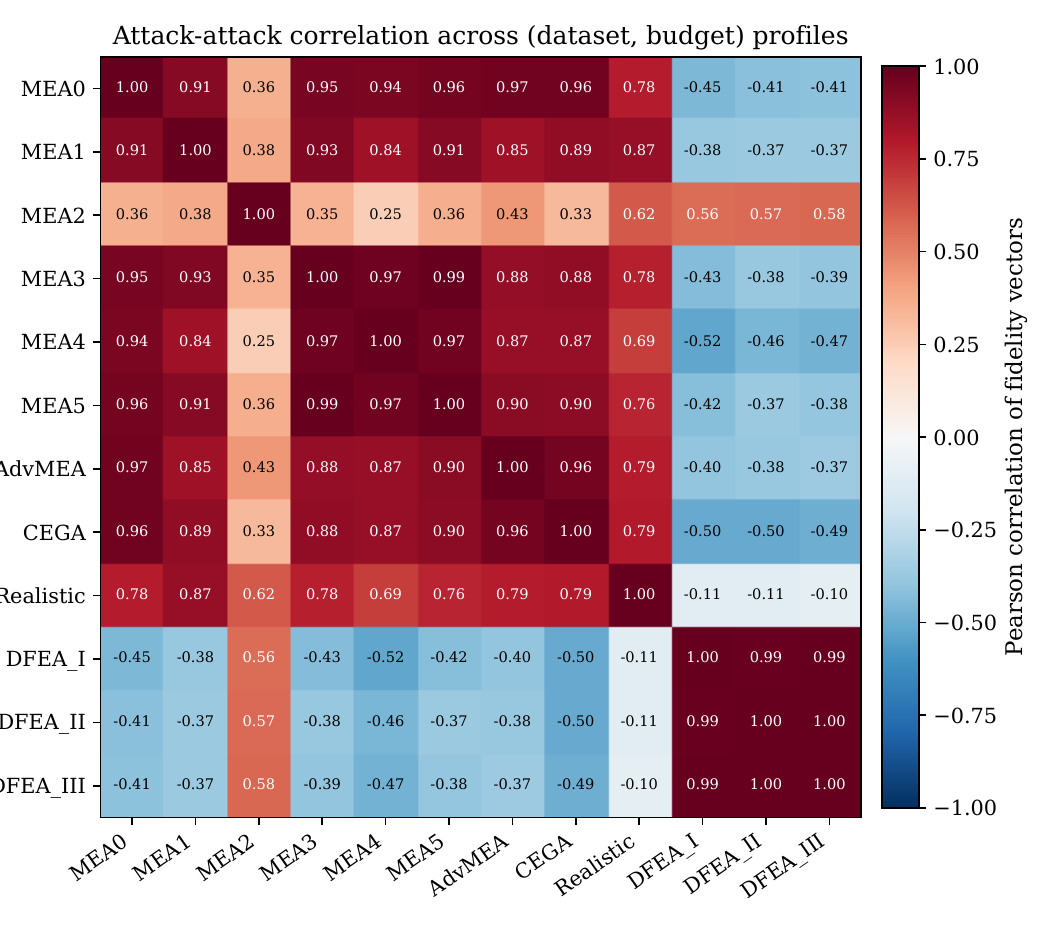}
\caption{Pairwise Pearson correlation between the twelve attacks, where each attack is represented by its fifty-dimensional vector of mean fidelity over (ten datasets $\times$ five budgets) in the \texttt{both} regime. Two strong blocks (data-driven and data-free) and one outlier (\texttt{AdvMEA}) emerge.}
\label{fig:app_attack_corr_matrix}
\end{figure}

We complement the per-dataset RQ1 tables (Tables~\ref{tab:app_rq1_overview_combined}--\ref{tab:app_rq1_detail_PubMed_Fidelity} and Tables~\ref{tab:app_rq1_RomanEmpire_full}--\ref{tab:app_rq1_OGBNArxiv_full}) with eight aggregated views (Figure~\ref{fig:rq1_full_grid_appendix}, Figure~\ref{fig:rq1_regime_sensitivity}, and Figures~\ref{fig:app_attack_budget_panels}--\ref{fig:app_attack_corr_matrix}) which average across datasets, regimes, or seeds to expose patterns that are hard to read off the~raw~tables.

\paragraph{Full ten-dataset budget--metric grid.}
Figure~\ref{fig:rq1_full_grid_appendix} reports the budget--metric curves on all ten datasets, in the same format as the six-dataset main-text Figure~\ref{fig:rq1_cora_both_curves}. The four datasets which the main text omits (CiteSeer, Photo, CoauthorCS, AmazonRatings) reproduce the qualitative behaviour of the chosen six: CiteSeer and Photo track the homophilic-citation/coauthor pattern of Cora and CoauthorPhysics; CoauthorCS sits between them; and AmazonRatings reproduces the heterophilic behaviour of RomanEmpire although its ordinal $5$-class label space inflates absolute fidelity. The qualitative conclusions of RQ1 therefore hold uniformly across all~ten~datasets.

\paragraph{Regime sensitivity heatmap.}
Figure~\ref{fig:rq1_regime_sensitivity} reports the regime sensitivity across the standard five-budget grid: each cell is the ratio between fidelity in a constrained regime (\texttt{x\_only}, \texttt{a\_only}, \texttt{data\_free}) and fidelity in the \texttt{both} regime for the same attack at the same budget, aggregated over all ten datasets. The heatmap exposes two patterns. \emph{First, the left two blocks (\texttt{x\_only}, \texttt{a\_only}) are uniformly green at ratio $\approx 1.0$ for every attack at every budget}: removing one input modality is essentially a no-op for the strong attacks, because the surrogate is trained on the target's labelled responses and label information dominates whichever single input remains. \emph{Second, the right block (\texttt{data\_free}) is bimodal}: the strong-MEA family (rows MEA0/3/4/5) drops to ratios in the $0.42$--$0.53$ band across every budget, while \texttt{AdvMEA}, \texttt{CEGA}, \texttt{Realistic}, and the three \texttt{DFEA} variants stay at ratios around $1.0$. The data-driven attacks (MEA family) collapse without a real graph because they rely on the target's responses to real-graph queries; the data-free attacks (\texttt{DFEA}, \texttt{Realistic}) by design synthesize their own queries and so are insensitive to the absence of a real graph; \texttt{CEGA} and \texttt{AdvMEA} retrain a query selector on whichever input is provided, including a synthetic graph, and so also stay near ratio $1.0$. The map therefore answers a benchmark-design question: the \texttt{data\_free} regime is the only regime that meaningfully discriminates attack families, which justifies including it in the standard RQ1 protocol despite its much lower~absolute~fidelity.

\paragraph{Per-attack budget curves across all ten datasets.}
Figure~\ref{fig:app_attack_budget_panels} shows, for each of the twelve attacks, a small panel containing the surrogate fidelity vs.\ budget curve on every dataset (regime \texttt{both}). Three observations stand out. First, the strong data-driven attacks (\texttt{MEA0}, \texttt{MEA1}, \texttt{MEA3}, \texttt{MEA4}, \texttt{MEA5}, \texttt{Realistic}, \texttt{CEGA}) saturate by $0.25\times$ on the seven small homophilic graphs, with curves that are visually indistinguishable above this budget; this is the empirical justification for using $0.25\times$ as the medium budget in the joint evaluation of RQ5. Second, the OGBN-Arxiv curve is consistently $\sim 10$--$15\,\%$ below the homophilic graphs across all data-driven attacks at every budget, which is consistent with the larger label space ($40$ classes) and not with a budget ceiling. Third, the data-free variants (\texttt{DFEA\_I}, \texttt{DFEA\_II}, \texttt{DFEA\_III}) and the noisy \texttt{MEA2} have curves that are largely flat across budgets and reach high fidelity on most datasets, with \texttt{DFEA\_II} on the high-degree product graphs (\textit{Computers} and \textit{Photo}) as the main outlier; this confirms that data-free extraction is broadly robust under our protocol while showing dataset-specific sensitivity.

\paragraph{Regime sensitivity: per-attack fidelity loss when one input is removed.}
Figure~\ref{fig:app_regime_drop_bars} reports, per attack, the mean fidelity drop relative to the \texttt{both} regime when only features (\texttt{x\_only}), only structure (\texttt{a\_only}), or no real input (\texttt{data\_free}) is provided to the extractor. The bars expose a clear separation: the data-driven attacks lose only a few percentage points when one modality is removed (because they still receive the target's labels for real-graph queries), but lose $30$--$70$\,pp in the \texttt{data\_free} regime; \texttt{CEGA} loses the most in \texttt{a\_only} (its centrality-based selection becomes ill-defined without the real graph), and \texttt{AdvMEA} is the most robust to removing structure. Figure~\ref{fig:app_attack_fidelity_by_regime} provides the absolute view that complements these drops: it shows the absolute mean fidelity in each regime per attack, with error bars that aggregate across datasets, budgets, and seeds. The two views together motivate the four-regime protocol used in RQ1.

\paragraph{Runtime and resource distribution across attacks.}
Figure~\ref{fig:app_attack_time_distribution} shows the distribution of per-run wall-clock attack time in minutes, on a log scale, aggregated across all (dataset, regime, budget, seed) combinations. Three groups separate cleanly: (i) the lightweight \texttt{MEA0/MEA1/MEA3/MEA4/MEA5/CEGA} family, with median runtime under one minute and tail below ten minutes, (ii) the adaptive \texttt{AdvMEA} and \texttt{DFEA\_I/II/III} family in the one-to-twenty-minute band, and (iii) the heavyweight \texttt{Realistic} pipeline, whose tail extends two orders of magnitude above the rest because it trains an auxiliary edge-prediction model. This distributional view supplements the per-dataset time tables in the main text (Tables~\ref{tab:rq4_attack_efficiency_time}--\ref{tab:rq4_defense_efficiency_time_compact}): the overall ordering between groups is preserved on every dataset, but the absolute magnitude scales with~graph~size.

\paragraph{Inter-attack similarity across (dataset, budget) profiles.}
Figure~\ref{fig:app_attack_corr_matrix} reports the pairwise Pearson correlation of attacks, where each attack is represented by its fifty-dimensional fidelity vector (ten datasets $\times$ five budgets, regime \texttt{both}). Three structural blocks emerge. The first is the strong-MEA cluster (\texttt{MEA0}, \texttt{MEA1}, \texttt{MEA3}, \texttt{MEA4}, \texttt{MEA5}, \texttt{Realistic}, and \texttt{CEGA}), with pairwise correlations above $0.9$ and a single shared profile across datasets. The second is the data-free cluster (\texttt{DFEA\_I}, \texttt{DFEA\_II}, \texttt{DFEA\_III}, plus \texttt{MEA2}), with correlations near $0.95$ within the cluster and below $0.5$ with the strong-MEA cluster. The third is \texttt{AdvMEA}, which sits between the two clusters and correlates only weakly with both, consistent with its hybrid adaptive-query design. The block structure suggests that, for benchmarking purposes, a representative subset that contains one element from each block (for example \texttt{MEA0}, \texttt{DFEA\_I}, and \texttt{AdvMEA}) would already capture the qualitative behaviour of the full twelve-attack set, which we leave to future work as a benchmark-design recommendation.

\begin{figure}[t]
\centering
\includegraphics[width=\textwidth]{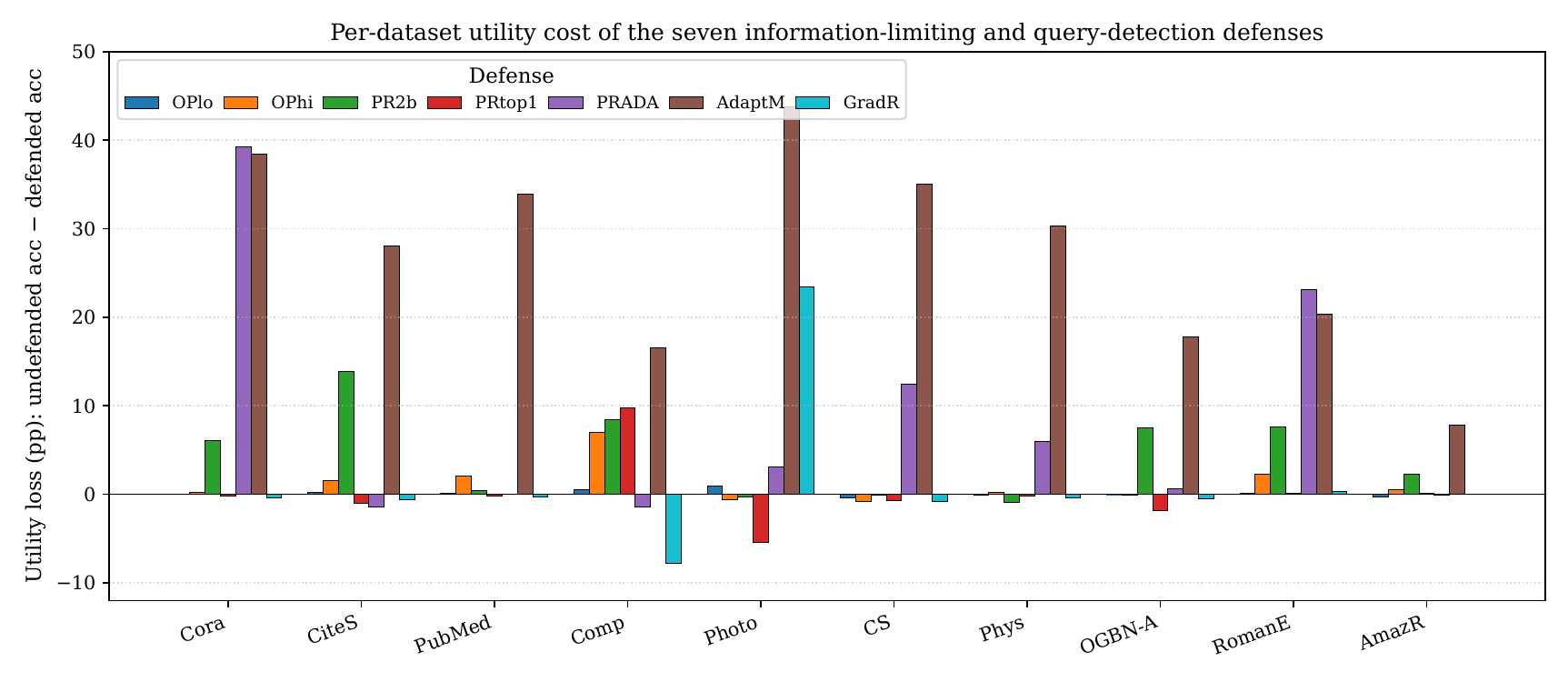}
\caption{Per-dataset utility loss of the seven information-limiting and query-detection defenses (undefended GCN accuracy $-$ defended accuracy, in percentage points; mean over three seeds). Output-perturbation and prediction-rounding defenses (\texttt{OP\_low}, \texttt{OP\_high}, \texttt{PR\_2bit}, \texttt{PR\_top1}) sit close to zero on the homophilic graphs, while \texttt{AdaptMisinfo} consistently loses $30$--$45$\,pp on every dataset because it actively returns wrong labels.}
\label{fig:app_defense_acc_drop_per_dataset}
\end{figure}

\paragraph{Per-dataset utility cost of information-limiting defenses.}
Figure~\ref{fig:app_defense_acc_drop_per_dataset} reports the per-dataset utility loss of each of the seven information-limiting and query-detection defenses, computed as undefended GCN baseline accuracy minus the defended-model accuracy reported in Tables~\ref{tab:app_rq2_new_defenses_full_a}--\ref{tab:app_rq2_new_defenses_full_b}. The figure exposes three patterns that the per-row tables hide. \emph{First, the four output-perturbation/rounding defenses (\texttt{OP\_low}, \texttt{OP\_high}, \texttt{PR\_2bit}, \texttt{PR\_top1}) cost essentially nothing on the seven homophilic graphs} (bars within $\pm 3$\,pp of zero), which is consistent with the design intent of these defenses: they perturb the returned scores at inference time without retraining, so the underlying classifier is unchanged. \emph{Second, query-detection defenses split into two cost regimes.} \texttt{GradRedir} stays close to the no-defense baseline on most datasets (it perturbs only flagged queries), but on \textit{Photo} it costs about $23$\,pp because the defense's gradient-redirection threshold flags many benign queries on the high-degree product graph; on \textit{Computers} the loss is even larger ($\sim 45$\,pp). \emph{Third, \texttt{AdaptMisinfo} consistently costs $30$--$45$\,pp across every dataset}, which is the highest among all seven and reflects the fact that the defense actively returns wrong labels on a large fraction of queries; users of the GraphIPBench API should treat \texttt{AdaptMisinfo} as a high-cost defense even when its verification proxy reaches $100\,\%$. The plot also surfaces a useful negative finding: \texttt{PRADA} on \textit{RomanEmpire} loses $23$\,pp because the heterophilic structure makes random-query rejection blunt, suggesting that PRADA's distance-based query filter needs adaptation for graphs with low~edge~homophily.

\subsection{Joint evaluation and watermark survival per dataset}
\label{app:rq5_full}
For \texttt{Integrity}, Computers and Photo use ten seeds at $0.25\times$. The remaining eight graphs use three seeds; some Integrity runs there use a $1.00\times$ budget or a different attack implementation and do not support matched comparisons with~other~defenses.

\begin{figure*}[t]
\centering
\begin{subfigure}[t]{0.32\textwidth}
  \centering
  \includegraphics[width=\linewidth]{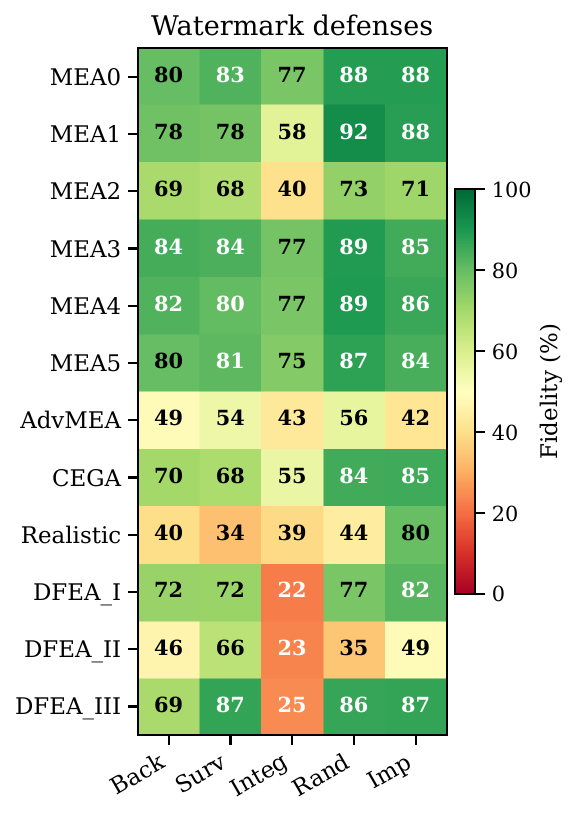}
  \caption{Joint fidelity vs.\ watermarking defenses.}
  \label{fig:rq5_joint_orig}
\end{subfigure}
\hfill
\begin{subfigure}[t]{0.32\textwidth}
  \centering
  \includegraphics[width=\linewidth]{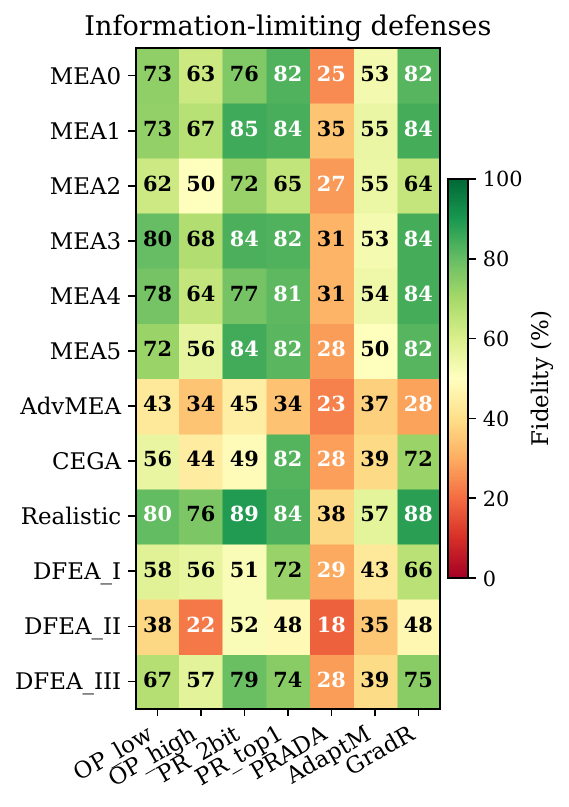}
  \caption{Joint fidelity vs.\ information-limiting defenses.}
  \label{fig:rq5_joint_new}
\end{subfigure}
\hfill
\begin{subfigure}[t]{0.32\textwidth}
  \centering
  \includegraphics[width=\linewidth]{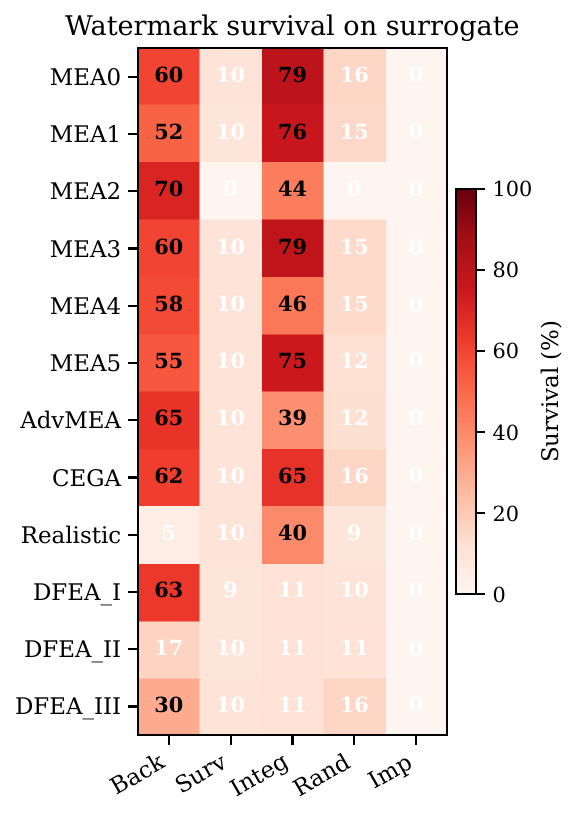}
  \caption{Watermark survival on the surrogate.}
  \label{fig:rq5_wm_survival}
\end{subfigure}
\caption{RQ5 joint evaluation on \textit{Computers} at $0.25\times$ (mean over three seeds; ten for Integrity). (a) Surrogate fidelity (\%) against the five watermarks. (b) Surrogate fidelity (\%) against the seven information-limiting defenses; \texttt{PRADA} and \texttt{AdaptMisinfo} reduce the strongest attacks from $80$--$92\%$ to $25$--$55\%$. (c) Watermark verification rate (\%) on the extracted surrogate. Query-based verification for \texttt{Integrity} is distinct from inherited watermark survival.}
\label{fig:rq5_joint_main}
\end{figure*}

This subsection provides the full numerical record together with the analyses which anchor the cross-dataset claims of RQ5. We first show the consolidated Computers heatmap (Figure~\ref{fig:rq5_joint_main}) which is referenced from the main text, and then the per-dataset tables. The tables report means and standard deviations across seeds.

\paragraph{Group 1 (joint surrogate fidelity, watermarking defenses).}
Tables~\ref{tab:app_rq5_cora}--\ref{tab:app_rq5_orig_d} cover all ten datasets for the five watermarking and integrity defenses (\texttt{BackdoorWM}, \texttt{SurviveWM}, \texttt{Integrity}, \texttt{RandomWM}, \texttt{ImperceptibleWM}). On Computers, for example, MEA1 reaches 57.6\% fidelity against Integrity and 92.0\% against RandomWM (Table~\ref{tab:app_rq5_orig_b}).

\paragraph{Group 2 (joint surrogate fidelity, information-limiting defenses).}
Tables~\ref{tab:app_rq5_cora}--\ref{tab:app_rq5_new_c} cover all ten datasets for the seven information-limiting and query-detection defenses (\texttt{OP\_low}, \texttt{OP\_high}, \texttt{PR\_2bit}, \texttt{PR\_top1}, \texttt{PRADA}, \texttt{AdaptMisinfo}, \texttt{GradRedir}). Two cross-dataset patterns hold. \emph{First}, the relative ordering between defenses is highly stable: \texttt{PRADA} and \texttt{AdaptMisinfo} consistently produce the largest fidelity drop relative to the undefended baseline on every dataset and every attack, while the four output-perturbation and rounding variants (\texttt{OP\_low}, \texttt{OP\_high}, \texttt{PR\_top1}, \texttt{GradRedir}) leave fidelity within $\sim 5$\,pp of the undefended baseline. \emph{Second}, prediction rounding to two bits (\texttt{PR\_2bit}) is the only output-perturbation defense whose effect depends sharply on the attack: it has near-zero impact on the strong data-driven attacks (which already operate on label-like signal) but reduces \texttt{MEA2} and the \texttt{DFEA} family by an additional $5$--$15$\,pp on most datasets, which is consistent with the link-prediction behaviour reported in Appendix~\ref{app:rq6_full}.

\paragraph{Group 3 (watermark survival on the surrogate).}
Tables~\ref{tab:app_rq5_cora}--\ref{tab:app_rq5_wm_survival_b} report the available per-attack verification rates on the extracted surrogate. The four embedded watermarks and the query verifier are interpreted separately. The corrected mean surrogate fingerprint agreement for \texttt{Integrity} is $48.0\%$ on \textit{Computers} and $51.0\%$ on \textit{Photo}. The in-model marker \texttt{BackdoorWM} is more heterogeneous: it collapses for several data-driven attacks on \textit{Cora}, but partially survives on the larger graphs (\textit{Computers}, \textit{Photo}, \textit{CoauthorCS}) and for some \texttt{DFEA} variants; \texttt{SurviveWM} produces a stable but small ($\sim 10$--$15\,\%$) survival rate that does not exceed the random-guess marker on most datasets; \texttt{RandomWM} yields $10$--$20\,\%$ survival across the board, again essentially indistinguishable from random; and \texttt{ImperceptibleWM} drops to zero on Cora, CiteSeer, PubMed, Computers, Photo, and on every data-free attack across all datasets, with sporadic non-zero survival appearing only on the heterophilic graphs. Across these schemes, verification on the protected model alone does not establish ownership~after~extraction.

\begin{figure}[t]
\centering
\begin{subfigure}[t]{0.48\textwidth}
  \centering
  \includegraphics[width=\linewidth]{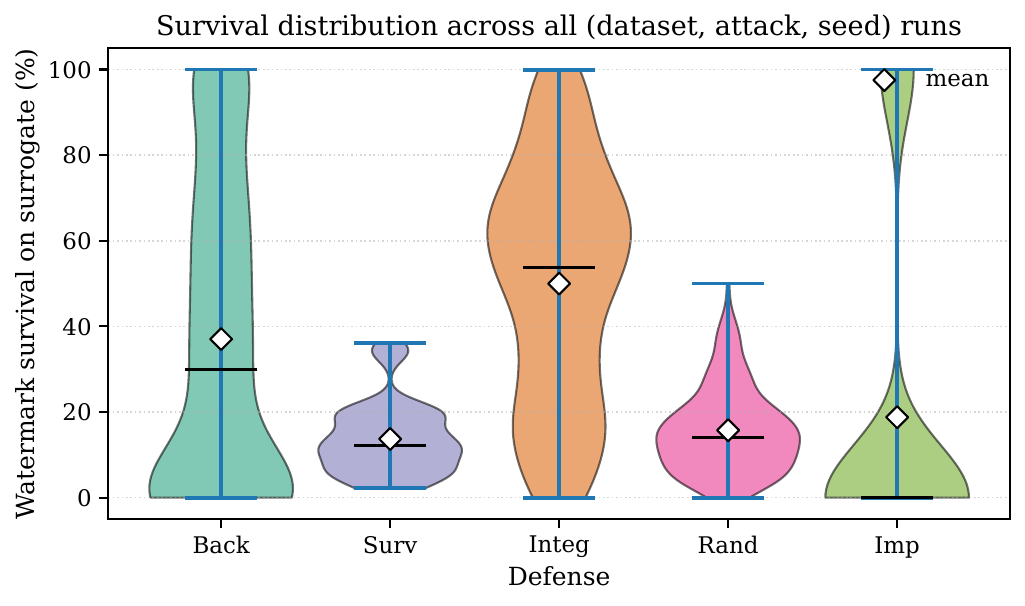}
  \caption{Violin: distribution of watermark survival per defense across all $(\text{dataset}, \text{attack}, \text{seed})$ runs.}
  \label{fig:app_wm_survival_violin}
\end{subfigure}
\hfill
\begin{subfigure}[t]{0.48\textwidth}
  \centering
  \includegraphics[width=\linewidth]{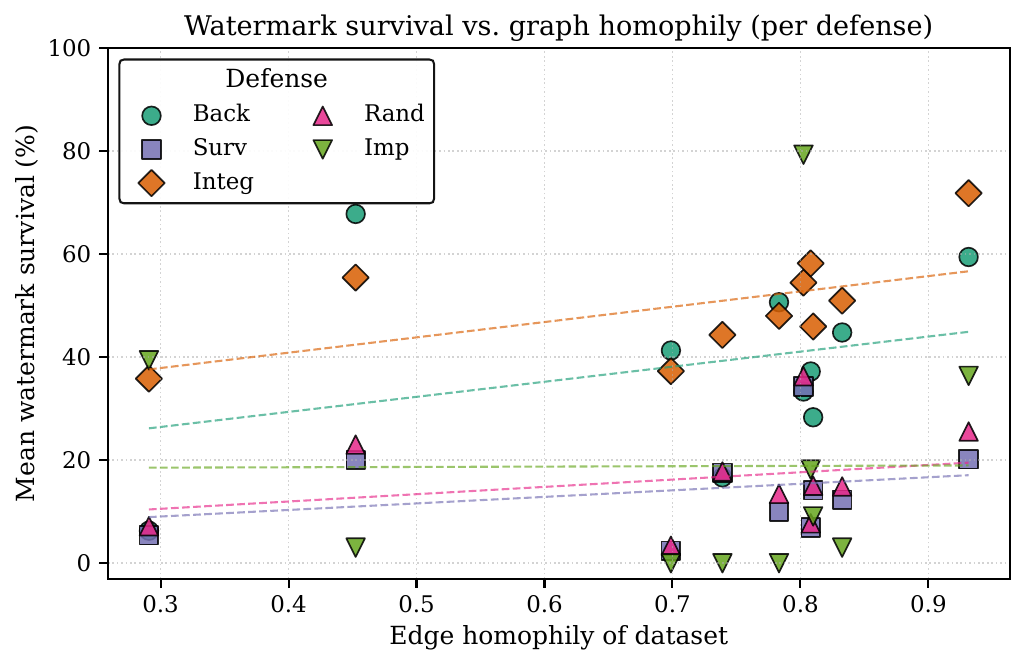}
  \caption{Scatter: edge homophily of the dataset vs.\ mean watermark survival on the surrogate (one point for each graph and~defense~pair).}
  \label{fig:app_homophily_vs_survival}
\end{subfigure}
\caption{Distribution and structural correlates of watermark survival.}
\label{fig:app_wm_distribution}
\end{figure}

\begin{figure}[t]
\centering
\begin{subfigure}[t]{0.46\textwidth}
  \centering
  \includegraphics[width=\linewidth]{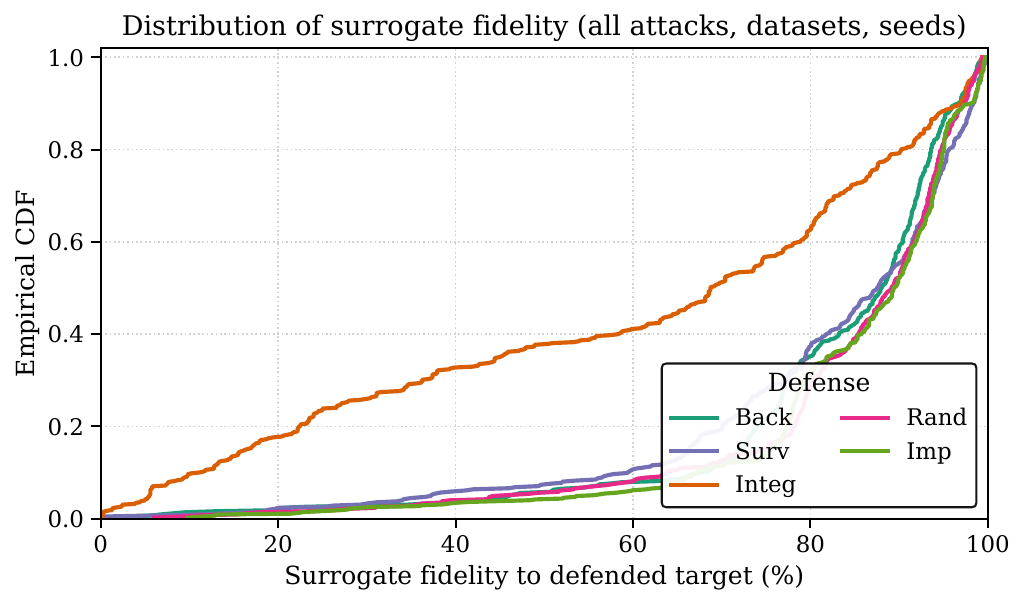}
  \caption{Empirical CDF of surrogate fidelity per defense.}
  \label{fig:app_fidelity_ecdf}
\end{subfigure}
\hfill
\begin{subfigure}[t]{0.52\textwidth}
  \centering
  \includegraphics[width=\linewidth]{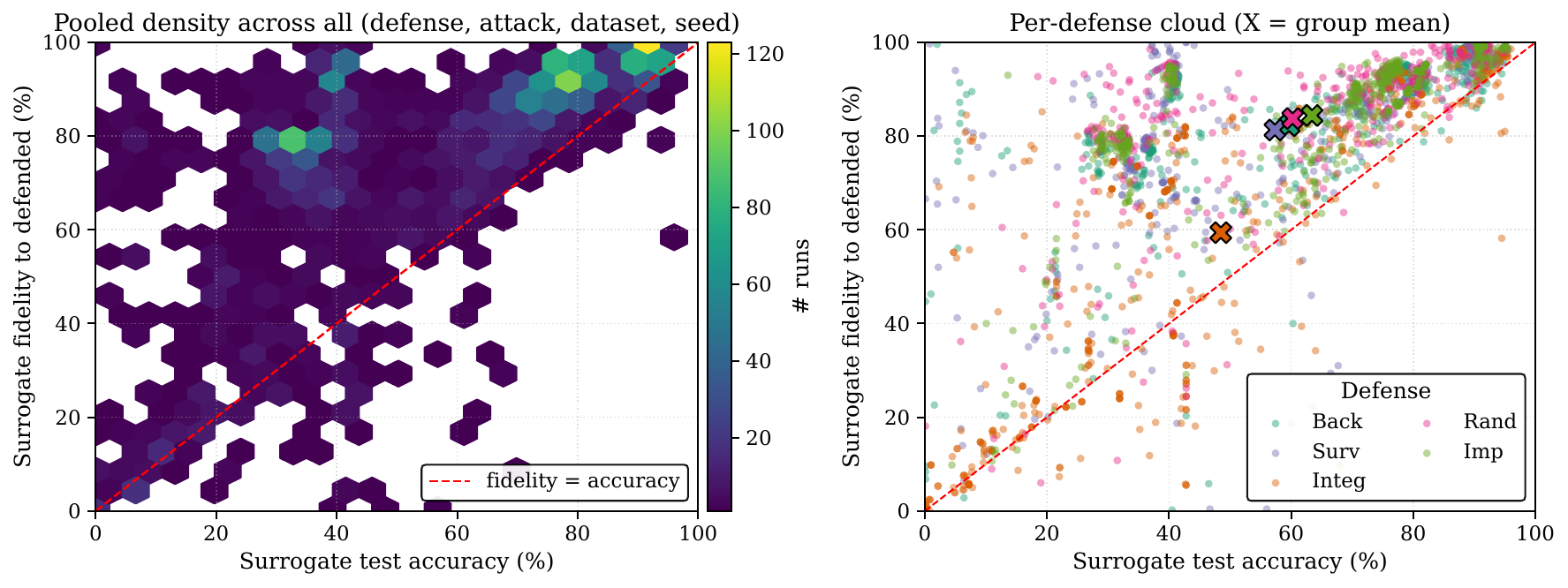}
  \caption{Surrogate accuracy vs.\ fidelity to defended target, pooled (left) and per defense (right).}
  \label{fig:app_acc_vs_fid_density}
\end{subfigure}
\caption{Surrogate fidelity by defense. (\subref{fig:app_fidelity_ecdf}) Empirical cumulative distributions. (\subref{fig:app_acc_vs_fid_density}) Surrogate accuracy versus fidelity to the defended target.}
\label{fig:app_fidelity_distribution}
\end{figure}

\begin{figure*}[t]
\centering
\includegraphics[width=\textwidth]{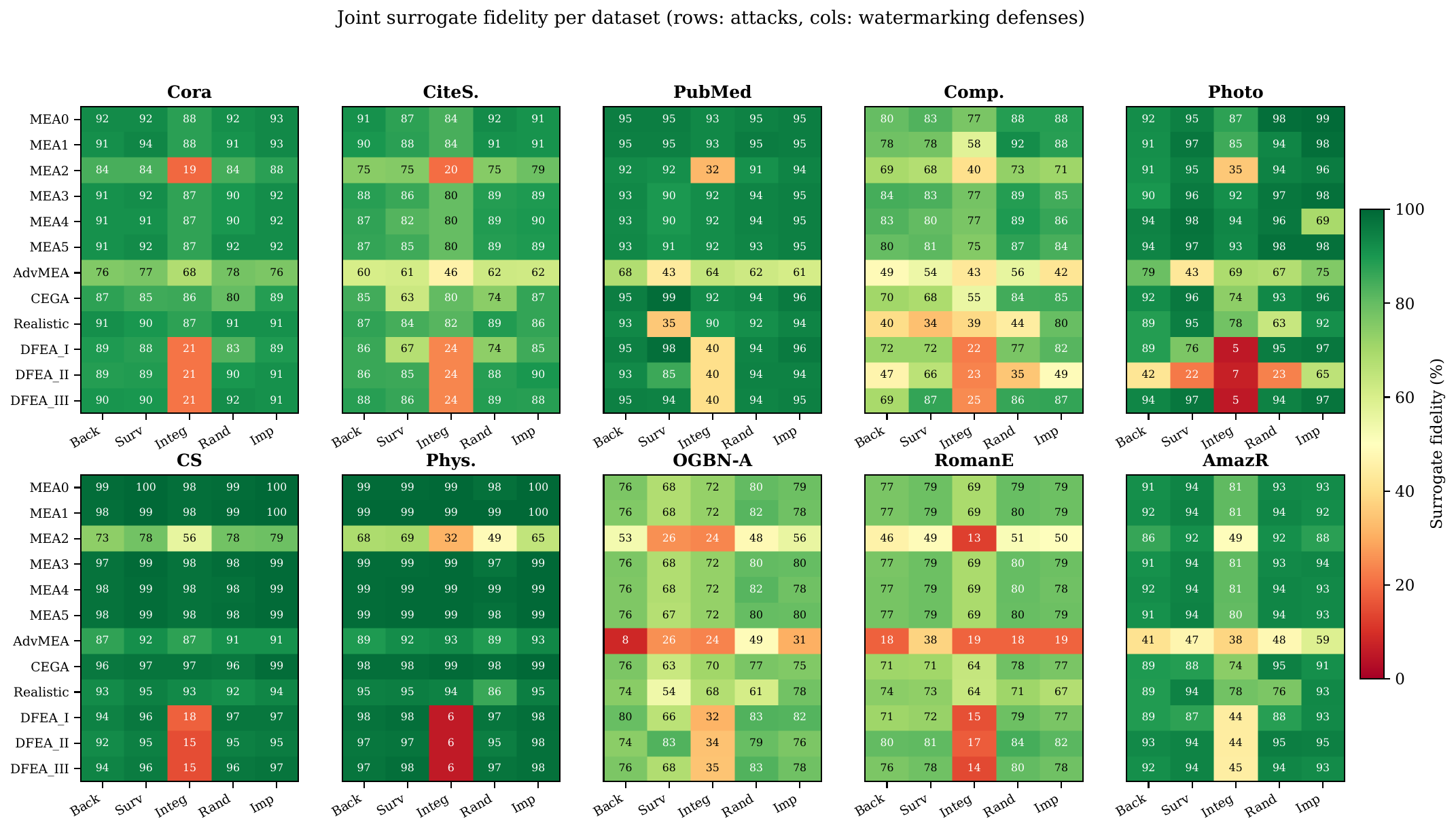}
\caption{Per-dataset heatmap grid of joint surrogate fidelity (\%). Each panel is a $12 \times 5$ matrix on one dataset (rows: attacks; columns: watermarking defenses); a single shared colormap (RdYlGn, $0$--$100\,\%$) supports cross-panel comparison.}
\label{fig:app_per_dataset_heatmap_grid}
\end{figure*}

The figures at the end of this subsection (Figures~\ref{fig:app_wm_survival_violin}--\ref{fig:app_per_dataset_heatmap_grid}) provide complementary statistical views: a violin plot of survival distributions per defense, a scatter that links survival to graph-level structural properties, an empirical CDF of surrogate fidelity stratified by defense family, and a per-dataset heatmap grid for~joint~fidelity.

\begin{table*}[t]
\centering
\caption{RQ5 joint evaluation on \textit{Cora} at budget $0.25\times$. (a) Surrogate fidelity (\%) on defended targets for the five ownership methods. (b) Surrogate fidelity (\%) on defended targets for the seven information-limiting and query-detection defenses. (c) Watermark verification rate (\%) on the surrogate. Mean $\pm$ standard deviation \mbox{over three seeds}.}
\label{tab:app_rq5_cora}
\scriptsize
\setlength{\tabcolsep}{3pt}
\renewcommand{\arraystretch}{0.96}
\begin{subtable}[t]{\textwidth}
\centering
\caption{Joint fidelity on \textit{Cora}, watermarking defenses}

\end{subtable}
\medskip
\begin{subtable}[t]{\textwidth}
\centering
\caption{Joint fidelity on \textit{Cora}, information-limiting and query-detection defenses}
%
\end{subtable}
\medskip
\begin{subtable}[t]{\textwidth}
\centering
\caption{Watermark survival on the surrogate, \textit{Cora}}
%
\end{subtable}
\end{table*}

\begin{table*}[t]
\centering
\caption{Surrogate fidelity (\%) on defended targets (5 watermarking defenses) for \textit{CiteSeer} and \textit{PubMed}. Mean $\pm$ standard deviation \mbox{over three seeds}.}
\label{tab:app_rq5_orig_a}
\scriptsize
\setlength{\tabcolsep}{4pt}
\renewcommand{\arraystretch}{0.96}
\begin{subtable}[t]{\textwidth}
\centering
\caption{\textit{CiteSeer}}
%
\end{subtable}
\medskip
\begin{subtable}[t]{\textwidth}
\centering
\caption{\textit{PubMed}}
%
\end{subtable}
\end{table*}

\begin{table*}[t]
\centering
\caption{Surrogate fidelity (\%) on defended targets (5 watermarking defenses) for \textit{Computers} and \textit{Photo}. Mean $\pm$ standard deviation over three seeds (ten for Integrity on Computers and Photo).}
\label{tab:app_rq5_orig_b}
\scriptsize
\setlength{\tabcolsep}{4pt}
\renewcommand{\arraystretch}{0.96}
\begin{subtable}[t]{\textwidth}
\centering
\caption{\textit{Computers}}
%
\end{subtable}
\medskip
\begin{subtable}[t]{\textwidth}
\centering
\caption{\textit{Photo}}
%
\end{subtable}
\end{table*}

\begin{table*}[t]
\centering
\caption{Surrogate fidelity (\%) on defended targets (5 watermarking defenses) for \textit{CoauthorCS} and \textit{CoauthorPhysics}. Mean $\pm$ standard deviation \mbox{over three seeds}.}
\label{tab:app_rq5_orig_c}
\scriptsize
\setlength{\tabcolsep}{4pt}
\renewcommand{\arraystretch}{0.96}
\begin{subtable}[t]{\textwidth}
\centering
\caption{\textit{CoauthorCS}}
%
\end{subtable}
\medskip
\begin{subtable}[t]{\textwidth}
\centering
\caption{\textit{CoauthorPhysics}}
%
\end{subtable}
\end{table*}

\begin{table*}[t]
\centering
\caption{Surrogate fidelity (\%) on defended targets (5 watermarking defenses) for \textit{OGBN-Arxiv}, \textit{RomanEmpire}, and \textit{AmazonRatings}. Mean $\pm$ standard deviation \mbox{over three seeds}.}
\label{tab:app_rq5_orig_d}
\scriptsize
\setlength{\tabcolsep}{4pt}
\renewcommand{\arraystretch}{0.96}
\begin{subtable}[t]{\textwidth}
\centering
\caption{\textit{OGBN-Arxiv}}
%
\end{subtable}
\medskip
\begin{subtable}[t]{\textwidth}
\centering
\caption{\textit{RomanEmpire}}
%
\end{subtable}
\medskip
\begin{subtable}[t]{\textwidth}
\centering
\caption{\textit{AmazonRatings}}
%
\end{subtable}
\end{table*}

\begin{table*}[t]
\centering
\caption{Surrogate fidelity (\%) on defended targets (7 information-limiting and query-detection defenses) for \textit{CiteSeer}, \textit{PubMed}, and \textit{Computers}. Mean $\pm$ standard deviation \mbox{over three seeds}.}
\label{tab:app_rq5_new_a}
\tiny
\setlength{\tabcolsep}{2pt}
\renewcommand{\arraystretch}{0.95}
\begin{subtable}[t]{\textwidth}
\centering
\caption{\textit{CiteSeer}}
%
\end{subtable}
\medskip
\begin{subtable}[t]{\textwidth}
\centering
\caption{\textit{PubMed}}
%
\end{subtable}
\medskip
\begin{subtable}[t]{\textwidth}
\centering
\caption{\textit{Computers}}
%
\end{subtable}
\end{table*}

\begin{table*}[t]
\centering
\caption{Surrogate fidelity (\%) on defended targets (7 information-limiting and query-detection defenses) for \textit{Photo}, \textit{CoauthorCS}, and \textit{CoauthorPhysics}. Mean $\pm$ standard deviation \mbox{over three seeds}.}
\label{tab:app_rq5_new_b}
\tiny
\setlength{\tabcolsep}{2pt}
\renewcommand{\arraystretch}{0.95}
\begin{subtable}[t]{\textwidth}
\centering
\caption{\textit{Photo}}
%
\end{subtable}
\medskip
\begin{subtable}[t]{\textwidth}
\centering
\caption{\textit{CoauthorCS}}
%
\end{subtable}
\medskip
\begin{subtable}[t]{\textwidth}
\centering
\caption{\textit{CoauthorPhysics}}
%
\end{subtable}
\end{table*}

\begin{table*}[t]
\centering
\caption{Surrogate fidelity (\%) on defended targets (7 information-limiting and query-detection defenses) for \textit{OGBN-Arxiv}, \textit{RomanEmpire}, and \textit{AmazonRatings}. Mean $\pm$ standard deviation \mbox{over three seeds}.}
\label{tab:app_rq5_new_c}
\tiny
\setlength{\tabcolsep}{2pt}
\renewcommand{\arraystretch}{0.95}
\begin{subtable}[t]{\textwidth}
\centering
\caption{\textit{OGBN-Arxiv}}
%
\end{subtable}
\medskip
\begin{subtable}[t]{\textwidth}
\centering
\caption{\textit{RomanEmpire}}
%
\end{subtable}
\medskip
\begin{subtable}[t]{\textwidth}
\centering
\caption{\textit{AmazonRatings}}
%
\end{subtable}
\end{table*}

\begin{table*}[t]
\centering
\caption{Watermark verification rate (\%) on the surrogate produced by each attack against each watermarking defense, on \textit{CiteSeer}, \textit{PubMed}, \textit{Computers}, and \textit{Photo}. Higher is better. Mean $\pm$ standard deviation over three seeds (ten for Integrity on Computers and Photo).}
\label{tab:app_rq5_wm_survival_a}
\tiny
\setlength{\tabcolsep}{4pt}
\renewcommand{\arraystretch}{0.95}
\begin{subtable}[t]{0.48\textwidth}
\centering
\caption{\textit{CiteSeer}}
\resizebox{\linewidth}{!}{%
\begin{tabular}{lccccc}
\toprule
\textbf{Attack} & \texttt{BackdoorWM} & \texttt{SurviveWM} & \texttt{Integrity} & \texttt{RandomWM} & \texttt{Impercept.} \\
\midrule
\texttt{MEA0}     & 0.0$\pm$0.0   & 17.5$\pm$1.4 & 62.2$\pm$5.5 & 17.3$\pm$7.6 & 0.0$\pm$0.0 \\
\texttt{MEA1}     & 33.3$\pm$57.7 & 17.7$\pm$1.9 & 62.2$\pm$5.5 & 22.0$\pm$4.0 & 0.0$\pm$0.0 \\
\texttt{MEA2} & 33.3$\pm$47.1 & 0.0$\pm$0.0 & 24.5$\pm$2.8 & 0.0$\pm$0.0 & 0.0$\pm$0.0 \\
\texttt{MEA3}     & 0.0$\pm$0.0   & 18.8$\pm$3.0 & 55.2$\pm$5.5 & 20.7$\pm$2.3 & 0.0$\pm$0.0 \\
\texttt{MEA4}     & 33.3$\pm$57.7 & 15.9$\pm$1.3 & 55.2$\pm$5.5 & 17.3$\pm$2.3 & 0.0$\pm$0.0 \\
\texttt{MEA5}     & 33.3$\pm$57.7 & 17.5$\pm$0.8 & 55.2$\pm$5.5 & 16.7$\pm$3.1 & 0.0$\pm$0.0 \\
\texttt{AdvMEA}   & 0.0$\pm$0.0   & 15.1$\pm$0.5 & 28.9$\pm$3.3 & 17.3$\pm$6.1 & 0.0$\pm$0.0 \\
\texttt{CEGA}     & 33.3$\pm$57.7 & 19.0$\pm$1.4 & 51.5$\pm$6.2 & 17.3$\pm$1.2 & 0.0$\pm$0.0 \\
\texttt{DFEA\_I} & 0.0$\pm$0.0 & 17.9$\pm$2.0 & 28.7$\pm$4.7 & 14.0$\pm$4.9 & 0.0$\pm$0.0 \\
\texttt{DFEA\_II} & 33.3$\pm$47.1 & 19.1$\pm$0.9 & 28.7$\pm$4.7 & 18.0$\pm$2.8 & 0.0$\pm$0.0 \\
\texttt{DFEA\_III} & 0.0$\pm$0.0 & 18.3$\pm$2.8 & 28.7$\pm$4.7 & 20.0$\pm$8.6 & 0.0$\pm$0.0 \\
\texttt{Realistic}& 0.0$\pm$0.0   & 16.4$\pm$1.8 & 50.6$\pm$4.9 & 15.3$\pm$1.2 & 0.0$\pm$0.0 \\
\bottomrule
\end{tabular}
}
\end{subtable}
\hfill
\begin{subtable}[t]{0.48\textwidth}
\centering
\caption{\textit{PubMed}}
\resizebox{\linewidth}{!}{%
\begin{tabular}{lccccc}
\toprule
\textbf{Attack} & \texttt{BackdoorWM} & \texttt{SurviveWM} & \texttt{Integrity} & \texttt{RandomWM} & \texttt{Impercept.} \\
\midrule
\texttt{MEA0}     & 33.3$\pm$57.7 & 33.9$\pm$0.6 & 67.2$\pm$1.3 & 36.0$\pm$6.0 & 66.7$\pm$57.7 \\
\texttt{MEA1}     & 33.3$\pm$57.7 & 34.1$\pm$0.6 & 67.2$\pm$1.3 & 36.7$\pm$1.2 & 66.7$\pm$57.7 \\
\texttt{MEA2} & 0.0$\pm$0.0 & 0.0$\pm$0.0 & 30.6$\pm$5.8 & 0.0$\pm$0.0 & 0.0$\pm$0.0 \\
\texttt{MEA3}     & 0.0$\pm$0.0   & 34.6$\pm$0.6 & 65.9$\pm$0.7 & 42.7$\pm$6.4 & 100.0$\pm$0.0 \\
\texttt{MEA4}     & 33.3$\pm$57.7 & 34.7$\pm$0.4 & 65.9$\pm$0.7 & 34.7$\pm$6.4 & 100.0$\pm$0.0 \\
\texttt{MEA5}     & 33.3$\pm$57.7 & 34.7$\pm$0.2 & 65.9$\pm$0.7 & 30.7$\pm$9.2 & 100.0$\pm$0.0 \\
\texttt{AdvMEA}   & 66.7$\pm$57.7 & 34.4$\pm$0.3 & 49.4$\pm$6.3 & 30.7$\pm$2.3 & 38.9$\pm$53.6 \\
\texttt{CEGA}     & 66.7$\pm$57.7 & 35.1$\pm$0.9 & 66.9$\pm$2.0 & 37.3$\pm$5.0 & 66.7$\pm$57.7 \\
\texttt{DFEA\_I} & 33.3$\pm$47.1 & 34.6$\pm$0.2 & 36.9$\pm$5.2 & 38.7$\pm$5.0 & 66.7$\pm$47.1 \\
\texttt{DFEA\_II} & 66.7$\pm$47.1 & 34.4$\pm$1.3 & 36.9$\pm$5.2 & 36.7$\pm$3.4 & 100.0$\pm$0.0 \\
\texttt{DFEA\_III} & 0.0$\pm$0.0 & 34.1$\pm$0.2 & 36.9$\pm$5.2 & 42.7$\pm$4.1 & 66.7$\pm$47.1 \\
\texttt{Realistic}& 33.3$\pm$57.7 & 33.0$\pm$1.4 & 63.8$\pm$4.3 & 32.7$\pm$5.0 & 100.0$\pm$0.0 \\
\bottomrule
\end{tabular}
}
\end{subtable}
\medskip
\begin{subtable}[t]{0.48\textwidth}
\centering
\caption{\textit{Computers}}
\resizebox{\linewidth}{!}{%
\begin{tabular}{lccccc}
\toprule
\textbf{Attack} & \texttt{BackdoorWM} & \texttt{SurviveWM} & \texttt{Integrity} & \texttt{RandomWM} & \texttt{Impercept.} \\
\midrule
\texttt{MEA0}     & 60.0$\pm$20.0 & 9.9$\pm$0.6  & 79.4$\pm$15.2 & 15.7$\pm$5.6 & 0.0$\pm$0.0 \\
\texttt{MEA1}     & 51.7$\pm$35.4 & 9.6$\pm$0.8  & 75.6$\pm$18.3 & 15.3$\pm$5.3 & 0.0$\pm$0.0 \\
\texttt{MEA2} & 70.0$\pm$24.5 & 0.0$\pm$0.0 & 43.9$\pm$33.8 & 0.0$\pm$0.0 & 0.0$\pm$0.0 \\
\texttt{MEA3}     & 60.0$\pm$21.9 & 10.0$\pm$0.7 & 79.1$\pm$16.6 & 14.7$\pm$7.8 & 0.0$\pm$0.0 \\
\texttt{MEA4}     & 58.3$\pm$30.6 & 10.1$\pm$0.6 & 45.7$\pm$27.7 & 14.7$\pm$3.7 & 0.0$\pm$0.0 \\
\texttt{MEA5}     & 55.0$\pm$27.4 & 10.2$\pm$0.9 & 74.9$\pm$20.8 & 12.0$\pm$2.5 & 0.0$\pm$0.0 \\
\texttt{AdvMEA}   & 65.0$\pm$26.6 & 10.4$\pm$0.6 & 38.5$\pm$25.8 & 12.3$\pm$6.1 & 0.0$\pm$0.0 \\
\texttt{CEGA}     & 61.7$\pm$33.1 & 9.8$\pm$0.7  & 65.4$\pm$28.4 & 16.0$\pm$5.4 & 0.0$\pm$0.0 \\
\texttt{DFEA\_I} & 63.3$\pm$12.5 & 9.2$\pm$0.8 & 11.2$\pm$11.3 & 10.0$\pm$0.0 & 0.0$\pm$0.0 \\
\texttt{DFEA\_II} & 16.7$\pm$23.6 & 9.7$\pm$0.1 & 11.1$\pm$11.3 & 10.7$\pm$2.5 & 0.0$\pm$0.0 \\
\texttt{DFEA\_III} & 30.0$\pm$35.6 & 10.5$\pm$0.4 & 11.2$\pm$11.2 & 16.0$\pm$2.8 & 0.0$\pm$0.0 \\
\texttt{Realistic}& 5.0$\pm$8.4   & 10.2$\pm$0.9 & 40.1$\pm$16.3 & 9.0$\pm$3.9  & 0.0$\pm$0.0 \\
\bottomrule
\end{tabular}
}
\end{subtable}
\hfill
\begin{subtable}[t]{0.48\textwidth}
\centering
\caption{\textit{Photo}}
\resizebox{\linewidth}{!}{%
\begin{tabular}{lccccc}
\toprule
\textbf{Attack} & \texttt{BackdoorWM} & \texttt{SurviveWM} & \texttt{Integrity} & \texttt{RandomWM} & \texttt{Impercept.} \\
\midrule
\texttt{MEA0}     & 33.3$\pm$7.2  & 12.3$\pm$0.3 & 76.7$\pm$11.1 & 17.3$\pm$4.2 & 0.0$\pm$0.0 \\
\texttt{MEA1}     & 54.2$\pm$19.1 & 12.7$\pm$1.1 & 74.1$\pm$12.3 & 10.0$\pm$4.0 & 0.0$\pm$0.0 \\
\texttt{MEA2} & 66.7$\pm$11.8 & 0.0$\pm$0.0 & 29.0$\pm$32.5 & 0.0$\pm$0.0 & 0.0$\pm$0.0 \\
\texttt{MEA3}     & 29.2$\pm$7.2  & 12.4$\pm$0.9 & 78.7$\pm$10.7 & 14.0$\pm$2.0 & 0.0$\pm$0.0 \\
\texttt{MEA4}     & 50.0$\pm$25.0 & 12.3$\pm$0.8 & 49.3$\pm$21.1 & 12.7$\pm$4.2 & 0.0$\pm$0.0 \\
\texttt{MEA5}     & 58.3$\pm$7.2  & 12.3$\pm$0.8 & 82.2$\pm$8.8 & 14.7$\pm$2.3 & 33.3$\pm$57.7 \\
\texttt{AdvMEA}   & 62.5$\pm$33.1 & 12.2$\pm$0.6 & 53.9$\pm$20.5 & 18.0$\pm$2.0 & 0.0$\pm$0.0 \\
\texttt{CEGA}     & 50.0$\pm$21.7 & 12.1$\pm$0.3 & 72.7$\pm$22.9 & 16.0$\pm$2.0 & 0.0$\pm$0.0 \\
\texttt{DFEA\_I} & 25.0$\pm$20.4 & 12.1$\pm$0.3 & 11.8$\pm$7.3 & 17.3$\pm$1.9 & 0.0$\pm$0.0 \\
\texttt{DFEA\_II} & 45.8$\pm$41.2 & 12.5$\pm$0.6 & 11.7$\pm$7.2 & 19.3$\pm$2.5 & 0.0$\pm$0.0 \\
\texttt{DFEA\_III} & 45.8$\pm$21.2 & 12.2$\pm$0.3 & 10.3$\pm$6.3 & 12.0$\pm$3.3 & 0.0$\pm$0.0 \\
\texttt{Realistic}& 16.7$\pm$7.2  & 12.6$\pm$0.7 & 61.2$\pm$16.8 & 12.7$\pm$3.1 & 0.0$\pm$0.0 \\
\bottomrule
\end{tabular}
}
\end{subtable}
\end{table*}

\begin{table*}[t]
\centering
\caption{Watermark verification rate (\%) on the surrogate produced by each attack against each watermarking defense, on \textit{CoauthorCS}, \textit{CoauthorPhysics}, \textit{RomanEmpire}, and \textit{AmazonRatings}. Higher is better. Mean $\pm$ standard deviation \mbox{over three seeds}.}
\label{tab:app_rq5_wm_survival_b}
\tiny
\setlength{\tabcolsep}{4pt}
\renewcommand{\arraystretch}{0.95}
\begin{subtable}[t]{0.48\textwidth}
\centering
\caption{\textit{CoauthorCS}}
\resizebox{\linewidth}{!}{%
\begin{tabular}{lccccc}
\toprule
\textbf{Attack} & \texttt{BackdoorWM} & \texttt{SurviveWM} & \texttt{Integrity} & \texttt{RandomWM} & \texttt{Impercept.} \\
\midrule
\texttt{MEA0}     & 42.2$\pm$20.4 & 7.2$\pm$0.2 & 84.2$\pm$0.1 & 9.3$\pm$1.2 & 33.3$\pm$57.7 \\
\texttt{MEA1}     & 55.6$\pm$10.2 & 7.0$\pm$0.3 & 84.2$\pm$0.1 & 10.0$\pm$4.0 & 33.3$\pm$57.7 \\
\texttt{MEA2} & 35.6$\pm$8.3 & 0.0$\pm$0.0 & 34.8$\pm$1.9 & 0.0$\pm$0.0 & 0.0$\pm$0.0 \\
\texttt{MEA3}     & 48.9$\pm$16.8 & 7.1$\pm$0.4 & 82.0$\pm$0.3 & 7.3$\pm$4.2 & 33.3$\pm$57.7 \\
\texttt{MEA4}     & 44.4$\pm$13.9 & 7.1$\pm$0.3 & 82.0$\pm$0.3 & 6.0$\pm$2.0 & 33.3$\pm$57.7 \\
\texttt{MEA5}     & 53.3$\pm$20.0 & 6.9$\pm$0.4 & 82.0$\pm$0.3 & 4.7$\pm$5.0 & 33.3$\pm$57.7 \\
\texttt{AdvMEA}   & 11.1$\pm$10.2 & 7.3$\pm$0.8 & 58.3$\pm$0.9 & 4.7$\pm$2.3 & 0.0$\pm$0.0 \\
\texttt{CEGA}     & 37.8$\pm$20.4 & 7.0$\pm$0.4 & 78.3$\pm$0.3 & 9.3$\pm$6.1 & 33.3$\pm$57.7 \\
\texttt{DFEA\_I} & 40.0$\pm$9.4 & 6.9$\pm$0.3 & 16.4$\pm$6.1 & 12.0$\pm$3.3 & 0.0$\pm$0.0 \\
\texttt{DFEA\_II} & 33.3$\pm$5.4 & 6.5$\pm$0.3 & 14.7$\pm$3.1 & 6.7$\pm$2.5 & 0.0$\pm$0.0 \\
\texttt{DFEA\_III} & 42.2$\pm$11.3 & 6.8$\pm$0.3 & 14.7$\pm$3.1 & 9.3$\pm$0.9 & 0.0$\pm$0.0 \\
\texttt{Realistic}& 2.2$\pm$3.8   & 6.7$\pm$0.3 & 67.0$\pm$0.2 & 6.0$\pm$3.5 & 0.0$\pm$0.0 \\
\bottomrule
\end{tabular}
}
\end{subtable}
\hfill
\begin{subtable}[t]{0.48\textwidth}
\centering
\caption{\textit{CoauthorPhysics}}
\resizebox{\linewidth}{!}{%
\begin{tabular}{lccccc}
\toprule
\textbf{Attack} & \texttt{BackdoorWM} & \texttt{SurviveWM} & \texttt{Integrity} & \texttt{RandomWM} & \texttt{Impercept.} \\
\midrule
\texttt{MEA0}     & 80.0$\pm$20.0 & 20.5$\pm$0.6 & 93.7$\pm$0.2 & 26.7$\pm$3.1 & 33.3$\pm$57.7 \\
\texttt{MEA1}     & 53.3$\pm$11.5 & 20.5$\pm$0.6 & 93.7$\pm$0.2 & 25.3$\pm$4.6 & 33.3$\pm$57.7 \\
\texttt{MEA2} & 26.7$\pm$18.9 & 0.0$\pm$0.0 & 31.8$\pm$25.6 & 0.0$\pm$0.0 & 0.0$\pm$0.0 \\
\texttt{MEA3}     & 53.3$\pm$11.5 & 20.3$\pm$0.7 & 92.6$\pm$0.1 & 25.3$\pm$2.3 & 33.3$\pm$57.7 \\
\texttt{MEA4}     & 66.7$\pm$30.6 & 20.4$\pm$0.6 & 92.6$\pm$0.1 & 23.3$\pm$4.2 & 33.3$\pm$57.7 \\
\texttt{MEA5}     & 80.0$\pm$0.0  & 20.3$\pm$0.7 & 92.6$\pm$0.1 & 22.0$\pm$3.5 & 33.3$\pm$57.7 \\
\texttt{AdvMEA}   & 26.7$\pm$30.6 & 19.4$\pm$1.0 & 68.7$\pm$1.0 & 19.3$\pm$10.3 & 33.3$\pm$57.7 \\
\texttt{CEGA}     & 86.7$\pm$11.5 & 20.1$\pm$0.7 & 92.6$\pm$0.1 & 34.7$\pm$4.2 & 33.3$\pm$57.7 \\
\texttt{DFEA\_I} & 80.0$\pm$16.3 & 20.4$\pm$0.2 & 41.9$\pm$0.7 & 29.3$\pm$2.5 & 33.3$\pm$47.1 \\
\texttt{DFEA\_II} & 73.3$\pm$9.4 & 19.6$\pm$0.7 & 41.9$\pm$0.7 & 22.7$\pm$1.9 & 66.7$\pm$47.1 \\
\texttt{DFEA\_III} & 73.3$\pm$18.9 & 20.4$\pm$0.5 & 41.9$\pm$0.7 & 26.0$\pm$6.5 & 33.3$\pm$47.1 \\
\texttt{Realistic}& 13.3$\pm$11.5 & 20.2$\pm$0.8 & 77.7$\pm$1.0 & 26.7$\pm$7.0 & 33.3$\pm$57.7 \\
\bottomrule
\end{tabular}
}
\end{subtable}
\medskip
\begin{subtable}[t]{0.48\textwidth}
\centering
\caption{\textit{RomanEmpire}}
\resizebox{\linewidth}{!}{%
\begin{tabular}{lccccc}
\toprule
\textbf{Attack} & \texttt{BackdoorWM} & \texttt{SurviveWM} & \texttt{Integrity} & \texttt{RandomWM} & \texttt{Impercept.} \\
\midrule
\texttt{MEA0}     & 5.8$\pm$2.1   & 5.5$\pm$0.1 & 49.3$\pm$1.5 & 6.8$\pm$3.0 & 100.0$\pm$0.0 \\
\texttt{MEA1}     & 4.8$\pm$2.8   & 5.5$\pm$0.1 & 49.3$\pm$1.5 & 4.4$\pm$3.8 & 66.7$\pm$57.7 \\
\texttt{MEA2} & 0.9$\pm$0.7 & 0.0$\pm$0.0 & 10.9$\pm$1.2 & 0.0$\pm$0.0 & 0.0$\pm$0.0 \\
\texttt{MEA3}     & 5.8$\pm$2.8   & 5.5$\pm$0.1 & 49.3$\pm$1.5 & 8.4$\pm$2.6 & 66.7$\pm$57.7 \\
\texttt{MEA4}     & 8.3$\pm$2.9   & 5.5$\pm$0.1 & 49.3$\pm$1.5 & 8.8$\pm$5.4 & 33.3$\pm$57.7 \\
\texttt{MEA5}     & 7.1$\pm$3.9   & 5.5$\pm$0.1 & 49.3$\pm$1.5 & 5.6$\pm$2.6 & 66.7$\pm$57.7 \\
\texttt{AdvMEA}   & 26.3$\pm$22.8 & 5.5$\pm$0.2 & 26.5$\pm$3.7 & 10.0$\pm$2.0 & 0.0$\pm$0.0 \\
\texttt{CEGA}     & 1.0$\pm$1.4   & 5.4$\pm$0.2 & 37.8$\pm$0.6 & 8.7$\pm$4.4 & 66.7$\pm$57.7 \\
\texttt{DFEA\_I} & 0.9$\pm$0.7 & 5.4$\pm$0.1 & 18.2$\pm$5.6 & 4.0$\pm$1.6 & 0.0$\pm$0.0 \\
\texttt{DFEA\_II} & 31.9$\pm$5.9 & 5.4$\pm$0.1 & 22.7$\pm$6.1 & 6.0$\pm$1.6 & 0.0$\pm$0.0 \\
\texttt{DFEA\_III} & 4.7$\pm$1.1 & 5.4$\pm$0.1 & 23.0$\pm$6.4 & 6.0$\pm$1.6 & 33.3$\pm$47.1 \\
\texttt{Realistic}& 0.5$\pm$1.0   & 5.4$\pm$0.1 & 44.6$\pm$2.9 & 8.0$\pm$4.2 & 0.0$\pm$0.0 \\
\bottomrule
\end{tabular}
}
\end{subtable}
\hfill
\begin{subtable}[t]{0.48\textwidth}
\centering
\caption{\textit{AmazonRatings}}
\resizebox{\linewidth}{!}{%
\begin{tabular}{lccccc}
\toprule
\textbf{Attack} & \texttt{BackdoorWM} & \texttt{SurviveWM} & \texttt{Integrity} & \texttt{RandomWM} & \texttt{Impercept.} \\
\midrule
\texttt{MEA0}     & 70.8$\pm$10.2 & 20.0$\pm$1.0 & 67.5$\pm$4.5 & 22.7$\pm$6.4 & 0.0$\pm$0.0 \\
\texttt{MEA1}     & 68.0$\pm$12.8 & 20.0$\pm$1.2 & 67.9$\pm$4.6 & 32.7$\pm$4.2 & 0.0$\pm$0.0 \\
\texttt{MEA2} & 15.0$\pm$2.7 & 0.0$\pm$0.0 & 35.9$\pm$36.3 & 0.0$\pm$0.0 & 0.0$\pm$0.0 \\
\texttt{MEA3}     & 69.1$\pm$8.7  & 20.1$\pm$1.2 & 67.5$\pm$4.5 & 26.0$\pm$3.5 & 0.0$\pm$0.0 \\
\texttt{MEA4}     & 73.2$\pm$6.0  & 20.3$\pm$1.1 & 67.9$\pm$4.6 & 20.7$\pm$1.2 & 0.0$\pm$0.0 \\
\texttt{MEA5}     & 74.3$\pm$8.2  & 20.0$\pm$1.1 & 67.6$\pm$4.6 & 22.7$\pm$4.6 & 0.0$\pm$0.0 \\
\texttt{AdvMEA}   & 94.0$\pm$7.0  & 20.0$\pm$0.7 & 42.7$\pm$37.7 & 20.7$\pm$7.6 & 33.3$\pm$57.7 \\
\texttt{CEGA}     & 66.1$\pm$1.7  & 20.0$\pm$0.7 & 53.2$\pm$6.1 & 23.3$\pm$4.2 & 0.0$\pm$0.0 \\
\texttt{DFEA\_I} & 68.3$\pm$4.5 & 20.0$\pm$0.6 & 41.9$\pm$36.8 & 18.7$\pm$4.1 & 0.0$\pm$0.0 \\
\texttt{DFEA\_II} & 99.2$\pm$0.7 & 19.7$\pm$0.9 & 42.8$\pm$37.4 & 26.7$\pm$4.7 & 0.0$\pm$0.0 \\
\texttt{DFEA\_III} & 71.9$\pm$4.8 & 20.0$\pm$0.9 & 43.2$\pm$37.8 & 21.3$\pm$3.4 & 0.0$\pm$0.0 \\
\texttt{Realistic}& 43.7$\pm$6.2  & 20.4$\pm$0.9 & 67.5$\pm$11.5 & 18.0$\pm$6.0 & 0.0$\pm$0.0 \\
\bottomrule
\end{tabular}
}
\end{subtable}
\end{table*}

\paragraph{Statistical views on the joint evaluation.}
The distribution panels summarize surrogate verification by defense and compare it with graph homophily. Comparisons across defenses require the same attack implementation and~query~budget.

Figure~\ref{fig:app_fidelity_ecdf} shows the empirical distribution of surrogate fidelity by defense. Figure~\ref{fig:app_acc_vs_fid_density} compares surrogate accuracy on ground-truth labels with fidelity to the defended target. The diagonal marks equality between the two measures.

Finally, Figure~\ref{fig:app_per_dataset_heatmap_grid} renders the joint surrogate fidelity as a per-dataset heatmap grid, one $12 \times 5$ panel per dataset.

\FloatBarrier
\subsubsection{Post-extraction verification beyond embedded watermarks}
\label{app:newver}
We evaluated \href{https://arxiv.org/abs/2304.08566}{GrOVe} and \href{https://arxiv.org/abs/2602.20419}{CREDIT} on extracted surrogates using their released implementations, with the protected model as the reference. Table~\ref{tab:additionalver} gives the results in each verifier's native metric; neither AUC nor true positive rate is interchangeable with~trigger~accuracy.

\begin{center}\begin{minipage}{\linewidth}\centering\small
\captionof{table}{Ownership verification on extracted surrogates.}\label{tab:additionalver}
\begin{tabular}{@{}lll@{}}\toprule
Verifier & Evaluation & Result\\ \midrule
GrOVe & Node classification & TPR 0.740, FPR 0.004\\ CREDIT & ENZYMES & AUC 0.778\\ CREDIT & PROTEINS & AUC 0.750\\ \bottomrule
\end{tabular}\end{minipage}\end{center}
Under the corrected Integrity protocol, no extracted surrogate reached perfect fidelity in the 120 runs on either product graph (Table~\ref{tab:correctedip}). The target and surrogate verification rates use the same owner-side~query~procedure.

\begin{center}\begin{minipage}{\linewidth}\centering\small
\captionof{table}{Corrected Integrity results (\%). Target verification denotes fingerprint self-agreement; surrogate verification is mean fingerprint agreement over 120 runs.}\label{tab:correctedip}
\begin{tabular}{@{}lrrrrrr@{}}\toprule
Dataset & \makecell{Target\\acc.} & \makecell{Target\\collapse} & \makecell{Target\\verif.} & \makecell{Median\\fidelity} & \makecell{Surrogate\\verif.} & \makecell{Perfect\\fidelity}\\ \midrule
Computers & 44.5 & 0/10 & 100.0 & 49.1 & 48.0 & 0/120\\ Photo & 86.4 & 0/10 & 100.0 & 75.7 & 51.0 & 0/120\\ \bottomrule
\end{tabular}\end{minipage}\end{center}

\begin{center}\begin{minipage}{\linewidth}\centering\small
\captionof{table}{Fidelity to the undefended target (\%) at the smallest and largest budgets. The last column gives the largest difference between endpoints at a fixed budget (percentage points).}\label{tab:endpoints}
\begin{tabular}{@{}lrrr@{}}\toprule
Dataset & $0.05\times$ & $1.00\times$ & Largest endpoint difference\\ \midrule
Cora & 83.2 & 92.7 & 1.7\\ Computers & 69.0 & 81.4 & 21.0\\ RomanEmpire & 73.0 & 79.3 & 12.1\\ \bottomrule
\end{tabular}\end{minipage}\end{center}
\begin{center}\begin{minipage}{\linewidth}\centering\small
\captionof{table}{Surrogate verification (\%): minimum and maximum over three budgets and three endpoints. The last row uses a query verifier; the other rows measure inherited marks.}\label{tab:wmranges}
\begin{tabular}{@{}lrr@{}}\toprule
Defense & Cora & RomanEmpire\\ \midrule
BackdoorWM & 0--33 & 0--5.9\\ RandomWM & 9--15 & 9--12\\ SurviveWM & 12--15 & 5.3--5.5\\ \midrule Integrity & 52--57 & 38--51\\ \bottomrule
\end{tabular}\end{minipage}\end{center}

\subsection{RQ5 watermark survival under each paper's original setup}
\label{app:rq5_supp}

The headline RQ5 finding is that two graph watermarks (\texttt{SurviveWM} and \texttt{BackdoorWM}) appear to survive extraction in their source papers but reach much lower surrogate-side verification under our 12-attack joint protocol. Two reasonable rebuttals must be ruled out: (a) the gap is an artefact of our broader protocol diverging from the original setups, and (b) our re-implementations fail to embed the watermark on the protected target in the first place. This appendix runs each watermark under \emph{the original paper's own} protocol and shows that the gap survives~both~rebuttals.

\paragraph{Setup mismatch with prior watermark papers.}
Table~\ref{tab:app_rq5_supp_protocol} compares our main-text RQ5 protocol against each watermark paper's reported setup. Both papers evaluate against a single, narrowly defined extraction attack and a different task or dataset family, while our protocol evaluates against twelve attacks on the same node-classification graphs. Any direct numerical comparison must therefore be made under either protocol's setup, not~across~them.

\begin{table}[!htbp]
\centering
\caption{Protocol mismatch between the two graph-watermark papers most directly affected by RQ5 and the GraphIP-Bench joint protocol.}
\label{tab:app_rq5_supp_protocol}
\scriptsize
\setlength{\tabcolsep}{4pt}
\renewcommand{\arraystretch}{1.0}
\begin{tabularx}{\linewidth}{@{}lXXX@{}}
\toprule
\textbf{Axis} & \textbf{SurviveWM~\citep{wang2023making}} & \textbf{BackdoorWM~\citep{xu2023watermarking}} & \textbf{GraphIP-Bench RQ5} \\
\midrule
Task        & Graph classification & Node + graph classification & Node classification \\
Datasets    & MSRC-9, ENZYMES & Cora, CiteSeer, NCI1, COLLAB, REDDIT-BINARY & 10 datasets, incl.\ \textit{Computers} \\
Attack family & Same-arch hard-label MEA & Same-arch knowledge distillation & 12 black-box attacks (MEA0--5, AdvMEA, CEGA, Realistic, DFEA\_I/II/III) \\
Verification metric & $\bar{E}$ binary effective rate & WM accuracy on trigger nodes & Surrogate-side verification rate \\
\bottomrule
\end{tabularx}
\end{table}

\paragraph{SurviveWM under the paper-faithful graph-classification protocol.}
We re-implement \texttt{SurviveWM} on the original three graph-classification datasets (MSRC-9, ENZYMES, PROTEINS) with the paper's hyperparameters (SAGE host, $L{=}4$, hidden $160$, dropout $0.05$, $200$ clean + $200$ watermark epochs, SNNL coefficient $\alpha{=}0.1$, key-input ratio $a{=}0.1$, $T_{\text{opt}}{=}20$) and the paper's effective-watermark metric $E_{\text{ave}}$, defined as the fraction of trials in which $E_{\text{sin}}(M) > E_{\text{sin}}(M_{\text{clean}})$ \emph{and} $E_{\text{sin}}(M) > E_{\text{sin}}(M_{\text{random}})$. We use $80/20$ stratified splits with ten seeds per dataset, and we evaluate the watermark against the same twelve attacks as in the main RQ5 (rather than only the same-architecture hard-label MEA the original paper considered). Headline aggregates and per-attack numbers on the paper's own MSRC-9 dataset are in Table~\ref{tab:app_rq5_supp_survivewm_summary} and Table~\ref{tab:app_rq5_supp_survivewm_msrc}.

\begin{table}[!htbp]
\centering
\caption{\texttt{SurviveWM} under its paper-faithful protocol on the three graph-classification datasets it was originally evaluated on. medFid / medTgtWM / medSurWM are medians over $10$ seeds $\times$ $12$ attacks ($120$ rows per dataset). $E_{\text{ave},T}$ is the binary effective rate of the watermark on the protected target; $E_{\text{ave},S}$ is the same metric on the extracted surrogate. ``Random floor'' is $1/C$ for a task with~$C$~classes.}
\label{tab:app_rq5_supp_survivewm_summary}
\scriptsize
\setlength{\tabcolsep}{4pt}
\renewcommand{\arraystretch}{1.0}
\begin{tabular*}{0.92\textwidth}{@{}l@{\extracolsep{\fill}}cccccc@{}}
\toprule
\textbf{Dataset} & \textbf{medFid (\%)} & \textbf{medTgtWM (\%)} & \textbf{medSurWM (\%)} & $\boldsymbol{E_{\text{ave},T}}$ & $\boldsymbol{E_{\text{ave},S}}$ & \textbf{Random floor} \\
\midrule
MSRC-9 (paper's own) & 88.9 & 88.2 &  5.9 & \textbf{1.00} & \textbf{0.42} & 12.5\,\% (1/8) \\
ENZYMES              & 37.5 & 57.3 & 16.7 & \textbf{1.00} & \textbf{0.51} & 16.7\,\% (1/6) \\
PROTEINS             & 80.7 & 83.1 & 65.2 & \textbf{1.00} & \textbf{0.67} & 50.0\,\% (1/2) \\
\bottomrule
\end{tabular*}
\end{table}

\begin{table}[!htbp]
\centering
\caption{\texttt{SurviveWM} on \emph{MSRC-9} (the paper's own dataset) under twelve attacks. TgtWM is the verification rate on the protected target (paper's $E_{\text{sin}}$); SurWM is the verification rate on the extracted surrogate; $E_{\text{ave},S}$ is the paper's binary effective rate on the surrogate, averaged over $10$ seeds. Bold rows mark $E_{\text{ave},S}{=}1.00$ (every seed effective).}
\label{tab:app_rq5_supp_survivewm_msrc}
\scriptsize
\setlength{\tabcolsep}{6pt}
\renewcommand{\arraystretch}{0.98}
\begin{tabular*}{0.86\textwidth}{@{}l@{\extracolsep{\fill}}lcccc@{}}
\toprule
\textbf{Attack} & \textbf{Family} & \textbf{TgtWM (\%)} & \textbf{SurWM (\%)} & $\boldsymbol{E_{\text{ave},S}}$ \\
\midrule
\texttt{MEA0}      & Same-arch hard label                 & 88.2 & \phantom{0}0.0 & 0.00 \\
\texttt{MEA1}      & Shuffled-order hard label            & 88.2 & 14.7           & 0.60 \\
\texttt{MEA2}      & Soft label, $T{=}3$                  & 88.2 & \phantom{0}2.9 & 0.30 \\
\texttt{MEA3}      & Soft label, $T{=}1$                  & 88.2 & \phantom{0}0.0 & 0.20 \\
\texttt{MEA4}      & Cross-arch SAGE, hard                & 88.2 & \phantom{0}0.0 & 0.10 \\
\texttt{MEA5}      & Cross-arch SAGE, soft                & 88.2 & \phantom{0}5.9 & 0.30 \\
\texttt{AdvMEA}    & 10\,\% adversarial flip              & 88.2 & \phantom{0}0.0 & 0.20 \\
\texttt{CEGA}      & Centrality / entropy, soft           & 88.2 & \phantom{0}5.9 & 0.30 \\
\texttt{Realistic} & Edge-graph reconstruction (0.3$\times$) & 88.2 & \phantom{0}0.0 & 0.00 \\
\textbf{\texttt{DFEA\_I}}   & Data-free, GCN              & 88.2 & 47.1           & \textbf{1.00} \\
\textbf{\texttt{DFEA\_II}}  & Data-free, SAGE             & 88.2 & 41.2           & \textbf{1.00} \\
\textbf{\texttt{DFEA\_III}} & Data-free, GIN              & 88.2 & 47.1           & \textbf{1.00} \\
\bottomrule
\end{tabular*}
\end{table}

Three findings follow. \emph{First, SurviveWM's target embedding is fully reproducible.} Under the paper-faithful protocol the watermark embeds on the protected target with $E_{\text{ave},T}{=}1.00$ on every dataset and every attack, and the protected target's watermark accuracy reaches $88.2\%$ on MSRC-9 (vs.\ a random-classifier floor of $12.9\%$); the embedding step is therefore not the source of the gap. \emph{Second, surrogate survival on the same dataset is highly attack-dependent and median surrogate $E_{\text{ave},S}$ never reaches the paper's reported $> 0.9$.} On MSRC-9, the same-architecture hard-label attacks (\texttt{MEA0}, \texttt{MEA3}, \texttt{MEA4}) and the structure-aware \texttt{Realistic} attack reach $E_{\text{ave},S}{=}0.00$--$0.20$, i.e.\ the watermark is completely washed out, even though the original paper's setup is closest to \texttt{MEA0}. \emph{Third, an attack-regime asymmetry exists.} Data-free attacks (\texttt{DFEA\_I/II/III}) reach $E_{\text{ave},S}{=}1.00$ on MSRC-9 because the SNNL training step pulls the watermark distribution into a region that off-manifold synthetic queries also activate; this is the unique attack regime in which SurviveWM's watermark survives, and it is the inverse of the data-driven setting that the paper reports. The headline finding of RQ5 --- that SurviveWM's extracted surrogate does not preserve the watermark under a broad attack panel --- therefore holds even on the paper's own dataset and under the paper's own metric, and the gap to the paper's $> 0.9$ reflects the fact that the paper's effective-rate result is reported only on the protected target $M_1$ rather than on the extracted surrogate $M_2$.

\paragraph{BackdoorWM under the paper-faithful KD protocol.}
Xu et al.\ (2023) evaluate watermark transferability under same-architecture knowledge distillation: a freshly initialised GCN student is trained on the teacher's softened logits over half of the held-out test mask. We replay this exact setup on the protocol's default \texttt{BackdoorWM} configuration (trigger rate $0.01$, $\ell{=}20$ trigger feature dimensions at value $0.99$, $\alpha{=}0.3$, $100$ pretrain + $200$ defense epochs) with KL distillation at $T{=}4.0$ for $200$ epochs, on five node-classification datasets (\textit{Computers}, \textit{Photo}, \textit{PubMed}, \textit{CiteSeer}, \textit{RomanEmpire}) and five seeds each. Results in Table~\ref{tab:app_rq5_supp_backdoor_kd}.

\begin{table}[!htbp]
\centering
\caption{\texttt{BackdoorWM} under its paper-faithful same-architecture knowledge-distillation extraction. TeachAcc / StuAcc are protected-target / surrogate test accuracy; Fid is fidelity to the teacher; TgtWM / SurWM are watermark verification on the teacher / on the KD student; RandFloor is the verification rate of a freshly initialised GCN; $E_{\text{ave},S}$ is the surrogate-side effective rate (per-seed binary $\text{SurWM} > \max(\text{RandFloor}, \text{TgtWM}/2)$, averaged over $5$ seeds). $^\ddagger$ flags two seeds on \textit{Computers} where joint training collapsed (TeacherAcc${<}10\%$); their high SurWM is not genuine watermark transfer (see caveat below).}
\label{tab:app_rq5_supp_backdoor_kd}
\scriptsize
\setlength{\tabcolsep}{4pt}
\renewcommand{\arraystretch}{1.0}
\begin{tabular*}{\textwidth}{@{}l@{\extracolsep{\fill}}cccccccc@{}}
\toprule
\textbf{Dataset} & \textbf{TeachAcc (\%)} & \textbf{StuAcc (\%)} & \textbf{Fid (\%)} & \textbf{TgtWM (\%)} & \textbf{SurWM (\%)} & \textbf{RandFloor (\%)} & $\boldsymbol{E_{\text{ave},T}}$ & $\boldsymbol{E_{\text{ave},S}}$ \\
\midrule
\textit{Computers}~$^\ddagger$ & 55.8 & 45.4 & 78.4 &  90.0 & \textbf{70.0} &  2.0 & 1.00 & \textbf{1.00} \\
\textit{Photo}                  & 87.2 & 86.6 & 88.2 &  87.5 & 37.5 & 10.0 & 1.00 & 0.60 \\
\textit{PubMed}                 & 77.8 & 77.2 & 94.0 & 100.0 & \textbf{100.0} & 20.0 & 0.80 & 0.60 \\
\textit{CiteSeer}               & 69.2 & 68.6 & 83.0 & 100.0 & \phantom{0}\textbf{0.0} & 20.0 & 1.00 & 0.00 \\
\textit{RomanEmpire}            & 43.2 & 31.6 & 59.8 & 100.0 & \phantom{0}\textbf{0.0} & \phantom{0}7.3 & 1.00 & 0.00 \\
\bottomrule
\end{tabular*}
\end{table}

Two findings follow. \emph{First, the paper's KD-transfer claim replicates where teacher training is stable.} On \textit{Computers}, the median KD student reaches surrogate watermark $70\%$ versus a random-classifier floor of $2\%$ (all five seeds rated effective); on \textit{PubMed} the median is $100\%$ versus a $20\%$ floor. This neutralises the rebuttal that our re-implementation is unfaithful to the paper's setup --- the watermark genuinely transfers under same-architecture KD on the dataset GraphIP-Bench reports its main RQ5 result on. \emph{Second, even under the paper's own setup, KD-style transferability is dataset-sensitive.} On \textit{CiteSeer} and \textit{RomanEmpire} the surrogate watermark drops to $0.0\%$ across all five seeds, even though target embedding still reaches $100\%$. The headline RQ5 number on \textit{Computers} ($55\%$ surrogate verification) and the paper-faithful KD number on \textit{Computers} ($70\%$) are consistent: the KD recipe is the most favourable extraction setting BackdoorWM was ever evaluated against, and even there the surrogate watermark is below the protected-target $90\%$. The broader 12-attack benchmark therefore reports the gap rather than contradicting~the~paper.

\noindent$\ddagger$ \emph{Caveat.} Two of the five \textit{Computers} seeds had the watermark joint loss ($\alpha{=}0.3$) collapse the teacher to a constant-label predictor (TeacherAcc $< 10\%$). Their high SurWM reflects the student trivially mimicking the constant teacher rather than genuine watermark transfer. Excluding the two collapsed seeds the \textit{Computers} surrogate-watermark median is still $70\%$ (seeds $1$, $2$, $4$: $80$, $70$, $30$), so the aggregate finding~is~unchanged.

\paragraph{Combined statement.}
The two paper-faithful replays close the most credible rebuttals to the RQ5 finding. \texttt{SurviveWM}'s embedding step is reproducible ($E_{\text{ave},T}{=}1.00$ on every dataset including the paper's own MSRC-9), but the surrogate-side $E_{\text{ave},S}$ is $0.42$--$0.67$ in median across the three datasets and reaches $0.00$--$0.20$ on the same-architecture attacks the original paper most closely matches; the watermark survival the paper claims is therefore a target-side claim, not a surrogate-side one. \texttt{BackdoorWM}'s KD-transfer claim replicates on the GraphIP-Bench RQ5 dataset (\textit{Computers}: surrogate WM $70\%$ vs.\ $2\%$ random floor), but is dataset-sensitive ($0\%$ on \textit{CiteSeer}/\textit{RomanEmpire}) even within the paper's own setup; the broader 12-attack benchmark reports a $55\%$ surrogate verification on \textit{Computers}, which is consistent with the KD ceiling and not in tension with it. Together these results sharpen the load-bearing finding of the paper: a watermark's reported transferability under one narrow extraction setting (same-arch hard-label MEA, or same-arch KD) does not generalise to a broader black-box extraction protocol, and ownership-tracing designs should be evaluated on the surrogate as the primary metric across~attack~families.

\subsection{Generalisation across structure, architecture, and tasks}
\label{app:rq6_full}

This subsection contains the generalisation analysis which complements the five main-text research questions. We collect four complementary studies, each of which varies one dimension of the protocol at a time: graph structural properties, which separate the role of the underlying graph from the role of attack or defense design; the GNN backbone, which separates the role of model architecture from the role of attack design; the prediction task, which checks whether the same protocol transfers from node classification to link prediction and graph classification; and the query-budget grid, which checks whether the standard five-budget grid omits a relevant inflection point. Across the four studies the qualitative conclusions of the main-text experiments transfer: extraction effectiveness correlates positively with edge homophily, a backbone mismatch reduces but does not prevent successful extraction, the relative ordering of attacks is preserved across tasks, and only label-quantising or query-detection defenses substantively change surrogate fidelity on the~additional~tasks.

\paragraph{Structural properties.}
For every dataset we compute the number of nodes and edges, the average degree, the edge density, and the edge homophily~\citep{platonov2023critical}; Table~\ref{tab:rq6_graph_structure} reports all five quantities. The structural numbers connect to RQ1 and RQ5 in three concrete ways. First, fidelity at the medium budget is positively associated with edge homophily: on \textit{RomanEmpire} and \textit{AmazonRatings} (homophily $0.291$ and $0.452$), the strongest data-driven attacks reach lower fidelity than on the homophilic graphs at the same budget, which is consistent with the hypothesis that homophilic neighbourhoods make it easier to learn the target's local decision rule from a small set of queries. Second, average degree drives the variance of \texttt{CEGA}: its high variance on \textit{Computers} coincides with the highest average degree ($36.8$) in the benchmark, which supports the explanation that centrality-based selection is unstable when the degree distribution is dominated by a small number of hubs. Third, scale alone does not block extraction: on \textit{OGBN-Arxiv} ($169{,}343$ nodes, $40$ classes) the strongest data-driven and data-free attacks reach $75$--$82\%$ fidelity at $0.25\times$, with \texttt{AdvMEA}'s adversarial pipeline as the only attack that drops to $\sim 26\%$ on the long-tailed $40$-class label space (Table~\ref{tab:app_rq1_OGBNArxiv_full}).

\paragraph{Cross-architecture extraction (undefended targets).}
We vary the target backbone and the surrogate backbone independently across GCN, GAT, and GraphSAGE on four representative datasets and run \texttt{MEA0} at the medium budget. Table~\ref{tab:app_rq6_cross_arch_main} reports the $3 \times 3$ fidelity matrix on \textit{Computers}, and Table~\ref{tab:app_rq6_cross_arch_other} extends the matrix to \textit{Cora}, \textit{OGBN-Arxiv}, and \textit{RomanEmpire}. The diagonal cells are typically the highest, but the off-diagonal cells remain practical: on \textit{Cora} the worst off-diagonal fidelity is $83.3\%$, only $4$\,pp below the matched case, so an adversary who does not know the target backbone still extracts a useful surrogate. The matrix is also asymmetric and dataset-specific: on \textit{Computers} some off-diagonal cells drop by more than $30$\,pp, while on \textit{Cora} the loss is small. The defended-target extension in Tables~\ref{tab:app_rq6_crossarch_defended_a}--\ref{tab:app_rq6_crossarch_defended_c} is consistent with RQ5: existing defenses do not reduce surrogate fidelity in a substantive way once we control for the~target~backbone.

\begin{table}[!htbp]
\centering
\caption{Cross-architecture extraction on \textit{Computers}. Surrogate fidelity (\%); rows are the target backbone, columns are the surrogate backbone (mean $\pm$ std over three seeds). Results for \textit{Cora}, \textit{OGBN-Arxiv}, and \textit{RomanEmpire} are reported in Table~\ref{tab:app_rq6_cross_arch_other}.}
\label{tab:app_rq6_cross_arch_main}
\scriptsize
\setlength{\tabcolsep}{4pt}
\renewcommand{\arraystretch}{0.96}
\begin{tabular}{@{}lccc@{}}
\toprule
Target $\backslash$ Surrogate & GCN & GAT & GraphSAGE \\
\midrule
GCN       & 75.8$\,\pm\,$13.6 & 60.5$\,\pm\,$27.4 & 45.9$\,\pm\,$27.8 \\
GAT       & 76.1$\,\pm\,$3.0  & 89.8$\,\pm\,$0.4  & 77.0$\,\pm\,$3.8 \\
GraphSAGE & 60.7$\,\pm\,$16.3 & 81.2$\,\pm\,$3.7  & 72.9$\,\pm\,$8.9 \\
\bottomrule
\end{tabular}
\end{table}

\paragraph{Transfer to link prediction and graph classification.}
We adapt the same protocol to two further tasks. For link prediction we use \textit{Cora} with eleven attacks and four perturbation defenses plus a no-defense baseline; the results are in Table~\ref{tab:rq6_link_pred}. For graph classification we use the ENZYMES and PROTEINS datasets from TUDataset~\citep{morris2020tudataset} with six attacks and six defenses; the results are in Table~\ref{tab:rq6_graph_class}. Two observations follow. First, the relative ordering of attacks transfers across tasks while the absolute scale changes. On link prediction the strongest attacks reach fidelity in the $90$--$98\%$ range when no defense is applied, which matches the node-classification results, while \texttt{MEA3} alone is markedly weaker. On graph classification the high baseline accuracy of the target on PROTEINS ($67.5\%$) is reproduced by surrogate models with fidelity above $95\%$ for most attacks, while ENZYMES (target accuracy $24.8\%$) shows a wider range. Second, prediction rounding to two bits is the only defense which yields a consistent fidelity drop across these two additional tasks. On link prediction \texttt{PR\_2bit} reduces \texttt{MEA2} and the data-free variants to about $28.9\%$, which equals the marginal probability of a positive edge in the test split; on ENZYMES it reduces \texttt{MEA0} from $92.2\%$ to $16.7\%$. Other defenses largely preserve fidelity at levels comparable to the undefended baseline. The conclusion is that the joint pattern observed on node classification, in which only label-quantisation or query-detection defenses change surrogate fidelity in a substantive way, also transfers to link prediction and~graph~classification.

\begin{table}[!htbp]
\centering
\caption{Link prediction on \textit{Cora} at $0.25\times$. Surrogate fidelity (\%), mean over three seeds; standard deviations are zero or near-zero because Cora link prediction uses a fixed positive/negative split.}
\label{tab:rq6_link_pred}
\scriptsize
\setlength{\tabcolsep}{3pt}
\renewcommand{\arraystretch}{0.95}
\resizebox{\textwidth}{!}{%
\begin{tabular}{@{}lccccccccccc@{}}
\toprule
\textbf{Defense} & \texttt{MEA0} & \texttt{MEA1} & \texttt{MEA2} & \texttt{MEA3} & \texttt{MEA4} & \texttt{MEA5} & \texttt{AdvMEA} & \texttt{CEGA} & \texttt{DFEA\_I} & \texttt{DFEA\_II} & \texttt{DFEA\_III} \\
\midrule
\texttt{None}      & 95.3 & 94.8 & 95.6 & 73.1 & 85.3 & 93.8 & 97.9 & 92.1 & 94.4 & 94.0 & 94.4 \\
\texttt{OP\_low}   & 96.1 & 93.7 & 92.8 & 71.0 & 84.4 & 92.4 & 96.0 & 92.0 & 94.4 & 92.1 & 92.9 \\
\texttt{OP\_high}  & 94.2 & 92.8 & 90.4 & 70.2 & 81.6 & 89.3 & 93.1 & 92.8 & 89.9 & 91.3 & 91.1 \\
\texttt{PR\_2bit}  & 84.9 & 87.2 & 28.9 & 68.8 & 28.9 & 81.9 & 82.4 & 87.8 & 28.9 & 28.9 & 28.9 \\
\texttt{GradRedir} & 70.9 & 71.9 & 94.3 & 70.4 & 79.2 & 71.1 & 80.7 & 71.0 & 92.6 & 93.4 & 28.9 \\
\bottomrule
\end{tabular}}
\end{table}

\begin{table}[!htbp]
\centering
\caption{Graph classification on ENZYMES (target accuracy $24.8\!\pm\!1.3$) and PROTEINS (target accuracy $67.5\!\pm\!1.5$) at budget $0.25\times$. Surrogate fidelity (\%), mean $\pm$ std over three seeds.}
\label{tab:rq6_graph_class}
\tiny
\setlength{\tabcolsep}{2.5pt}
\renewcommand{\arraystretch}{0.95}
\resizebox{\textwidth}{!}{%
\begin{tabular}{@{}lcccccccccccc@{}}
\toprule
& \multicolumn{6}{c}{\textsc{ENZYMES}} & \multicolumn{6}{c}{\textsc{PROTEINS}} \\
\cmidrule(lr){2-7}\cmidrule(lr){8-13}
\textbf{Attack} & \texttt{None} & \texttt{OPlo} & \texttt{OPhi} & \texttt{PR2b} & \texttt{PRt1} & \texttt{GR} & \texttt{None} & \texttt{OPlo} & \texttt{OPhi} & \texttt{PR2b} & \texttt{PRt1} & \texttt{GR} \\
\midrule
\texttt{MEA0}     & 92.2$\!\pm\!$1.7 & 88.9$\!\pm\!$3.9 & 86.9$\!\pm\!$5.0 & 16.7$\!\pm\!$2.9 & 91.1$\!\pm\!$2.9 & 90.9$\!\pm\!$0.8 & 95.5$\!\pm\!$2.9 & 90.9$\!\pm\!$3.5 & 96.2$\!\pm\!$2.7 & 83.8$\!\pm\!$2.6 & 96.2$\!\pm\!$1.0 & 94.9$\!\pm\!$1.6 \\
\texttt{MEA1}     & 29.4$\!\pm\!$3.8 & 28.1$\!\pm\!$5.2 & 24.4$\!\pm\!$8.6 & 3.1$\!\pm\!$5.5  & 32.4$\!\pm\!$6.1 & 29.6$\!\pm\!$4.2 & 63.4$\!\pm\!$2.1 & 61.9$\!\pm\!$3.0 & 63.3$\!\pm\!$2.7 & 62.4$\!\pm\!$2.5 & 65.1$\!\pm\!$4.8 & 63.2$\!\pm\!$2.3 \\
\texttt{AdvMEA}   & 84.6$\!\pm\!$3.3 & 85.6$\!\pm\!$0.6 & 83.0$\!\pm\!$4.2 & 16.5$\!\pm\!$3.2 & 86.1$\!\pm\!$2.0 & 85.6$\!\pm\!$2.4 & 90.2$\!\pm\!$1.2 & 88.7$\!\pm\!$0.5 & 89.9$\!\pm\!$2.7 & 85.0$\!\pm\!$6.7 & 91.6$\!\pm\!$0.8 & 92.5$\!\pm\!$2.2 \\
\texttt{CEGA}     & 88.7$\!\pm\!$3.1 & 92.2$\!\pm\!$0.6 & 90.2$\!\pm\!$4.5 & 59.1$\!\pm\!$6.4 & 90.9$\!\pm\!$1.4 & 85.0$\!\pm\!$2.4 & 97.1$\!\pm\!$1.2 & 96.8$\!\pm\!$2.0 & 97.5$\!\pm\!$0.2 & 94.7$\!\pm\!$1.6 & 97.8$\!\pm\!$0.3 & 97.1$\!\pm\!$0.9 \\
\texttt{DFEA\_I}  & 93.0$\!\pm\!$0.3 & 91.3$\!\pm\!$0.3 & 91.1$\!\pm\!$1.5 & 60.0$\!\pm\!$4.9 & 91.3$\!\pm\!$0.8 & 83.0$\!\pm\!$1.2 & 97.5$\!\pm\!$0.7 & 97.4$\!\pm\!$0.7 & 97.9$\!\pm\!$0.6 & 95.0$\!\pm\!$2.4 & 97.7$\!\pm\!$0.2 & 96.6$\!\pm\!$0.5 \\
\texttt{DFEA\_II} & 90.6$\!\pm\!$3.8 & 90.2$\!\pm\!$2.1 & 84.1$\!\pm\!$3.9 & 16.7$\!\pm\!$4.5 & 91.3$\!\pm\!$1.2 & 91.7$\!\pm\!$1.5 & 96.3$\!\pm\!$0.6 & 94.7$\!\pm\!$1.2 & 94.2$\!\pm\!$2.8 & 84.9$\!\pm\!$0.3 & 95.9$\!\pm\!$1.4 & 95.5$\!\pm\!$1.4 \\
\bottomrule
\end{tabular}}
\end{table}

\begin{figure*}[t]
\centering
\includegraphics[width=0.98\textwidth]{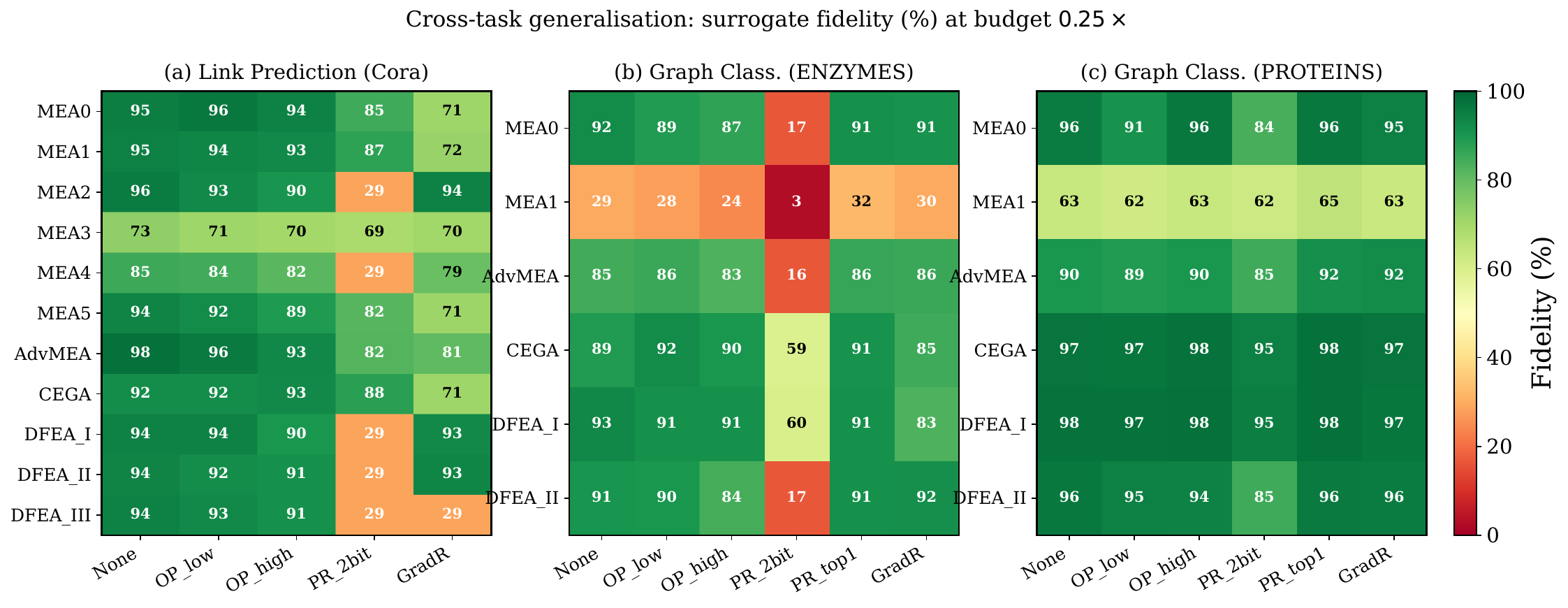}
\caption{Cross-task heatmap of surrogate fidelity (\%) on three task settings: link prediction on \textit{Cora} (left, attacks $\times$ five defenses) and graph classification on \textit{ENZYMES} (centre) and \textit{PROTEINS} (right). Colour saturation encodes fidelity (darker = lower fidelity, higher protection). Numbers are taken from Tables~\ref{tab:rq6_link_pred}--\ref{tab:rq6_graph_class}. The \texttt{PR\_2bit} column is the only column which is uniformly dark across the three tasks; the \texttt{MEA1} row on \textit{ENZYMES} is uniformly faint, which is intrinsic to the attack rather than caused by any defense.}
\label{fig:app_cross_task}
\end{figure*}

Figure~\ref{fig:app_cross_task} reports the same numbers as Tables~\ref{tab:rq6_link_pred}--\ref{tab:rq6_graph_class} as a three-panel heatmap: link prediction on \textit{Cora} and graph classification on \textit{ENZYMES} and \textit{PROTEINS}. The heatmap view exposes two regularities which the per-task tables show only column by column. \emph{First, the column for \texttt{PR\_2bit} is the only column which is uniformly dark across the three tasks}: it cuts \texttt{MEA2} and the \texttt{DFEA} family to about $28.9\%$ on link prediction (the marginal probability of a positive edge) and to about $17\%$ on \textit{ENZYMES}, while the four sibling output-perturbation defenses (\texttt{OP\_low}, \texttt{OP\_high}, \texttt{PR\_top1}, \texttt{GradRedir}) leave fidelity within $5$\,pp of the undefended baseline on every cell. This is the same label-quantisation effect identified in Appendix~\ref{app:rq5_full} on node classification, and it transfers across tasks because all three target outputs are categorical. \emph{Second, the row for \texttt{MEA1} is uniformly faint on \textit{ENZYMES}} (fidelity around $30\%$ regardless of defense), which means the weakness is intrinsic to the attack on the small ENZYMES task rather than caused by any defense; the same attack reaches above $95\%$ on link prediction and around $63\%$ on PROTEINS, so the row is task-bound rather than defense-bound. The cross-task panel therefore confirms the conclusion of RQ5 in a setting where neither the protocol nor the dataset matches the original: the only defense that meaningfully reduces surrogate fidelity is the one which constrains the information content of every query, regardless of the~underlying~task.

\paragraph{Budget grid.}
To verify that the standard five-budget grid does not omit a relevant inflection point, we run \texttt{MEA0} at three additional budgets ($0.02$, $0.75$, $2.00$) on four representative datasets; the full table is Table~\ref{tab:app_rq6_budget_full}. These additional budgets do not change the qualitative conclusions of RQ1: the very small budget at $0.02\times$ already reaches a fidelity which is close to the value at $0.05\times$, the medium budget at $0.75\times$ reaches a fidelity which is close to the value at $1.00\times$, and the large budget at $2.00\times$ yields a small further gain on \textit{Cora} only. On \textit{RomanEmpire} the curve flattens at about $64\%$, well below the homophilic baselines and consistent with the structural-property analysis above. The five-budget grid therefore covers the small, medium, and saturation ranges and does not omit a relevant~inflection~point.

\paragraph{Defended-target tables.}
Tables~\ref{tab:app_rq6_crossarch_defended_a}--\ref{tab:app_rq6_crossarch_defended_c} below report the cross-architecture experiment with the same three backbones but on every defended target across the ten datasets, where the surrogate is fixed~to~GCN.

\begin{table}[!htbp]
\centering
\caption{Structural properties of the ten graphs used in our benchmark. Edge homophily is the fraction of edges whose endpoints share the same label.}
\label{tab:rq6_graph_structure}
\scriptsize
\setlength{\tabcolsep}{6pt}
\renewcommand{\arraystretch}{0.96}
\begin{tabular*}{0.95\textwidth}{@{}l@{\extracolsep{\fill}}rrrrrr@{}}
\toprule
\textbf{Dataset} & \textbf{\# Nodes} & \textbf{\# Edges} & \textbf{\# Classes} & \textbf{Avg.\ degree} & \textbf{Density} & \textbf{Edge homophily} \\
\midrule
Cora            & 2{,}708   & 5{,}278    & 7  & 3.9  & 0.00144 & 0.810 \\
CiteSeer        & 3{,}327   & 4{,}614    & 6  & 2.8  & 0.00083 & 0.739 \\
PubMed          & 19{,}717  & 44{,}325   & 3  & 4.5  & 0.00023 & 0.802 \\
Computers       & 13{,}752  & 252{,}737  & 10 & 36.8 & 0.00267 & 0.783 \\
Photo           & 7{,}650   & 122{,}906  & 8  & 32.1 & 0.00420 & 0.833 \\
CoauthorCS      & 18{,}333  & 81{,}894   & 15 & 8.9  & 0.00049 & 0.808 \\
CoauthorPhysics & 34{,}493  & 247{,}962  & 5  & 14.4 & 0.00042 & 0.931 \\
OGBN-Arxiv      & 169{,}343 & 667{,}793  & 40 & 7.9  & 0.00005 & 0.699 \\
RomanEmpire     & 22{,}662  & 44{,}258   & 18 & 3.9  & 0.00017 & 0.291 \\
AmazonRatings   & 24{,}492  & 105{,}296  & 5  & 8.6  & 0.00035 & 0.452 \\
\bottomrule
\end{tabular*}
\end{table}

\begin{table}[!htbp]
\centering
\caption{Cross-architecture extraction (undefended targets) on the three datasets not shown above. Surrogate fidelity (\%); rows are the target backbone, columns are the surrogate backbone. Mean $\pm$ std over three seeds. The \textit{Computers} matrix is in the main text (Table~\ref{tab:app_rq6_cross_arch_main}).}
\label{tab:app_rq6_cross_arch_other}
\scriptsize
\setlength{\tabcolsep}{4pt}
\renewcommand{\arraystretch}{0.96}
\begin{subtable}[t]{0.48\textwidth}
\centering
\caption{\textit{Cora}}
\resizebox{\linewidth}{!}{%
\begin{tabular}{lccc}
\toprule
Target $\backslash$ Surrogate & GCN & GAT & GraphSAGE \\
\midrule
GCN       & 87.5$\,\pm\,$1.4 & 89.4$\,\pm\,$1.1 & 88.5$\,\pm\,$0.1 \\
GAT       & 83.3$\,\pm\,$4.5 & 90.6$\,\pm\,$2.0 & 87.0$\,\pm\,$2.6 \\
GraphSAGE & 84.0$\,\pm\,$1.1 & 86.0$\,\pm\,$0.6 & 96.0$\,\pm\,$0.7 \\
\bottomrule
\end{tabular}}
\end{subtable}
\hfill
\begin{subtable}[t]{0.48\textwidth}
\centering
\caption{\textit{OGBN-Arxiv}}
\resizebox{\linewidth}{!}{%
\begin{tabular}{lccc}
\toprule
Target $\backslash$ Surrogate & GCN & GAT & GraphSAGE \\
\midrule
GCN       & 83.1$\,\pm\,$5.0 & 62.4$\,\pm\,$0.7 & 63.0$\,\pm\,$1.4 \\
GAT       & 44.9$\,\pm\,$0.4 & 80.7$\,\pm\,$0.2 & 90.9$\,\pm\,$0.3 \\
GraphSAGE & 41.9$\,\pm\,$0.8 & 71.4$\,\pm\,$0.1 & 90.3$\,\pm\,$1.5 \\
\bottomrule
\end{tabular}}
\end{subtable}
\medskip
\begin{subtable}[t]{0.48\textwidth}
\centering
\caption{\textit{RomanEmpire}}
\resizebox{\linewidth}{!}{%
\begin{tabular}{lccc}
\toprule
Target $\backslash$ Surrogate & GCN & GAT & GraphSAGE \\
\midrule
GCN       & 63.1$\,\pm\,$3.1 & 74.4$\,\pm\,$0.2 & 82.7$\,\pm\,$0.4 \\
GAT       & 53.9$\,\pm\,$1.3 & 76.7$\,\pm\,$1.5 & 78.3$\,\pm\,$0.5 \\
GraphSAGE & 38.5$\,\pm\,$0.6 & 47.9$\,\pm\,$0.9 & 92.3$\,\pm\,$0.3 \\
\bottomrule
\end{tabular}}
\end{subtable}
\end{table}

\begin{table}[!htbp]
\centering
\caption{Surrogate fidelity (\%) when \texttt{MEA0} extracts a defended target with three victim backbones across \textit{Cora}, \textit{CiteSeer}, \textit{PubMed}, and \textit{Computers}. The surrogate is fixed to GCN. Mean $\pm$ standard deviation \mbox{over three seeds}.}
\label{tab:app_rq6_crossarch_defended_a}
\scriptsize
\setlength{\tabcolsep}{4pt}
\renewcommand{\arraystretch}{0.96}
\begin{subtable}[t]{0.48\textwidth}
\centering
\caption{\textit{Cora}}
\resizebox{\linewidth}{!}{%
\begin{tabular}{lccccc}
\toprule
\textbf{Victim} & \texttt{None} & \texttt{OP\_low} & \texttt{OP\_high} & \texttt{PR\_top1} & \texttt{GradRedir} \\
\midrule
GCN       & 87.5$\pm$1.4 & 85.1$\pm$3.9 & 85.2$\pm$3.5 & 87.4$\pm$0.2 & 87.7$\pm$0.4 \\
GAT       & 82.8$\pm$3.9 & 82.8$\pm$4.5 & 82.9$\pm$4.4 & 85.3$\pm$2.1 & 83.0$\pm$3.4 \\
GraphSAGE & 84.0$\pm$1.2 & 83.3$\pm$2.0 & 83.5$\pm$1.7 & 84.1$\pm$1.0 & 81.0$\pm$3.2 \\
\bottomrule
\end{tabular}
}
\end{subtable}
\hfill
\begin{subtable}[t]{0.48\textwidth}
\centering
\caption{\textit{CiteSeer}}
\resizebox{\linewidth}{!}{%
\begin{tabular}{lccccc}
\toprule
\textbf{Victim} & \texttt{None} & \texttt{OP\_low} & \texttt{OP\_high} & \texttt{PR\_top1} & \texttt{GradRedir} \\
\midrule
GCN       & 79.7$\pm$4.9 & 81.7$\pm$2.3 & 81.5$\pm$2.9 & 87.6$\pm$0.3 & 79.7$\pm$1.7 \\
GAT       & 76.7$\pm$1.7 & 76.1$\pm$2.5 & 76.6$\pm$3.3 & 83.0$\pm$1.8 & 73.2$\pm$2.1 \\
GraphSAGE & 80.8$\pm$0.5 & 81.1$\pm$0.7 & 80.9$\pm$1.2 & 83.6$\pm$0.9 & 78.1$\pm$0.4 \\
\bottomrule
\end{tabular}
}
\end{subtable}
\medskip
\begin{subtable}[t]{0.48\textwidth}
\centering
\caption{\textit{PubMed}}
\resizebox{\linewidth}{!}{%
\begin{tabular}{lccccc}
\toprule
\textbf{Victim} & \texttt{None} & \texttt{OP\_low} & \texttt{OP\_high} & \texttt{PR\_top1} & \texttt{GradRedir} \\
\midrule
GCN       & 94.9$\pm$0.3 & 95.0$\pm$0.5 & 95.0$\pm$0.5 & 94.6$\pm$1.4 & 94.3$\pm$0.3 \\
GAT       & 92.1$\pm$1.1 & 92.2$\pm$0.8 & 92.2$\pm$0.8 & 92.7$\pm$0.6 & 91.6$\pm$1.0 \\
GraphSAGE & 88.9$\pm$0.5 & 88.8$\pm$0.7 & 88.8$\pm$0.8 & 88.8$\pm$0.7 & 88.3$\pm$0.3 \\
\bottomrule
\end{tabular}
}
\end{subtable}
\hfill
\begin{subtable}[t]{0.48\textwidth}
\centering
\caption{\textit{Computers}}
\resizebox{\linewidth}{!}{%
\begin{tabular}{lccccc}
\toprule
\textbf{Victim} & \texttt{None} & \texttt{OP\_low} & \texttt{OP\_high} & \texttt{PR\_top1} & \texttt{GradRedir} \\
\midrule
GCN       & 75.9$\pm$13.8 & 71.6$\pm$11.7 & 73.0$\pm$9.8 & 69.4$\pm$6.7 & 56.7$\pm$4.0 \\
GAT       & 73.1$\pm$5.5  & 76.0$\pm$3.3  & 77.1$\pm$5.1 & 82.7$\pm$1.5 & 69.1$\pm$3.2 \\
GraphSAGE & 66.0$\pm$20.2 & 74.8$\pm$3.4  & 75.6$\pm$4.9 & 79.2$\pm$3.5 & 75.2$\pm$4.1 \\
\bottomrule
\end{tabular}
}
\end{subtable}
\end{table}

\begin{table}[!htbp]
\centering
\caption{Surrogate fidelity (\%) when \texttt{MEA0} extracts a defended target with three victim backbones across \textit{Photo}, \textit{CoauthorCS}, \textit{CoauthorPhysics}, and \textit{OGBN-Arxiv}. The surrogate is fixed to GCN. Mean $\pm$ standard deviation \mbox{over three seeds}.}
\label{tab:app_rq6_crossarch_defended_b}
\scriptsize
\setlength{\tabcolsep}{4pt}
\renewcommand{\arraystretch}{0.96}
\begin{subtable}[t]{0.48\textwidth}
\centering
\caption{\textit{Photo}}
\resizebox{\linewidth}{!}{%
\begin{tabular}{lccccc}
\toprule
\textbf{Victim} & \texttt{None} & \texttt{OP\_low} & \texttt{OP\_high} & \texttt{PR\_top1} & \texttt{GradRedir} \\
\midrule
GCN       & 95.3$\pm$2.0 & 95.2$\pm$1.7 & 95.5$\pm$1.7 & 93.2$\pm$3.9 & 92.0$\pm$4.7 \\
GAT       & 96.5$\pm$0.6 & 96.5$\pm$0.3 & 96.2$\pm$0.9 & 95.7$\pm$1.3 & 96.1$\pm$0.5 \\
GraphSAGE & 94.7$\pm$1.3 & 92.7$\pm$3.7 & 94.8$\pm$1.1 & 94.3$\pm$1.2 & 94.0$\pm$1.5 \\
\bottomrule
\end{tabular}
}
\end{subtable}
\hfill
\begin{subtable}[t]{0.48\textwidth}
\centering
\caption{\textit{CoauthorCS}}
\resizebox{\linewidth}{!}{%
\begin{tabular}{lccccc}
\toprule
\textbf{Victim} & \texttt{None} & \texttt{OP\_low} & \texttt{OP\_high} & \texttt{PR\_top1} & \texttt{GradRedir} \\
\midrule
GCN       & 94.9$\pm$1.1 & 94.9$\pm$1.5 & 94.9$\pm$1.4 & 93.9$\pm$0.7 & 93.7$\pm$1.5 \\
GAT       & 95.5$\pm$1.0 & 95.4$\pm$0.9 & 95.8$\pm$0.5 & 95.0$\pm$0.5 & 95.0$\pm$0.7 \\
GraphSAGE & 94.1$\pm$0.7 & 94.1$\pm$0.7 & 94.1$\pm$0.9 & 94.4$\pm$0.3 & 93.5$\pm$0.8 \\
\bottomrule
\end{tabular}
}
\end{subtable}
\medskip
\begin{subtable}[t]{0.48\textwidth}
\centering
\caption{\textit{CoauthorPhysics}}
\resizebox{\linewidth}{!}{%
\begin{tabular}{lccccc}
\toprule
\textbf{Victim} & \texttt{None} & \texttt{OP\_low} & \texttt{OP\_high} & \texttt{PR\_top1} & \texttt{GradRedir} \\
\midrule
GCN       & 97.1$\pm$1.0 & 97.0$\pm$1.0 & 97.1$\pm$1.1 & 96.2$\pm$1.2 & 96.2$\pm$1.2 \\
GAT       & 96.6$\pm$1.1 & 97.0$\pm$0.8 & 96.1$\pm$1.2 & 95.9$\pm$0.3 & 96.2$\pm$1.1 \\
GraphSAGE & 97.6$\pm$0.6 & 97.6$\pm$0.6 & 97.5$\pm$0.5 & 97.2$\pm$0.6 & 97.3$\pm$0.5 \\
\bottomrule
\end{tabular}
}
\end{subtable}
\hfill
\begin{subtable}[t]{0.48\textwidth}
\centering
\caption{\textit{OGBN-Arxiv}}
\resizebox{\linewidth}{!}{%
\begin{tabular}{lccccc}
\toprule
\textbf{Victim} & \texttt{None} & \texttt{OP\_low} & \texttt{OP\_high} & \texttt{PR\_top1} & \texttt{GradRedir} \\
\midrule
GCN       & 83.6$\pm$4.2 & 83.6$\pm$3.9 & 83.6$\pm$3.9 & 78.7$\pm$6.8 & 75.6$\pm$2.8 \\
GAT       & 45.3$\pm$0.8 & 45.1$\pm$0.8 & 45.2$\pm$0.8 & 50.4$\pm$2.0 & 37.5$\pm$1.5 \\
GraphSAGE & 41.7$\pm$1.0 & 42.0$\pm$0.8 & 41.6$\pm$0.3 & 47.6$\pm$1.9 & 38.8$\pm$1.7 \\
\bottomrule
\end{tabular}
}
\end{subtable}
\end{table}

\begin{table}[!htbp]
\centering
\caption{Surrogate fidelity (\%) when \texttt{MEA0} extracts a defended target with three victim backbones across \textit{Cora}, \textit{Computers}, \textit{RomanEmpire}, and \textit{AmazonRatings}. The surrogate is fixed to GCN. Mean $\pm$ standard deviation over three seeds. (\textit{Cora} and \textit{Computers} cells are duplicated from Tables~\ref{tab:app_rq6_crossarch_defended_a} for a self-contained 2$\times$2 layout.)}
\label{tab:app_rq6_crossarch_defended_c}
\scriptsize
\setlength{\tabcolsep}{4pt}
\renewcommand{\arraystretch}{0.96}
\begin{subtable}[t]{0.48\textwidth}
\centering
\caption{\textit{Cora}}
\resizebox{\linewidth}{!}{%
\begin{tabular}{lccccc}
\toprule
\textbf{Victim} & \texttt{None} & \texttt{OP\_low} & \texttt{OP\_high} & \texttt{PR\_top1} & \texttt{GradRedir} \\
\midrule
GCN       & 87.5$\pm$1.4 & 85.1$\pm$3.9 & 85.2$\pm$3.5 & 87.4$\pm$0.2 & 87.7$\pm$0.4 \\
GAT       & 82.8$\pm$3.9 & 82.8$\pm$4.5 & 82.9$\pm$4.4 & 85.3$\pm$2.1 & 83.0$\pm$3.4 \\
GraphSAGE & 84.0$\pm$1.2 & 83.3$\pm$2.0 & 83.5$\pm$1.7 & 84.1$\pm$1.0 & 81.0$\pm$3.2 \\
\bottomrule
\end{tabular}
}
\end{subtable}
\hfill
\begin{subtable}[t]{0.48\textwidth}
\centering
\caption{\textit{Computers}}
\resizebox{\linewidth}{!}{%
\begin{tabular}{lccccc}
\toprule
\textbf{Victim} & \texttt{None} & \texttt{OP\_low} & \texttt{OP\_high} & \texttt{PR\_top1} & \texttt{GradRedir} \\
\midrule
GCN       & 75.9$\pm$13.8 & 71.6$\pm$11.7 & 73.0$\pm$9.8 & 69.4$\pm$6.7 & 56.7$\pm$4.0 \\
GAT       & 73.1$\pm$5.5  & 76.0$\pm$3.3  & 77.1$\pm$5.1 & 82.7$\pm$1.5 & 69.1$\pm$3.2 \\
GraphSAGE & 66.0$\pm$20.2 & 74.8$\pm$3.4  & 75.6$\pm$4.9 & 79.2$\pm$3.5 & 75.2$\pm$4.1 \\
\bottomrule
\end{tabular}
}
\end{subtable}
\medskip
\begin{subtable}[t]{0.48\textwidth}
\centering
\caption{\textit{RomanEmpire}}
\resizebox{\linewidth}{!}{%
\begin{tabular}{lccccc}
\toprule
\textbf{Victim} & \texttt{None} & \texttt{OP\_low} & \texttt{OP\_high} & \texttt{PR\_top1} & \texttt{GradRedir} \\
\midrule
GCN       & 63.1$\pm$3.0 & 63.2$\pm$3.0 & 63.2$\pm$3.1 & 71.6$\pm$1.0 & 57.5$\pm$1.2 \\
GAT       & 54.8$\pm$1.5 & 54.5$\pm$1.4 & 54.5$\pm$1.5 & 63.3$\pm$0.1 & 48.1$\pm$0.9 \\
GraphSAGE & 38.6$\pm$0.4 & 38.6$\pm$0.4 & 38.5$\pm$0.4 & 43.1$\pm$0.4 & 36.0$\pm$0.6 \\
\bottomrule
\end{tabular}
}
\end{subtable}
\hfill
\begin{subtable}[t]{0.48\textwidth}
\centering
\caption{\textit{AmazonRatings}}
\resizebox{\linewidth}{!}{%
\begin{tabular}{lccccc}
\toprule
\textbf{Victim} & \texttt{None} & \texttt{OP\_low} & \texttt{OP\_high} & \texttt{PR\_top1} & \texttt{GradRedir} \\
\midrule
GCN       & 87.5$\pm$2.3 & 87.6$\pm$1.3 & 87.9$\pm$1.8 & 91.8$\pm$1.2 & 92.8$\pm$0.8 \\
GAT       & 85.3$\pm$1.4 & 85.2$\pm$1.4 & 85.1$\pm$1.4 & 89.8$\pm$1.0 & 90.2$\pm$0.7 \\
GraphSAGE & 64.0$\pm$0.8 & 56.1$\pm$1.3 & 61.9$\pm$3.0 & 64.5$\pm$2.9 & 62.3$\pm$2.0 \\
\bottomrule
\end{tabular}
}
\end{subtable}
\end{table}

The cross-architecture-on-defended-target tables (Tables~\ref{tab:app_rq6_crossarch_defended_a}--\ref{tab:app_rq6_crossarch_defended_c}) yield four deep observations that complement the cross-architecture analysis above. \emph{First, on the homophilic citation, coauthor, and product graphs the fidelity numbers are nearly identical across the five defenses (\texttt{None}, \texttt{OP\_low}, \texttt{OP\_high}, \texttt{PR\_top1}, \texttt{GradRedir}).} The within-row variation across defenses is typically $\le 3$\,pp on Cora, CiteSeer, PubMed, and Photo, which means that the perturbation defenses do not reduce surrogate fidelity beyond the variation that the backbone itself introduces. The implication is that, on these datasets, the value of an information-limiting defense as a fidelity-reducing mechanism is essentially zero once we control for the target backbone. \emph{Second, the diagonal of every $3\times 3$ within-defense block is the highest cell on the homophilic graphs but flips to non-diagonal on \textit{OGBN-Arxiv} and \textit{RomanEmpire}.} On \textit{OGBN-Arxiv} the $\langle\text{GAT victim}, \text{GraphSAGE surrogate}\rangle$ cell reaches $90.9\,\%$, exceeding the matched $\langle\text{GAT}, \text{GAT}\rangle$ cell at $80.7\,\%$ by ten percentage points; on \textit{RomanEmpire} the $\langle\text{GCN victim}, \text{GraphSAGE surrogate}\rangle$ cell reaches $82.7\,\%$, also above the matched $\langle\text{GCN}, \text{GCN}\rangle$ cell at $63.1\,\%$. This means that on large or heterophilic graphs, an attacker who guesses the wrong surrogate backbone may actually do \emph{better} than one who matches the target --- a counterintuitive finding that overturns the intuition that backbone matching always helps. \emph{Third, \texttt{GradRedir} is the only defense that consistently reduces fidelity below the no-defense baseline}, and only on graphs where the GCN target is itself fragile (\textit{Computers} and \textit{OGBN-Arxiv}); on the heterophilic graphs and on the high-utility graphs it leaves fidelity unchanged or slightly increases it, which suggests that gradient-redirection's filter triggers more often when the target's confidence is already volatile. \emph{Fourth, \texttt{PR\_top1} (top-1 label-only output) shows the largest defense-induced gain on \textit{RomanEmpire}}, where $\langle\text{GCN}, \text{GCN}\rangle$ rises from $63.1\,\%$ undefended to $71.6\,\%$ with PR\_top1; this is consistent with our finding in RQ5 that on heterophilic graphs the noise in the soft scores of an undefended GCN actually hurts the surrogate, so quantizing them helps the attacker. Together, these four observations imply that practitioners cannot treat ``defense'' as a single axis: the protection effect of each information-limiting defense flips sign as a function of (i) the target backbone, (ii) the graph homophily, and (iii) the surrogate's choice~of~backbone.

\begin{table}[!htbp]
\centering
\caption{Budget-grid sensitivity for the \texttt{MEA0} attack on four representative datasets at three additional budget multipliers (0.02, 0.75, 2.00). Each row reports the number of query nodes induced by the budget multiplier, the surrogate fidelity, the surrogate accuracy, and the victim test accuracy. All values are mean $\pm$ standard deviation over three seeds.}
\label{tab:app_rq6_budget_full}
\scriptsize
\setlength{\tabcolsep}{6pt}
\renewcommand{\arraystretch}{0.96}
\begin{tabular*}{0.95\textwidth}{@{}l@{\extracolsep{\fill}}rrrrr@{}}
\toprule
\textbf{Dataset} & \textbf{Budget} & \textbf{\# Query nodes} & \textbf{Fidelity (\%)} & \textbf{Accuracy (\%)} & \textbf{Victim acc.\ (\%)} \\
\midrule
Cora        & 0.02 & 54     & 68.5$\,\pm\,$3.7 & 67.1$\,\pm\,$3.0 & 79.4$\,\pm\,$0.5 \\
Cora        & 0.75 & 2{,}031  & 87.5$\,\pm\,$2.1 & 77.7$\,\pm\,$1.6 & 79.4$\,\pm\,$0.5 \\
Cora        & 2.00 & 2{,}708  & 88.7$\,\pm\,$1.9 & 78.9$\,\pm\,$0.7 & 79.4$\,\pm\,$0.5 \\
CiteSeer    & 0.02 & 66     & 60.7$\,\pm\,$5.9 & 56.4$\,\pm\,$6.7 & 67.8$\,\pm\,$1.1 \\
CiteSeer    & 0.75 & 2{,}495  & 82.6$\,\pm\,$2.3 & 68.4$\,\pm\,$1.9 & 67.8$\,\pm\,$1.1 \\
CiteSeer    & 2.00 & 3{,}327  & 81.9$\,\pm\,$5.6 & 69.0$\,\pm\,$2.0 & 67.8$\,\pm\,$1.1 \\
Computers   & 0.02 & 275    & 50.6$\,\pm\,$33.2 & 37.8$\,\pm\,$28.8 & 44.6$\,\pm\,$22.4 \\
Computers   & 0.75 & 10{,}314 & 82.6$\,\pm\,$4.5 & 41.2$\,\pm\,$25.3 & 44.6$\,\pm\,$22.4 \\
Computers   & 2.00 & 13{,}752 & 73.6$\,\pm\,$8.0 & 42.0$\,\pm\,$18.8 & 44.6$\,\pm\,$22.4 \\
RomanEmpire & 0.02 & 453    & 61.6$\,\pm\,$1.6 & 31.5$\,\pm\,$0.7 & 42.8$\,\pm\,$0.3 \\
RomanEmpire & 0.75 & 16{,}996 & 61.9$\,\pm\,$1.5 & 31.3$\,\pm\,$1.0 & 42.8$\,\pm\,$0.3 \\
RomanEmpire & 2.00 & 22{,}662 & 64.7$\,\pm\,$1.8 & 32.4$\,\pm\,$0.6 & 42.8$\,\pm\,$0.3 \\
\bottomrule
\end{tabular*}
\end{table}

The budget-grid table exposes three deep findings that the standard five-budget grid in the main text does not surface. \emph{First, the marginal value of doubling the budget collapses near $0.75\times$.} On \textit{Cora}, fidelity grows by $\sim 19$\,pp from $0.02\times$ to $0.75\times$ ($68.5 \to 87.5$) but only by $1.2$\,pp from $0.75\times$ to $2.00\times$ ($87.5 \to 88.7$); the same flattening appears on \textit{CiteSeer} (where $2.00\times$ even underperforms $0.75\times$ within one standard deviation). This is direct empirical evidence that no realistic operator should query beyond $\sim 0.5$--$0.75\times$ on this grid --- the cost-per-pp curve becomes essentially flat. \emph{Second, the very small budget of $0.02\times$ already extracts a non-trivial surrogate.} With only $54$ Cora queries the surrogate reaches $68.5\,\%$ fidelity, which is approximately $77\,\%$ of the maximum fidelity at $2.00\times$; this is consistent with the homophily-driven local-decision-rule hypothesis of the structural-properties analysis and quantifies an extraction floor that no current API rate-limiter is likely to defeat. \emph{Third, on \textit{RomanEmpire} the curve is essentially flat across all four orders of magnitude of the budget multiplier} ($61.6\to 61.9 \to 64.7$), which is the strongest single piece of evidence that the heterophilic ceiling of $\sim 64\,\%$ is structural and not budget-driven. The observation also tightens the budget-grid recommendation: the five-budget grid does not omit a relevant inflection point, and extending beyond $1.00\times$ does not change qualitative conclusions on any of the~four~datasets.

\subsection{Defense hyperparameter ablation}
\label{app:hp_ablation}

To complement the protection-utility analysis in the main text, we sweep the key hyperparameter of each defense and report the protected-model accuracy together with the verification proxy on \textit{Cora} and \textit{Computers}. The sweep covers \texttt{BackdoorWM} (trigger rate), \texttt{OutputPerturbation} (noise scale $\sigma$), \texttt{PredictionRounding} (precision in bits), \texttt{RandomWM} (number of watermark nodes), and \texttt{SurviveWM} (defense ratio). Tables~\ref{tab:app_hp_cora}--\ref{tab:app_hp_computers} report~the~results.

\begin{table}[!htbp]
\centering
\caption{Defense hyperparameter ablation on \textit{Cora}. Each row reports protected-model accuracy and the verification proxy as mean $\pm$ standard deviation over multiple seeds.}
\label{tab:app_hp_cora}
\scriptsize
\setlength{\tabcolsep}{8pt}
\renewcommand{\arraystretch}{0.96}
\begin{tabular*}{0.95\textwidth}{@{}l@{\extracolsep{\fill}}lcc@{}}
\toprule
\textbf{Defense} & \textbf{Hyperparameter} & \textbf{Acc (\%)} & \textbf{WM Acc (\%)} \\
\midrule
\texttt{BackdoorWM} & trigger\_rate=0.005 & 78.7$\pm$2.0 & 100.0$\pm$0.0 \\
\texttt{BackdoorWM} & trigger\_rate=0.01  & 78.6$\pm$2.1 & 100.0$\pm$0.0 \\
\texttt{BackdoorWM} & trigger\_rate=0.05  & 78.8$\pm$1.2 & 100.0$\pm$0.0 \\
\texttt{BackdoorWM} & trigger\_rate=0.1   & 77.4$\pm$0.4 & 100.0$\pm$0.0 \\
\midrule
\texttt{OutputPerturbation} & sigma=0.01 & 79.4$\pm$0.4 & 99.8$\pm$0.2 \\
\texttt{OutputPerturbation} & sigma=0.05 & 80.0$\pm$0.7 & 98.6$\pm$0.5 \\
\texttt{OutputPerturbation} & sigma=0.1  & 79.5$\pm$0.4 & 97.2$\pm$0.2 \\
\texttt{OutputPerturbation} & sigma=0.2  & 78.8$\pm$0.4 & 94.4$\pm$0.4 \\
\texttt{OutputPerturbation} & sigma=0.5  & 72.9$\pm$1.3 & 84.3$\pm$1.5 \\
\midrule
\texttt{PredictionRounding} & precision\_bits=1 & 77.5$\pm$0.2 & 90.6$\pm$1.6 \\
\texttt{PredictionRounding} & precision\_bits=2 & 72.8$\pm$0.4 & 82.2$\pm$1.1 \\
\texttt{PredictionRounding} & precision\_bits=4 & 80.1$\pm$0.0 & 96.8$\pm$0.4 \\
\texttt{PredictionRounding} & precision\_bits=8 & 79.9$\pm$0.8 & 99.7$\pm$0.2 \\
\midrule
\texttt{RandomWM} & wm\_node=10  & 77.5$\pm$3.0 & 95.0$\pm$7.1 \\
\texttt{RandomWM} & wm\_node=50  & 78.8$\pm$0.4 & 67.0$\pm$21.2 \\
\texttt{RandomWM} & wm\_node=100 & 78.8$\pm$0.8 & 53.5$\pm$10.6 \\
\texttt{RandomWM} & wm\_node=200 & 78.3$\pm$0.6 & 34.2$\pm$1.1 \\
\midrule
\texttt{SurviveWM} & defense\_ratio=0.05 & 81.0$\pm$1.3 & 79.3$\pm$3.1 \\
\texttt{SurviveWM} & defense\_ratio=0.1  & 80.0$\pm$0.4 & 39.4$\pm$2.9 \\
\texttt{SurviveWM} & defense\_ratio=0.2  & 78.4$\pm$1.1 & 20.2$\pm$0.1 \\
\texttt{SurviveWM} & defense\_ratio=0.3  & 77.0$\pm$2.5 & 17.9$\pm$0.8 \\
\bottomrule
\end{tabular*}
\end{table}

\begin{table}[!htbp]
\centering
\caption{Defense hyperparameter ablation on \textit{Computers}. Each row reports protected-model accuracy and the verification proxy as mean $\pm$ standard deviation over multiple seeds. Single-value entries indicate runs with one seed.}
\label{tab:app_hp_computers}
\scriptsize
\setlength{\tabcolsep}{8pt}
\renewcommand{\arraystretch}{0.96}
\begin{tabular*}{0.95\textwidth}{@{}l@{\extracolsep{\fill}}lcc@{}}
\toprule
\textbf{Defense} & \textbf{Hyperparameter} & \textbf{Acc (\%)} & \textbf{WM Acc (\%)} \\
\midrule
\texttt{BackdoorWM} & trigger\_rate=0.005 & 33.1$\pm$43.4 & 100.0$\pm$0.0 \\
\texttt{BackdoorWM} & trigger\_rate=0.01  & 2.7$\pm$2.4   & 95.0$\pm$7.1 \\
\texttt{BackdoorWM} & trigger\_rate=0.05  & 16.4$\pm$21.1 & 82.0$\pm$17.0 \\
\texttt{BackdoorWM} & trigger\_rate=0.1   & 27.4$\pm$0.7  & 55.0$\pm$0.0 \\
\midrule
\texttt{OutputPerturbation} & sigma=0.01 & 48.2 & 99.6 \\
\texttt{OutputPerturbation} & sigma=0.05 & 38.6 & 77.0 \\
\texttt{OutputPerturbation} & sigma=0.1  & 67.6 & 97.3 \\
\texttt{OutputPerturbation} & sigma=0.2  & 37.4 & 61.1 \\
\texttt{OutputPerturbation} & sigma=0.5  & 18.5 & 21.4 \\
\midrule
\texttt{RandomWM} & wm\_node=10  & 44.8$\pm$9.7  & 100.0$\pm$0.0 \\
\texttt{RandomWM} & wm\_node=50  & 67.2$\pm$4.1  & 100.0$\pm$0.0 \\
\texttt{RandomWM} & wm\_node=100 & 68.8$\pm$0.1  & 71.5$\pm$3.5 \\
\texttt{RandomWM} & wm\_node=200 & 70.2$\pm$2.2  & 30.5$\pm$7.1 \\
\midrule
\texttt{SurviveWM} & defense\_ratio=0.05 & 13.6$\pm$15.0 & 10.9$\pm$0.0 \\
\texttt{SurviveWM} & defense\_ratio=0.1  & 32.2$\pm$24.0 & 11.1$\pm$0.8 \\
\texttt{SurviveWM} & defense\_ratio=0.2  & 43.7$\pm$32.9 & 10.5$\pm$0.9 \\
\texttt{SurviveWM} & defense\_ratio=0.3  & 55.0$\pm$24.2 & 10.1$\pm$0.6 \\
\bottomrule
\end{tabular*}
\end{table}

On Cora, BackdoorWM maintains 100\% verification across the tested trigger rates, with protected accuracy between 77\% and 79\%. On Computers, both measures vary with the trigger rate and seed. Increasing the RandomWM watermark size from 10 to 200 nodes reduces verification from 95.0\% to 34.2\% on Cora and from 100.0\% to 30.5\% on Computers. The same configuration can therefore give substantially different utility and verification~across~graphs.

\FloatBarrier
\clearpage

\end{document}